\documentclass[proof]{pasj00}
\draft
\newcommand{\hp}{\hspace*{-8mm}}
\newcommand{\one}[1]{\hp\includegraphics[width=40mm]{#1}}

\begin{document}
\SetRunningHead{K. Tomisaka}{Origin of Molecular Outflow Determined from Thermal Dust Polarization
}
\Received{2010/08/31}
\Accepted{2010/10/26}

\title{Origin of Molecular Outflow Determined from Thermal Dust Polarization
}

\author{Kohji \textsc{Tomisaka}}
\affil{National Astronomical Observatory of Japan\\
Osawa 2-21-1, Mitaka, Tokyo 181-8588, Japan}

%


\maketitle
\begin{abstract}
The observational expectation of polarization measurements of
 thermal dust radiation is investigated to find information on molecular outflows 
 based on magnetohydrodynamical (MHD) and radiation transfer simulations.
There are two major proposed models for the driving of molecular outflows:
 (1) molecular gas is accelerated by a magnetic pressure gradient
 or magnetocentrifugal wind mechanism before
 the magnetic field and molecular gas are decoupled,
 (2) the linear momentum of a highly collimated jet is transferred 
 to the ambient molecular gas.  
In order to distinguish between these two models, it is crucial to observe the configuration of the magnetic field.
An observation of a toroidal magnetic field is strong evidence
 that the first of the models is appropriate.
In this paper, we calculated the polarization distribution
 of thermal dust radiation due to the alignment of dust grains
 along the magnetic field 
using molecular outflow data calculated by two-dimensional axisymmetric 
 MHD simulations.
An asymmetric distribution around the $z$-axis is characteristic for
 magnetic fields composed of both poloidal and toroidal components.
We determined that the outflow has a low polarization degree compared with the envelope
and that the envelope and outflow have different polarization directions 
 (B-vector), namely, the magnetic field within the envelope is parallel
 to the global magnetic field lines
 while the magnetic field of the outflow is perpendicular to it. 
Thus we have demonstrated that the point-symmetric (rather than axisymmetric)
 distributions of low polarization regions indicate
 that molecular outflows are likely to be magnetically driven.
Observations of this polarization distribution with tools such as ALMA
 would confirm the origin of the molecular outflow. 
\end{abstract}

\section{Introduction}
Magnetic fields play a crucial role in the star formation process.
Magnetohydrodynamical (MHD) simulations have shown that molecular
 outflows and jets, which are ubiquitously observed in star forming
 regions, are launched by the magnetic Lorentz force \citep{tomisaka02,banerjee06,machida07,commercon10,tomida10}.
Furthermore, excess angular momentum of the molecular core of a star forming region is
 transferred by magnetic braking and the rotation rate is
 reduced to that observed for rotating stars \citep{tomisaka00,machida07}.
However, the origin of molecular outflows has not yet been determined observationally.   
There are at least two major models for explaining molecular outflows.
In the first model, the molecular outflow is accelerated by the magnetic
 Lorentz force, in the region in which the magnetic field and gas are well coupled
 (magnetically driven molecular outflow).
This mechanism begins to work after the formation of the first core, which
 is the first hydrostatic object made of hydrogen molecules in the process
 of star formation \citep{larson69},
 due to a combination of the magnetic field and 
 rotational motion around the first core. 
Before the formation of the first core, or in the isothermal runaway collapse phase,
 no molecular outflow is launched.  
The second model is based on the idea that the jet is a primary object
 and its linear momentum is transferred
 to the ambient gas to form a bipolar molecular outflow (entrainment model).
The well-collimated jet is thought to be magnetically accelerated
 \citep{shu94,kudoh98}.
In the entrainment model,
 the commonly observed wide opening angles in molecular outflows
 cannot be explained by the simple idea of momentum transfer \citep{stahler93},
 since the jet is well collimated and its width is much smaller than that of
 the molecular outflow.
A number of variations on the entrainment model have been proposed to address this problem, such as turbulent entrainment \citep{raga+93}
 and entrainment through a bow shock \citep{raga93,masson93}.
Comparisons are made between observed molecular outflows and models \citep{cabrit97,lee00},
 which indicate that a part of the molecular outflows have observational
 signatures consistent with jet-driven origin.
The magnetic drive model, however, can solve the angular
 momentum problem of
 newborn stars, as excess angular momentum, which must be greatly reduced
 to form stars, 
is removed to a distance by the magnetic torque and the molecular outflow
 \citep{tomisaka00,machida07}.

To explore the formation mechanism of molecular outflows,
 we need observational evidence to distinguish the above two models.
The magnetically driven molecular outflow model predicts 
 (1) rotation of the molecular outflow and
 (2) a toroidal magnetic field especially near the acceleration region.  
Although there are several observations of jet rotation
 \citep{chrysostomou00,davis00,bacciotti02,woitas05,coffey07}, only a few observations have been made
 on the rotation of molecular outflows.
\citet{launhardt09} observed a molecular outflow around a T-Tauri star
 in a dark cloud named CB26
 and found in an intensity-weighted velocity map of $^{12}{\rm CO}(J=2-1)$
 that two lobes of the molecular outflow have systematic rotation
 with the same orientation as the high-density circumstellar disk rotation. 
This seems to indicate a twisted magnetic field due to the rotation of
 the high-density disk exerting a torque on the molecular material. 

Further direct evidence of magnetic driven outflows is 
 the existence of a strong toroidal magnetic field.
The magnetic torque (or toroidal Lorentz force) arises from the
 combination of
 the poloidal current and poloidal magnetic field, and the poloidal current
 does not exist without a toroidal magnetic field.
Thus, the magnetic acceleration region 
 must be characterized by a toroidal magnetic field
 as well as a poloidal magnetic field.      
In the present paper,
 we demonstrate the characteristics of 
the configuration of a magnetic field
 driving the molecular outflow (i.e. a magnetic field consisting of both poloidal
 and toroidal components) in polarization observations 
 of thermal dust emission.

If we compare these predictions with observations,
 we can distinguish which model is suitable for molecular outflows.
This is done by post-processing the numerical results of an MHD simulation.
We call this procedure ``observational visualization,'' which
 is a visualization process of the numerical results to
 explain the undergoing physics but also emphasizing the observational 
 expectations from the simulation.
In the observations of the magnetic field,
 we focus on the polarization of thermal dust radiation,
 which gives information about the configuration of the 
 magnetic field.     
The strength of the magnetic field obtained from Zeeman splitting
 measurements will be presented in a future paper.

The plan of this paper is as follows:
 In section 2 we describe the molecular outflow model
 and summarize the results of MHD simulations.
We also describe the numerical method for calculating the polarization 
 of thermal dust radiation.
In section 3, we present the results of the observation visualization.
Section 4 is devoted to discussing the distribution of 
 the polarization degree and the characteristic features of
 the magnetically driven molecular outflow. 
 
\section{Model and Method}

\begin{figure}[h]
   \begin{center}
(a)\hspace*{70mm}(b)\\
      \includegraphics[width=70mm]{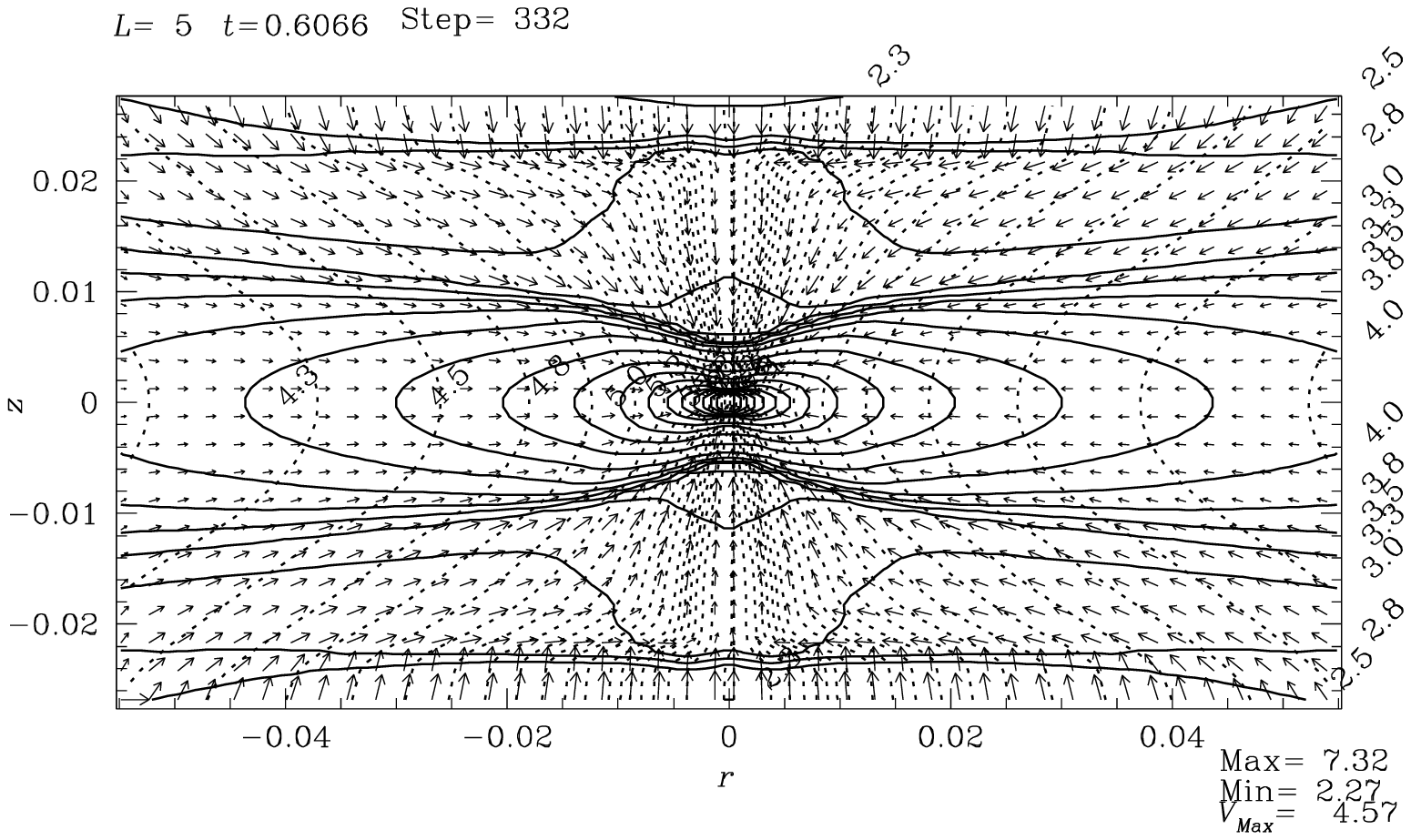}
      \hfil\includegraphics[width=70mm]{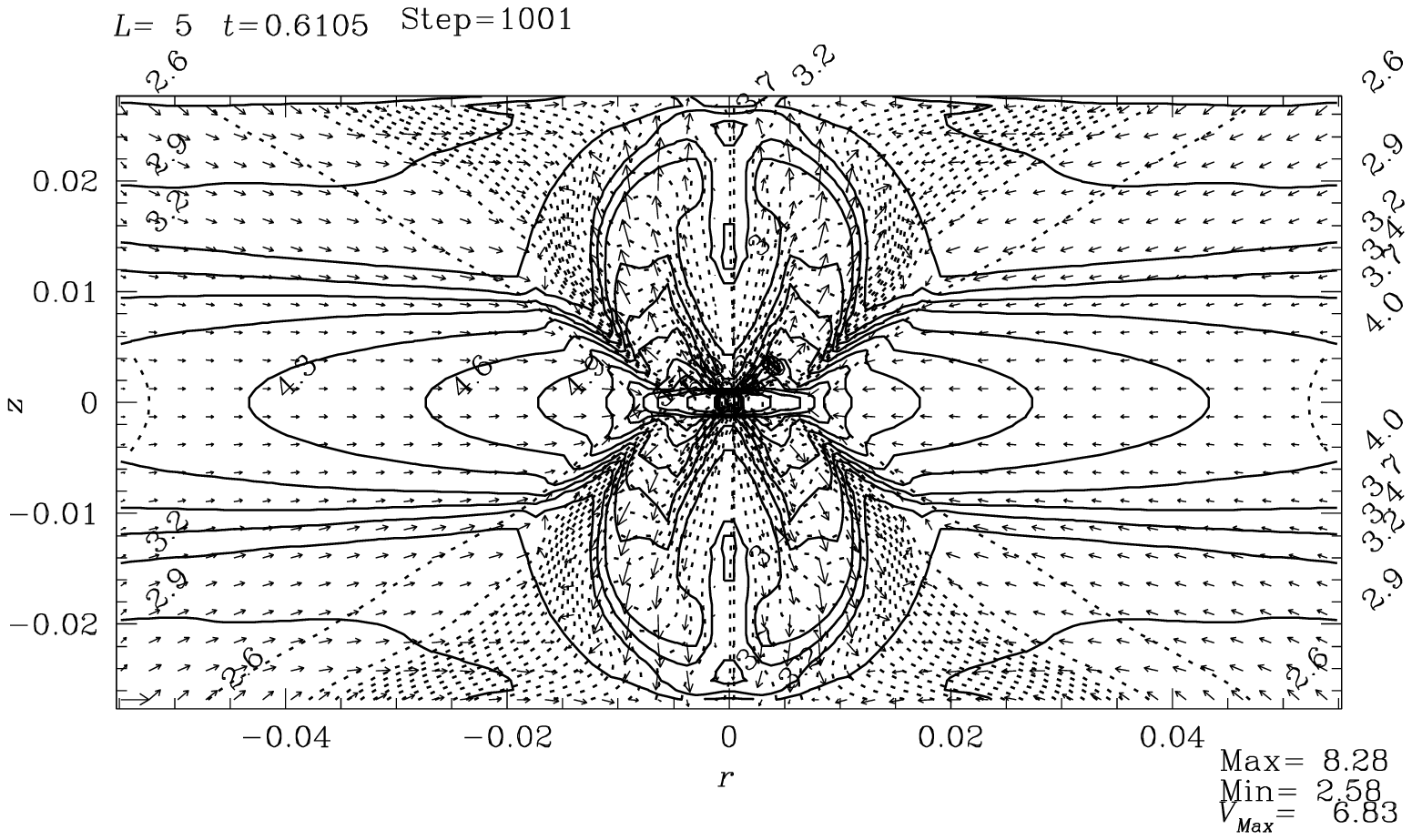}
   \end{center}
\caption{\small\label{fig:density}
Structure of contracting magnetized rotating cloud in
 the prestellar (a) and poststellar (b) phases.
 This model corresponds to model AH1 of Paper I  with $\alpha=1$ and $\Omega'=5$.
 The solid lines illustrate the density contour and the dashed lines
 represent the magnetic field lines.  
 The vectors show the velocity on the meridian plane.
}
\end{figure}

We have presented calculations of the evolution of a rotating magnetized axisymmetric 
 isothermal cloud in \citet{tomisaka02},
 which is hereafter referred to as Paper I.
In Paper I, we assumed a cloud in hydrostatic balance  
 characterized by dimensionless parameters to specify the magnetic
 field strength, 
 $\alpha=B_z^2/(4\pi \rho c_s^2)$ ($B_z$, $\rho$, and $c_s$ are
 the magnetic flux density, gas density, and isothermal sound speed),
 and the rotation rate, 
 $\Omega'=\Omega_0\tau_{\rm ff}=\Omega_0/(4\pi G \rho_s)^{1/2}$
 ($\tau_{\rm ff}=1/[4\pi G \rho_s]^{1/2}$ represents the characteristic free-fall
 time-scale for the initial surface density of the cloud
 $\rho_s$, $\Omega_0$ represents the angular rotation speed at the center).
Figure \ref{fig:density} illustrates the evolution of 
 model AH1 from Paper I  with $\alpha=1$ and $\Omega'=5$.
The parameters of this model correspond to 
 a central density of $\rho_c=10^4{\rm H_2\,cm^{-3}}$,
 surface density of $\rho_s=10^2{\rm H_2\,cm^{-3}}$,
 magnetic field strength at the center of
  $B_0=13.3(\rho_c/10^4{\rm H_2\,cm^{-3}})^{1/2}\alpha^{1/2}{\rm \mu G}$,
 rotation rate at the center of
  $\Omega_0=2.78{\rm km\,s^{-1}\,pc^{-1}}(\Omega'/5)$,
 and isothermal sound speed of
  $c_s=190{\rm m\,s^{-1}}$. 
The evolution is divided into two phases:
 before and after the first core formation.
The figure shows
 the structure just before the first core formation (Fig.\,\ref{fig:density}(a))
 and when $\tau=3.94\times 10^{-3}\tau_{\rm ff}
 =7000{\rm yr}(\rho_s/100{\rm H_2 cm^{-3}})^{-1/2}$
 has passed after the first core formation (Fig.\,\ref{fig:density}(b)).
Before the first core forms, we observe a nested disk system 
 composed of two disks,  cocentered and bounded
 by different accretion shocks (the half thickness of the thinner disk is 
 $z\simeq 0.01 H\sim 700 {\rm AU}(c_s/190{\rm m\,s^{-1}})(\rho_s/100{\rm H_2 cm^{-3}})^{-1/2}$ and that of the thicker disk is
 $z\simeq 0.02 H\sim 1400 {\rm AU}(c_s/190{\rm m\,s^{-1}})(\rho_s/100{\rm H_2 cm^{-3}})^{-1/2}$,
 where 
$H=7\times 10^4{\rm AU}
 (c_s/190{\rm m\,s^{-1}})(\rho_s/100{\rm H_2 cm^{-3}})^{-1/2}$ represents the scale height of the initial cloud). 
This disk is actually a pseudodisk which continues to contract
 supersonically \citep{galli93,tomisaka98}.
Just after the first core forms ($\tau \sim 1000{\rm yr}(\rho_s/100{\rm H_2 cm^{-3}})^{-1/2}$),
 gas begins to be ejected around the core due to the centrifugal
 force driven by the extra angular momentum transferred
 by the magnetic tension force (magnetocentrifugal wind mechanism
 \citep{blandford82}).   
At $\tau=7000{\rm yr}(\rho_s/100{\rm H_2 cm^{-3}})^{-1/2}$, 
 the outflow reaches $z\simeq 0.02H=1400{\rm AU}(c_s/190{\rm m\,s^{-1}})
(\rho_s/100{\rm H_2\, cm^{-3}})^{-1/2}$ (Fig.\,\ref{fig:density}(b)).
For the kinetic temperature, we assume a barotropic gas:
 an isothermal envelope is assumed with
 $T_K=10\,\rm K$ for $\rho < \rho_A = 10^{10}{\rm H_2 cm^{-3}}$,
 although a hydrostatic core (the first core)
 with a small volume has a higher temperature as
 $T_K=10(\rho / \rho_A)\,{\rm K}$
 for $\rho > \rho_A = 10^{10}{\rm H_2\, cm^{-3}}$. 

\begin{figure}[h]
   \begin{center}
\includegraphics[width=100mm]{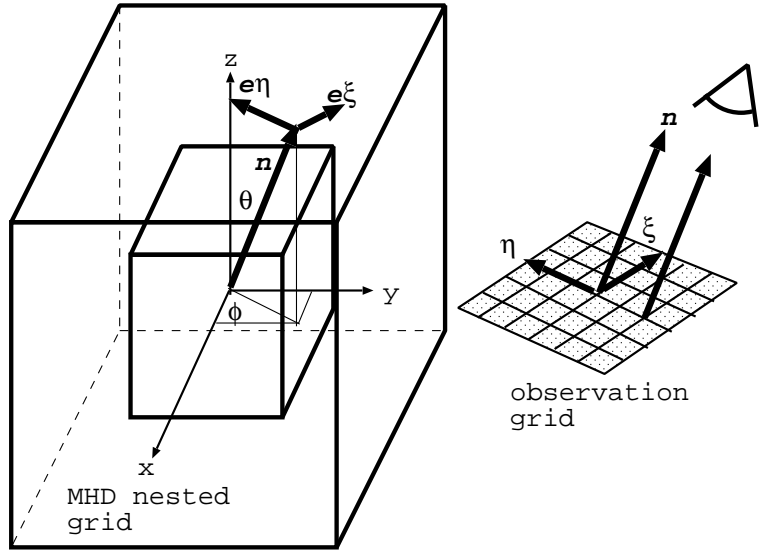}
   \end{center}
\caption{\label{fig:grid}Relationship between the grid
 used in the MHD simulations (left: nested grid)
 and the observation grid (right).
Observations were made by integrating along the normal vector
 $\boldsymbol{n}$.
That is, the direction of the observation is specified by $\boldsymbol{n}$.
The vertical and horizontal axes of the observation grid are chosen along
 the unit vectors 
 $\boldsymbol{e}_\eta =\left[\boldsymbol{e}_z-(\boldsymbol{e}_z\cdot \boldsymbol{n})\boldsymbol{n}\right]
 /\left|\boldsymbol{e}_z-(\boldsymbol{e}_z\cdot \boldsymbol{n})\boldsymbol{n}\right|$
 and
 $\boldsymbol{e}_\xi=\boldsymbol{e}_\eta\times \boldsymbol{n}$.
}
\end{figure}

We then calculate the expected polarization distribution for thermal dust 
 radiation observed from the direction $\boldsymbol{n}$. 
Figure \ref{fig:grid}
 shows the relationship between the grid used in the MHD simulation (left)
 and the observation grid (right).
The observation is made by integrating along the normal vector 
 $\boldsymbol{n}$. 
The vertical and horizontal axes of the observation grid are chosen along
 the unit vectors 
\begin{eqnarray}
\boldsymbol{e}_\eta &=&\frac{\boldsymbol{e}_z-(\boldsymbol{e}_z\cdot \boldsymbol{n})\boldsymbol{n}}
{\left|\boldsymbol{e}_z-(\boldsymbol{e}_z\cdot \boldsymbol{n})\boldsymbol{n}\right|},\\
\boldsymbol{e}_\xi&=&\boldsymbol{e}_\eta\times \boldsymbol{n}.
\end{eqnarray}
The vectors $\boldsymbol{n}$, $\boldsymbol{e}_\xi$, and $\boldsymbol{e}_\eta$
 form a right-handed coordinate system
 and the direction of $\boldsymbol{e}_\eta$ is chosen
 to be toward the $z$-axis.
The direction of $\boldsymbol{n}$ is specified by the angle $\theta$
 from the $z$-axis 
 of the simulation grid,
 along which the angular momentum and initial magnetic field
 are directed.
Since the model is axisymmetric around the $z$-axis, 
 the polarization is only dependent on the elevation angle
 $\theta$ and independent of the azimuth angle $\phi$.
We will discuss the case without axisymmetry in a future study.

\begin{figure}[h]
   \begin{center}
\includegraphics[width=100mm]{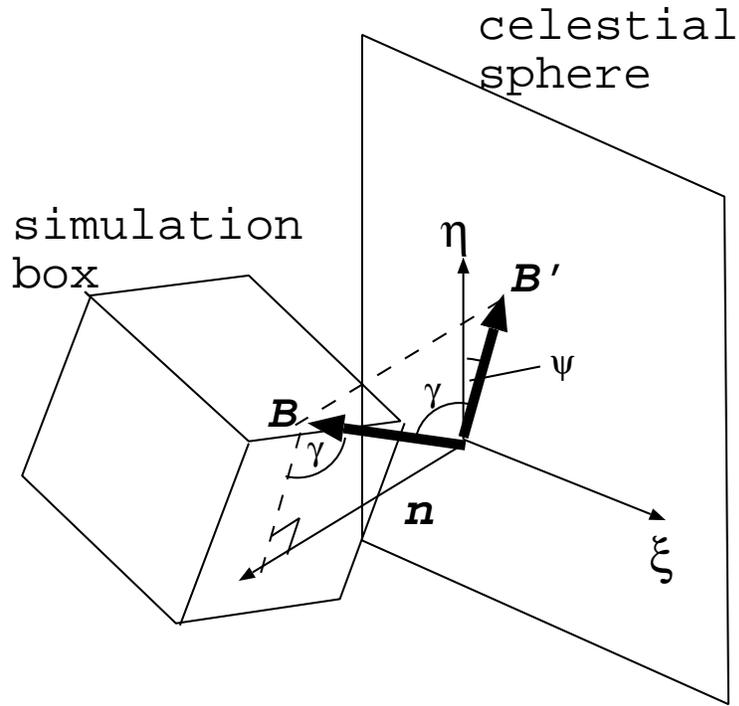}
   \end{center}
\caption{\label{fig:psigamma}
Relative Stokes' parameter $q$ and $u$ calculated by
 integrating eqs (\ref{eqn:q}) and (\ref{eqn:u}) along $\boldsymbol{n}$.
The angles used in eqs (\ref{eqn:q}) and (\ref{eqn:u}) are defined as follows:
 $\gamma$ and $\psi$ represent respectively
 the angle between the magnetic field and the plane of the sky
 and the angle between the projected magnetic field $\boldsymbol{B}'$ and
 the $\eta$-axis. 
}
\end{figure}

The polarization is calculated from the Stokes' $Q$ and $U$ parameters.
Assuming a constant emissivity per mass for dust,
 owing to the global isothermality in the molecular core and optically thin
 radiation, we substitute the {\em relative} Stokes' parameters $q$ and $u$ for
 $Q$ and $U$ (\citet{lee85,fiege00,matsumoto06}):
\begin{eqnarray}
q&=&\int\rho\cos 2\psi \cos^2 \gamma ds,\label{eqn:q}\\
u&=&\int\rho\sin 2\psi \cos^2 \gamma ds,\label{eqn:u}
\end{eqnarray}
where the integration is performed along the line-of-sight $\boldsymbol{n}$
 and the two angles $\gamma$ and $\psi$ represent respectively
 that between the magnetic field and the plane of the sky
 and that between the projected magnetic field $\boldsymbol{B}'$ and
 the $\eta$-axis (see Fig.\ref{fig:psigamma}).
From the relative Stokes' parameters, we derive the polarization direction 
 $\chi$ as
\begin{eqnarray}
\cos 2\chi &=& \frac{q}{\left(q^2+u^2\right)^{1/2}},\label{eqn:cos2chi}\\
\sin 2\chi &=& \frac{u}{\left(q^2+u^2\right)^{1/2}}.\label{eqn:sin2chi}
\end{eqnarray} 
Then, $\cos\chi$ and $\sin \chi$ are solved from equations (\ref{eqn:cos2chi})
 and (\ref{eqn:sin2chi}) as
\begin{eqnarray}
\cos \chi&=&\left(\frac{\cos2\chi+1}{2}\right)^{1/2}\\
\sin \chi&=&\frac{\sin 2\chi}{2\cos\chi}\ \ \  ({\rm for}\ \cos\chi \ne 0),
\end{eqnarray}
where $\sin\chi=1$ (for $\cos\chi = 0$) (see also \citet{fiege00}).
The polarization degree vector is calculated as
\begin{equation}
\boldsymbol{P}\equiv\left(\begin{array}{c}P_\xi\\P_\eta\end{array}\right)
=\left(\begin{array}{c}P\sin\chi\\
                       P\cos\chi\end{array}\right),
\end{equation} 
where $P$ represents the ratio of the polarized intensity to the total intensity
 and is given as  
\begin{equation}
P=p_0\frac{\left(q^2+u^2\right)^{1/2}}{\Sigma-p_0\Sigma_2},
\end{equation}
from the two integrated quantities 
\begin{eqnarray}
\Sigma &\equiv& \int \rho ds,\label{eqn:Sigma}\\
\Sigma_2 &\equiv& \int \rho \left(\frac{\cos^2\gamma}{2}-\frac{1}{3} \right)ds.
\label{eqn:Sigma2}
\end{eqnarray}
The numerical factor $p_0$ is chosen to be $p_0=0.15$
 to fit the maximum $P$ agreeing with the observations of typical dark clouds.
The timescale of the alignment of the dust in the magnetized interstellar
 medium is not established, especially near the molecular core center
 where the gas is accelerated\footnote{
The timescale necessary for alignment of dust grains was estimated quite long
 based on the damping timescale of rotation motion of paramagnetic dusts in 
 the interstellar magnetic field \citep{davis51}.
 However, at present, the alignment of angular momentum with the axis of maximum moment of
 inertia (internal alignment) is believed to be quite rapid \citep{lazarian99}.
And other mechanisms than the paramagnetic damping 
 such as the radiative torque mechanism and dynamical alignment mechanism
 are much more efficient for the alignment of the angular momentum with the magnetic field   
 (see for review \citet{lazarian07})}.
Therefore, we assume the dust grains are aligned in a similar way
 to the dark cloud.   

In conclusion, to calculate the polarization,
 we have to obtain four quantities ($u$, $q$, $\Sigma$, and $\Sigma_2$)
 integrating numerically along the line-of-sight 
 (eqs.[\ref{eqn:q}], [\ref{eqn:u}], [\ref{eqn:Sigma}],
  and [\ref{eqn:Sigma2}]).
This is done on the nested grid hierarchy, in which $N$ levels of
 grid with different spatial resolutions are placed concentrically
 (see Paper I and Fig.~\ref{fig:grid}).
Figure \ref{fig:density} illustrates the $L=5$ level, which has 
 $2^5=32$ times finer resolution than the $L=0$ level that covers the whole
 structure of the cloud. 
To make a polarization map with the same spatial resolution
 and the same spatial coverage as the $L=5$ grid, we integrate the above four equations
 (eqs.[\ref{eqn:u}], [\ref{eqn:q}], [\ref{eqn:Sigma}], and [\ref{eqn:Sigma2}])
 along the line-of-sight
 using all the data in grid levels $L=0,\ 1,\ \ldots,\ 5$.
This is done by the following procedure with target=5: 
\begin{verbatim}
    For L=0, target-1 do begin
      Integrate from outer to inner boundaries for grid level L
    End for
    Integrate for target grid level
    For L=target-1, 0, -1 do begin
      Integrate from inner to outer boundaries for grid level L
    End for
\end{verbatim} 
This procedure was tested by calculating the column density distribution obtained from
 a spherically symmetric density distribution $\rho(r)$, 
 with the corresponding column density obtained by a numerical integration. 

\section{Result}

\subsection{Runaway Collapse Phase}

\begin{figure}[h]
\begin{center}
\vspace*{-12mm}($L=3$)\hspace*{22mm}($L=5$)\hspace*{22mm}($L=7$)\hspace*{22mm}($L=9$)\\
\hspace*{24mm}
      \raisebox{-20mm}
{\one{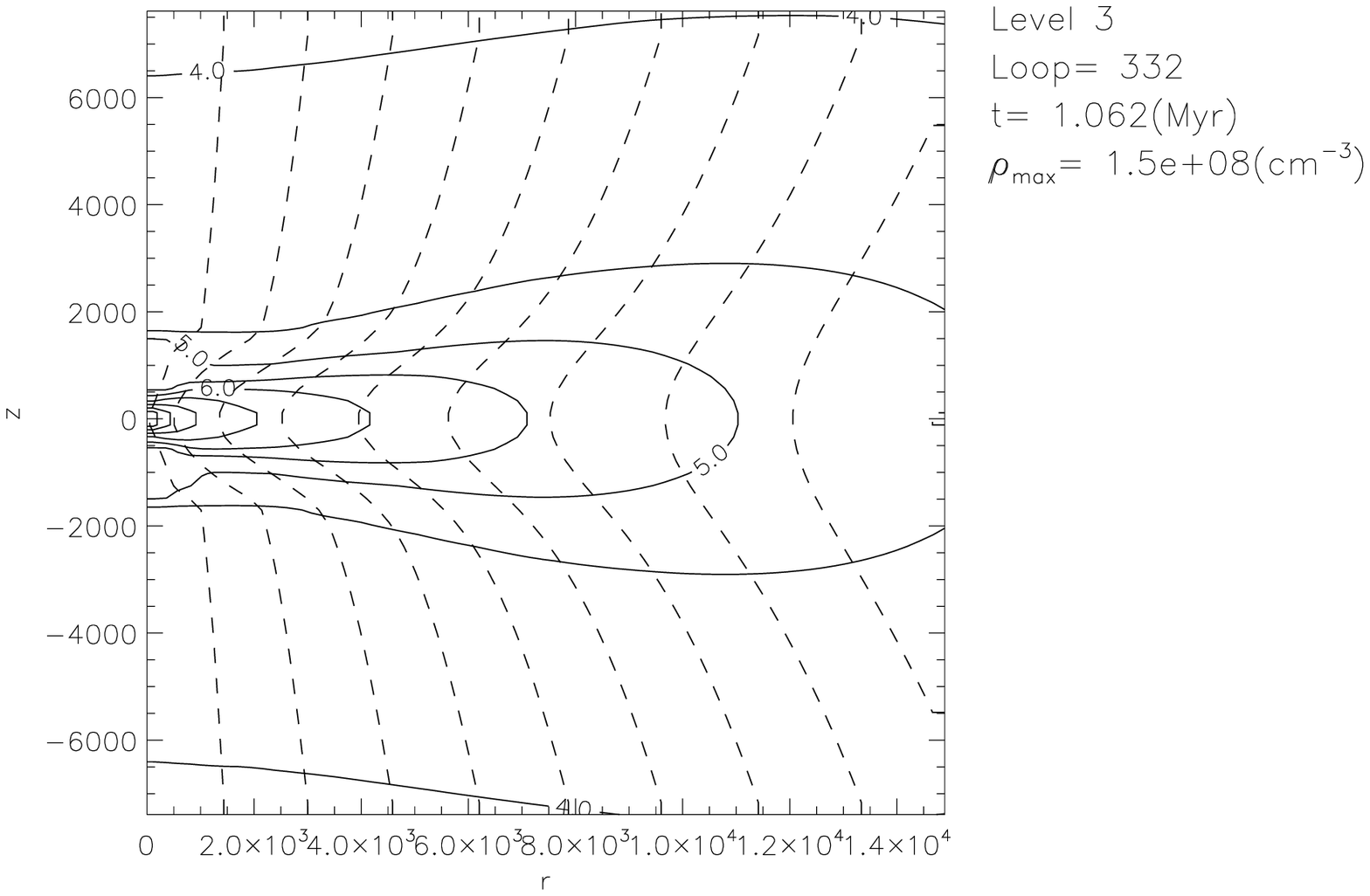}
      \one{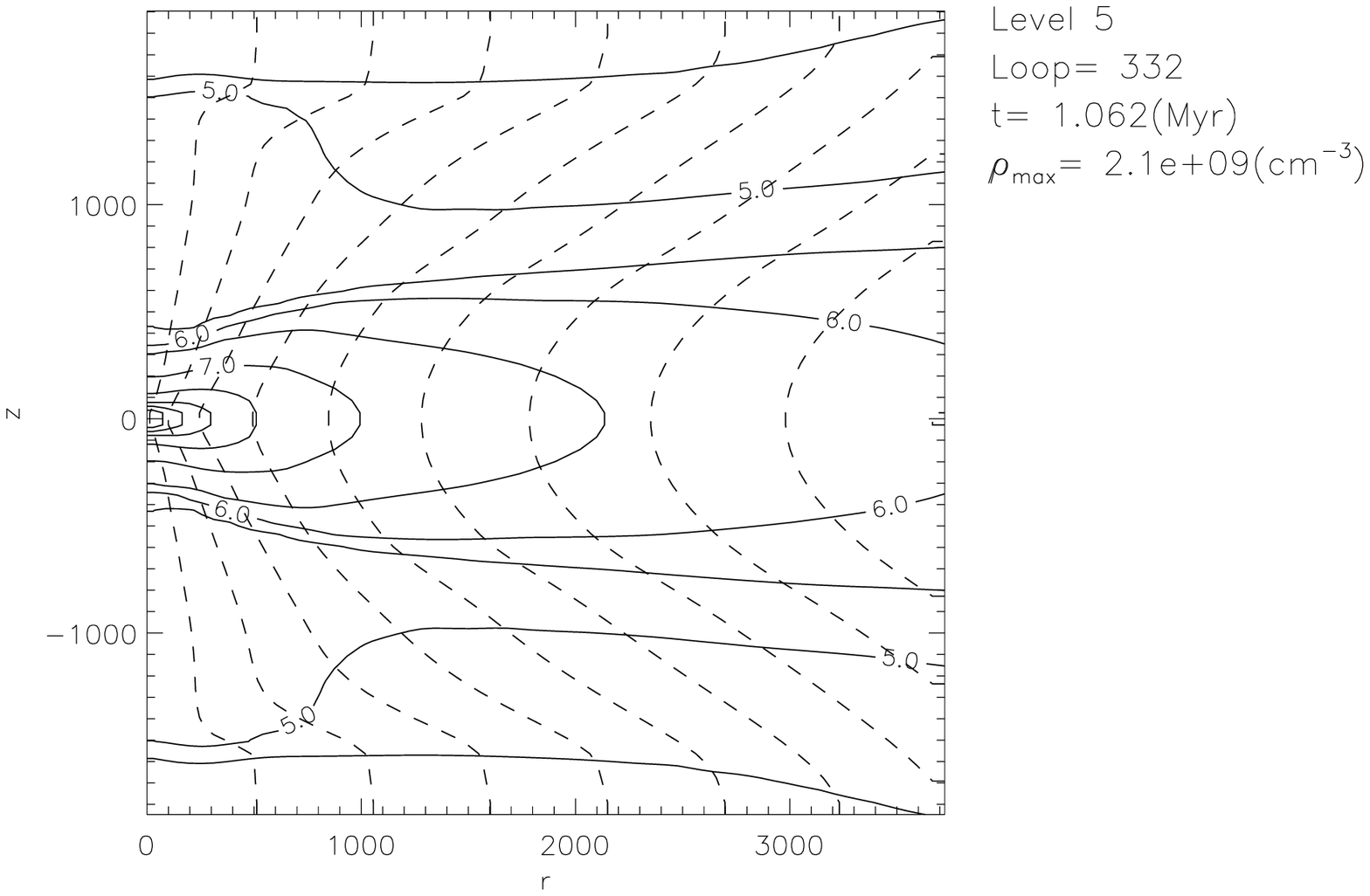}
      \one{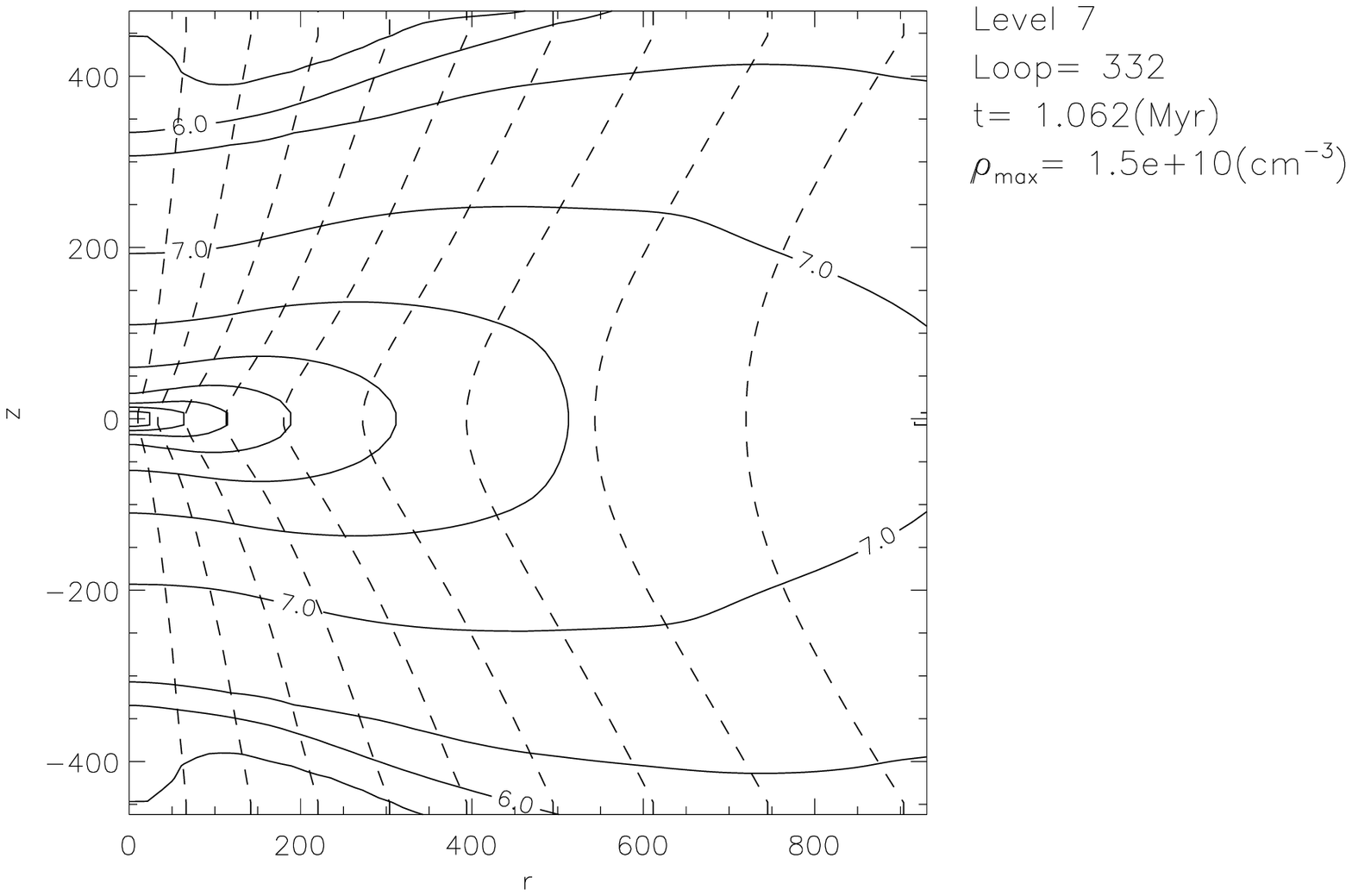}
      \one{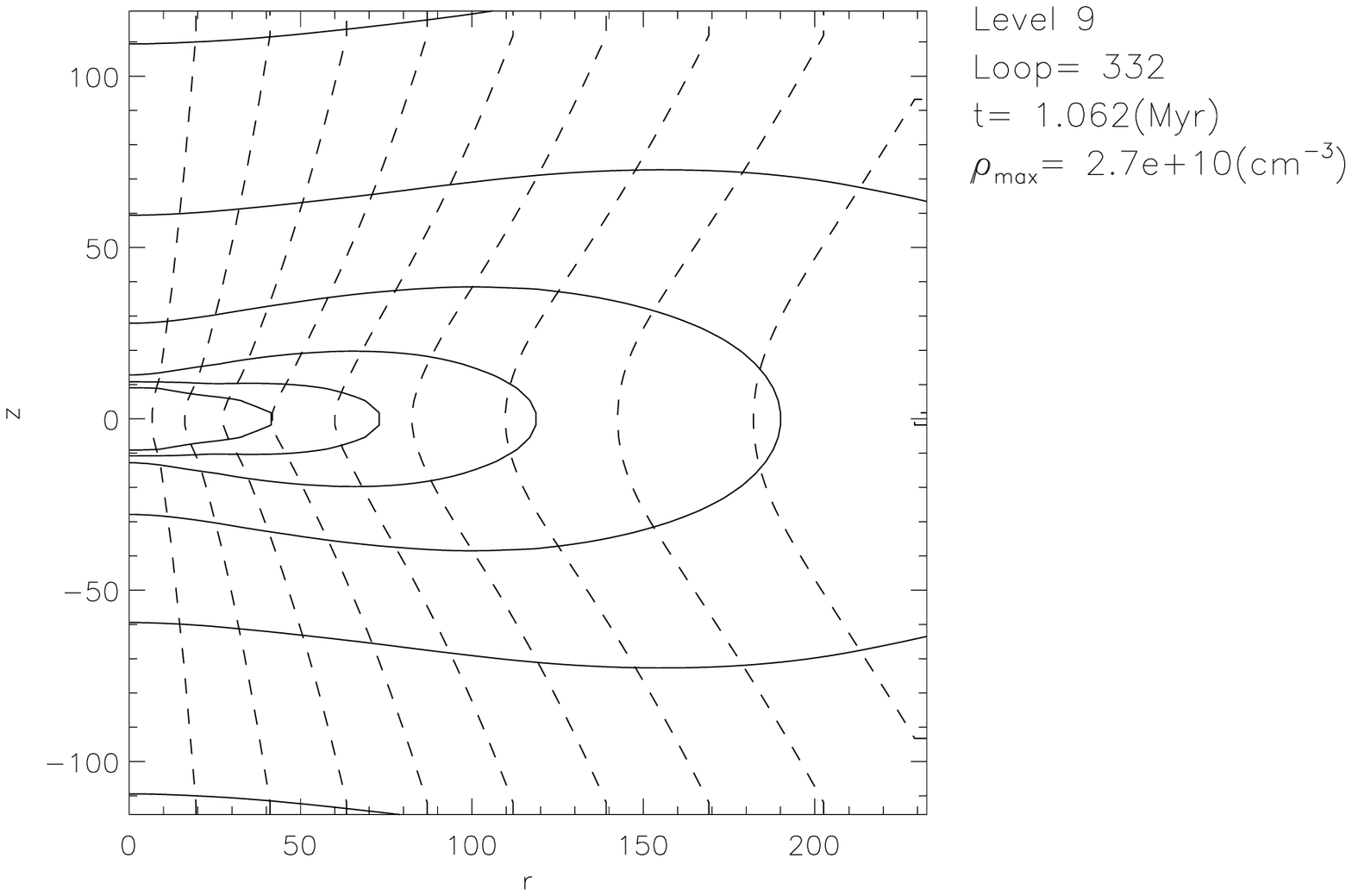}}\\

$\theta=0^\circ$\hspace*{12mm}
      \raisebox{-20mm}{\one{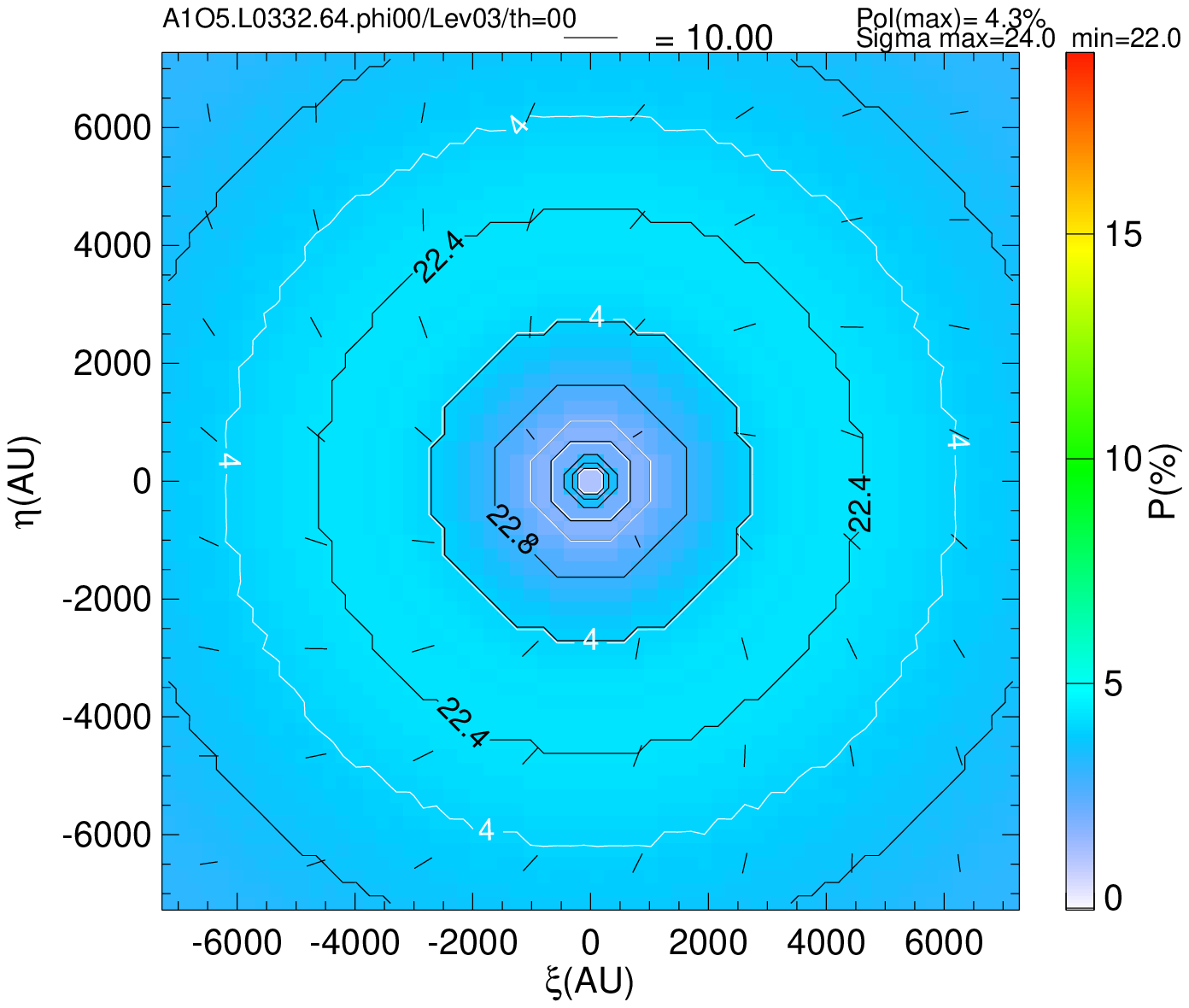}
      \one{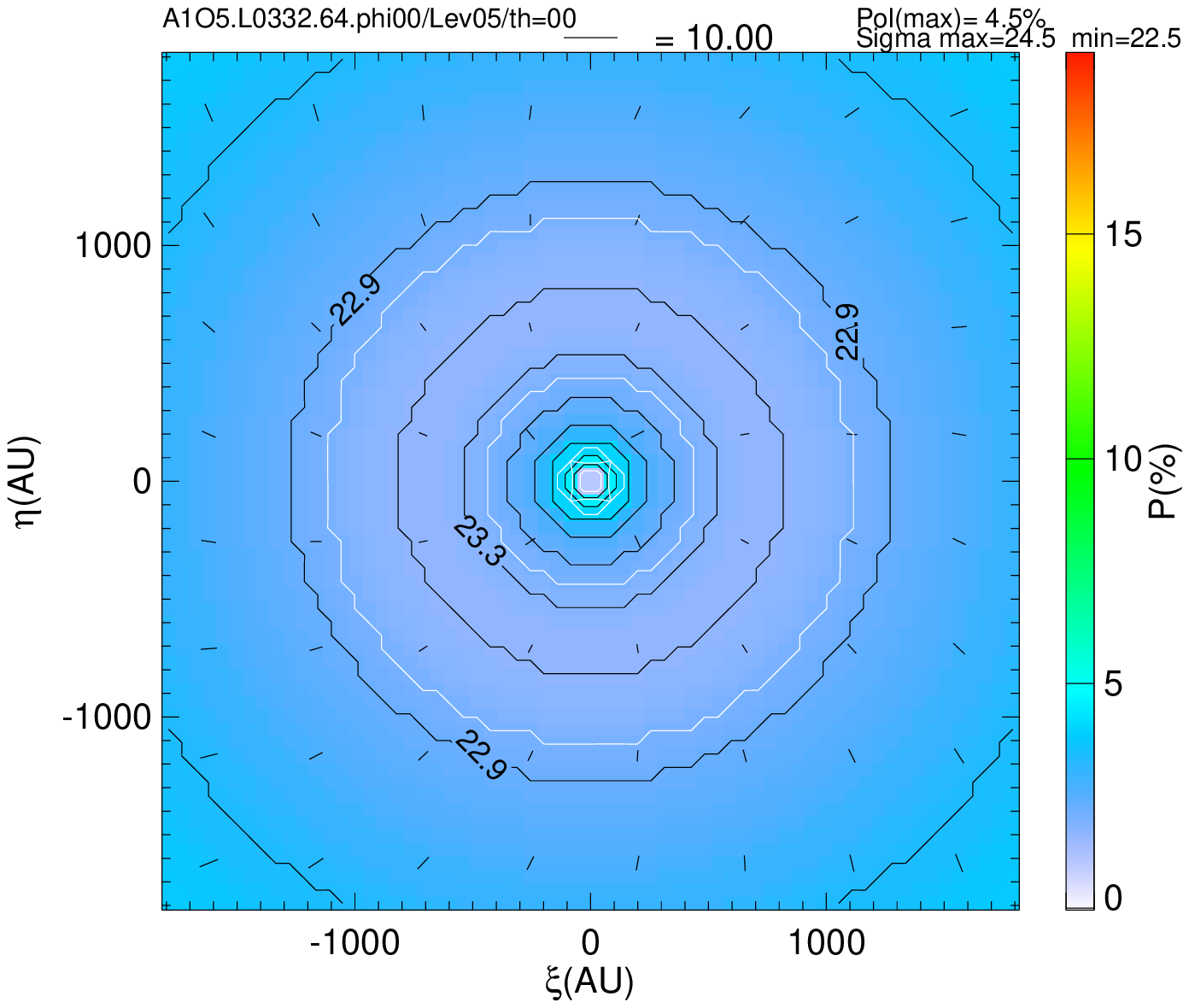}
      \one{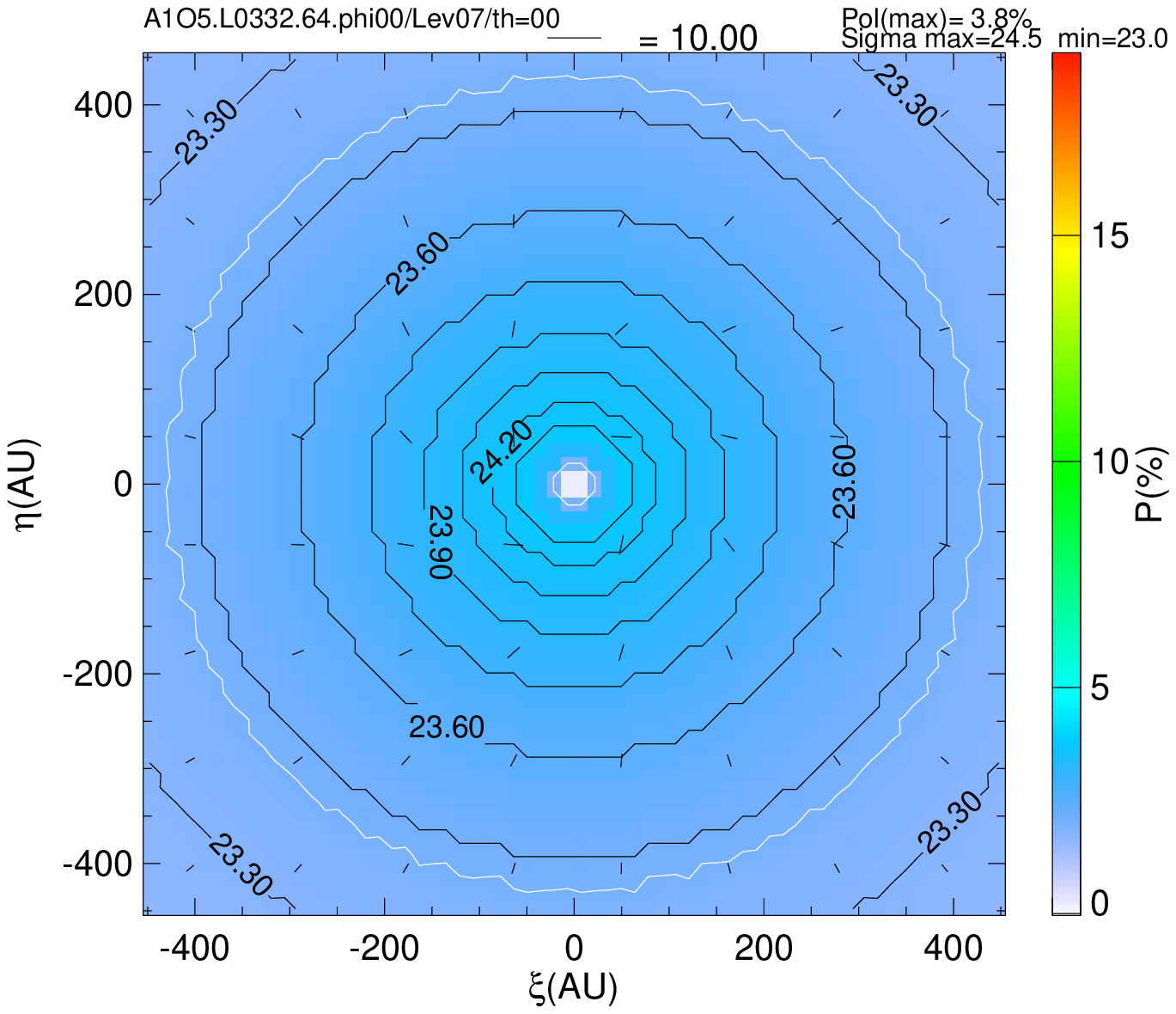}
      \one{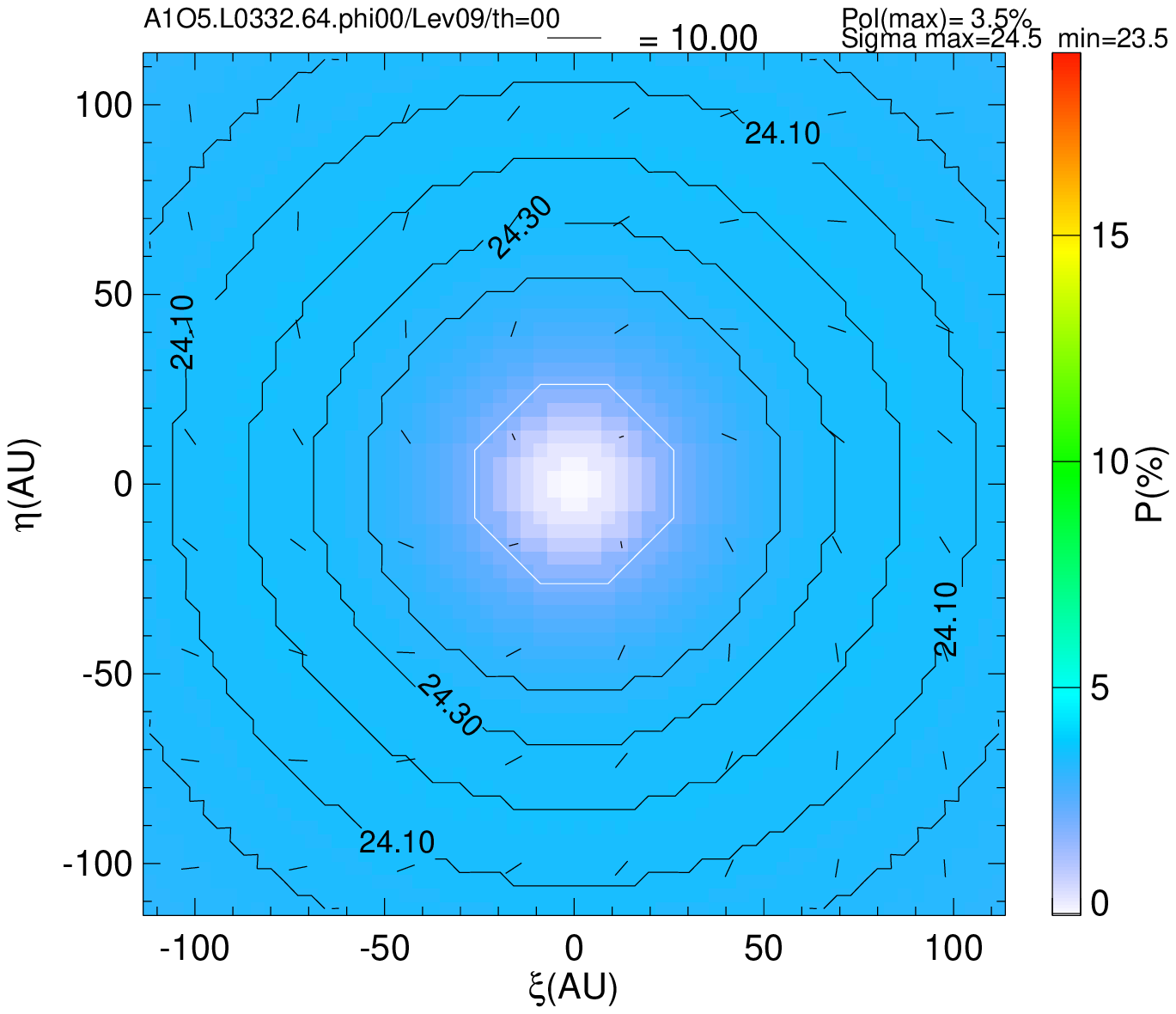}}\\
$\theta=30^\circ$\hspace*{10mm}
      \raisebox{-20mm}{\one{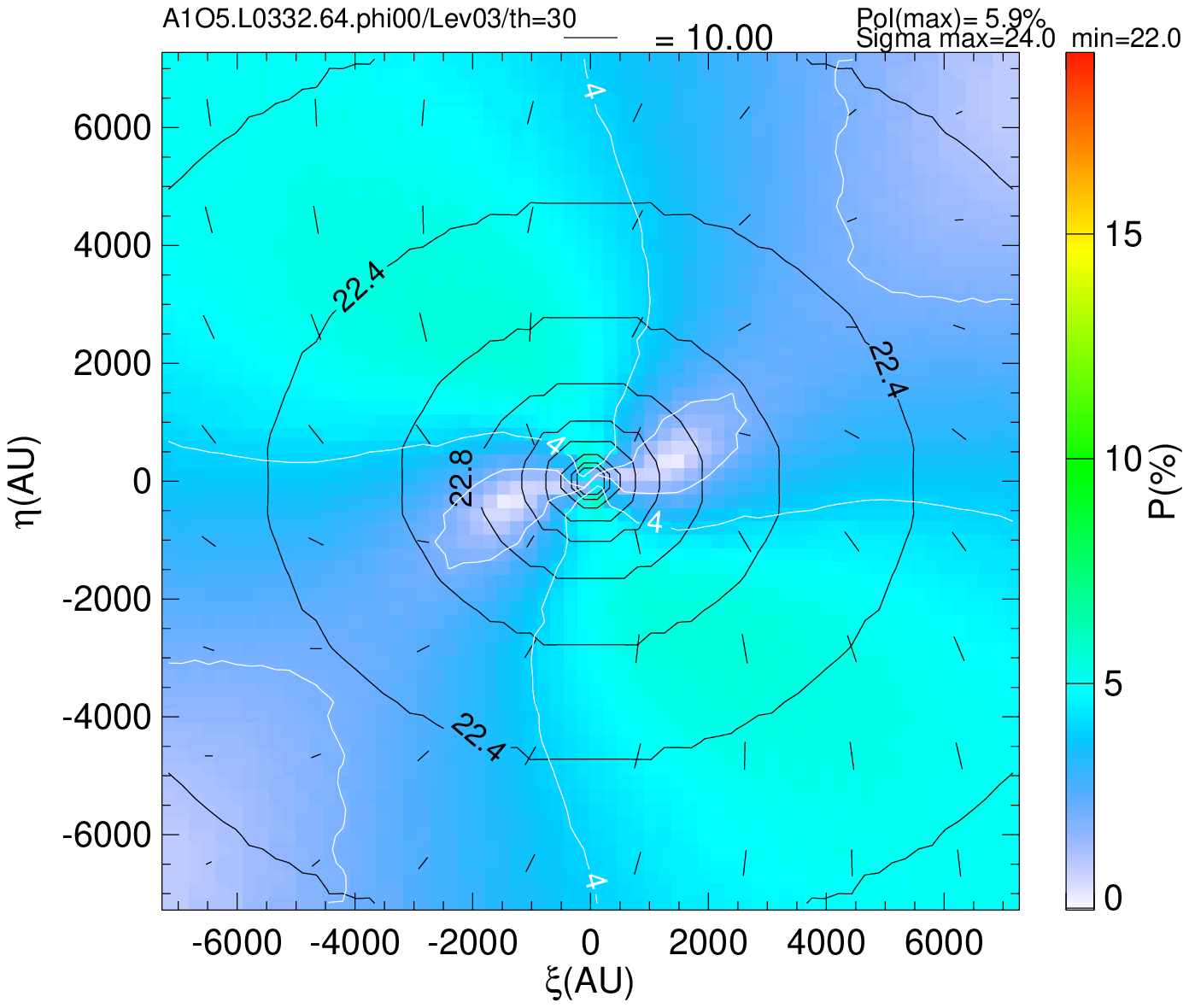}
      \one{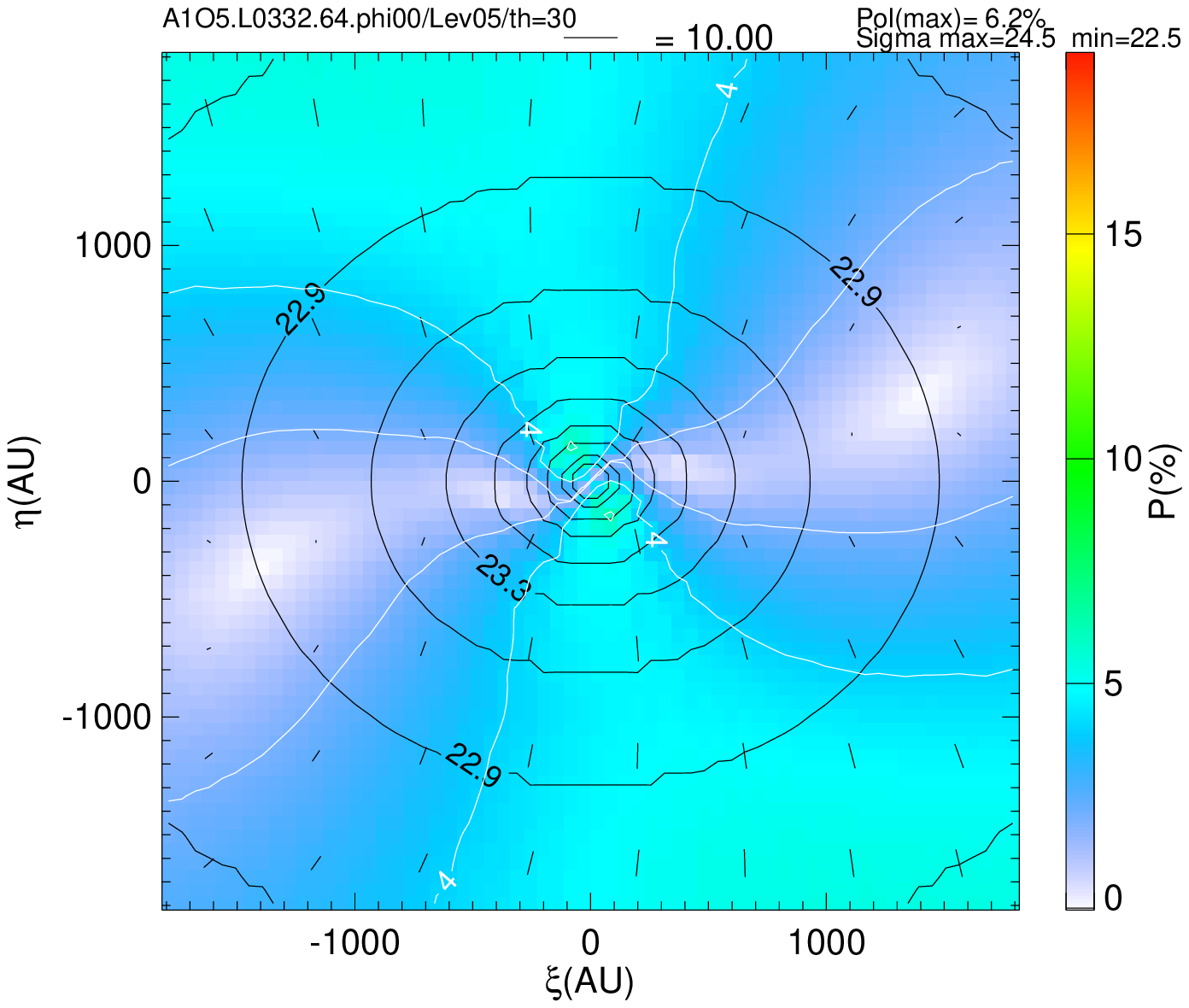}
      \one{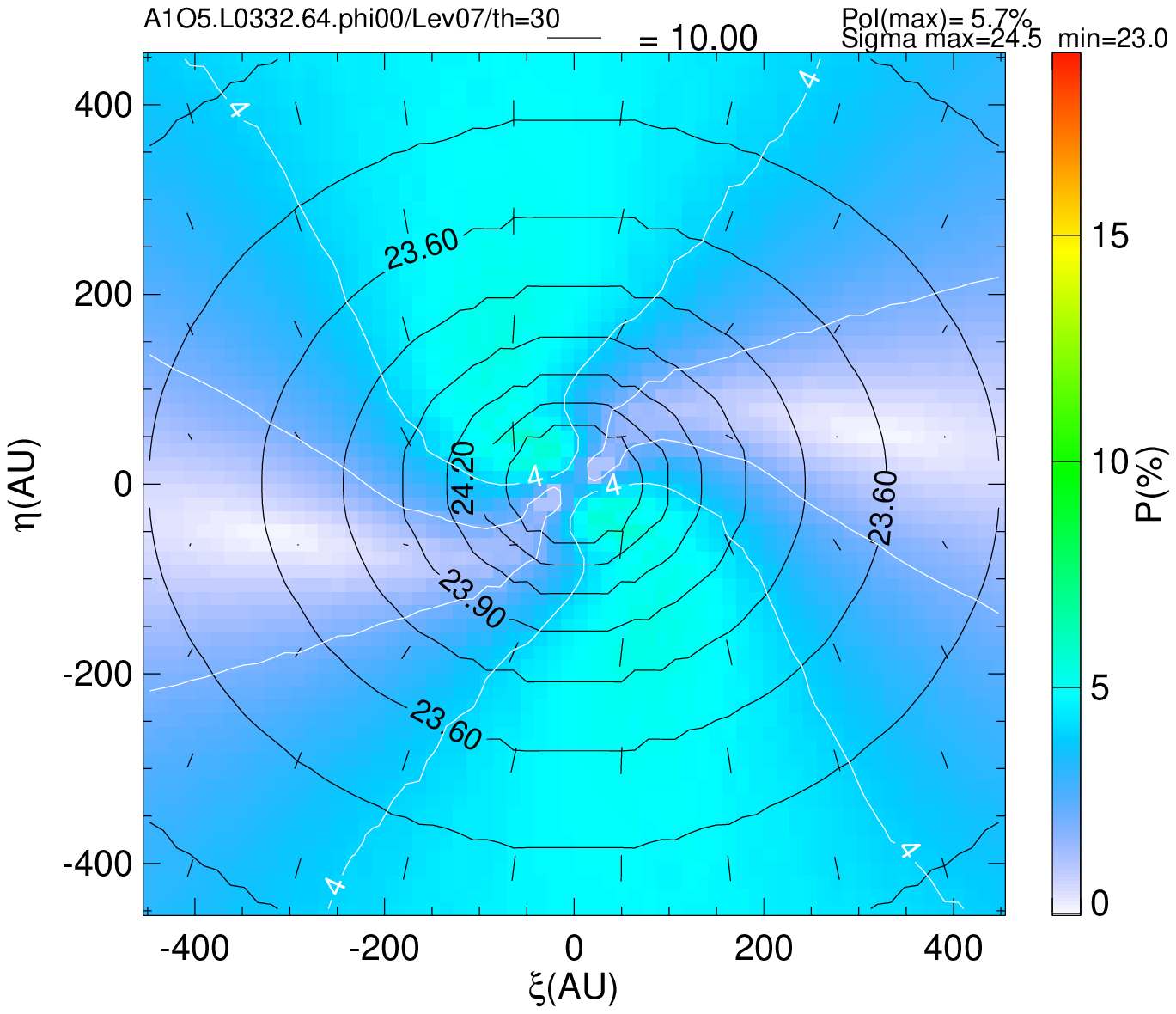}
      \one{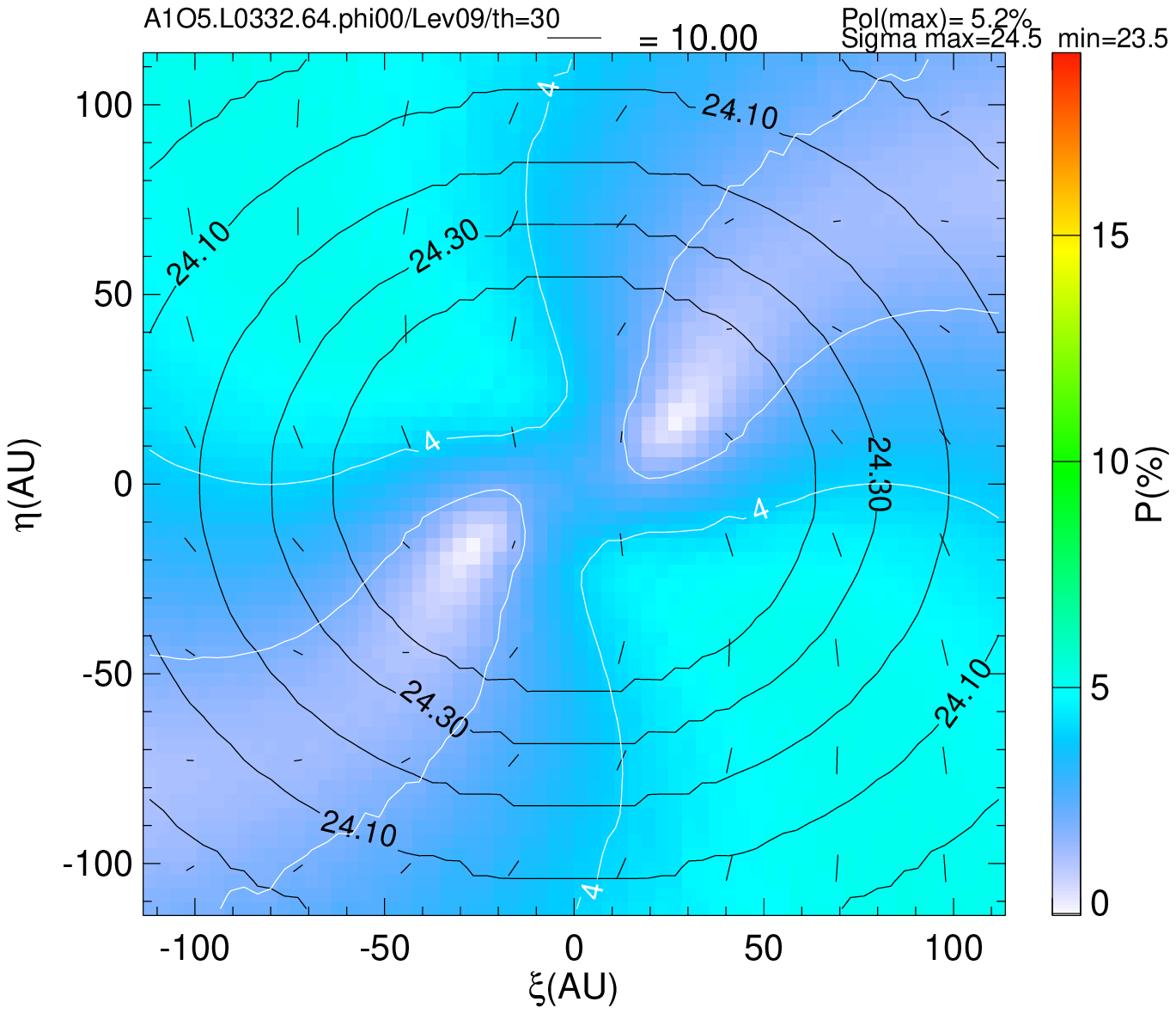}}\\
$\theta=45^\circ$\hspace*{10mm}
      \raisebox{-20mm}{\one{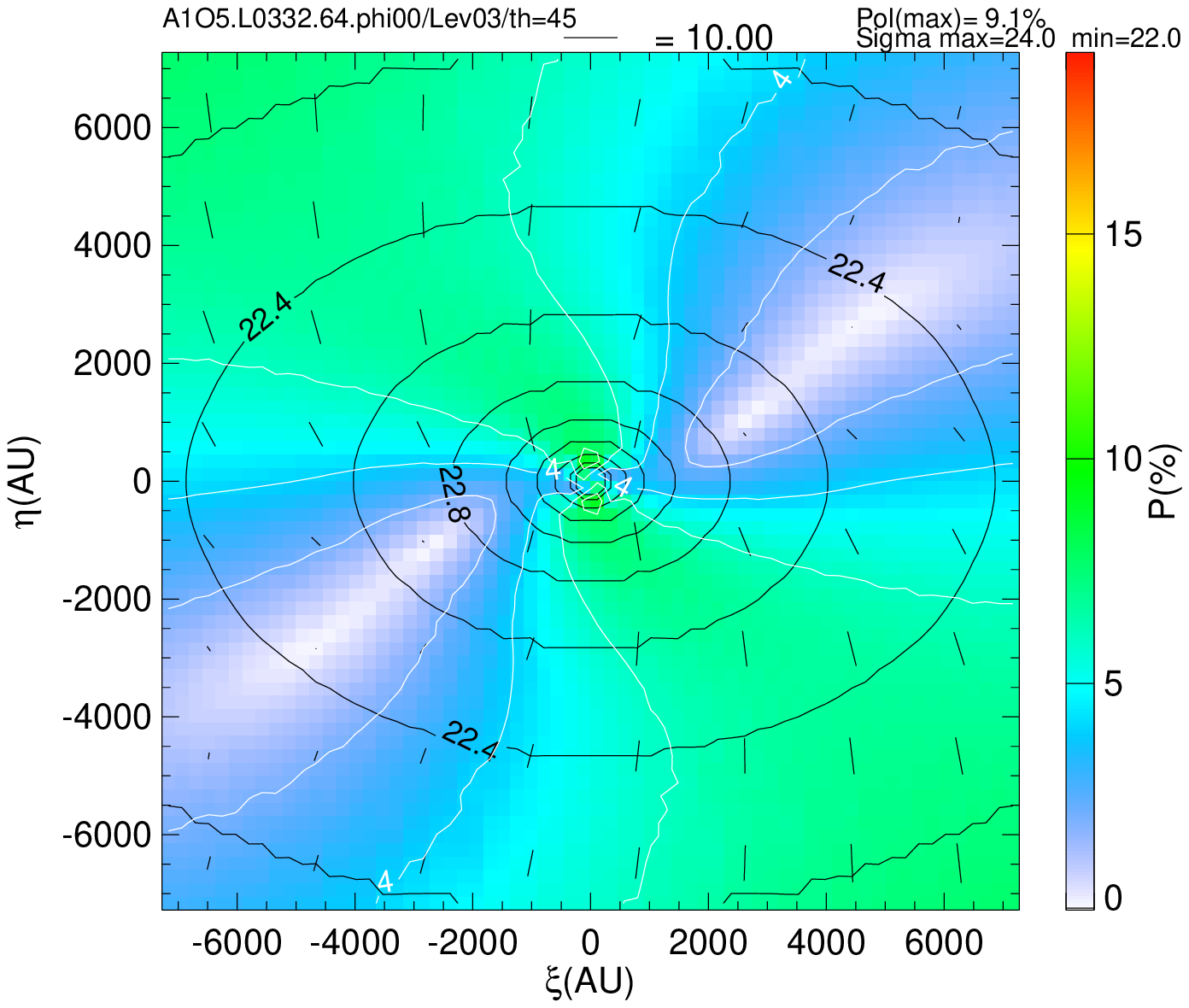}
      \one{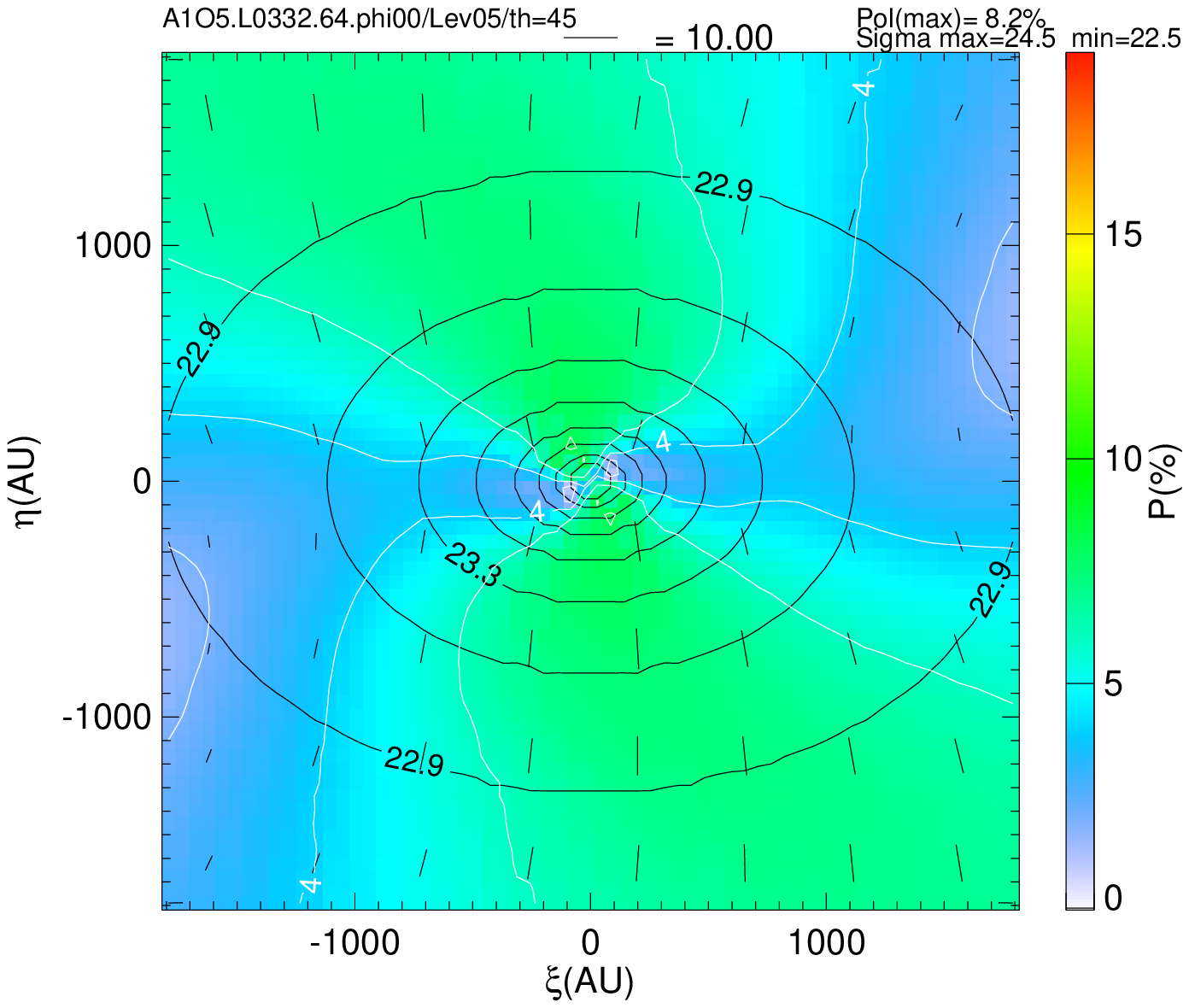}
      \one{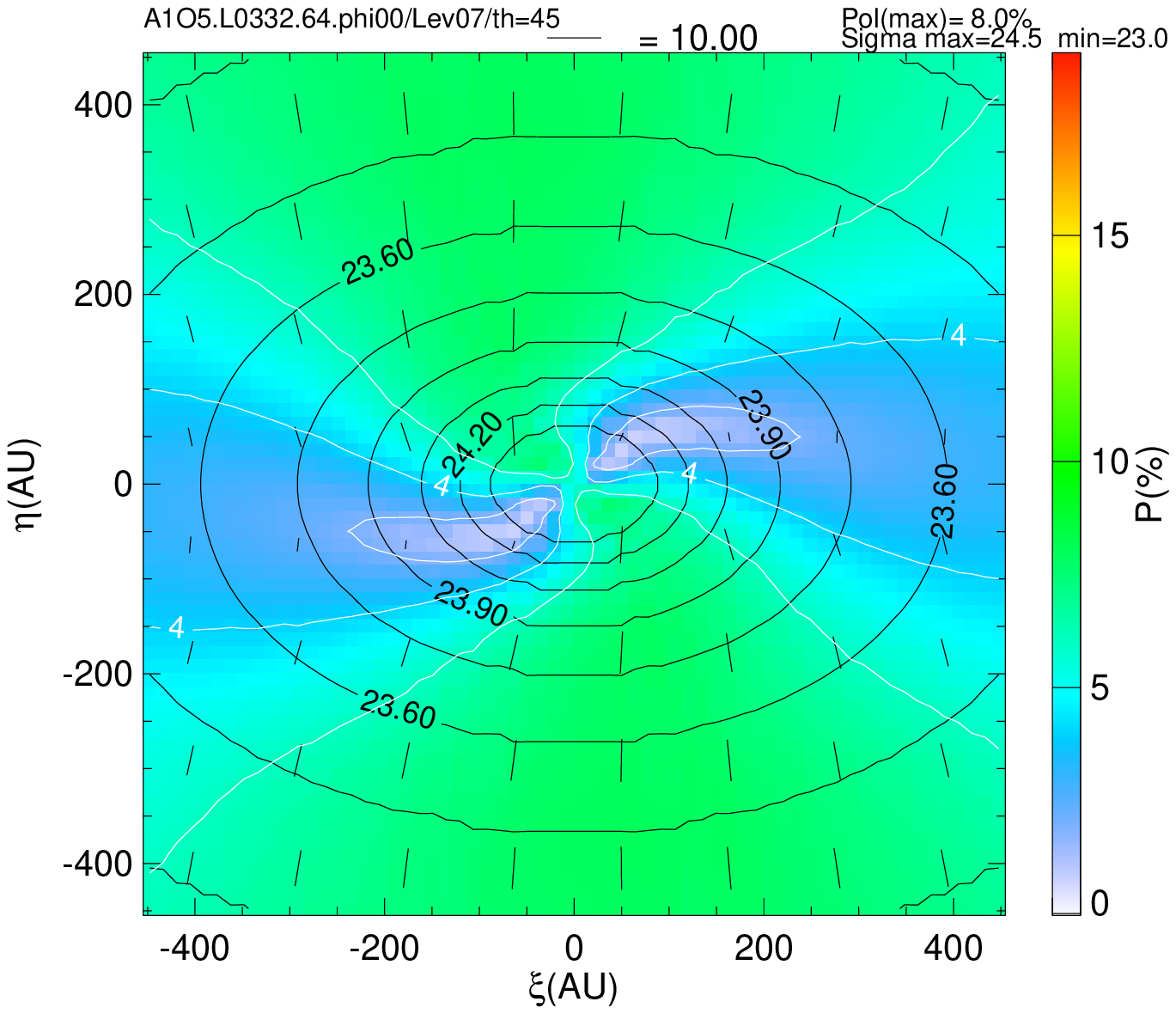}
      \one{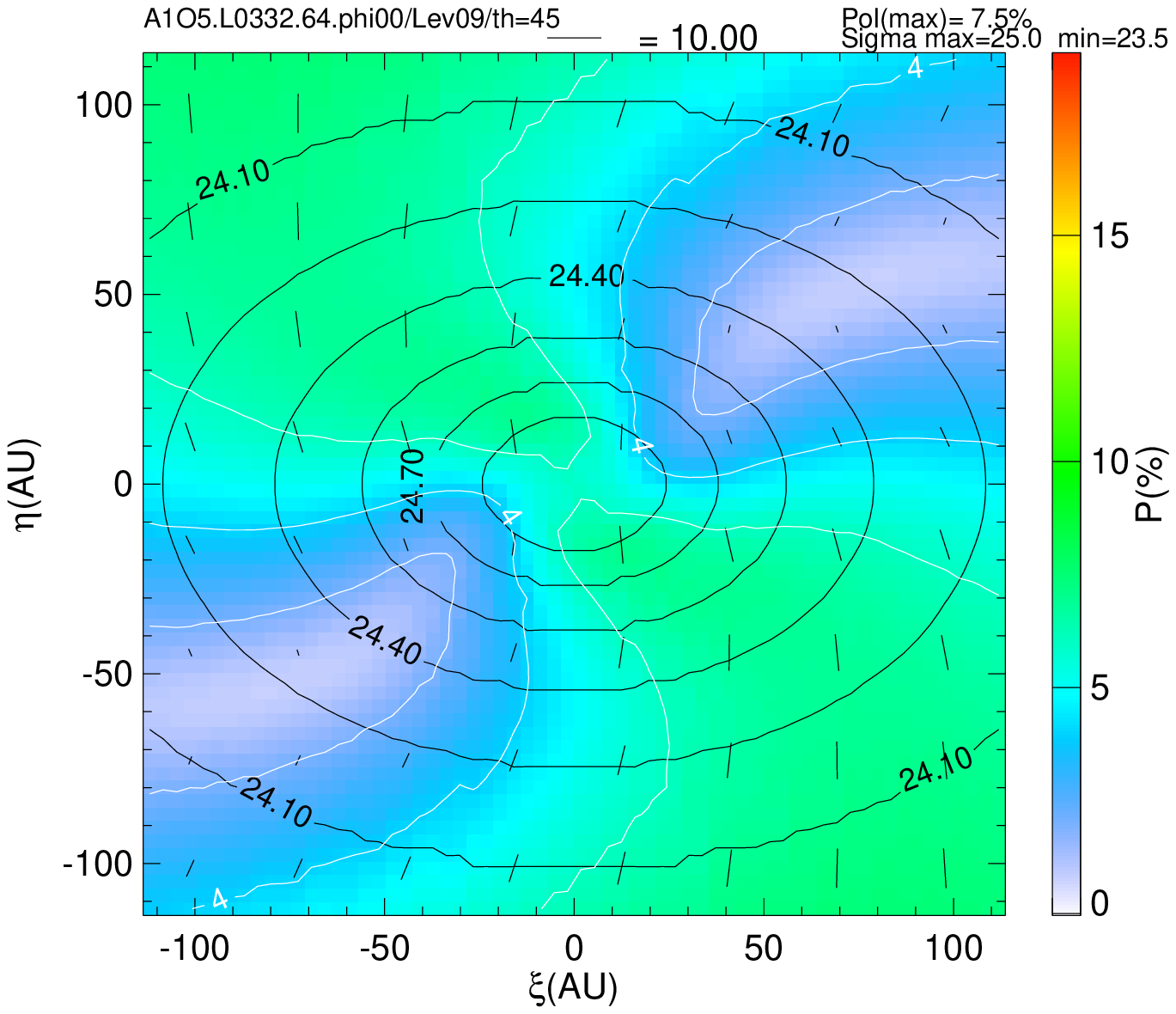}}\\
$\theta=60^\circ$\hspace*{10mm}
      \raisebox{-20mm}{\one{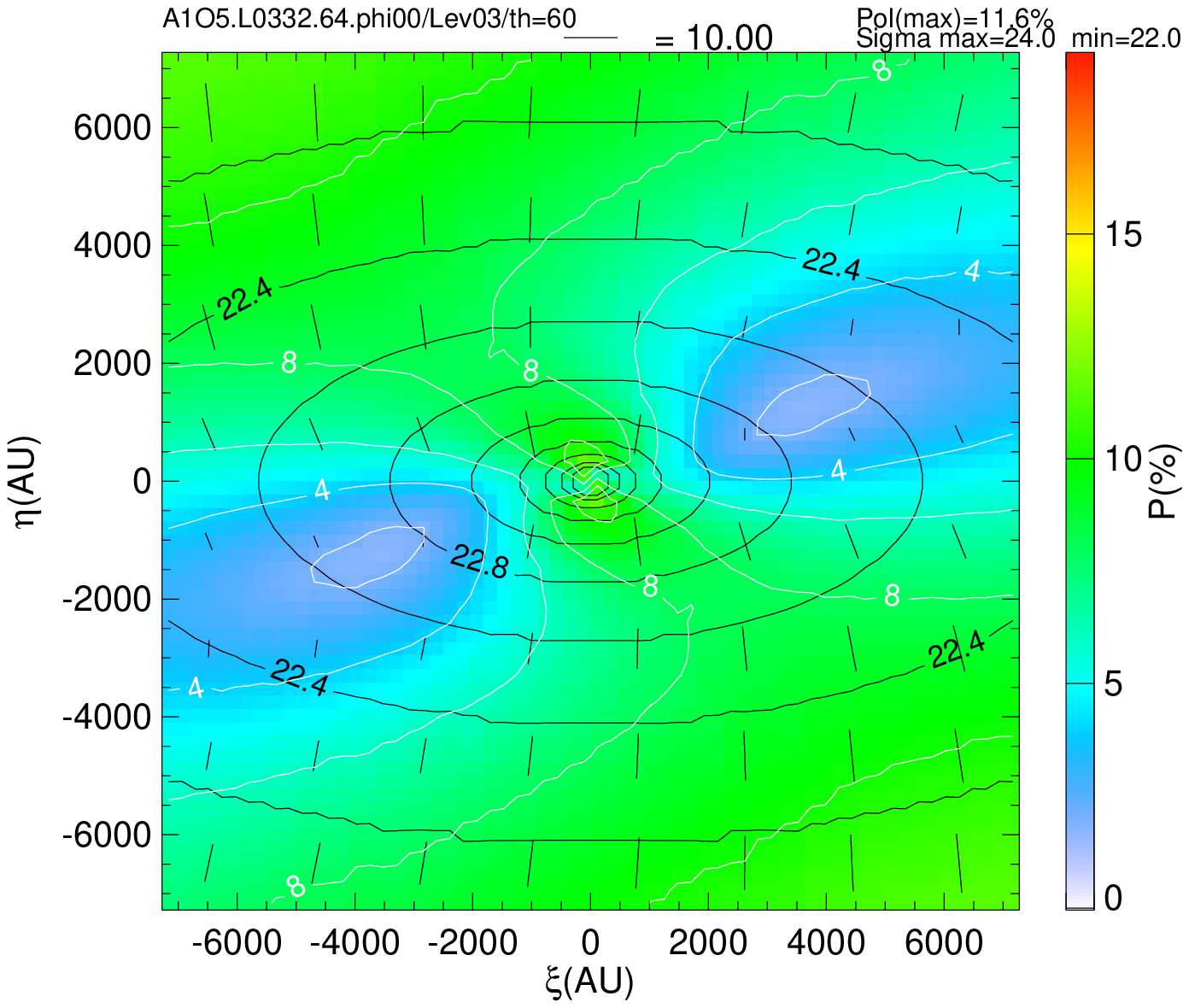}
      \one{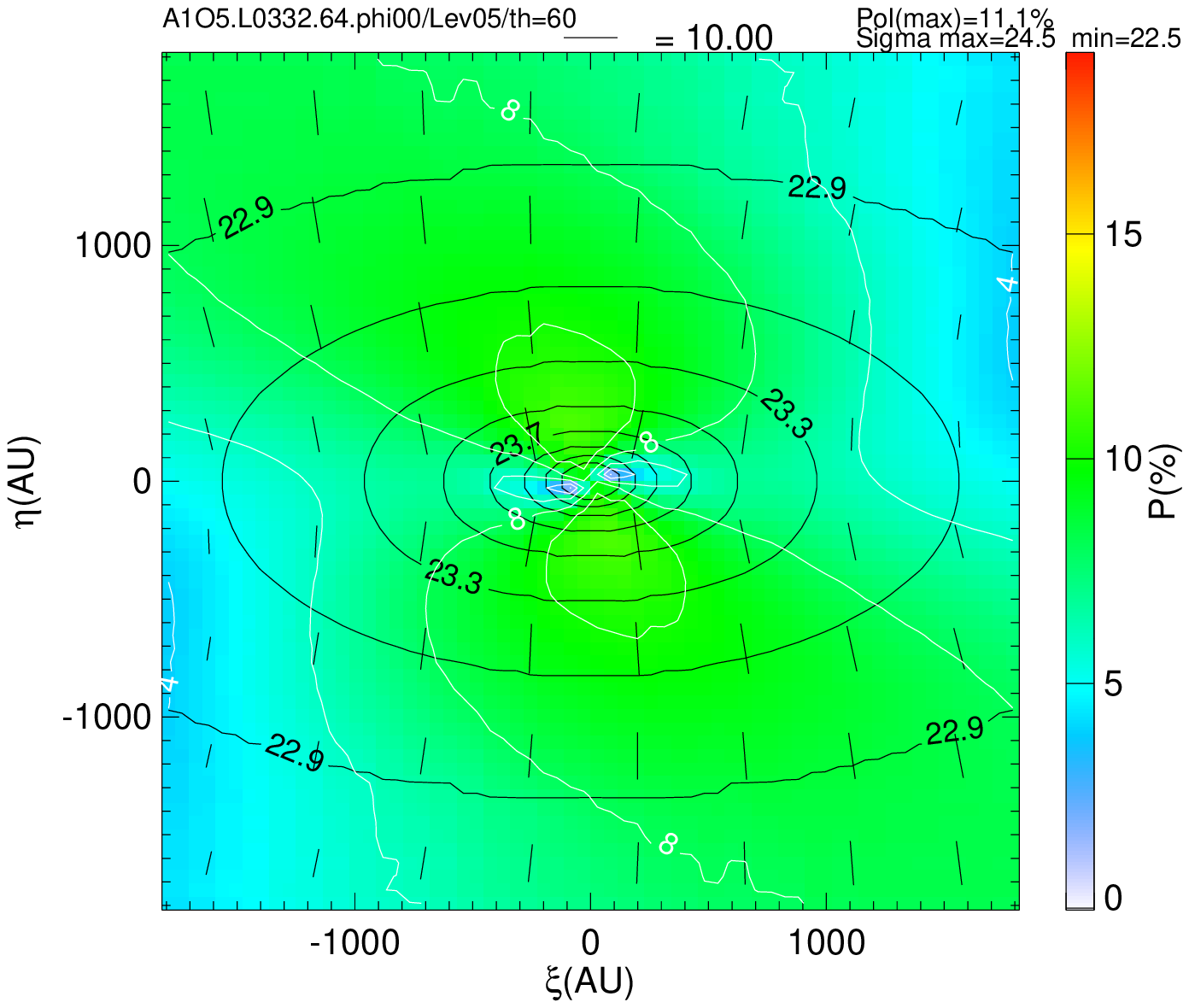}
      \one{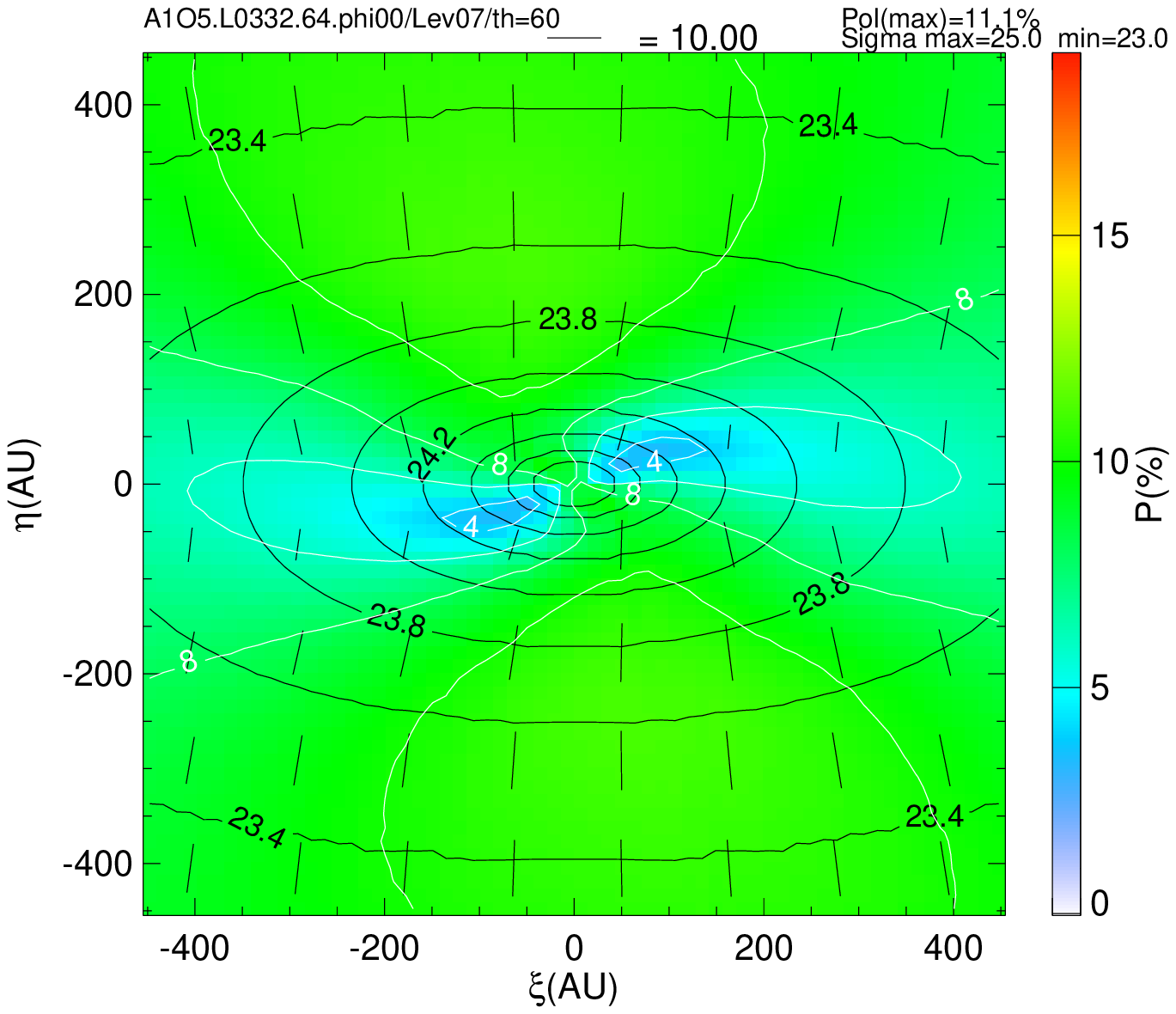}
      \one{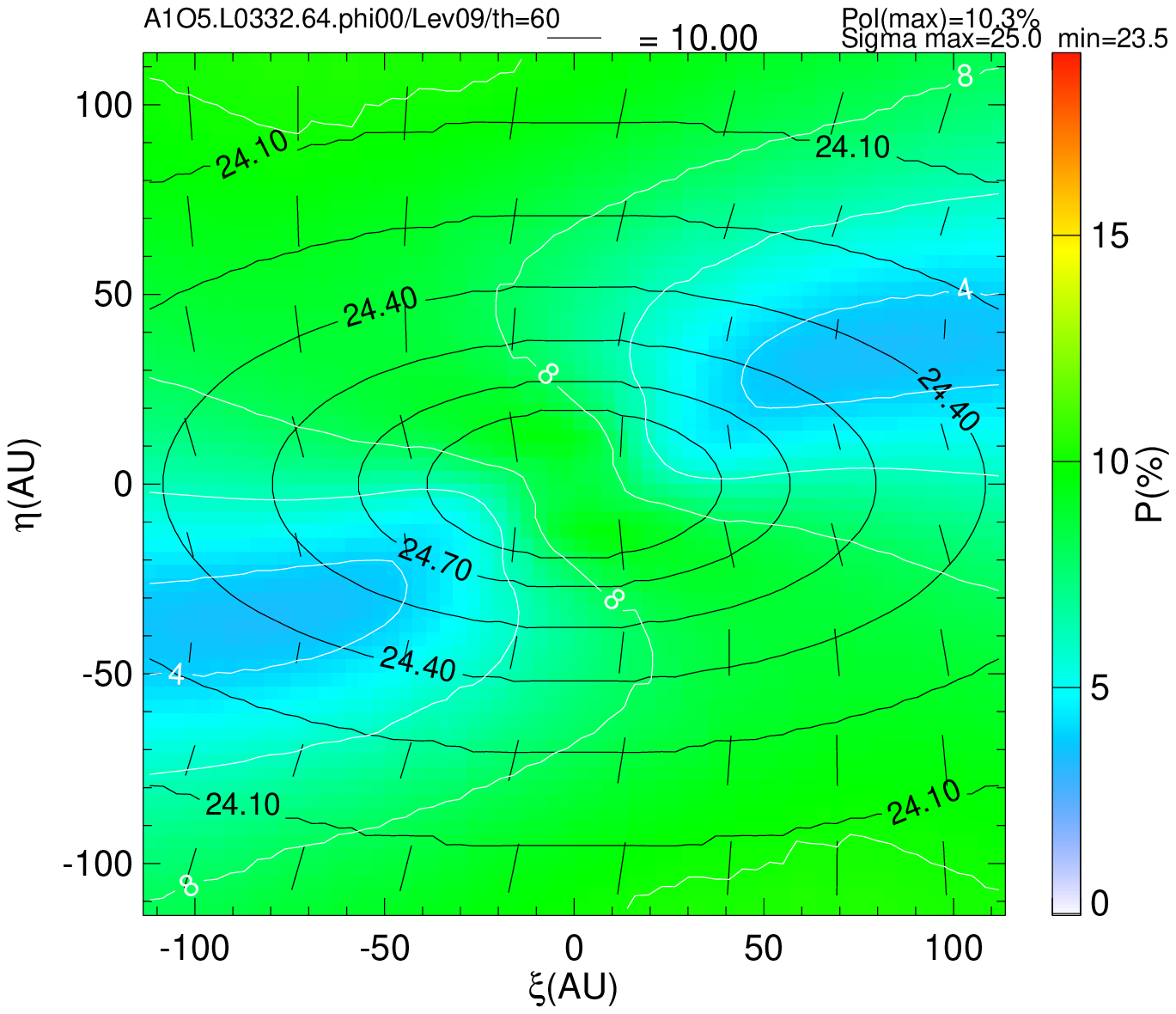}}\\
$\theta=80^\circ$\hspace*{10mm}
      \raisebox{-20mm}{\one{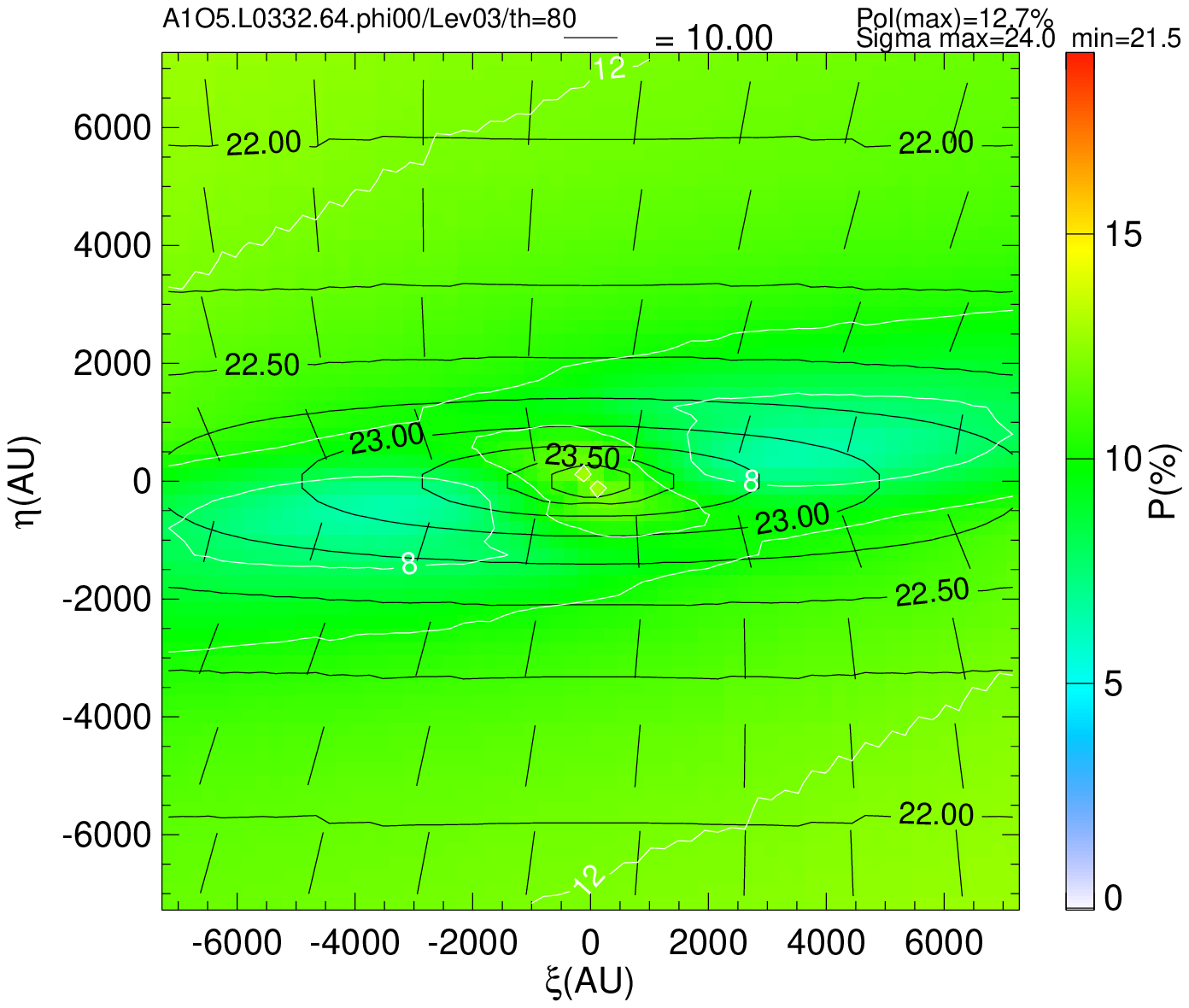}
      \one{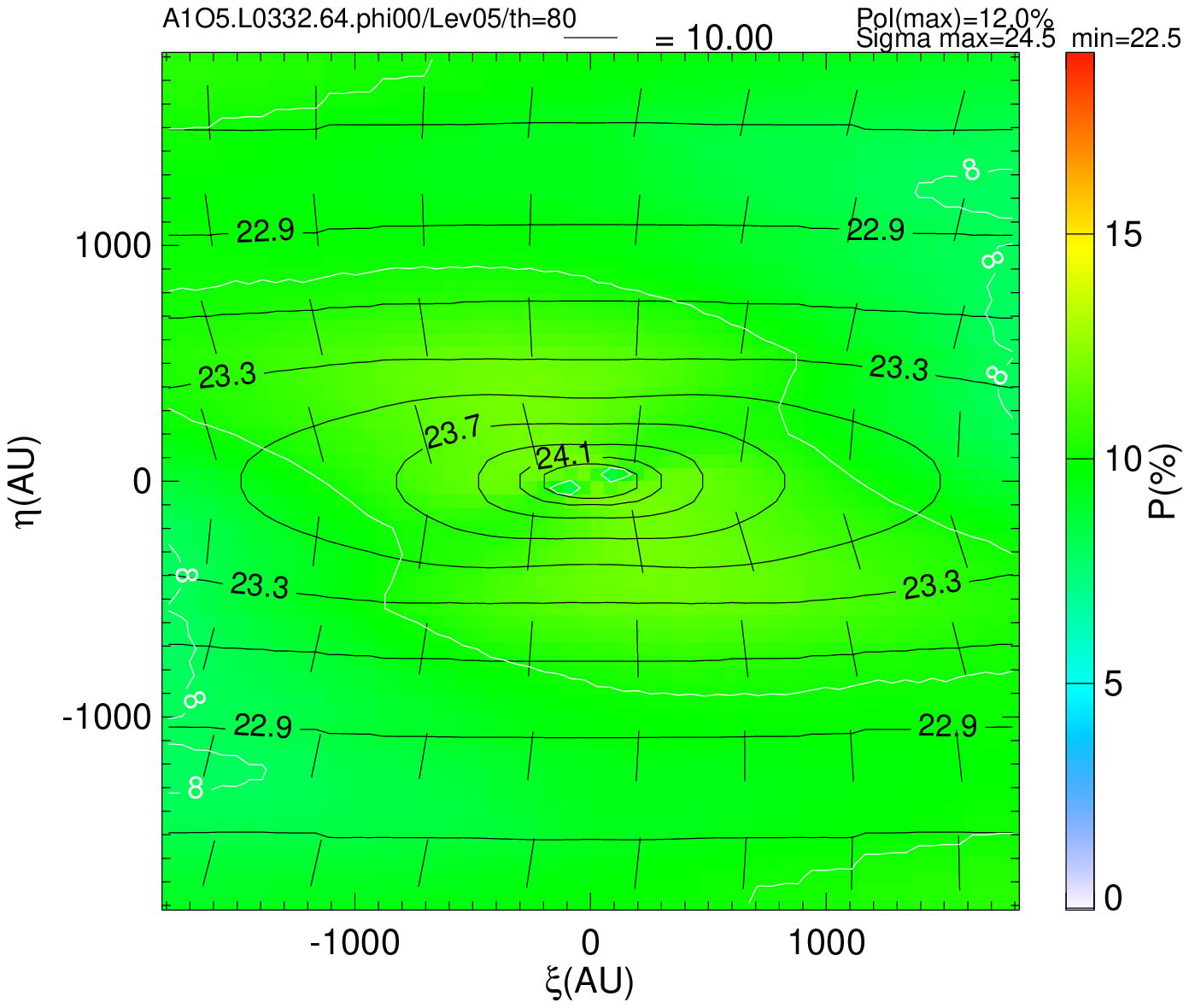}
      \one{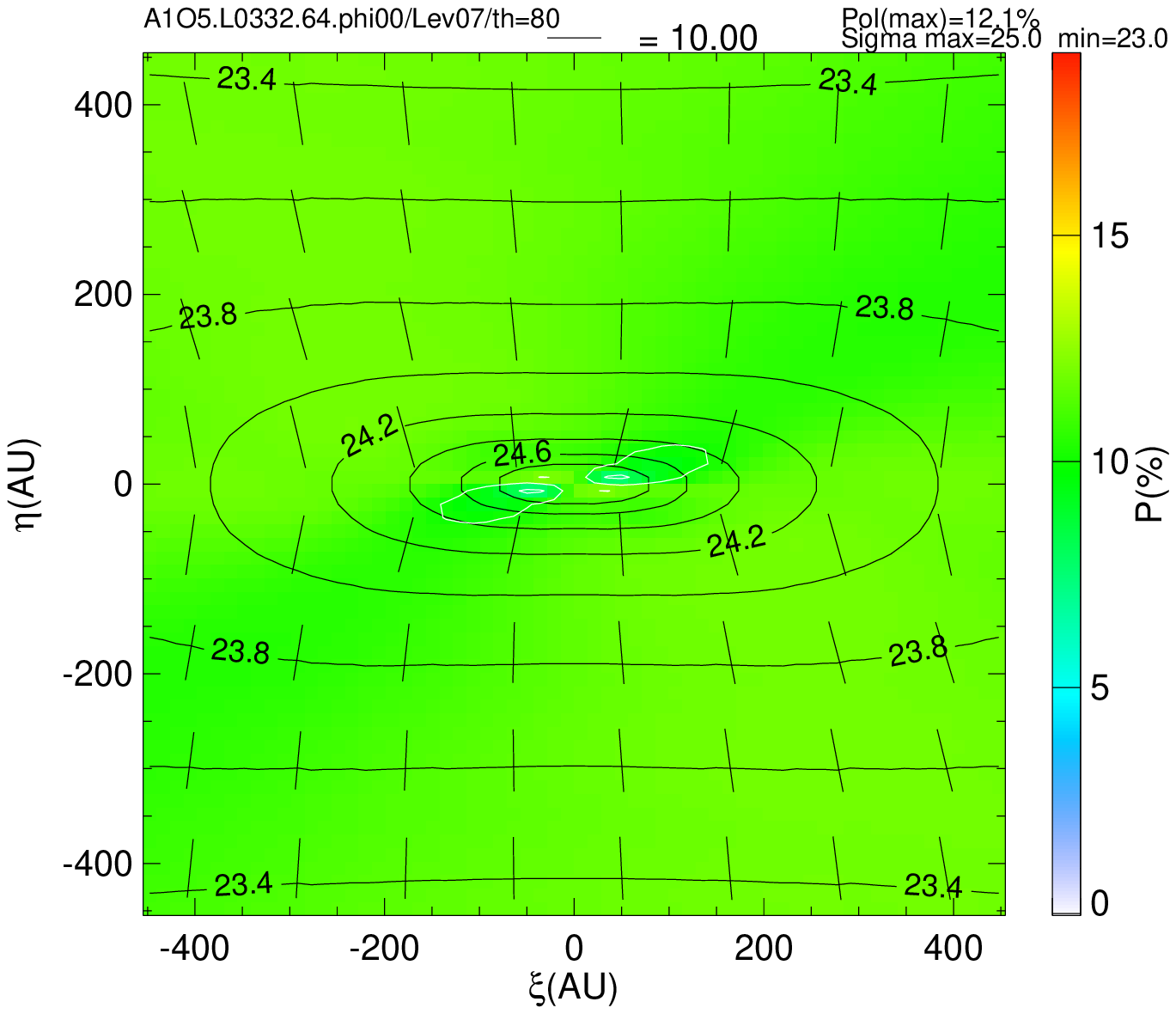}
      \one{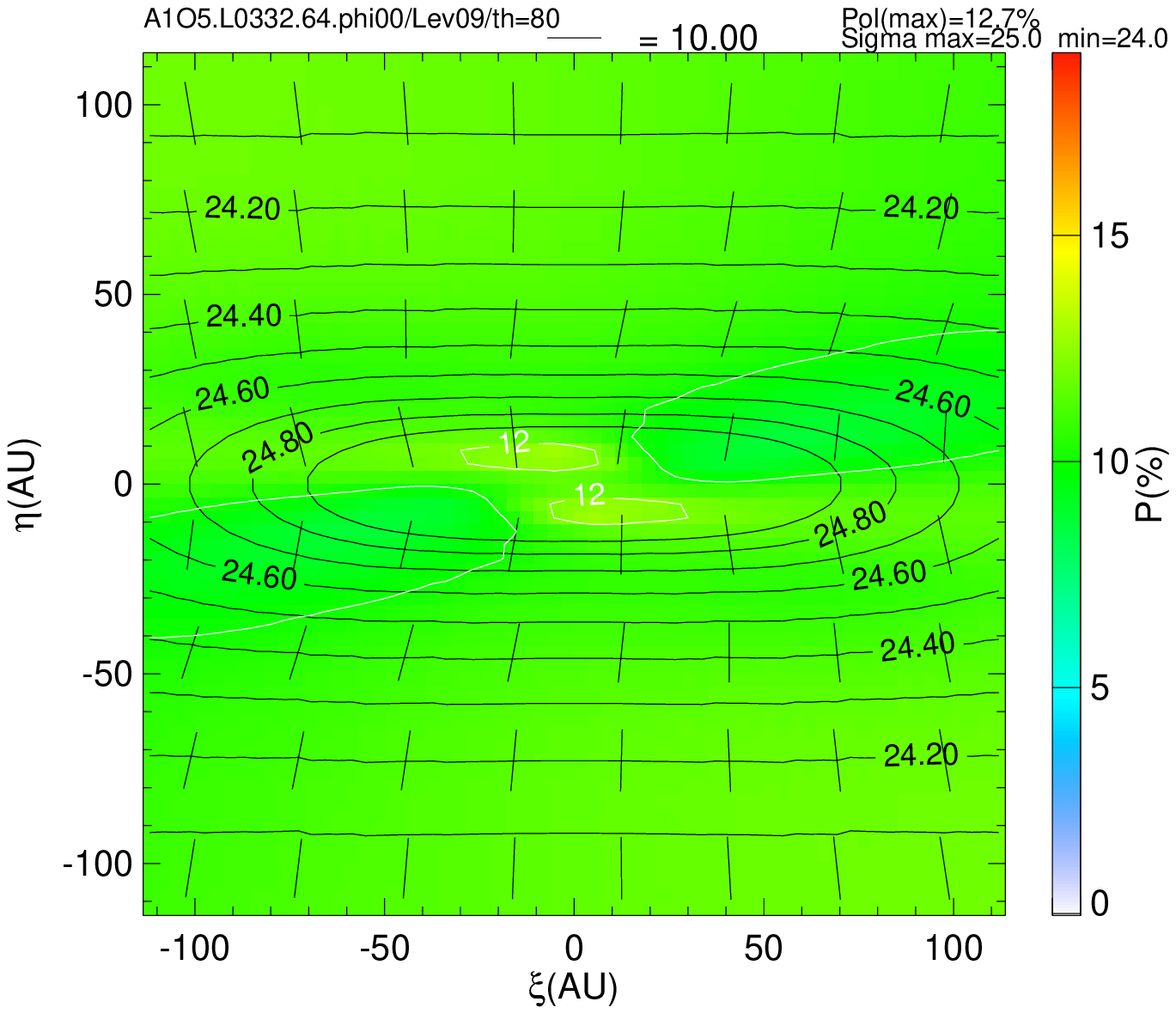}}\\
$\theta=90^\circ$\hspace*{10mm}
      \raisebox{-20mm}{\one{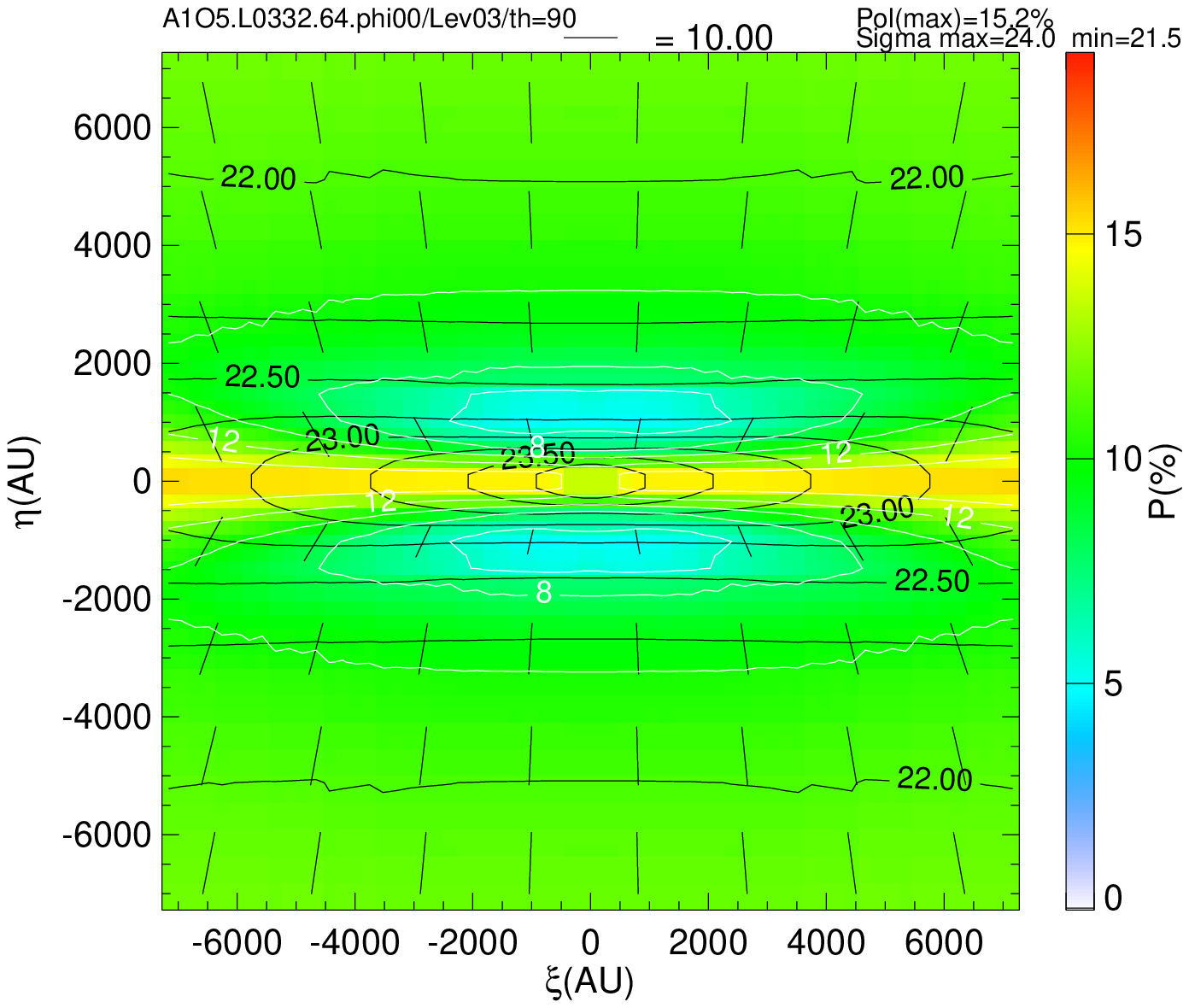}
      \one{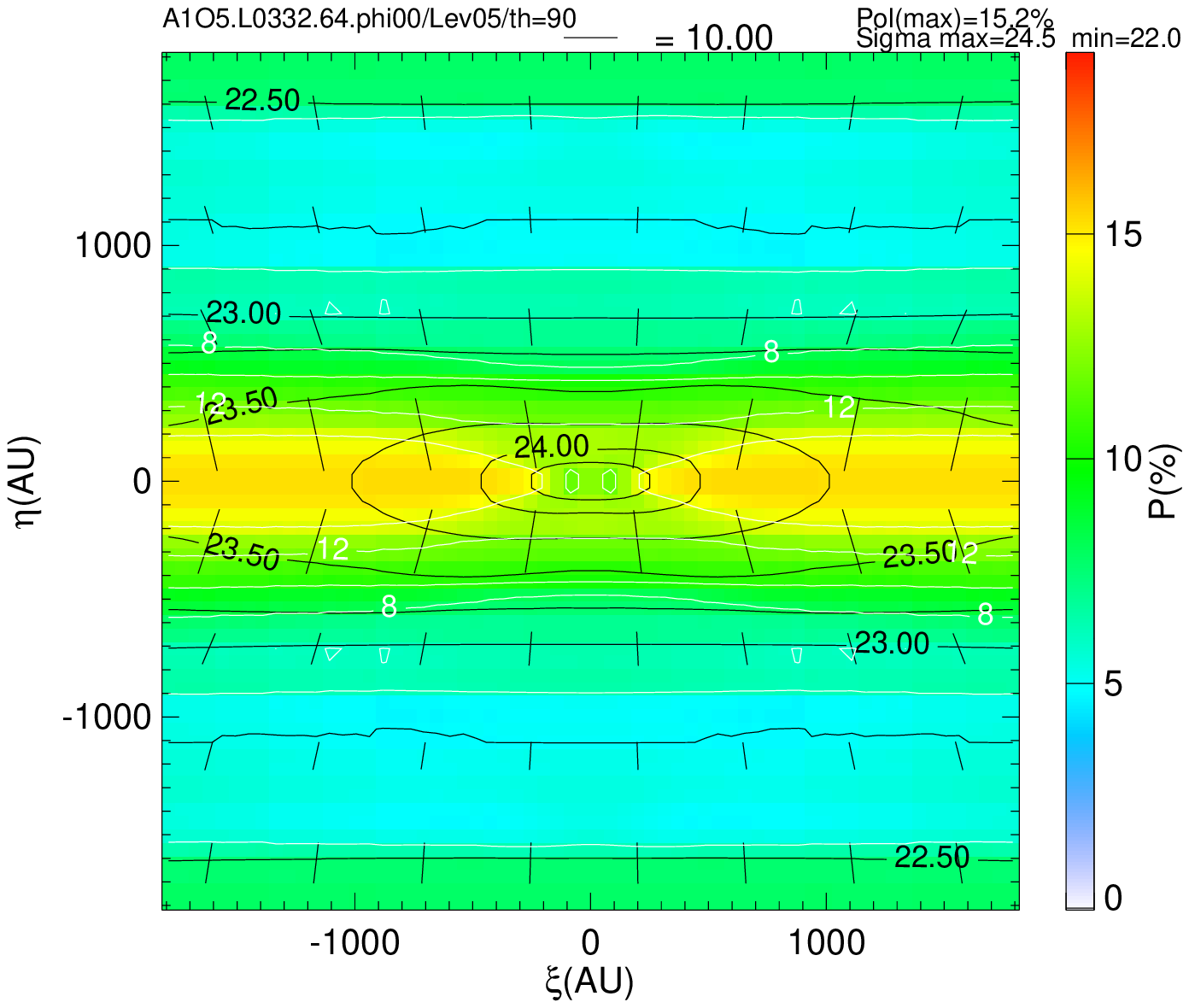}
      \one{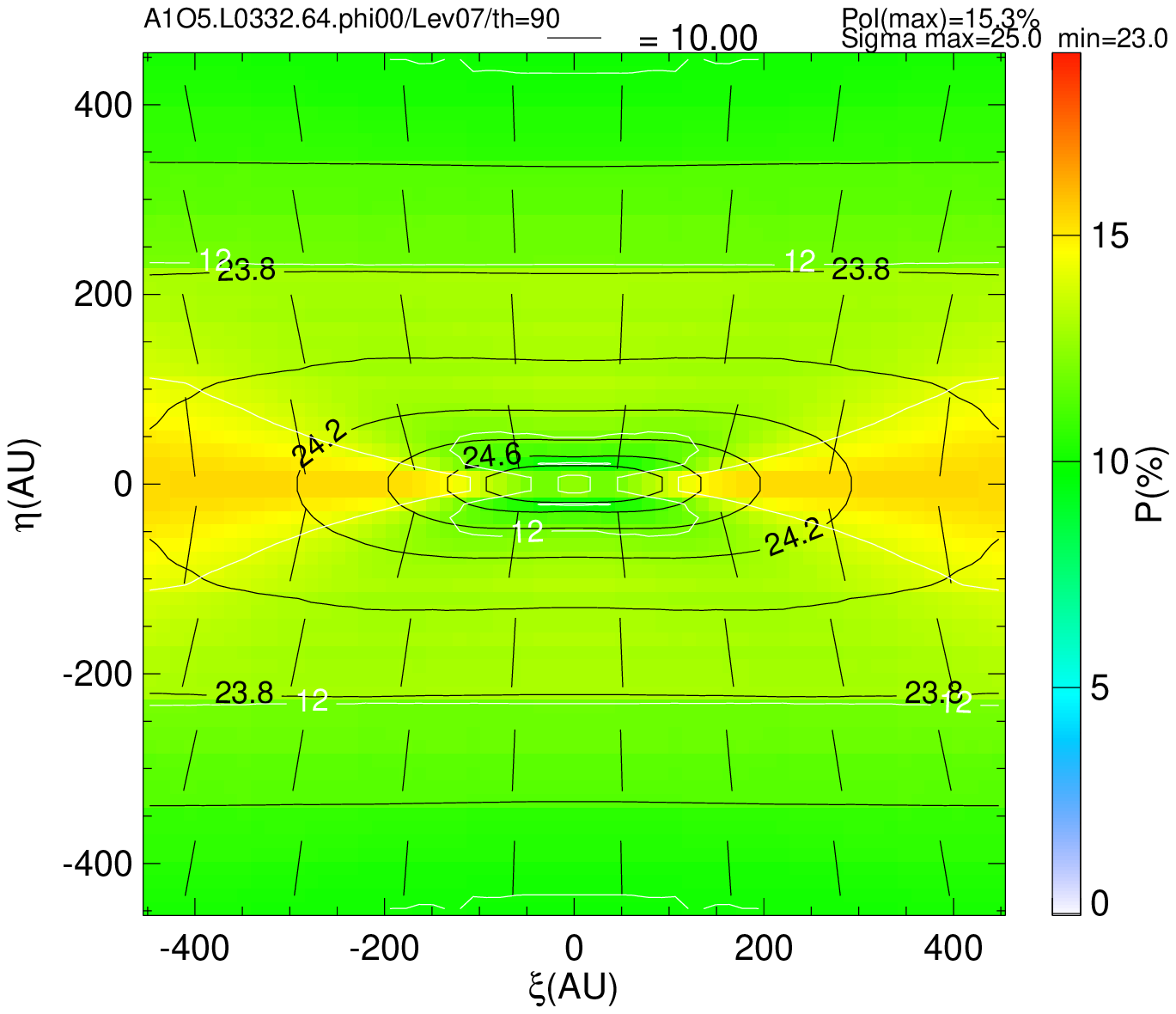}
      \one{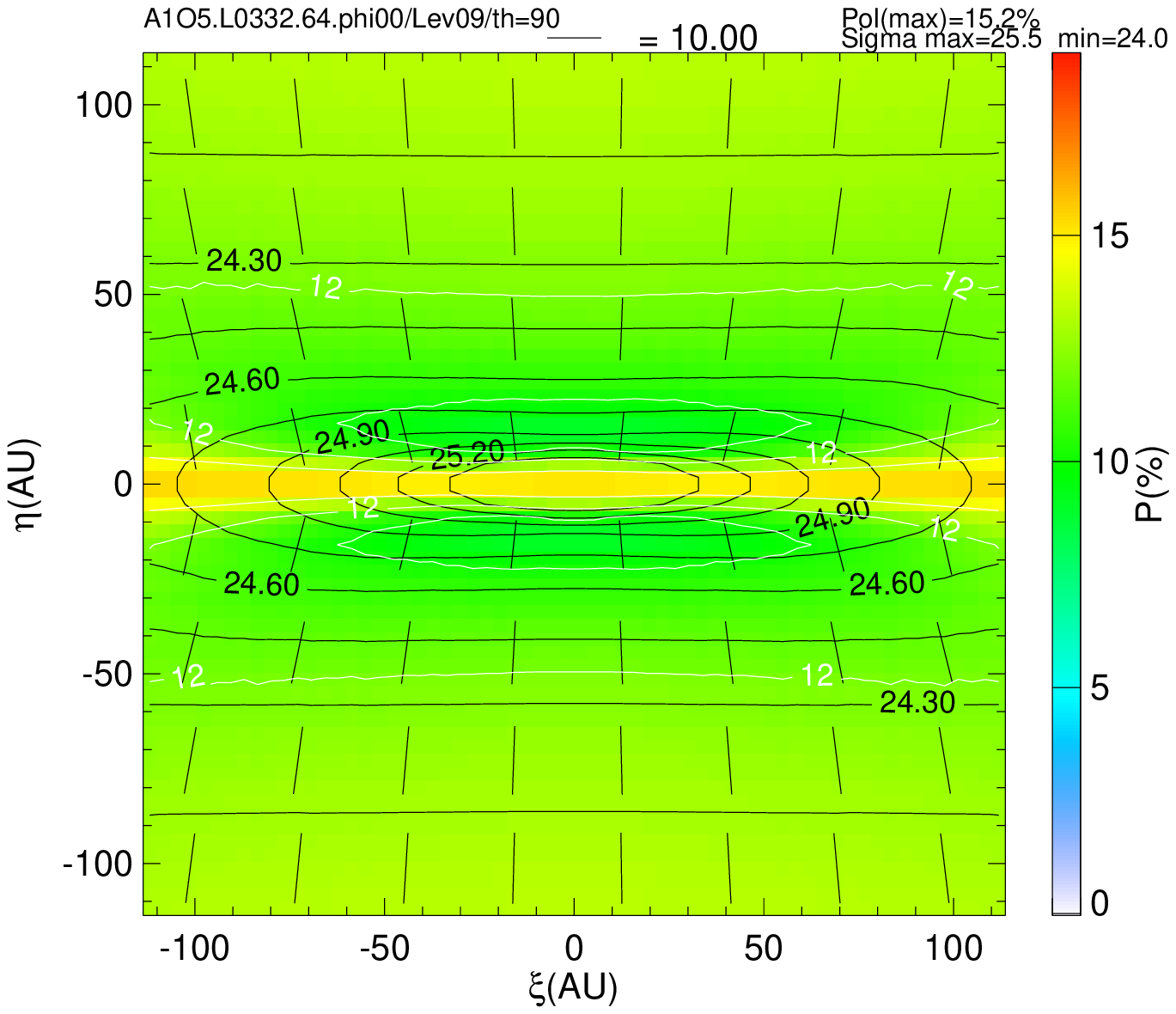}}
\end{center}
\caption{\label{fig:A1O5L332}Expected polarization for the runaway collapse
  phase. 
Column density (black contour lines),
 polarization vector (bar),
 and polarization degree (false color and white contour lines) are
 shown for Levels 3, 5, 7, and 9 of model AH1 with $\alpha=1$ and $\Omega'=5$.
In the uppermost panels, density (solid line contours) and magnetic field
 lines (dashed line) are plotted.}
\end{figure}

In Figure \ref{fig:A1O5L332},
 we plot the polarization at the final period of the isothermal runaway collapse
 phase just before the first core formation shown in Figure \ref{fig:density}(a).
The top row shows the distributions of the density and magnetic field
 seen in the nested grids for the $L=3$, $5$, $7$, and $9$ levels.
These show that a pseudodisk in the form of a contracting disk forms in the
 direction perpendicular to the magnetic field in this phase.
The other rows illustrate the polarization degree vectors,
 showing the directions of the B-vector and their polarization degree,
 as well as the column density (black contour lines) and polarization degree 
 (false color  and white contour lines).     
When a uniform magnetic field is deduced from the thermal dust emission,
 the radiation's B-vector gives the direction of the magnetic field.
The second to seventh rows represent the results for
 $\theta=0^\circ$ (along the $z$-axis or the pole-on view),
 $\theta=30^\circ$,  $\theta=45^\circ$,
 $\theta=60^\circ$, $\theta=80^\circ$, and $\theta=90^\circ$ (the edge-on view), respectively.

Figure \ref{fig:A1O5L332}
 shows that the polarization degree observed along the $z$-axis  
 is much lower than that observed for $\theta \gtrsim  45^\circ$.
The reason for this is clear: 
 observing along the $z$-axis, the magnetic field runs perpendicular
 to the celestial plane, which greatly reduces $q$ and $u$ due to
 the fact that $\gamma\simeq 90^\circ$.
In other words, dust grains do not align in a specific direction 
 on the celestial plane.

In the range $30^\circ \le \theta \le 60^\circ$,
 a pseudodisk which is seen in the middle of the top panels
 is observed as an elliptical distribution of the column density
 (black contour lines), which is an effect of the projection. 
Although the polarization degree has its minimum near the major axis of
 the column density distribution, the minimum direction does not coincide
 with the major axis with a difference of $20^\circ-40^\circ$.

In Figure \ref{fig:A1O5L332}, models with $\theta \ge 80^\circ$ (near edge-on)
 exhibit an hour-glass shape, in which the magnetic field is squeezed
 near the mid-plane.
However, it should be noted that
 the deviation from a straight magnetic field is small.
The bottom panels show that the polarization degree varies depending on the
 heights from the mid-plane.
Namely, a low polarization region extends over $500 {\rm AU} \lesssim z \lesssim 1500 {\rm AU}$ 
 ($L=3$ and $5$).
This corresponds to a post-shock region passing an accretion shock
 appearing at $z\sim 1500{\rm AU}$.
In the pre-shock region $z\gtrsim 1500{\rm AU}$, the magnetic field line is
 essentially perpendicular to the pseudodisk,
 while in the post-shock region $B_\phi$ and $B_r$ are amplified due to 
 compression. 
This configuration reduces the polarization degree in the post-shock region.
Another accretion shock appears near $z\sim 10$--$20{\rm AU}$,
 of which the post-shock region corresponds to a low polarization
 degree around $|\eta| \sim 10$--$20{\rm AU}$ in the $L=9$ grid.

\subsection{Accretion Phase}

In Figure \ref{fig:A1O5L1001},
 we show the polarization obtained after the first core formation.
As shown in the top row, the molecular outflow is ejected
 in this phase and the outflow reaches  
 $z\simeq 2000 {\rm AU}$ in $\tau \simeq 5000{\rm yr}(\rho_s/100{\rm H_2 cm^{-3}})^{-1/2}$
 after the core formation, which is seen in the $L=5$ grid.
In the $L=7$ grid, gas moves outward
 in the region 
 $200 {\rm AU} \lesssim r \lesssim 400 {\rm AU}$ at $z\sim 200{\rm AU}$,
 which is connected to a thick outflow lobe
 seen in the range $200 {\rm AU} \lesssim r \lesssim 800 {\rm AU}$
 at the height $z\sim 1000{\rm AU}$.

Although the pole-on view ($\theta=0^\circ$) exhibits low polarization
 in Figure \ref{fig:A1O5L332},
 the central part with radii $r \lesssim 500 {\rm AU}$
 has a larger polarization degree than the outer part,
 which is seen in the grids $L\ge 5$.
This is due to the fact that the toroidal magnetic field
 is amplified by rotation of the disk at the mid-plane
 in contrast to the former runaway collapse phase.

The molecular outflow is clearly seen as a region with
 low polarization degree ($\lesssim 5\%$) extending vertically
 in the plots of $L=5$ for $\theta = 60^\circ$--$90^\circ$
 (compare the panels in the second column from the left).
Outside the molecular outflow,
 the poloidal magnetic field is predominant, while inside the outflow
 the toroidal magnetic field is dominant.
Since the poloidal magnetic field induces larger polarization than the
 toroidal one if we observe the outflow edge-on,
 the outflow is seen as a region with low polarization degree.

Although the disk is seen in a high polarization degree region 
 in the edge-on view in the range $|\xi|\gtrsim 500~{\rm AU}$ ($L=3$ and $5$),
 the disk is composed of a region with low polarization degree
 looking further inside $|\xi|\lesssim 400~{\rm AU}$ ($L=7$).
Disk rotation generates the toroidal magnetic field in this region,
 while outside $|\xi|\sim 500~{\rm AU}$ the poloidal magnetic field is
 dominant. 

Between the pole-on and edge-on views ($30^\circ\le\theta\le 60^\circ$),
 a region with low polarization degree extends in the horizontal direction
 from the center,
 similar to the runaway collapse phase.
The panels of $L=5$ for $30^\circ\le\theta\le 60^\circ$ contain
 a compact $\sim 200{\rm AU}$-scale region with high polarization degree.
This is characteristic of the accretion phase,
 in contrast with the runaway collapse phase (Fig.\ref{fig:A1O5L332}).
This corresponds to the acceleration region where
 the toroidal magnetic field works to accelerate the gas.
This region forms a spiral feature with a radius of $\sim 500$ AU,
 as seen for $L=7$.
 
\subsection{Model with Weak Magnetic Field}

In Figure \ref{fig:A001O1L1322},
 we plot the density and magnetic field lines in the accretion phase
 of a model with a weak magnetic field 
 (model EH of Paper I; $\alpha=0.01$ and $\Omega'=1$),
 which has a ten-times weaker
 magnetic field than the previous model.
That is, the magnetic field strength at the center is equal to
  $B_0=1.33(\rho_c/10^4{\rm H_2\,cm^{-3}})^{1/2}(\alpha/0.01)^{1/2}{\rm \mu G}$
 and the rotation rate at the center is equal to 
  $\Omega_0=0.556{\rm km\,s^{-1}\,pc^{-1}}(\Omega'/1)$,
 if the other parameters are taken to be the same.
The rotation rate of this model is 1/5 that of the previous model.
The outflow reaches $z\simeq 1200 {\rm AU}$ in a time scale
 $\tau\sim 2\times 10^4\,{\rm yr}\,(\rho_s/100{\rm H_2\, cm^{-3}})^{-1/2}$.
Owing to the weaker magnetic field (and relatively slow rotation),
 the density distribution looks nearly spherical
 (see the panel of $L=3$ in the top row).
MHD simulations have already shown that 
 the relatively weak magnetic field gives an outflow
 driven by the magnetic pressure gradient in the toroidal component
 and this outflow is well collimated.
The strong magnetic field induces an outflow driven by
 the magnetocentrifugal wind, which has a wide opening angle
 \citep{kudoh98,tomisaka02}.
The model in this subsection corresponds to the model of outflows driven by the magnetic
 pressure gradient.
   
Similar to Figure \ref{fig:A1O5L1001}, in Figure \ref{fig:A001O1L1322}
 the outflow is seen as a low-polarization region with $\sim 3\%$
 ($L=5$ and $45^\circ \le\theta \le 90^\circ$).
However, the outflow has an interior structure,
 which is also seen in the third column from the left, i.e., the $L=7$ grid.
Near the rotation axis ($|\xi|\lesssim 200\, {\rm AU}$),
 a vertical structure with a relatively high polarization degree (10\%) 
 is seen.
Since the polarization B-vector in this region is in the horizontal direction,
 it appears to arise from the toroidal magnetic field compressed
 near the $z$-axis.
The toroidal field runs perpendicular
 to the celestial plane in the region
 $200~{\rm AU} \lesssim |\xi| \lesssim 600~{\rm AU}$,
 which reduces the polarization degree in this region.       
 
In the edge-on view,
 a disk is traced as a region with high polarization $\sim 15\%$ 
 ($\theta=90^\circ$ in the $L=3$ and $5$ levels).
Similar to the previous model,
 a high polarization disk is truncated inside $|\xi|\lesssim 800{\rm AU}$.
In this region, rotation and thus the toroidal magnetic field seem to
 predominate and reduce the polarization degree.

Between $\theta=30^\circ$ and $\theta=80^\circ$,
 a region with low polarization degree extends in the direction of
 the major axis
 of the column density distribution in the scale $L=3$ and $5$. 
Similar to the previous model, the direction of the low polarization degree
 and that of the major axis differ up to $\lesssim 30^\circ$.

\section{Discussion}

\subsection{Origin of the Asymmetry}

Although the distributions of both density and magnetic field are axisymmetric,
 the polarization distribution is not symmetric with respect to the $\eta$-axis.
This seems strange at first glance.
In this subsection, we consider this problem.
 
\begin{figure}[h]
   \begin{center}
     \hspace*{45mm}(poloidal)\hspace*{30mm}(toroidal)\\
$\theta=0^\circ$\hspace*{12mm}\raisebox{-20mm}{\one{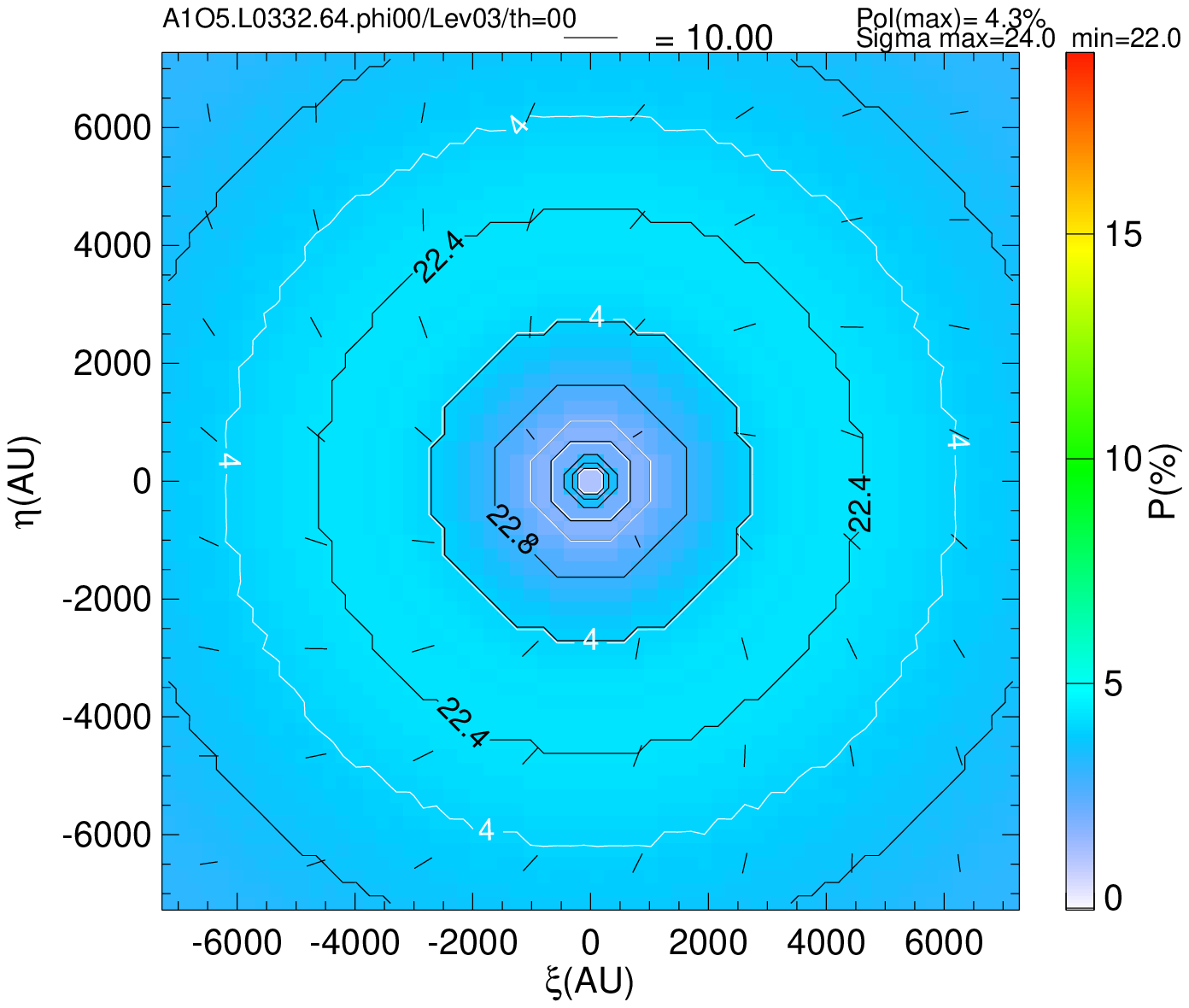}
      \hspace*{5mm}\one{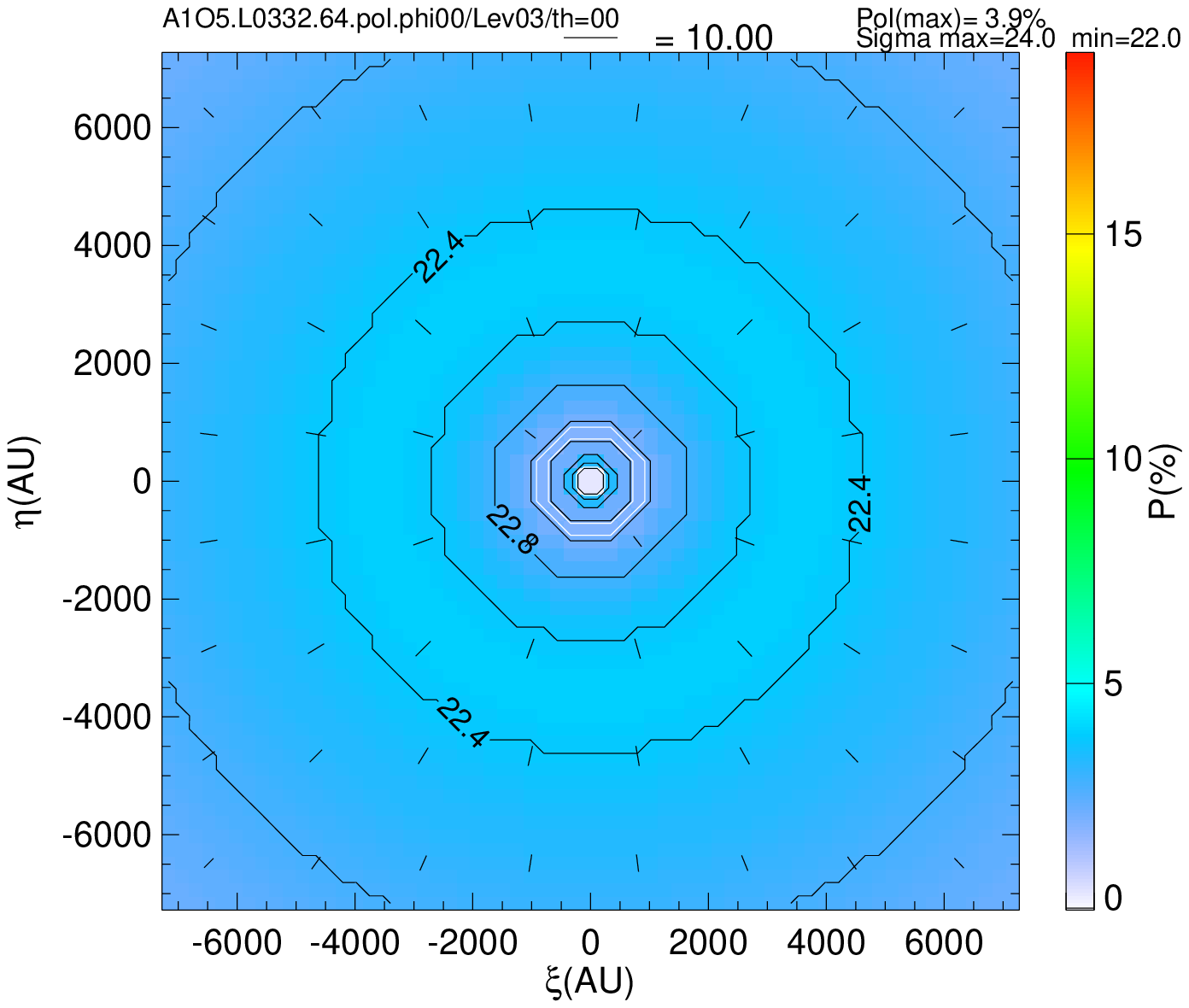} 
      \hspace*{5mm}\one{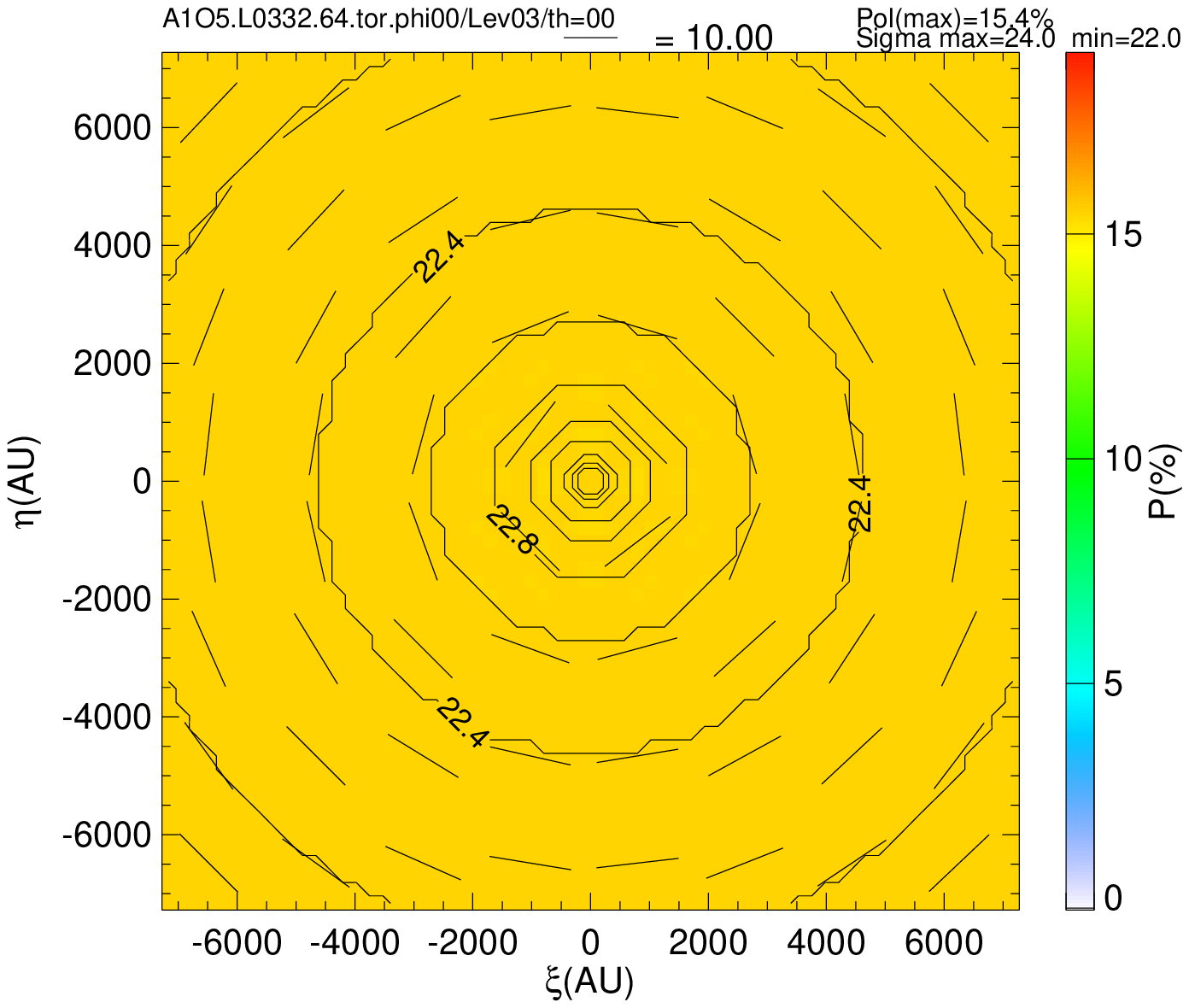}}\\[-0mm]
$\theta=45^\circ$\hspace*{10mm}\raisebox{-20mm}{\one{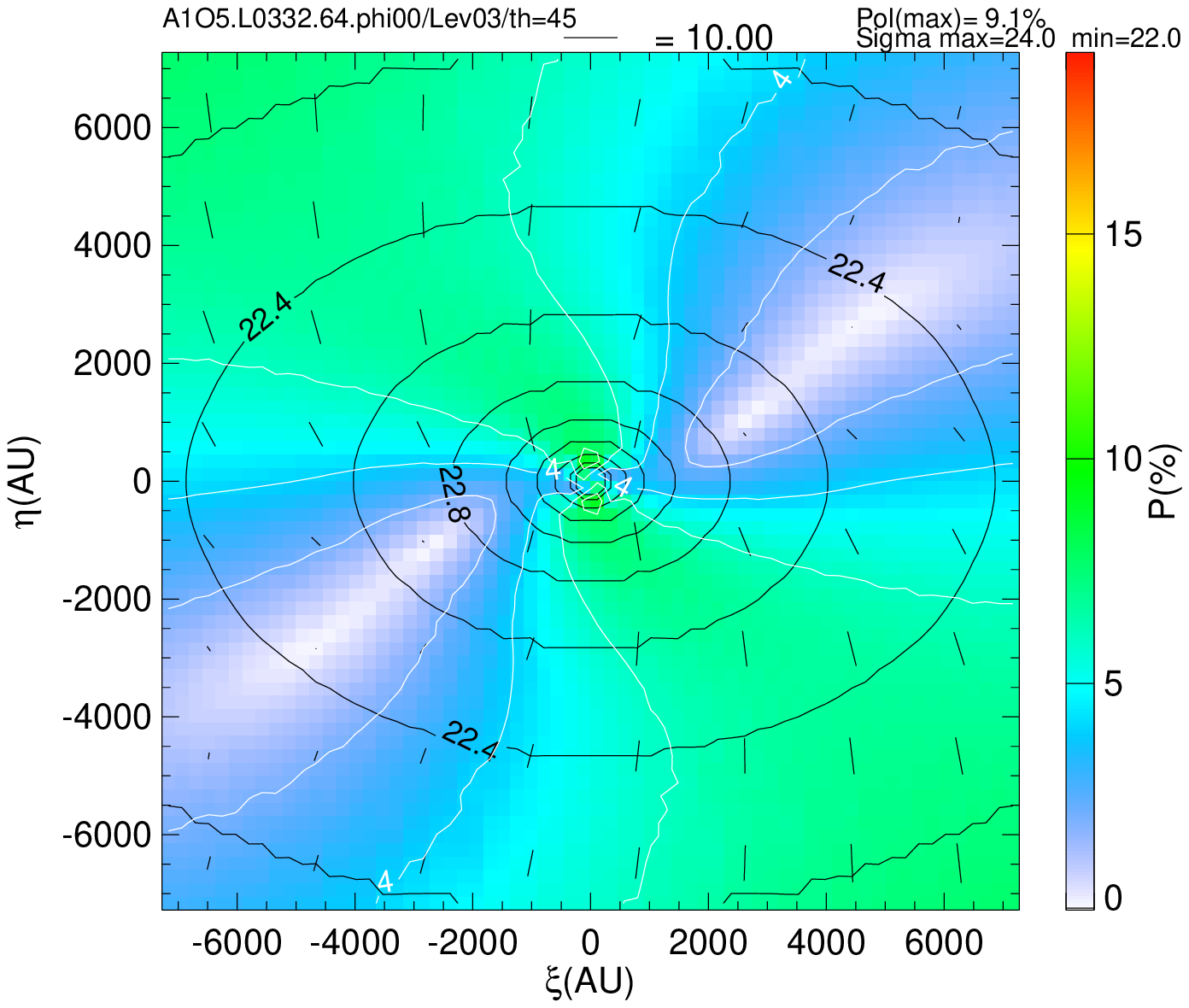}
      \hspace*{5mm}\one{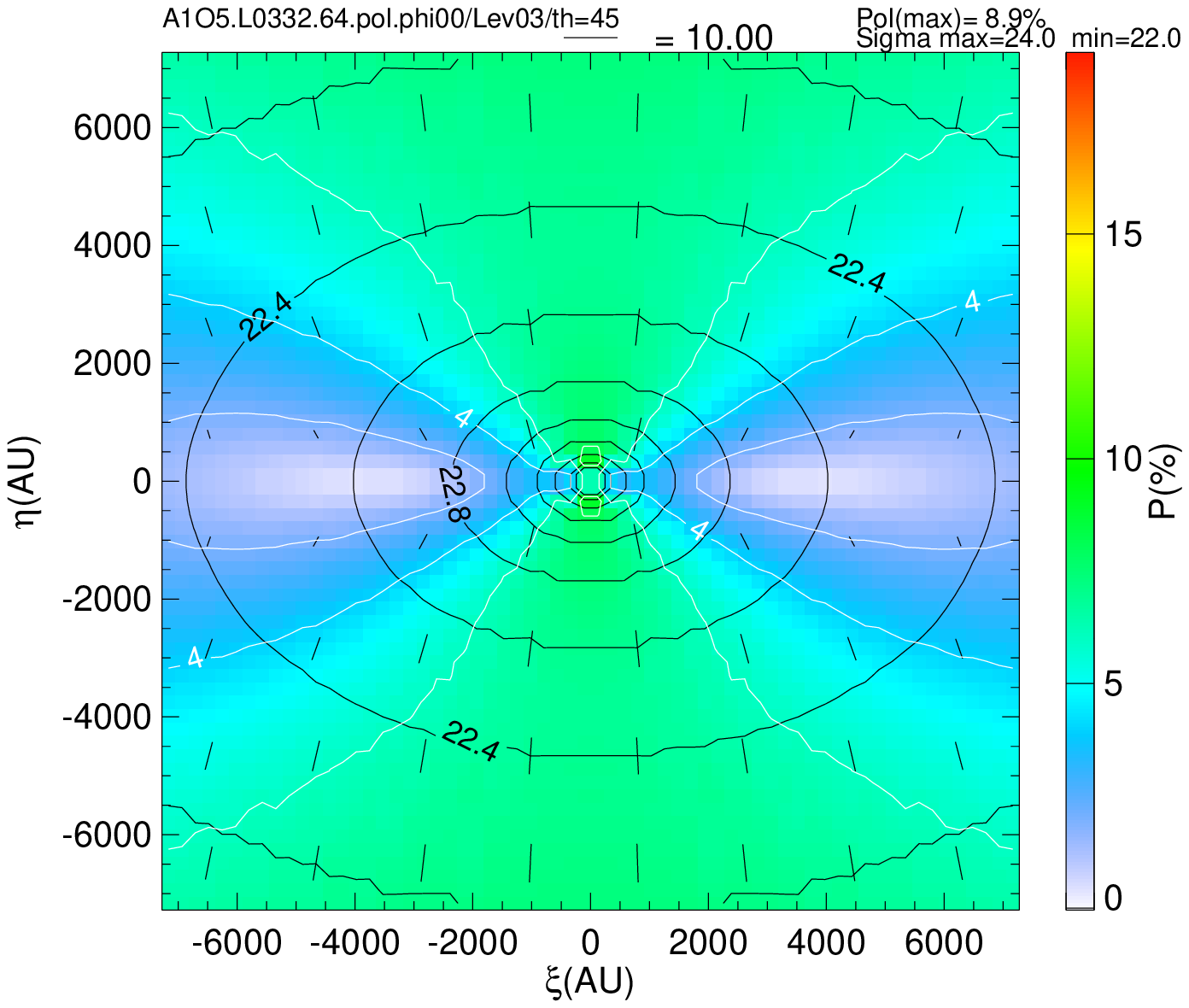}
      \hspace*{5mm}\one{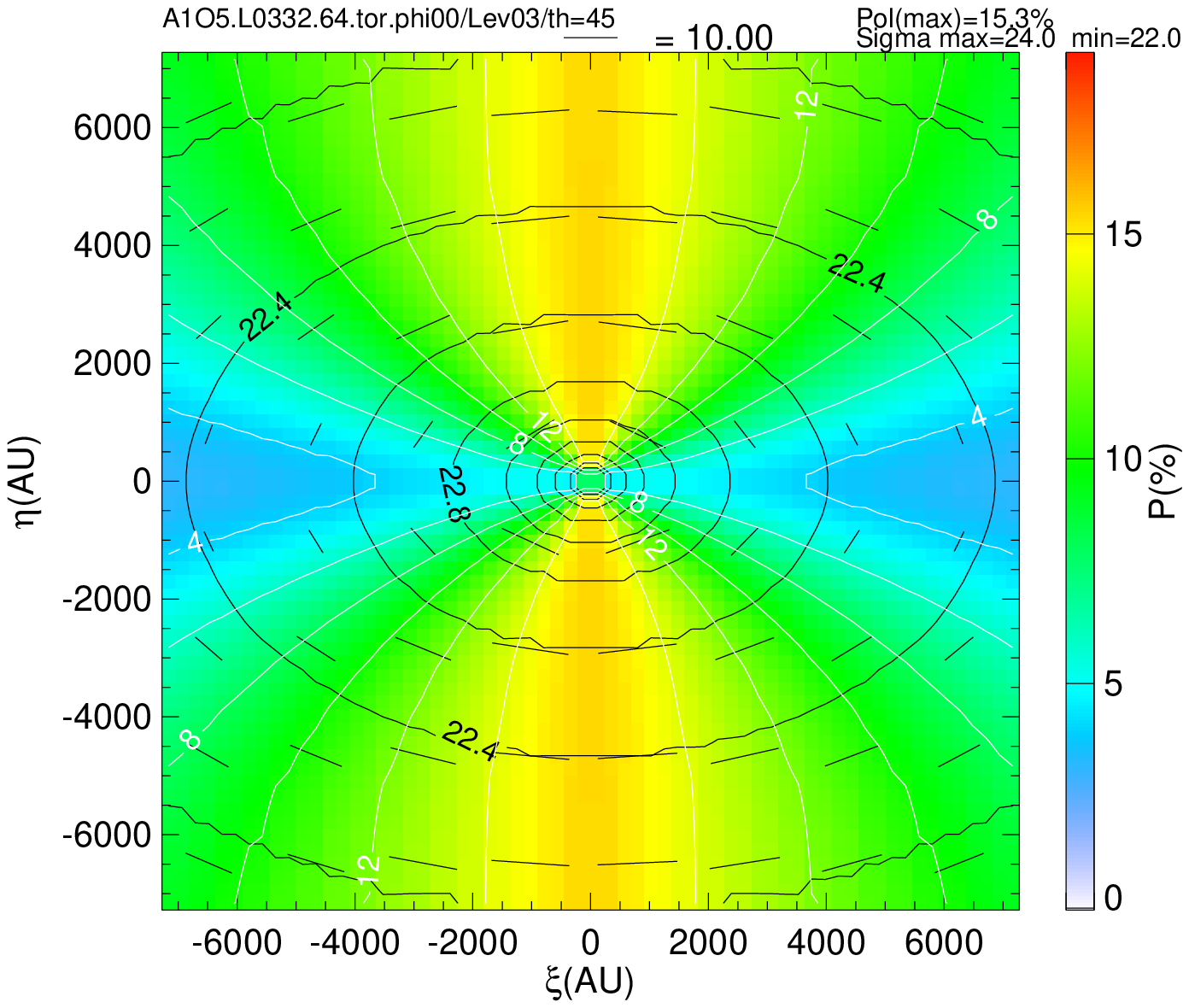}}\\[-0mm]
$\theta=90^\circ$\hspace*{10mm}\raisebox{-20mm}{\one{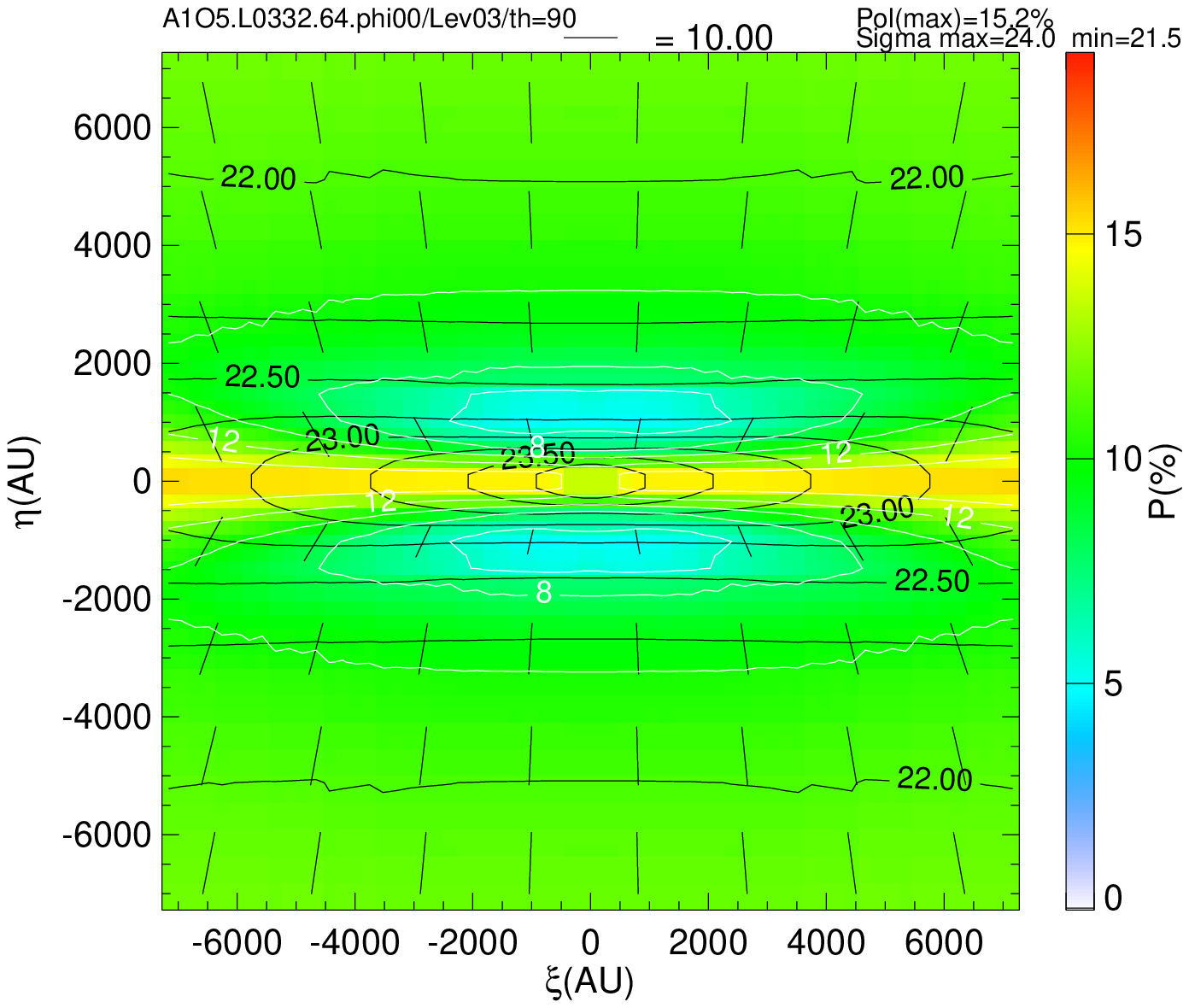}
      \hspace*{5mm}\one{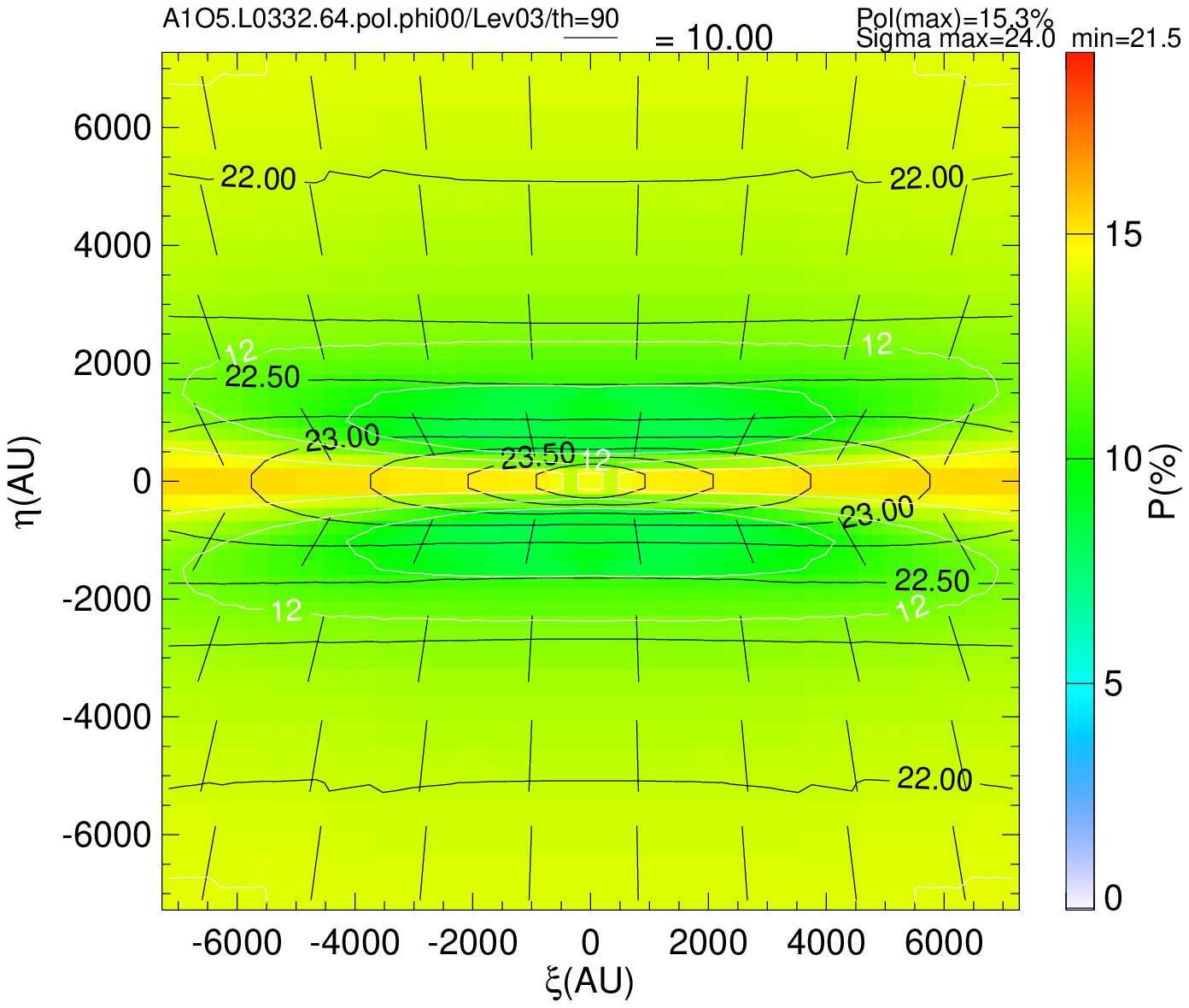}
      \hspace*{5mm}\one{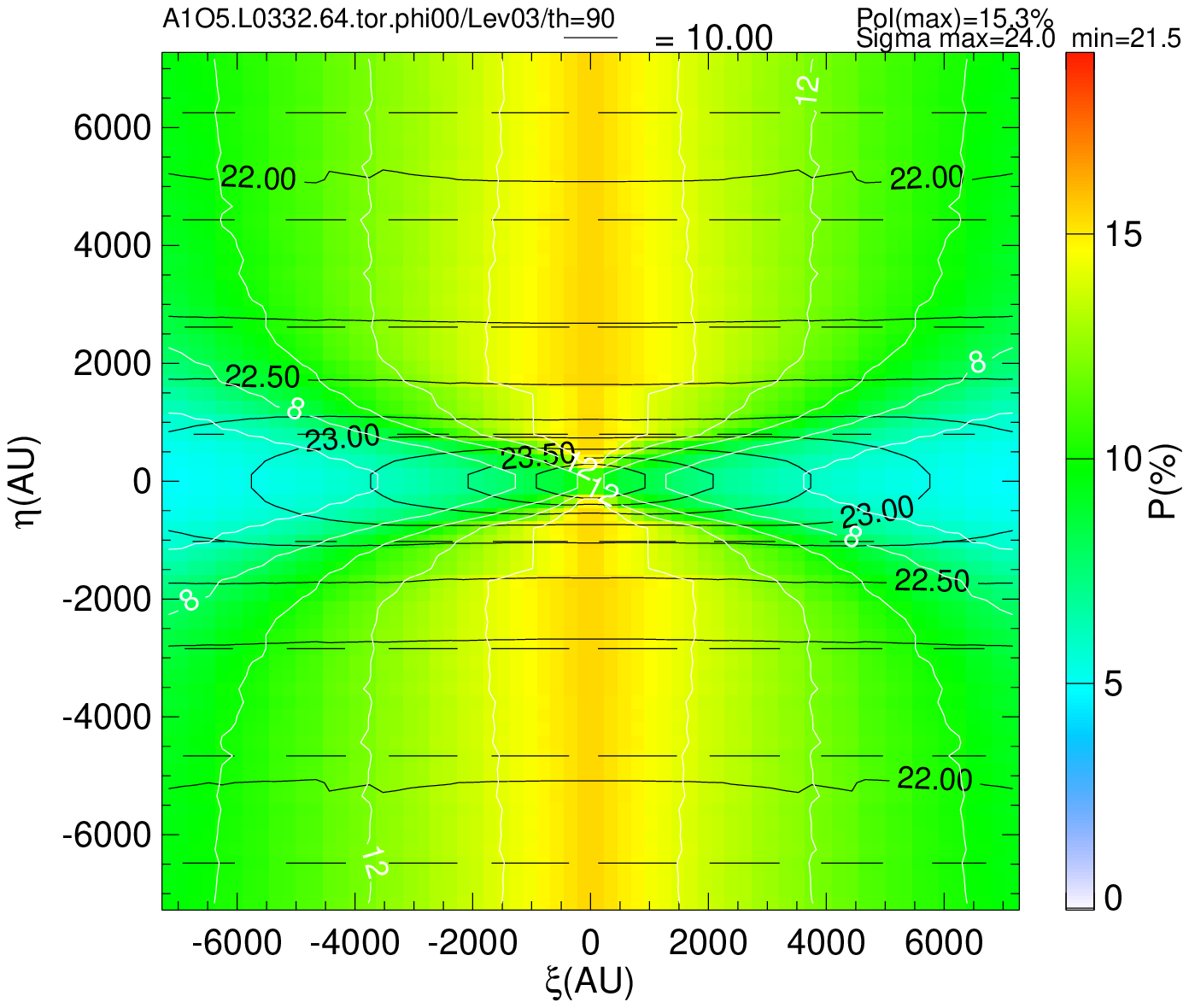}}
   \end{center}
\caption{\label{fig:0332A1O5L3poltor}As for Fig.~\ref{fig:A1O5L332} 
 (runaway collapse phase).
 However, the middle and right columns are results
 for artificial data consisting only of poloidal and toroidal magnetic fields,
 respectively.
 Level 3 of model AH1 with $\alpha=1$ and $\Omega'=5$.}
\end{figure}

In Figures \ref{fig:0332A1O5L3poltor} and \ref{fig:1001A1O5L5poltor},
 we plot the expected polarization map for the model shown in Figures
 \ref{fig:A1O5L332} and \ref{fig:A1O5L1001} (model AH1; $\alpha=1$ and $\Omega'=5$).
The left column is a plot of the results,
 while the middle and right columns are results
 obtained from artificial data consisting of the poloidal magnetic field
 without the toroidal field (middle: hereafter we call it a poloidal model)
 and from data consisting of the toroidal magnetic field without the
 poloidal field 
 (right: hereafter we call it a toroidal model).

In the case of neither pole-on nor edge-on (i.e., $0^\circ < \theta < 90^\circ$),
 the polarization degree distribution shown in the left column exhibits
 point symmetry rather than mirror symmetry. 
If we exclude the toroidal magnetic field (poloidal model) or
 exclude the poloidal magnetic field (toroidal model),  
 we observe mirror symmetry with respect to the $\eta$-axis.
Hence, only the true data, consisting of both the poloidal and toroidal magnetic fields,
 gives a polarization distribution without mirror symmetry.
Namely, co-existing poloidal and toroidal magnetic fields induce 
 this asymmetry.

The toroidal model (Figs. \ref{fig:0332A1O5L3poltor}
 and \ref{fig:1001A1O5L5poltor}) exhibits a vertical bar structure
 with a strong polarization degree 
 (the cases $\theta=45^\circ$ and $\theta=90^\circ$ of the toroidal model).
Integrating the toroidal magnetic field, the polarization in the horizontal 
 direction has a peak on the $\eta$-axis, since the magnetic field 
 is parallel to the celestial plane and the column density has
 a maximum there.     
The distributions in the left and middle columns show
 some similarity, in contrast to the right column.  
For example, for $\theta=90^\circ$,
 a pseudodisk is observed with a similar structure, exhibiting a polarization degree
 decrease followed by an increase, rising from the mid-plane.   
This indicates that the magnetic field can be regarded essentially as being poloidal
 in this run-away collapse phase.
Observing from $\theta=45^\circ$
 in the direction of a major axis of the disk,  
 the polarization degree is low.

The polarization distributions in the accretion phase are shown in  
 Figures \ref{fig:1001A1O5L5poltor} and \ref{fig:1322A001O1L6poltor},
 which correspond to models AH1 ($\alpha=1$ and $\Omega'=5$)
 and EH ($\alpha=0.01$ and $\Omega'=1$), respectively.
In Figure \ref{fig:1001A1O5L5poltor},
 an outflow lobe is seen in the left column $\theta > 0^\circ$
 and also in the middle column as a low-polarization region.
The fact that the lobe has a low polarization even in the poloidal
 model indicates that the low polarization is due to cancellation
 between the foreground and background.
That is, the toroidal field component $B_\phi>0$ is mapped to a negative $B_\xi<0$
 in the foreground but is mapped to a positive $B_\xi>0$ in the background.

\begin{figure}[h]
   \begin{center}
\includegraphics[width=100mm]{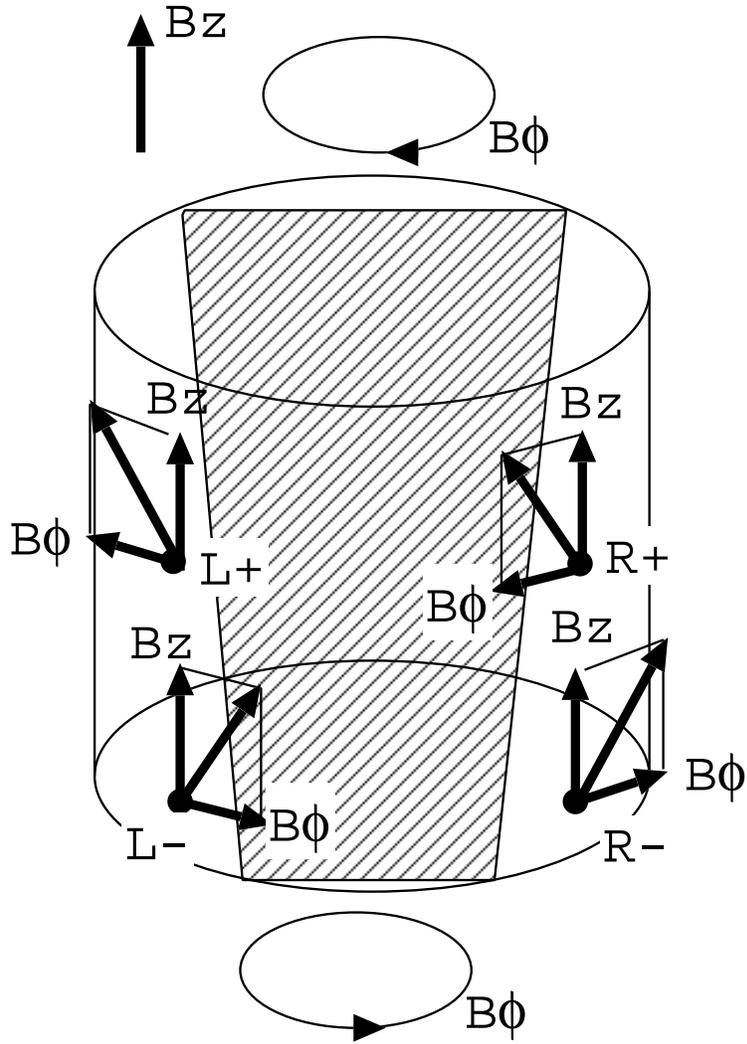}
   \end{center}
\caption{\label{fig:pol}Illustration of the reason why mirror symmetry with respect to 
 the $\eta$-axis is broken.
For two points R$+$ and R$-$
 at  symmetric positions with respect to the midplane,
if we assume that the magnetic field is composed of uniform $B_z$ and
 antisymmetric $B_\phi$ with respect to the midplane and
 the $\boldmath{B}$ vectors at R$+$ and R$-$ are projected on the celestial
 plane, 
 the projected two vectors $\boldmath{B}'$ have different amplitudes
 and are out of mirror symmetry.
Considering the symmetric points with respect to the $\eta$-axis, L$+$ and
 L$-$, the projected $\boldmath{B}'$ vectors at L$+$ and R$-$ are 
 in point symmetry.   
 }
\end{figure}

Figure \ref{fig:pol} shows how the mirror symmetry breaks in a 
 magnetic field consisting of poloidal and toroidal components. 
In this figure, we consider a magnetic field consisting of $B_z$ and $B_\phi$
 and assume $B_z=$ const and $B_\phi$ changes its sign for $z>0$ and $z<0$,
 which is obtained when the outflow is ejected by the magnetic Lorentz force.
Two points (R$+$ and R$-$) are taken at symmetric positions
 with respect to the mid-plane (and also L$+$ and L$-$).
R and L are chosen to be symmetric with respect to the $\eta$-axis.
Vectors of this magnetic field are projected onto a celestial plane
 for $\theta\simeq 45^\circ$.    
Looking from $0^\circ < \theta < 90^\circ$,
 the projected vector $\boldsymbol{B}'$ observed at R$+$ and R$-$ has
 different absolute values and points in different directions after
 the mirror reversal with respect to the $\xi$- or $\eta$-axes.
The situation is the same for the vectors at L$+$ and L$-$.
This explains how the mirror symmetry is broken. 
    
\subsection{Comparison with Observation}
\begin{figure}[h]
\begin{center}
\vspace*{-12mm}($L=3$)\hspace*{22mm}($L=5$)\hspace*{22mm}($L=7$)\hspace*{22mm}($L=9$)\\
\hspace*{24mm}
      \raisebox{-20mm}
{\one{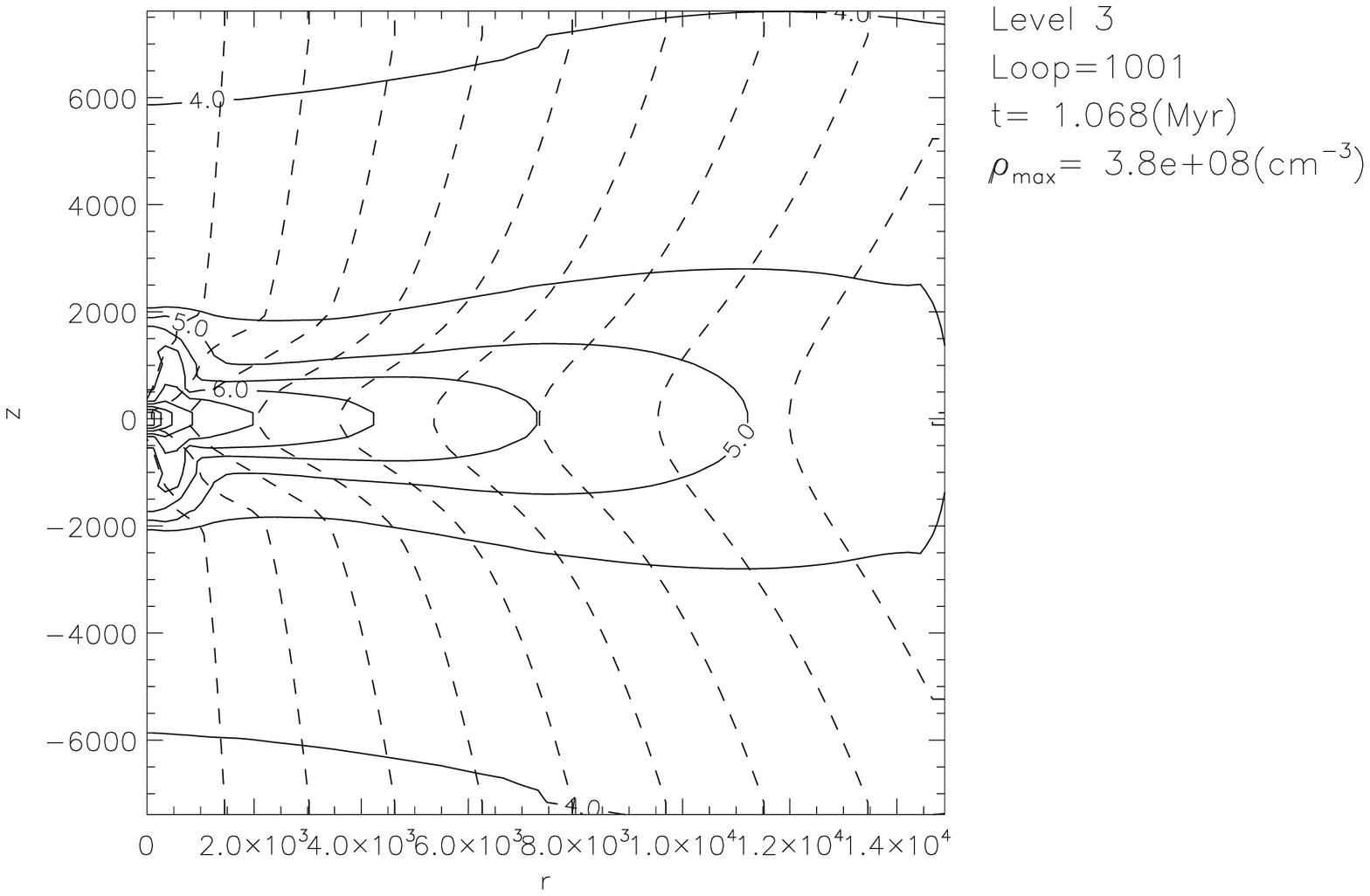}
      \one{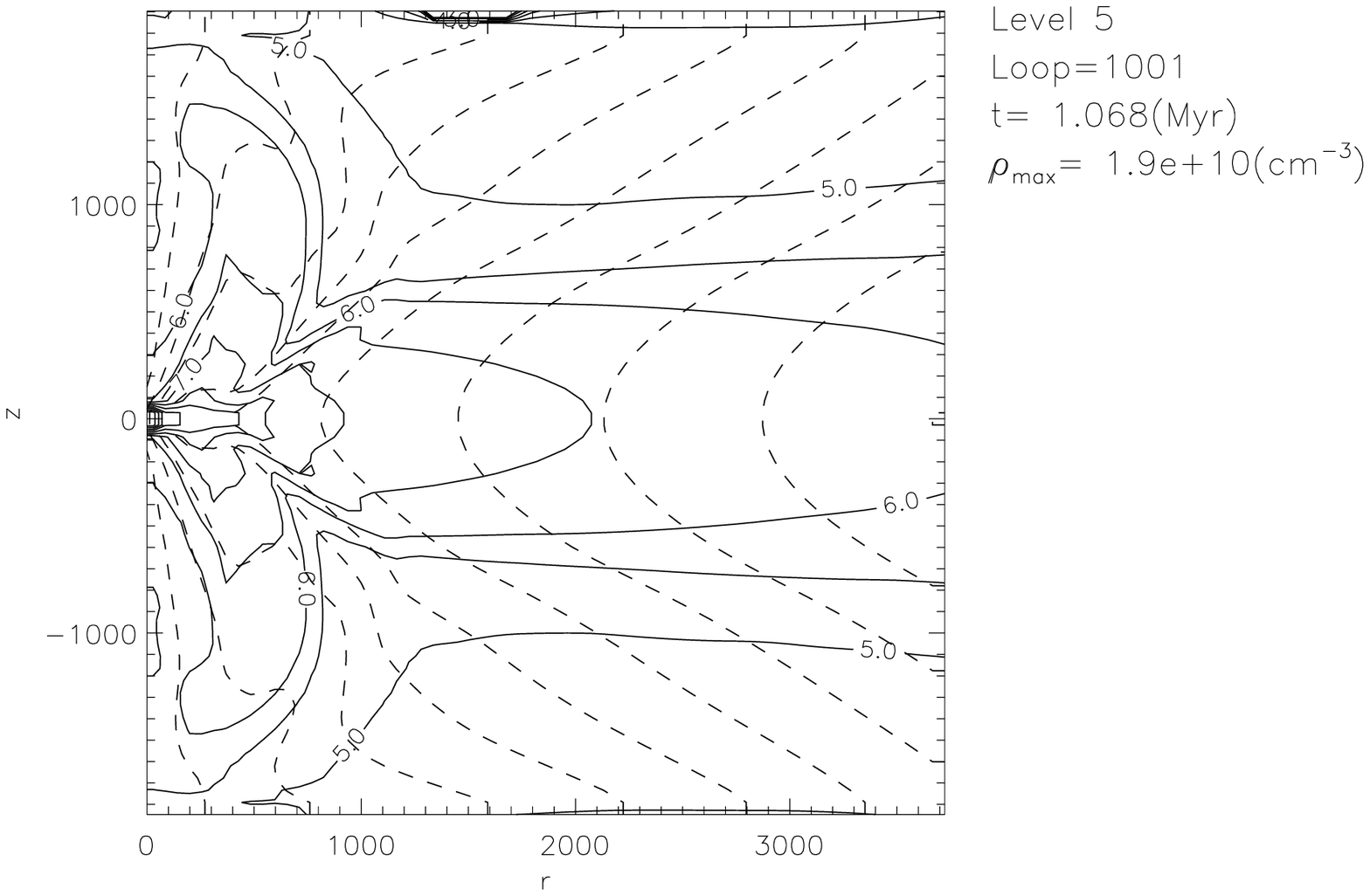}
      \one{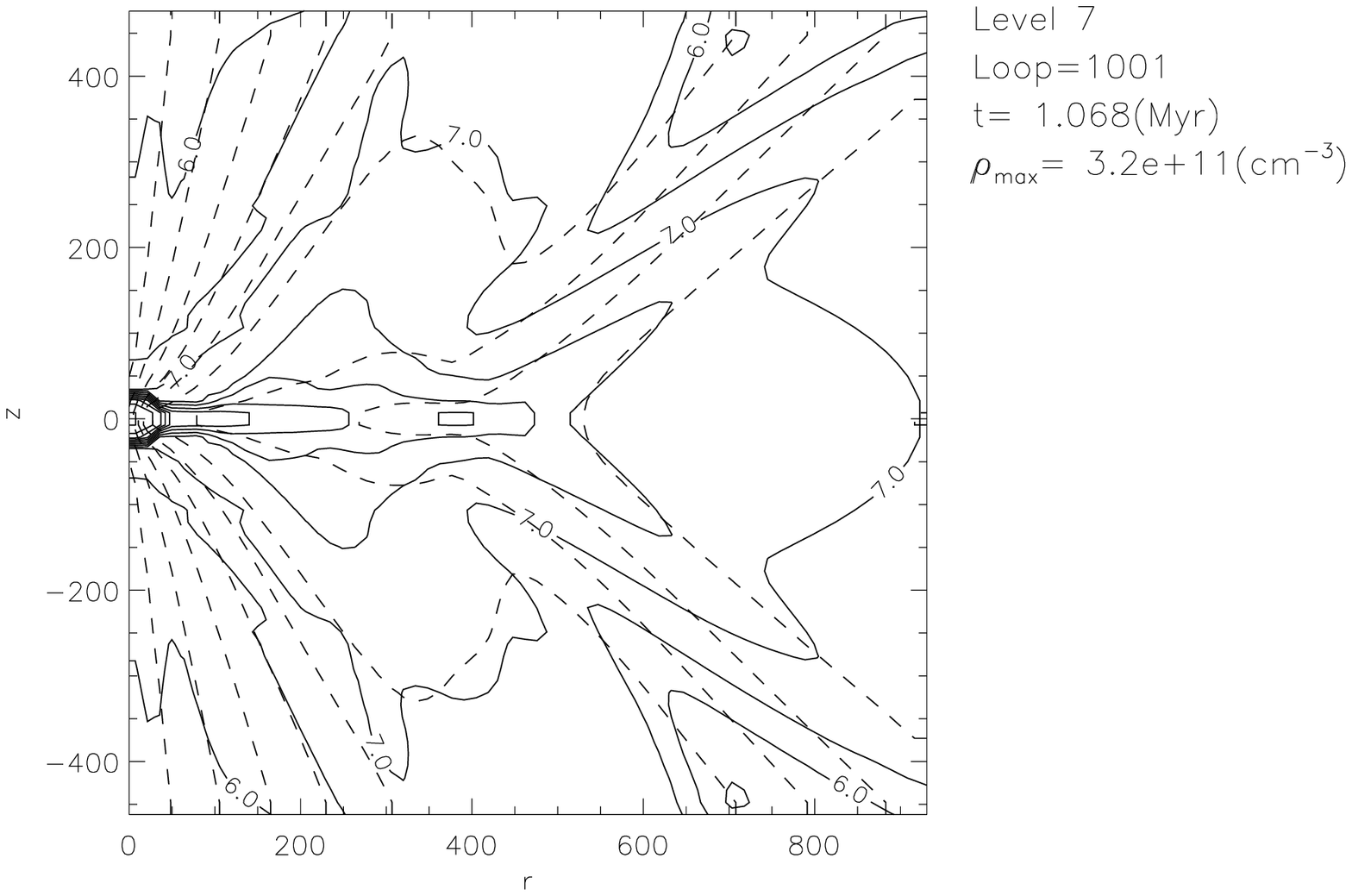}
      \one{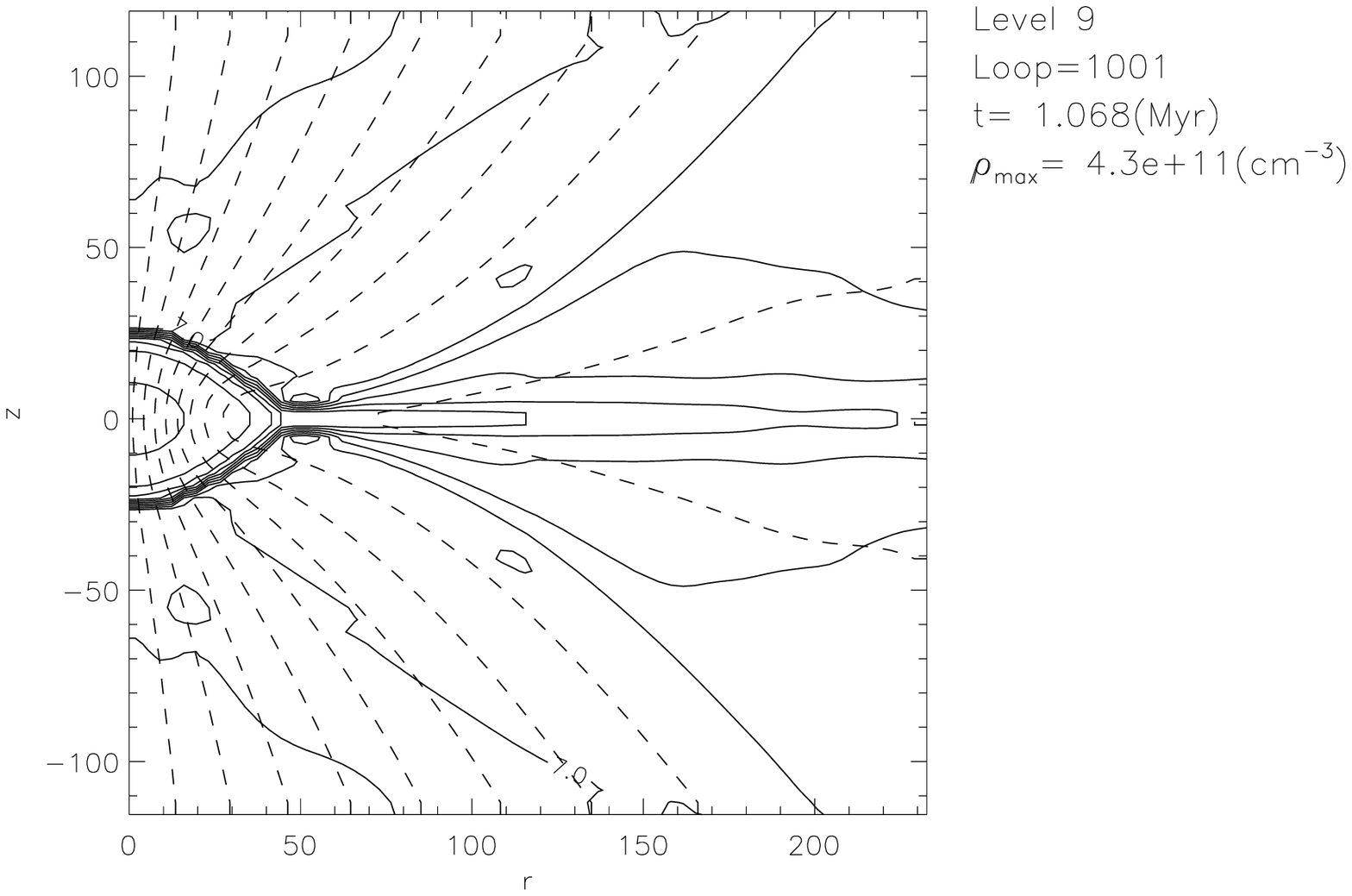}}\\

$\theta=0^\circ$\hspace*{12mm}
      \raisebox{-20mm}{\one{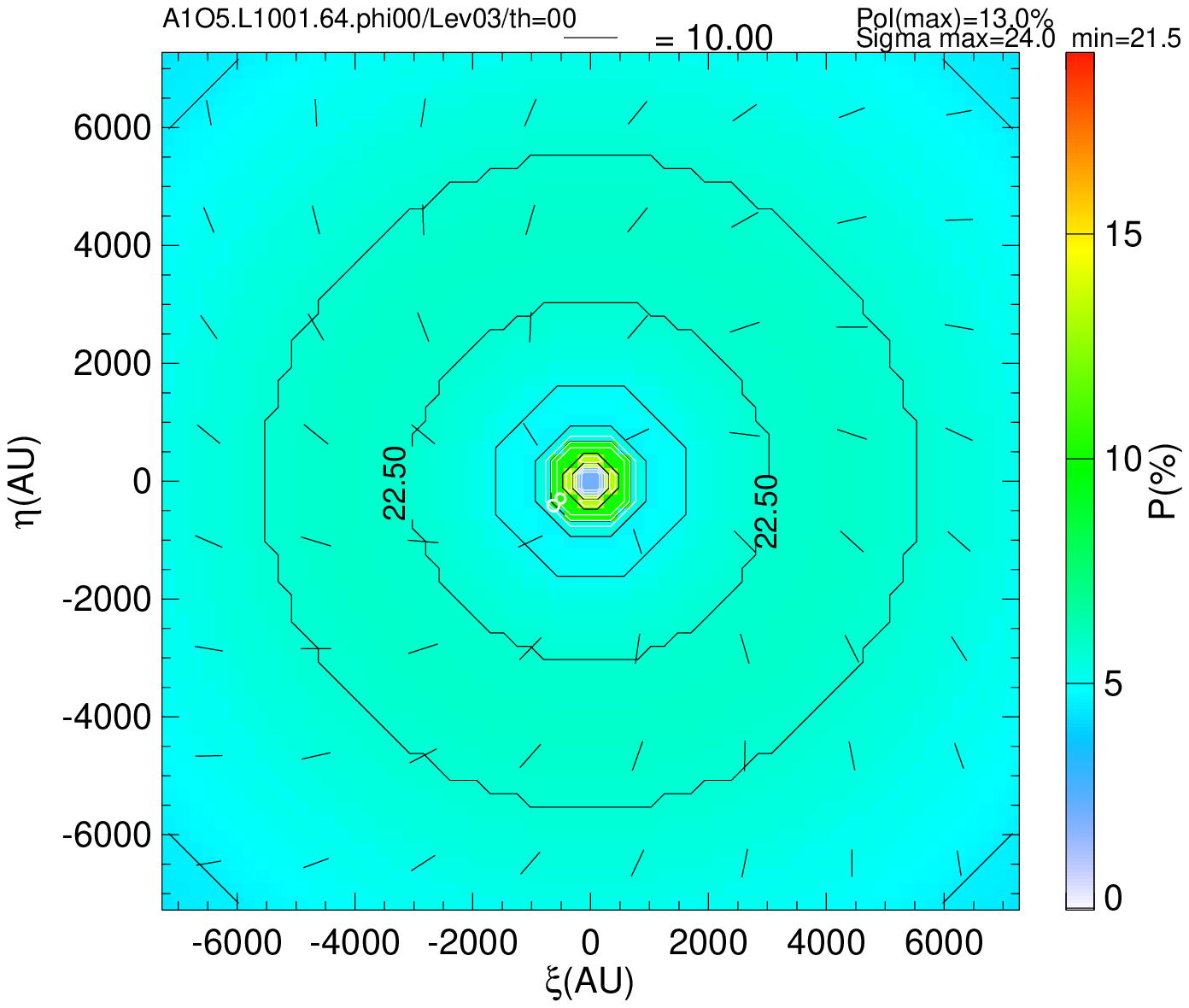}
      \one{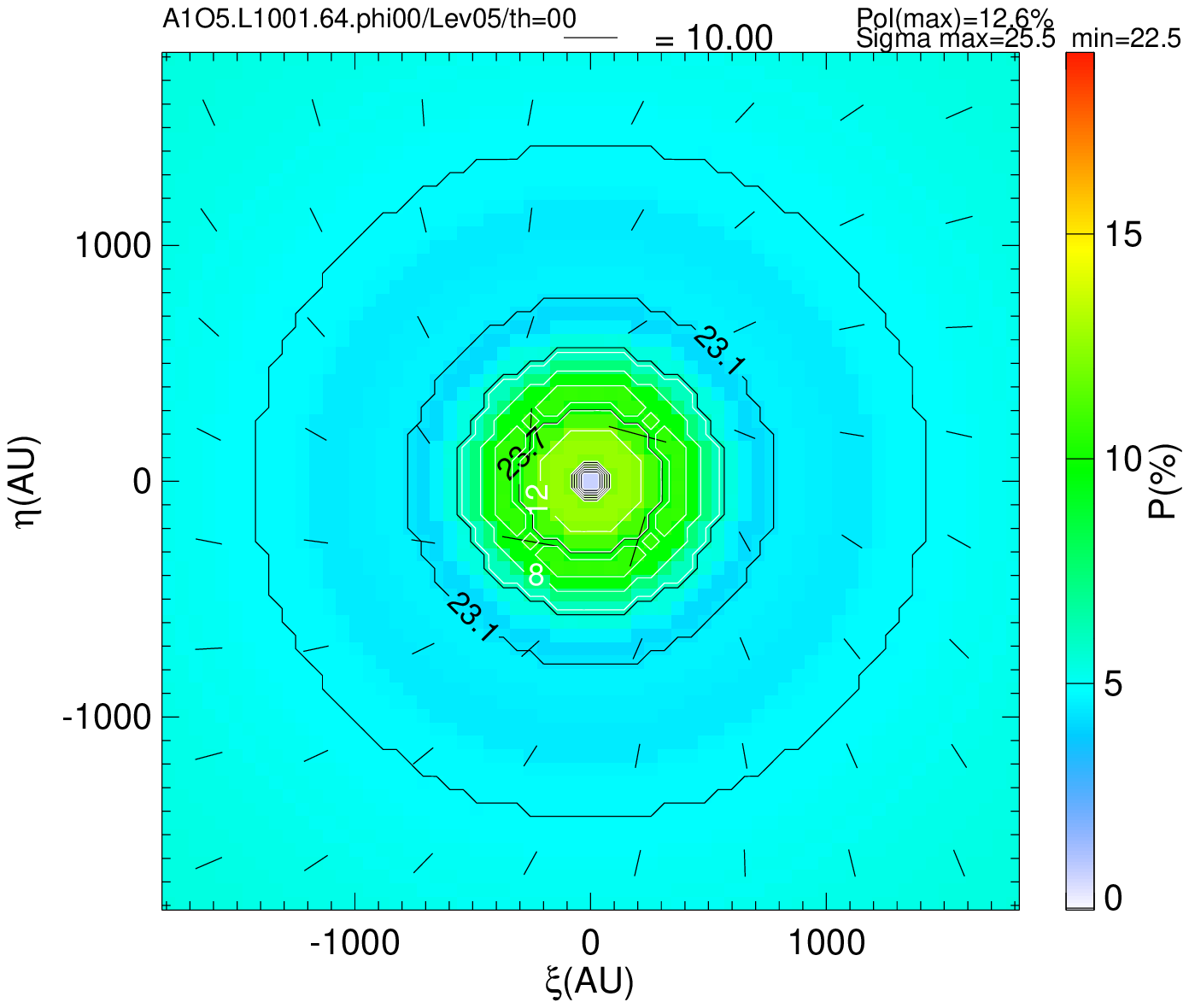}
      \one{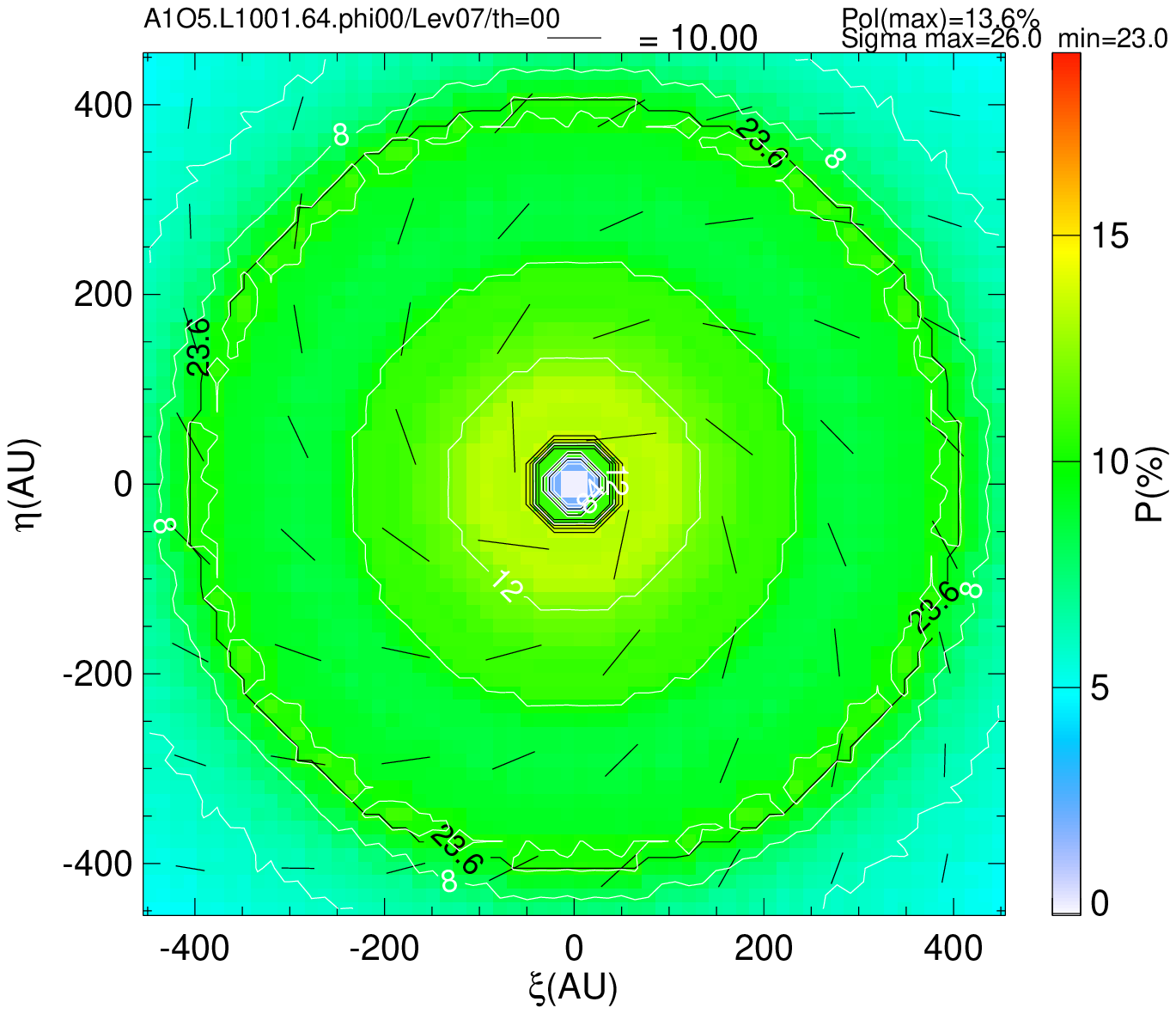}
      \one{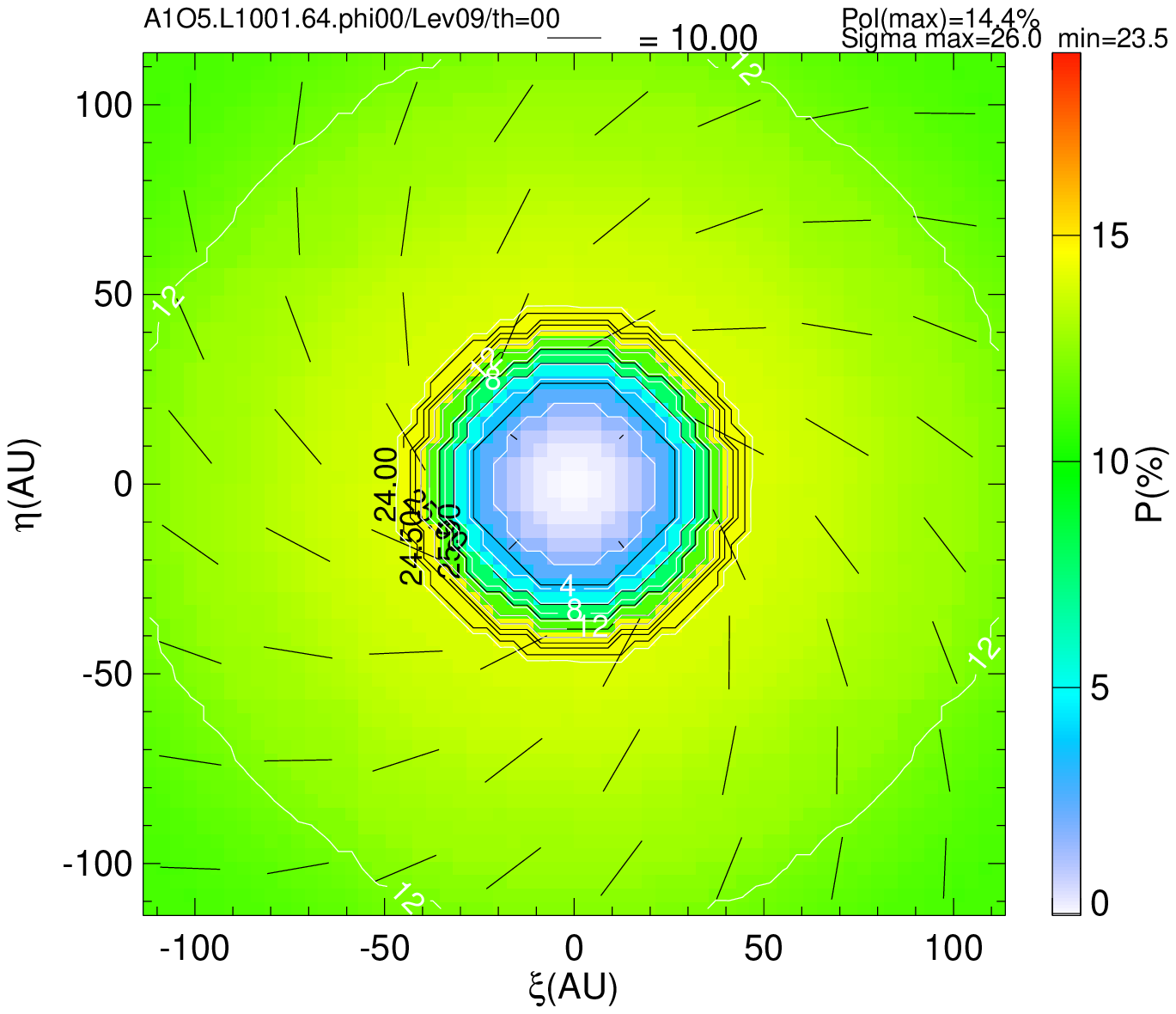}}\\
$\theta=30^\circ$\hspace*{10mm}
      \raisebox{-20mm}{\one{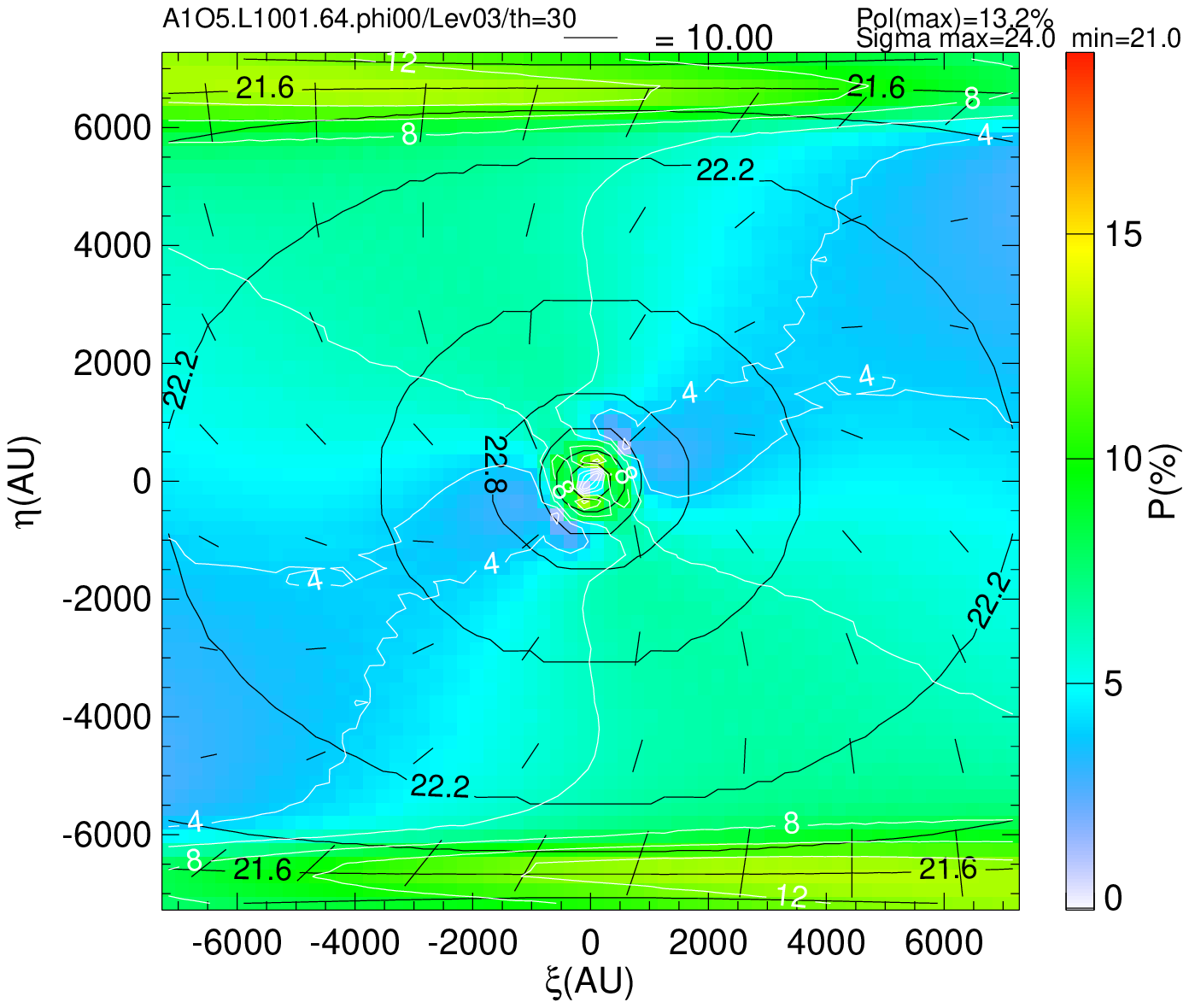}
      \one{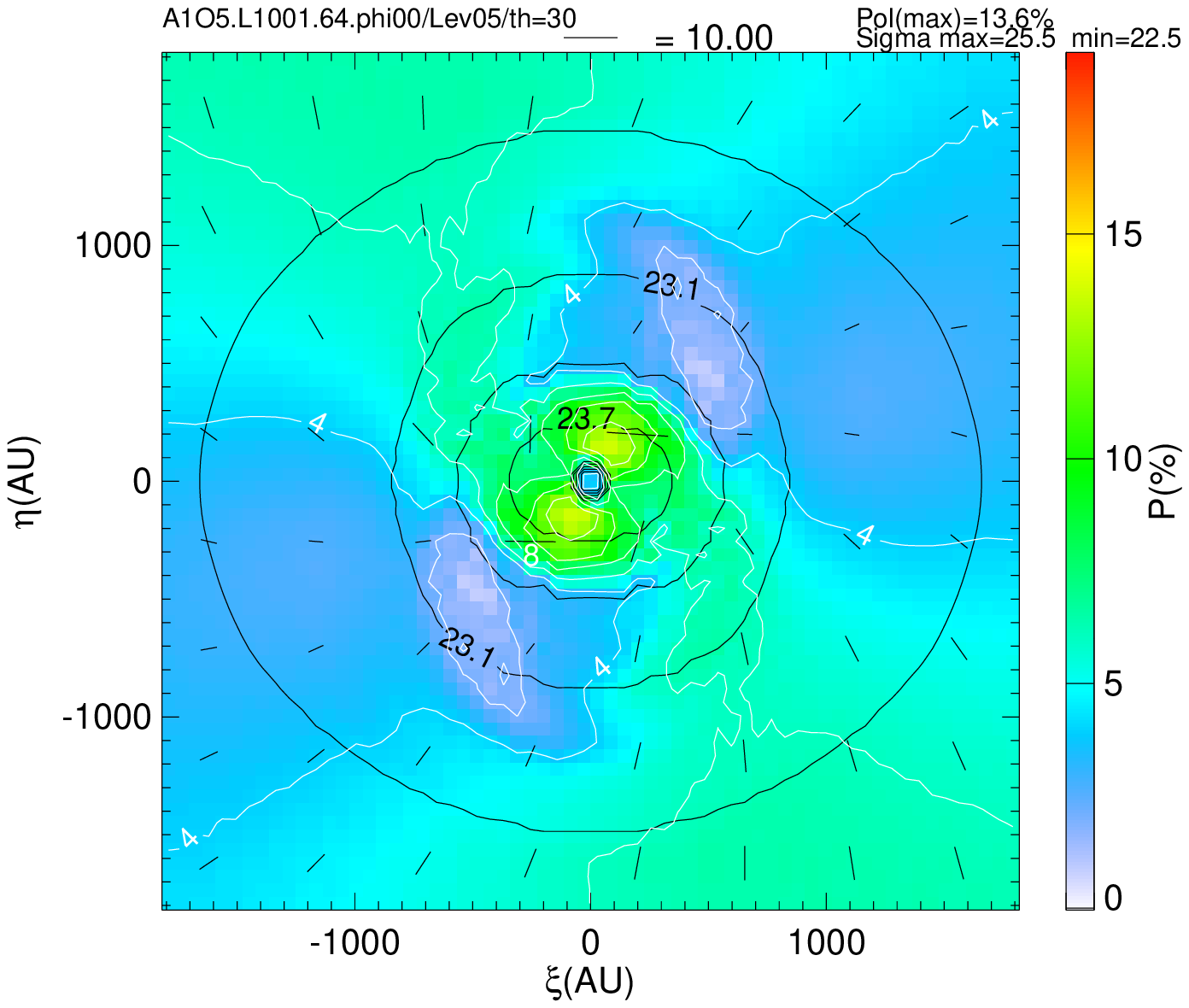}
      \one{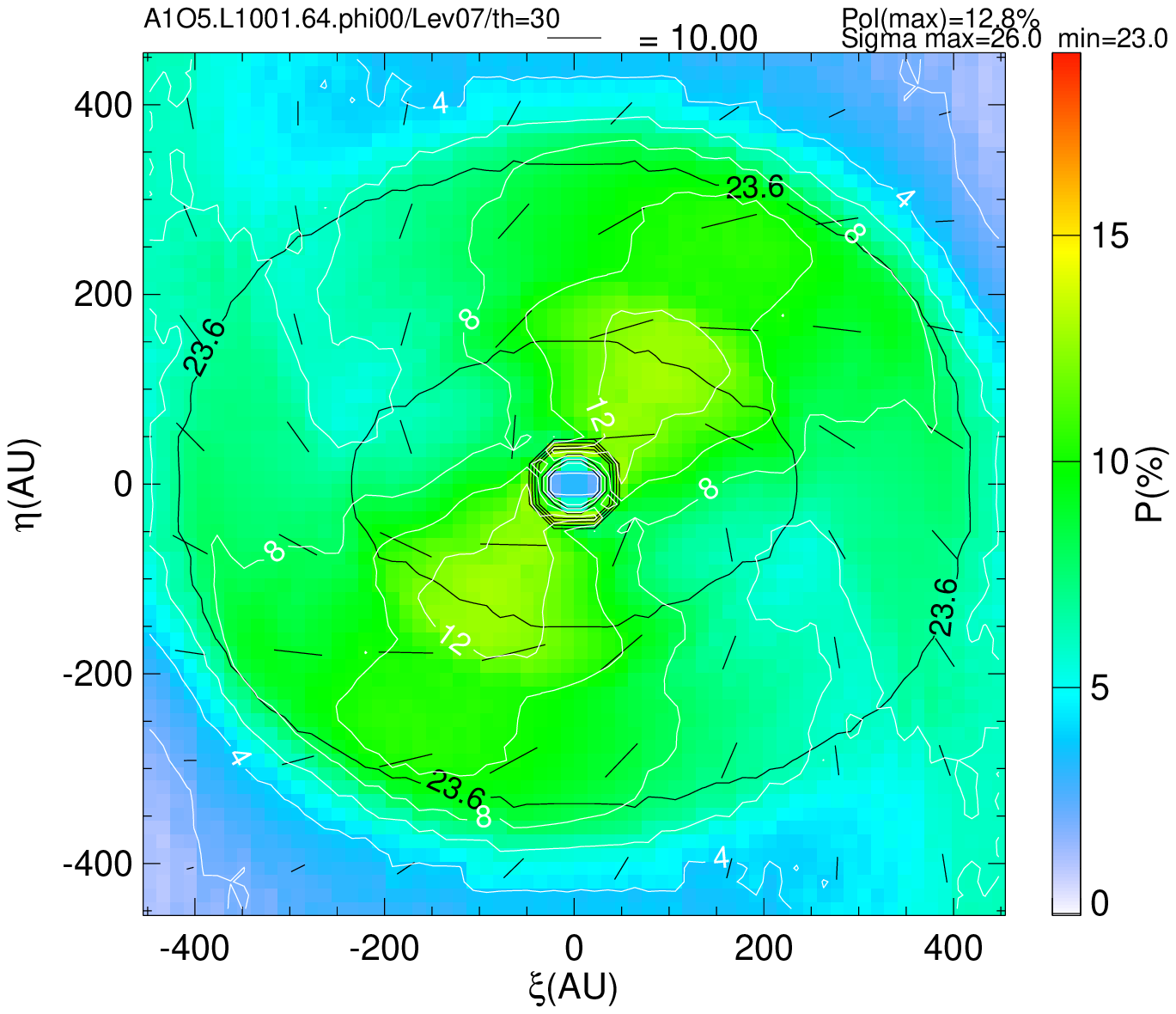}
      \one{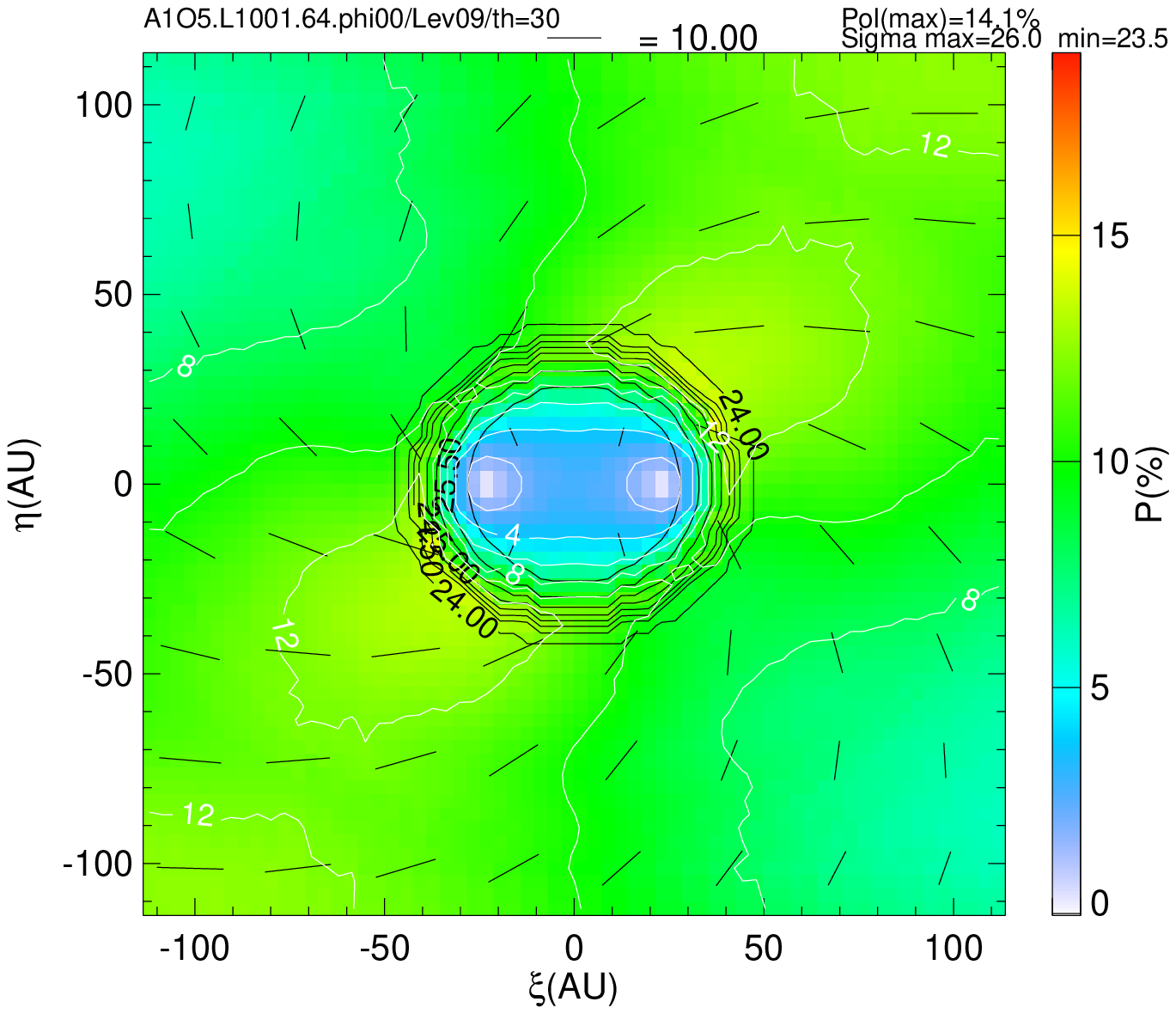}}\\
$\theta=45^\circ$\hspace*{10mm}
      \raisebox{-20mm}{\one{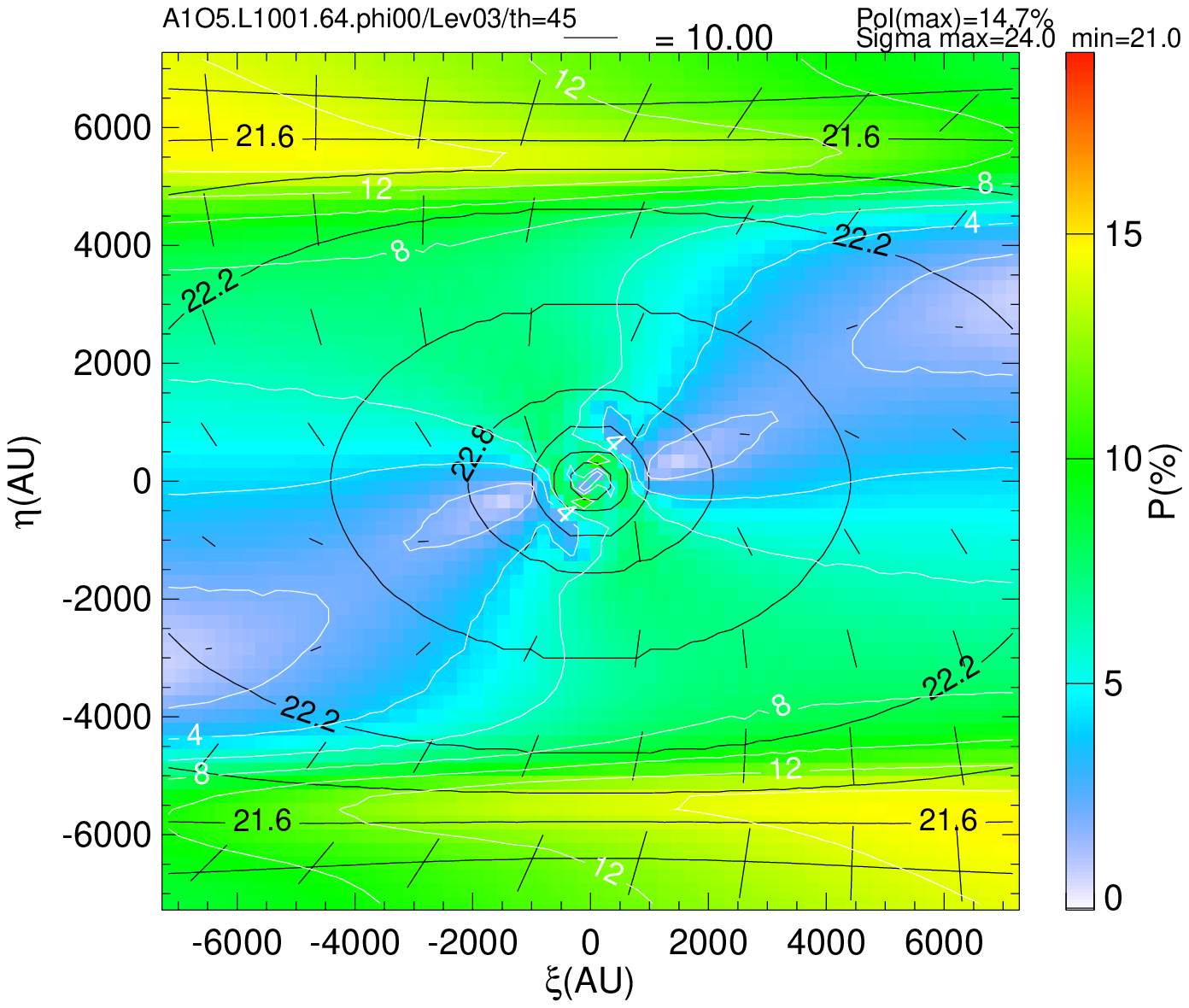}
      \one{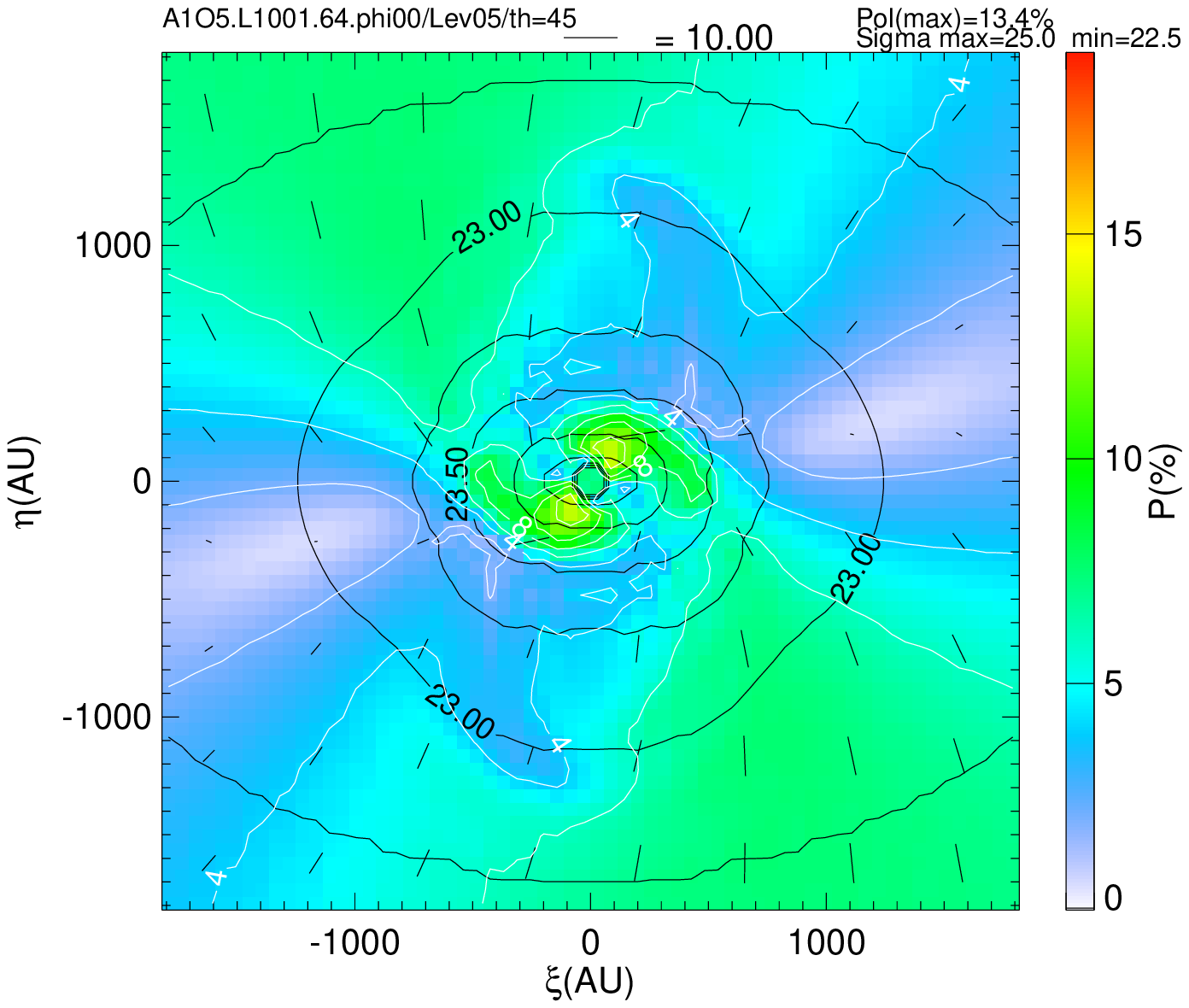}
      \one{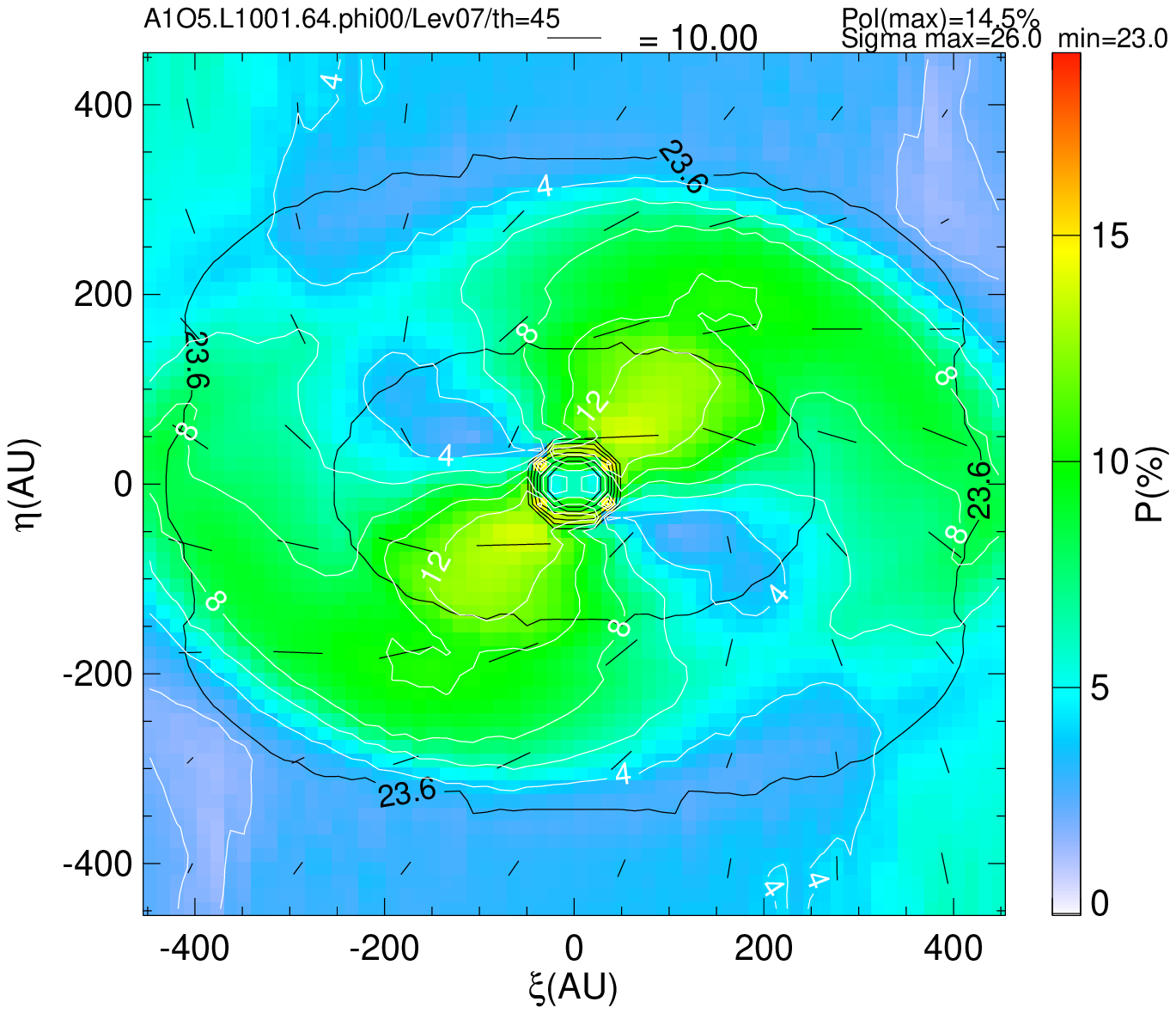}
      \one{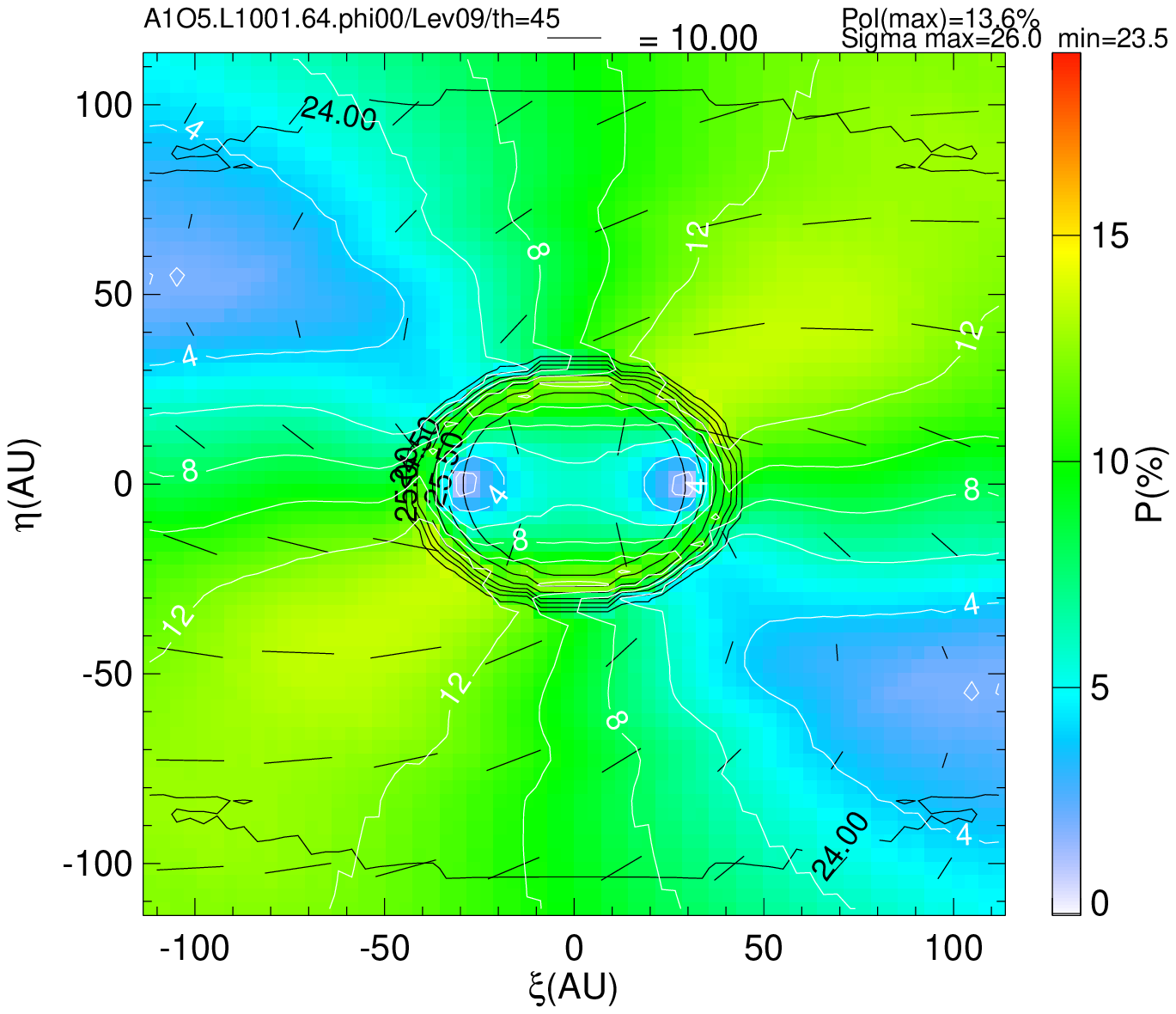}}\\
$\theta=60^\circ$\hspace*{10mm}
      \raisebox{-20mm}{\one{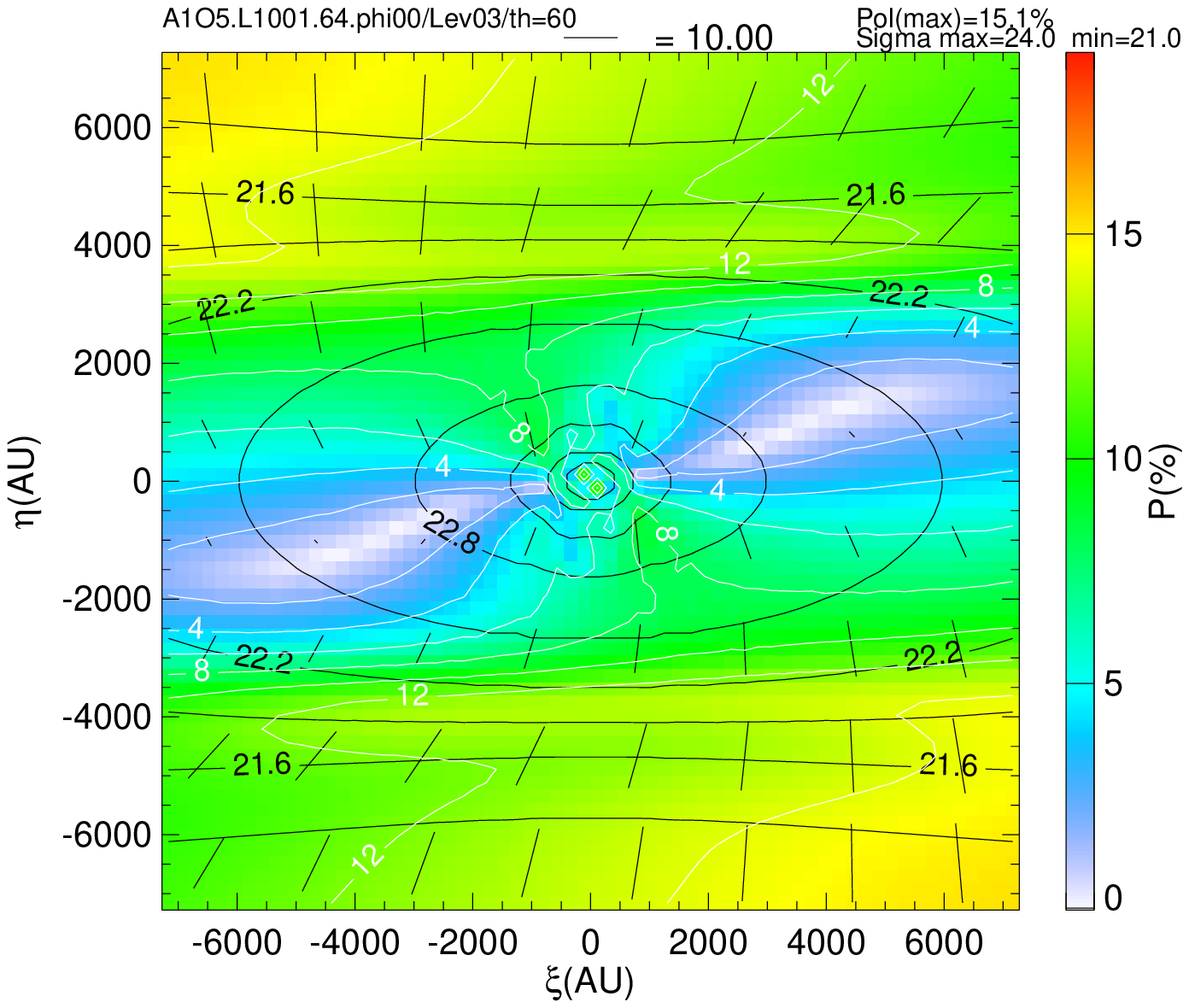}
      \one{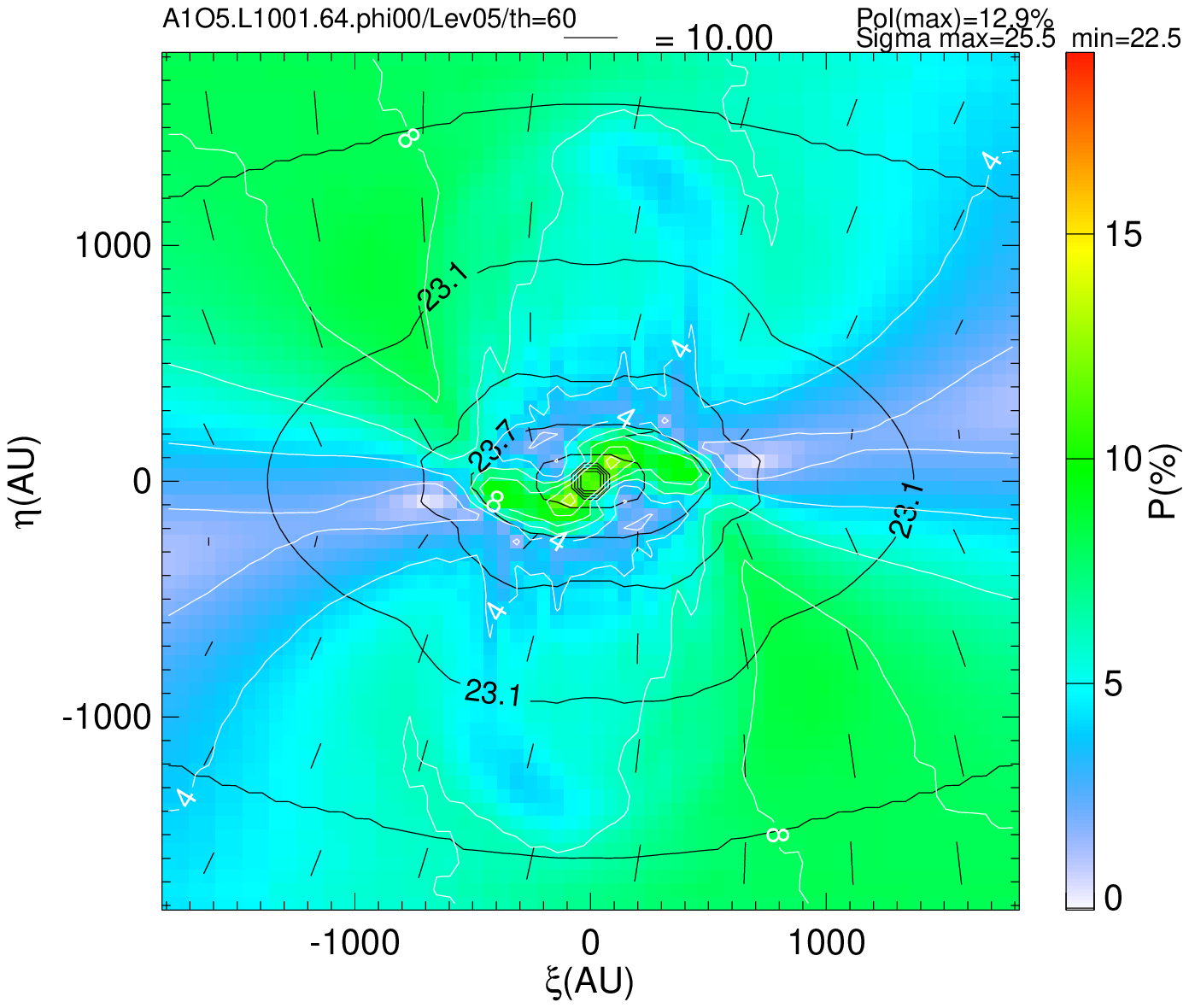}
      \one{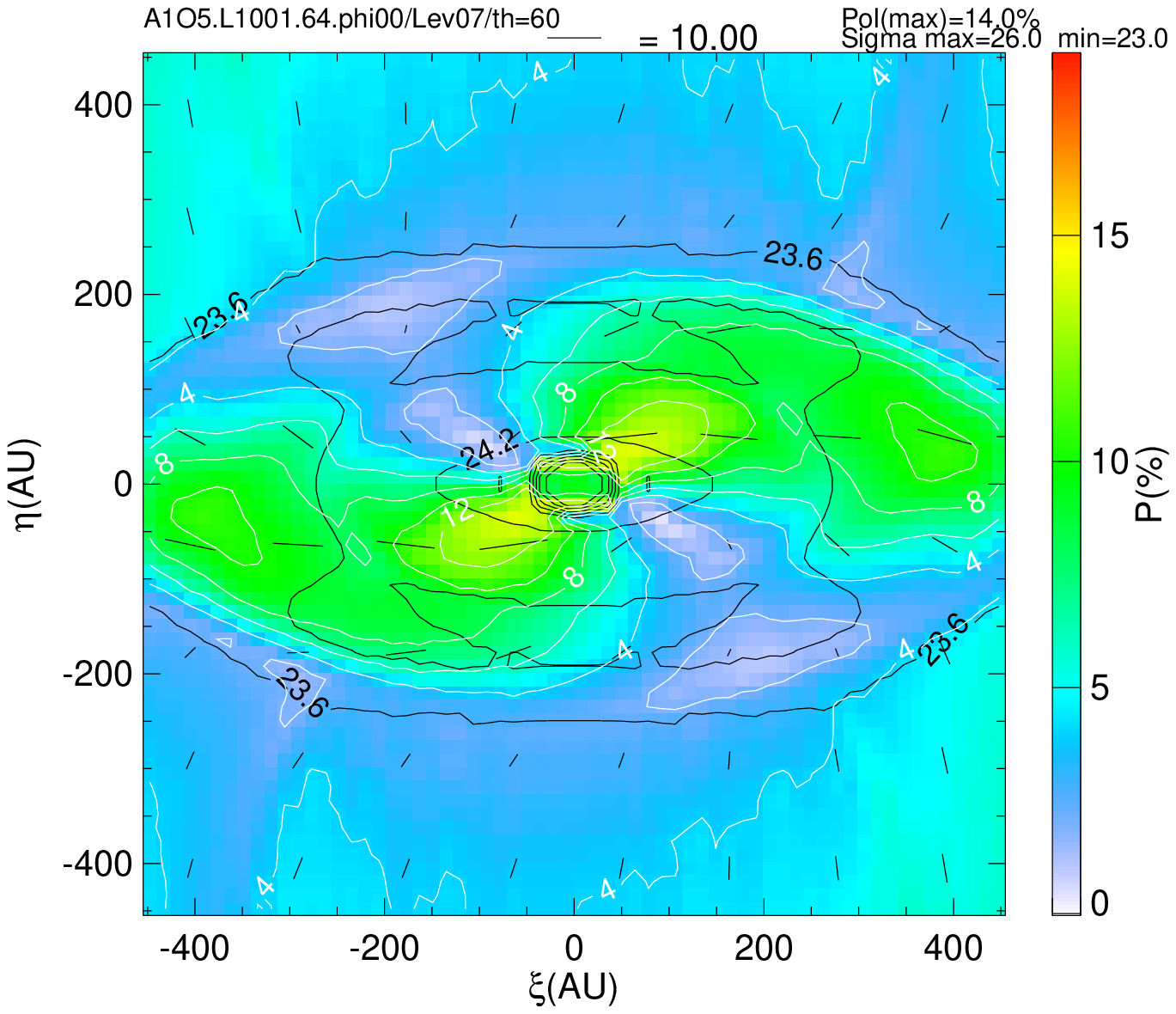}
      \one{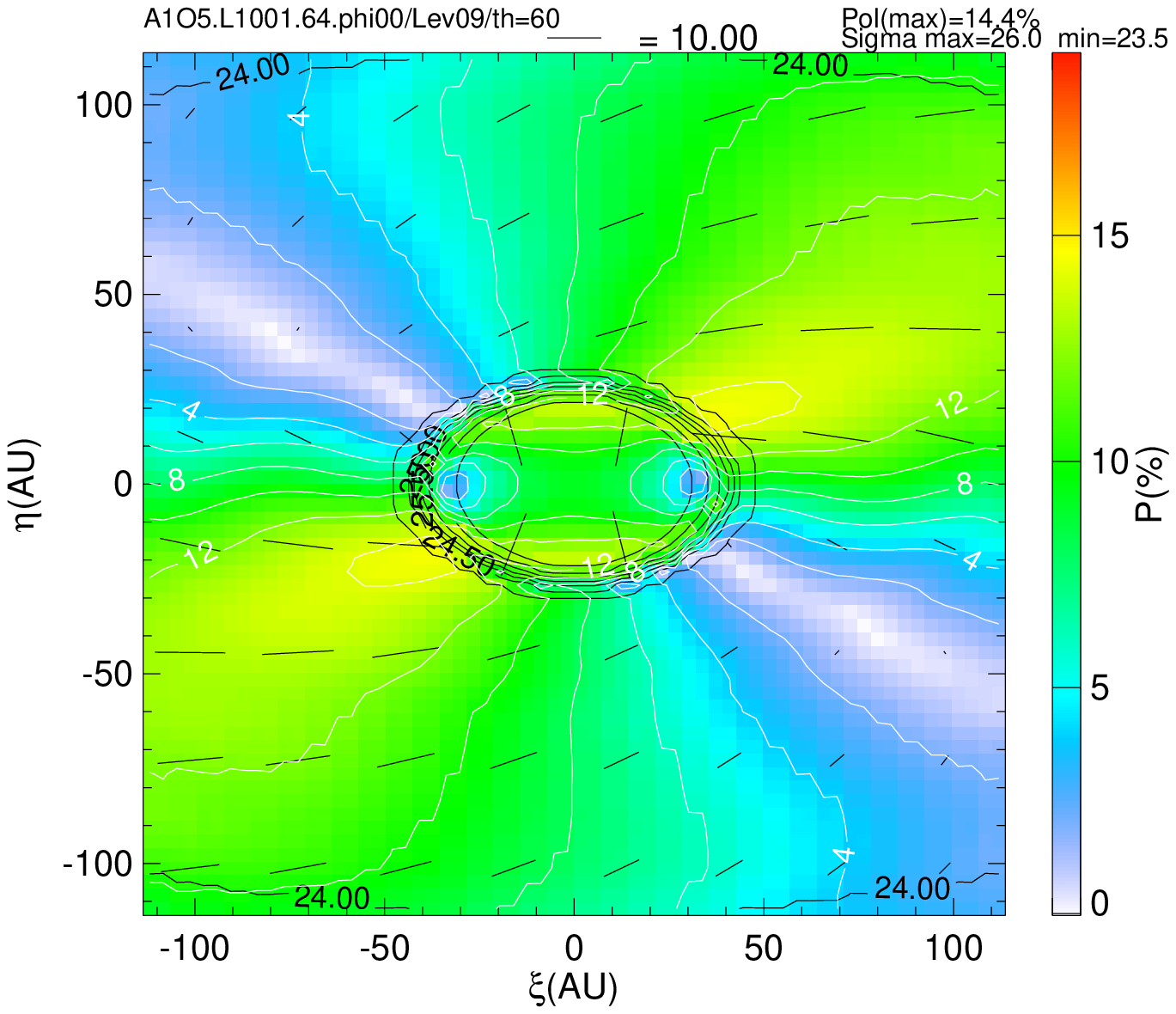}}\\
$\theta=80^\circ$\hspace*{10mm}
      \raisebox{-20mm}{\one{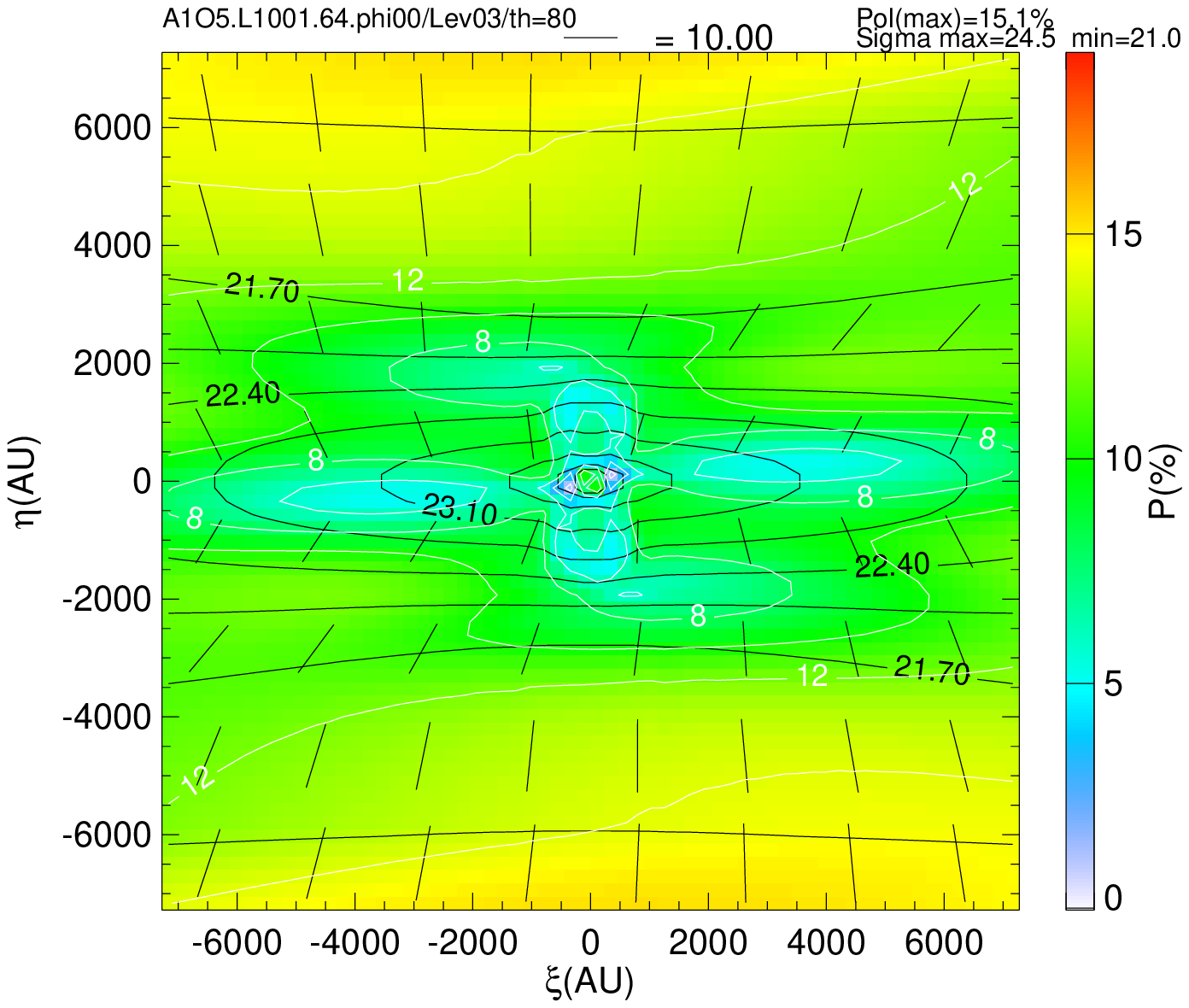}
      \one{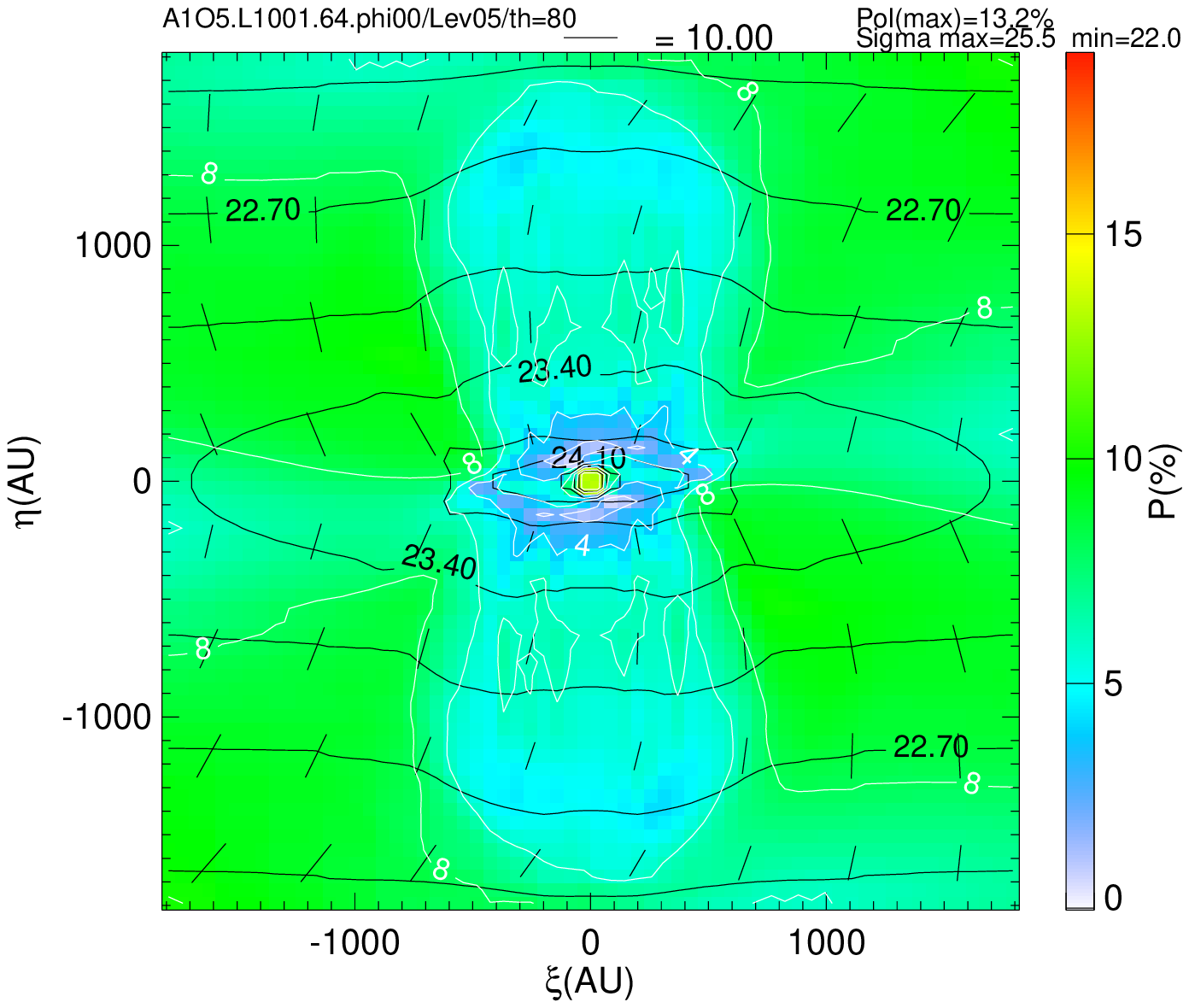}
      \one{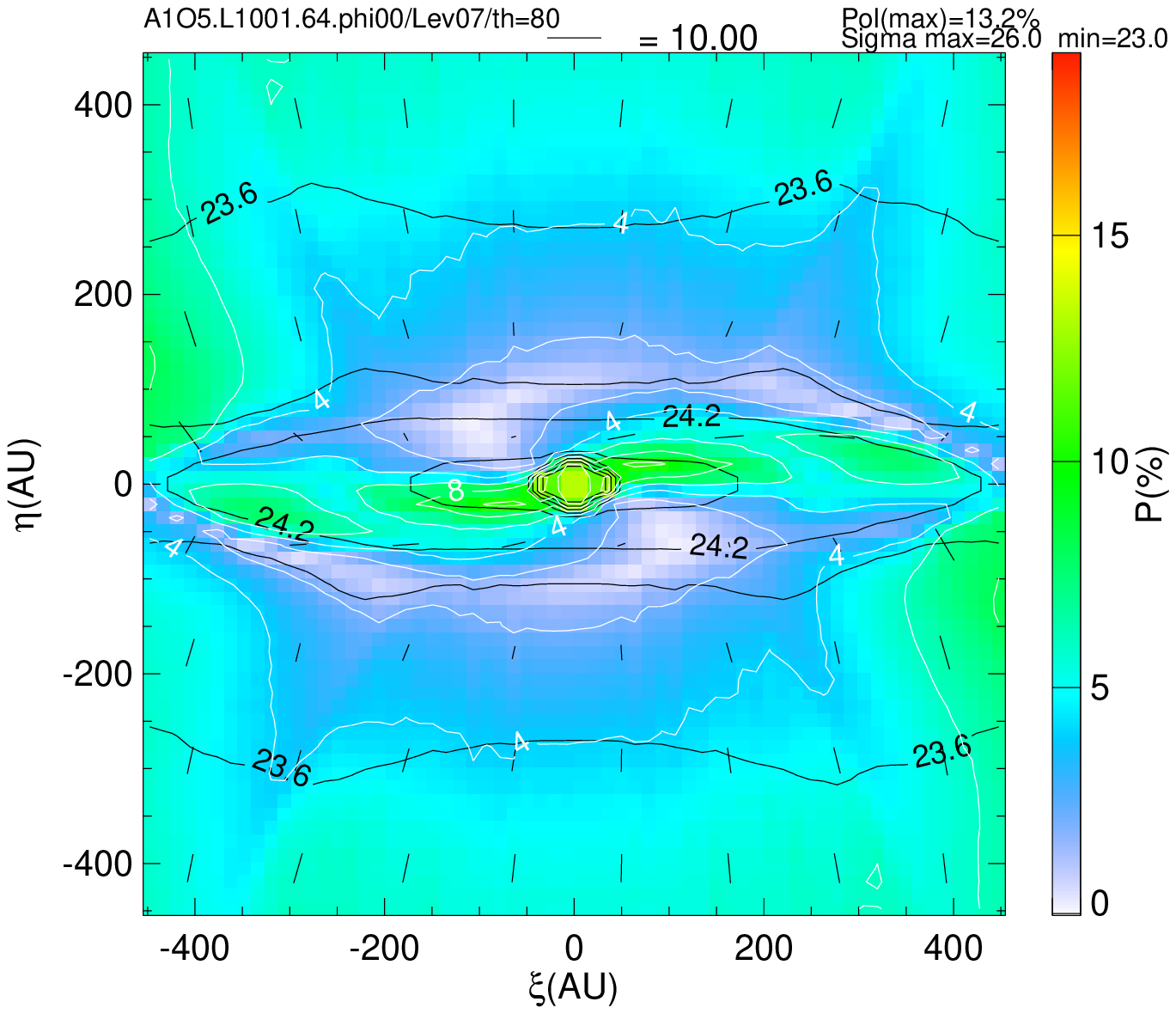}
      \one{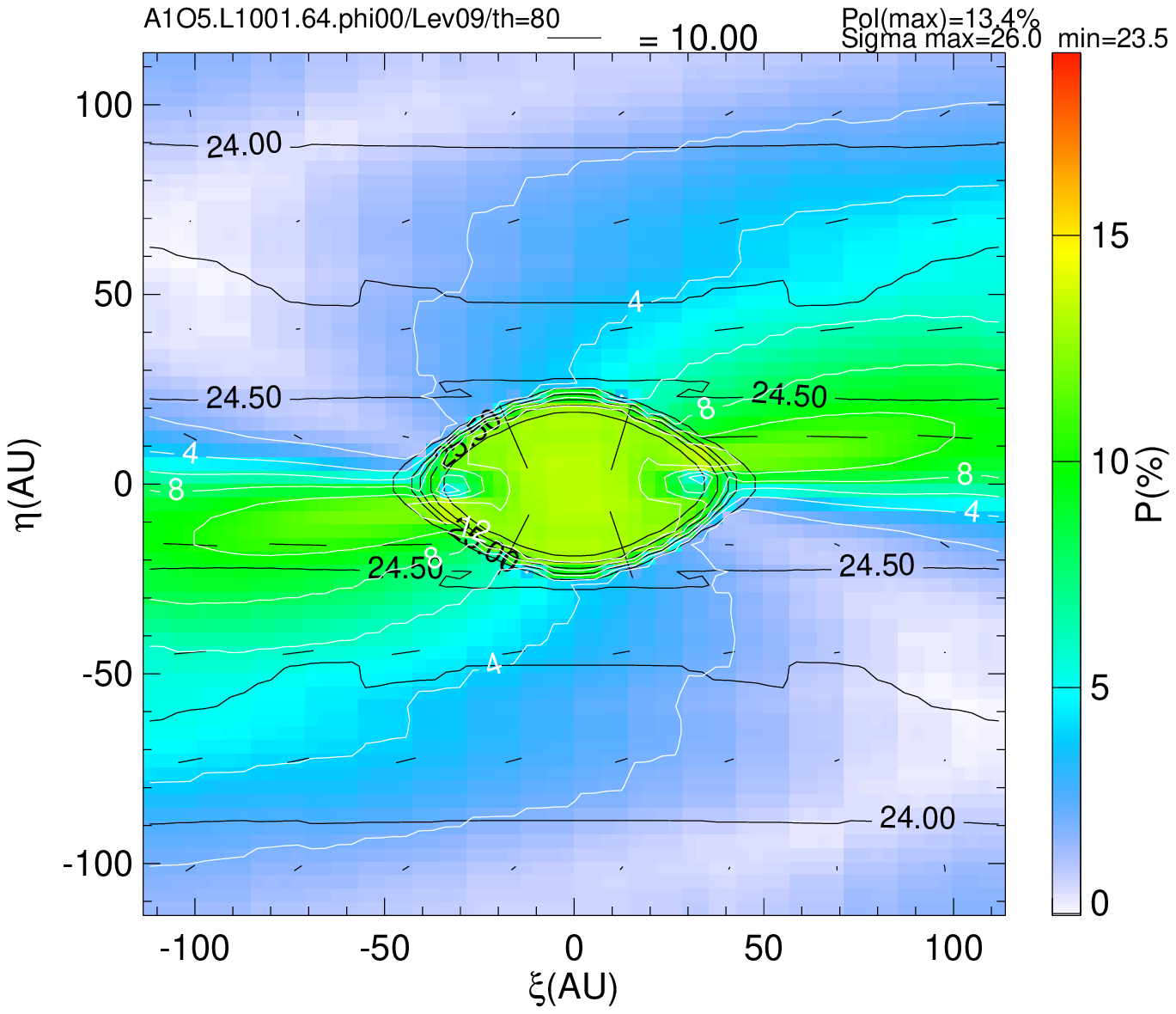}}\\
$\theta=90^\circ$\hspace*{10mm}
      \raisebox{-20mm}{\one{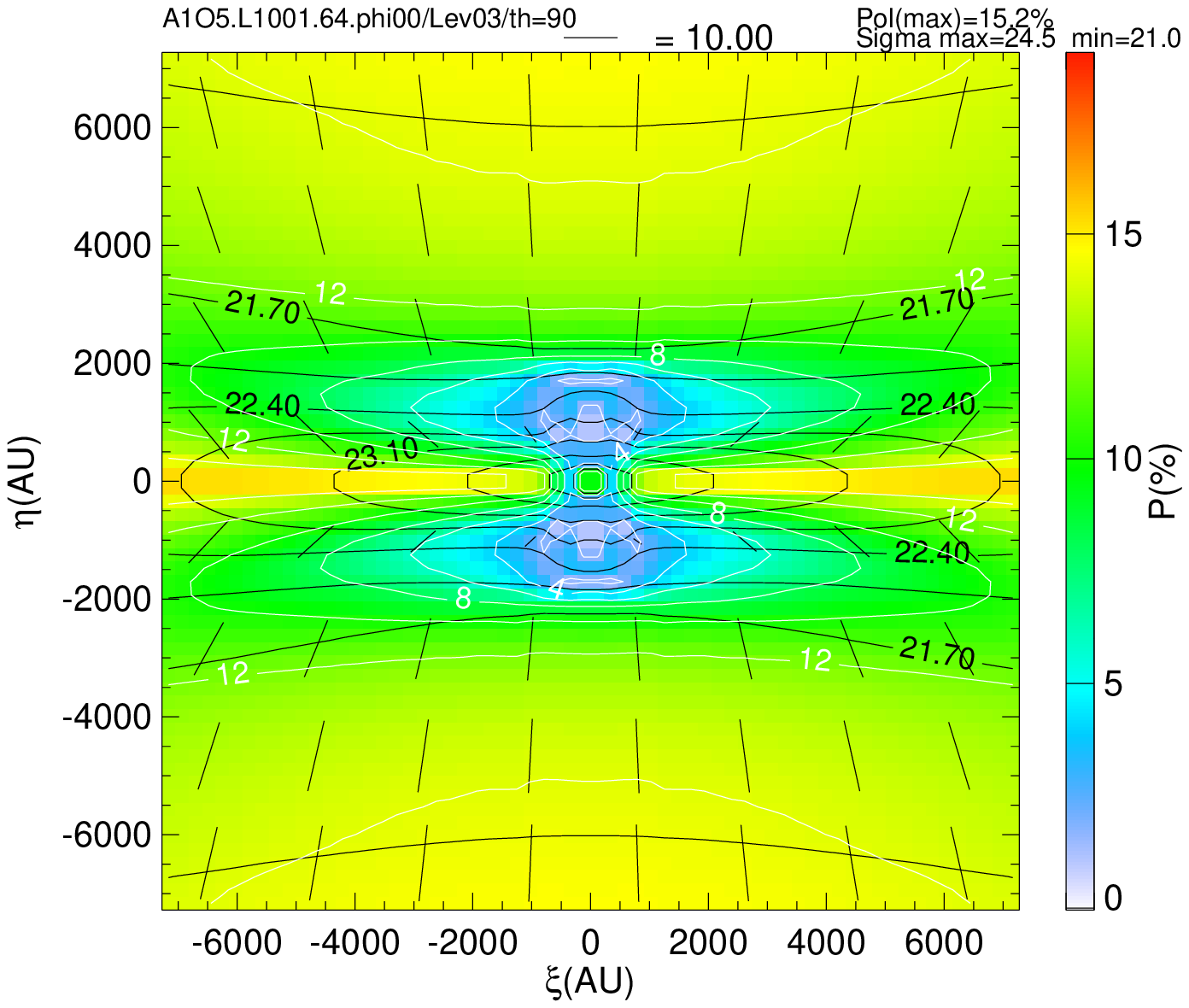}
      \one{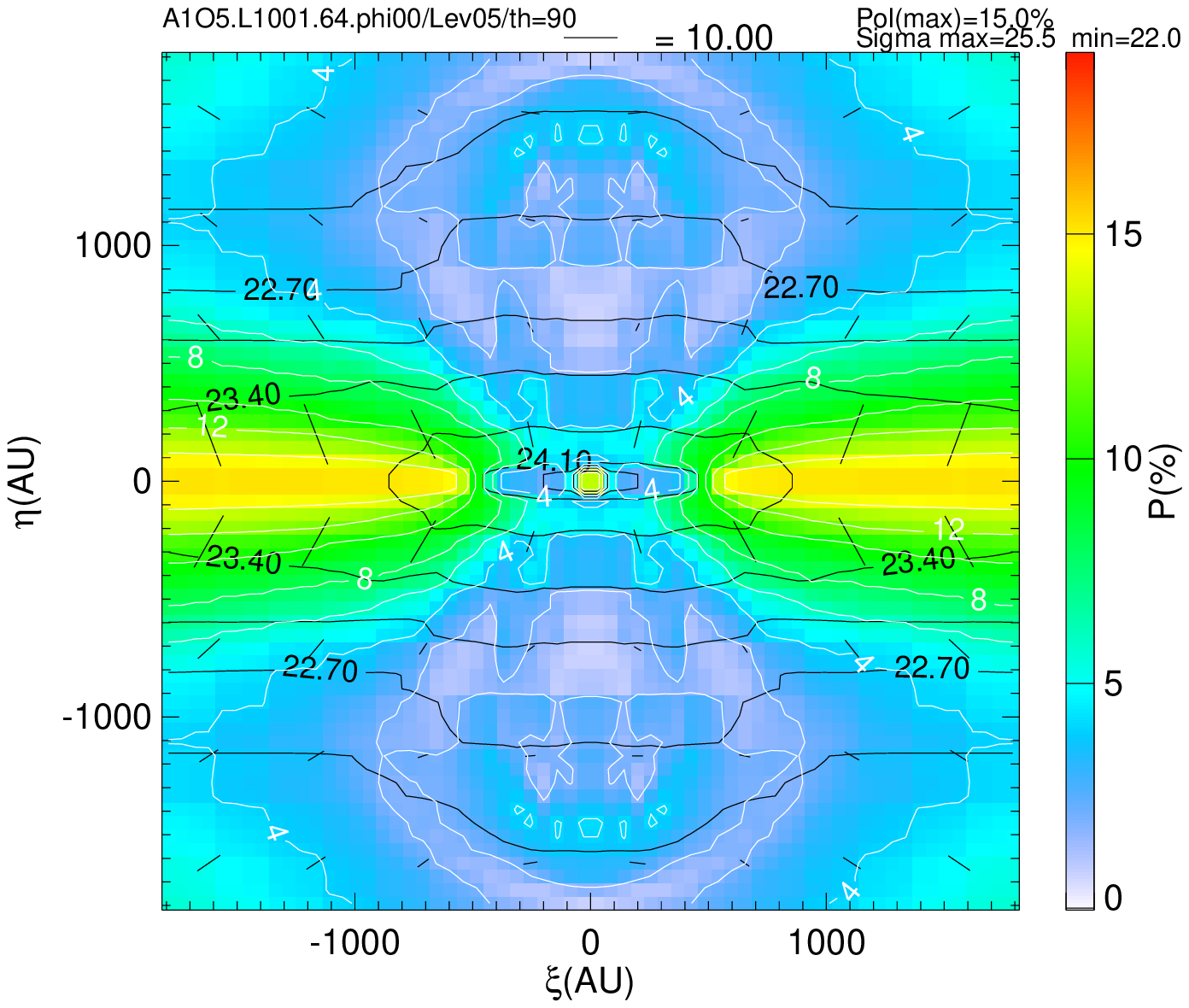}
      \one{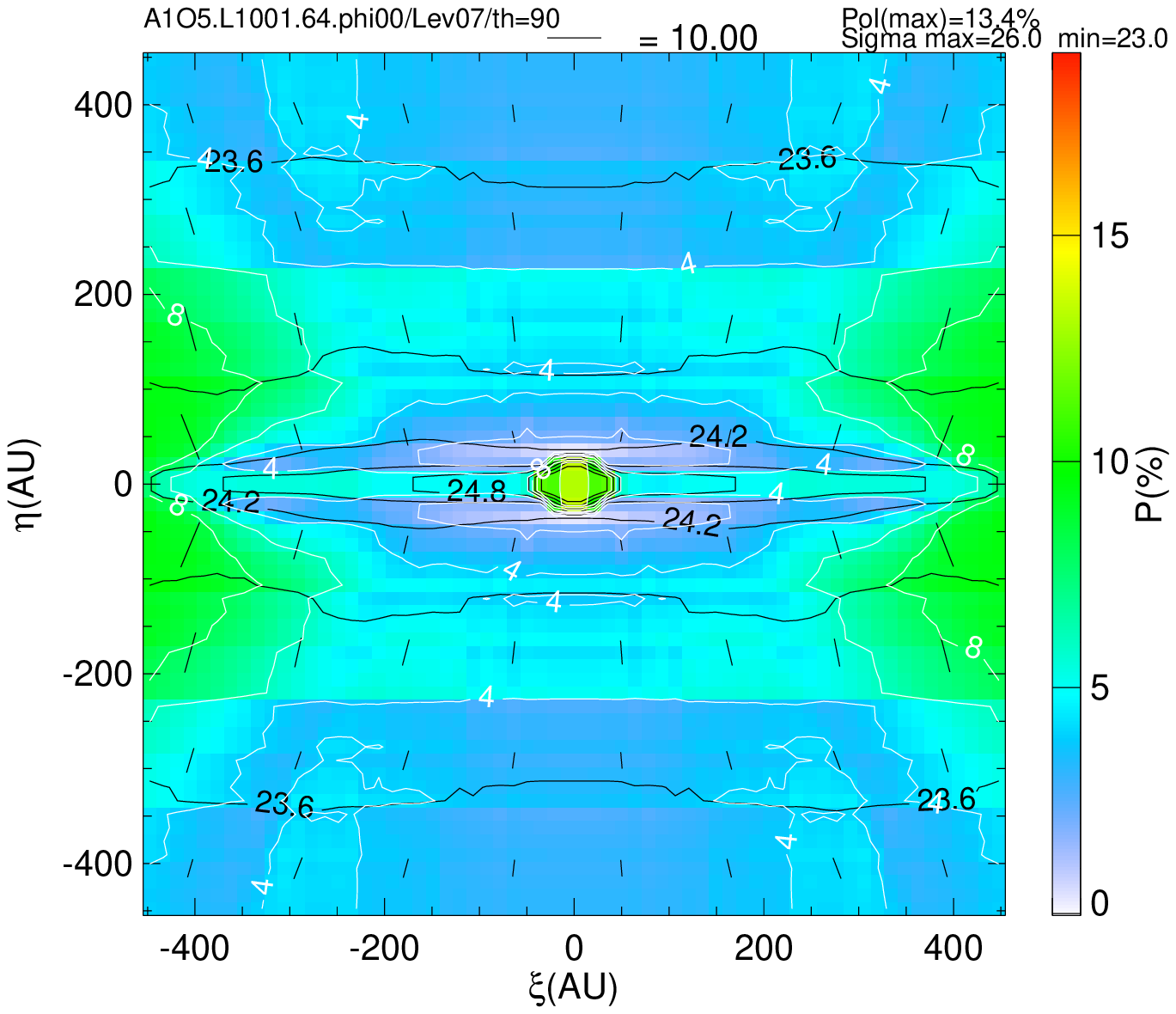}
      \one{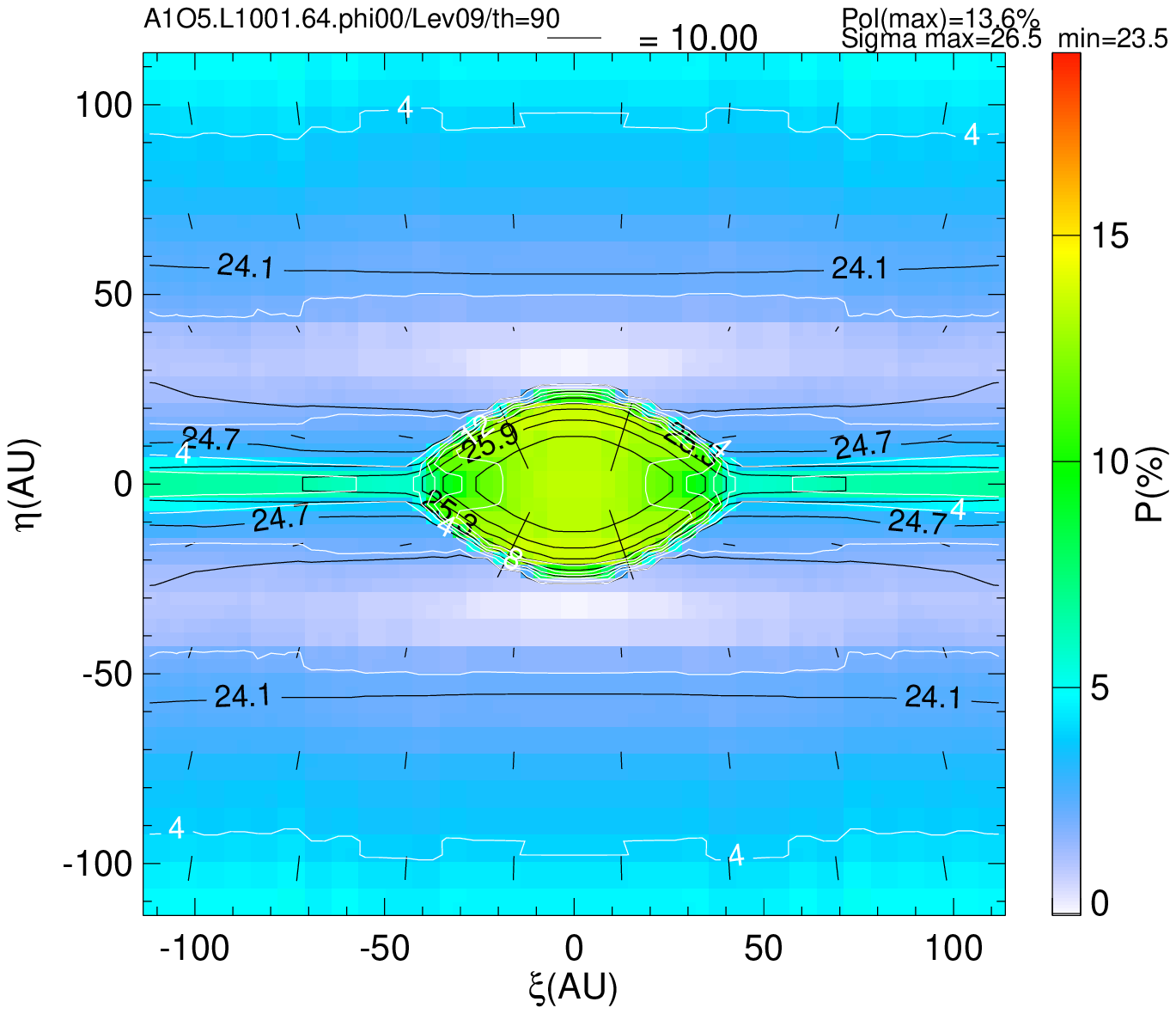}}
   \end{center}
\caption{\label{fig:A1O5L1001}As for Fig.\ref{fig:A1O5L332}
 but for the accretion phase, in which the outflow is driven by the
 magnetic force.}
\end{figure}

\begin{figure}[h]
   \begin{center}
\vspace*{-12mm}($L=3$)\hspace*{22mm}($L=5$)\hspace*{22mm}($L=7$)\hspace*{22mm}($L=9$)\\
\hspace*{24mm}
      \raisebox{-20mm}
{\one{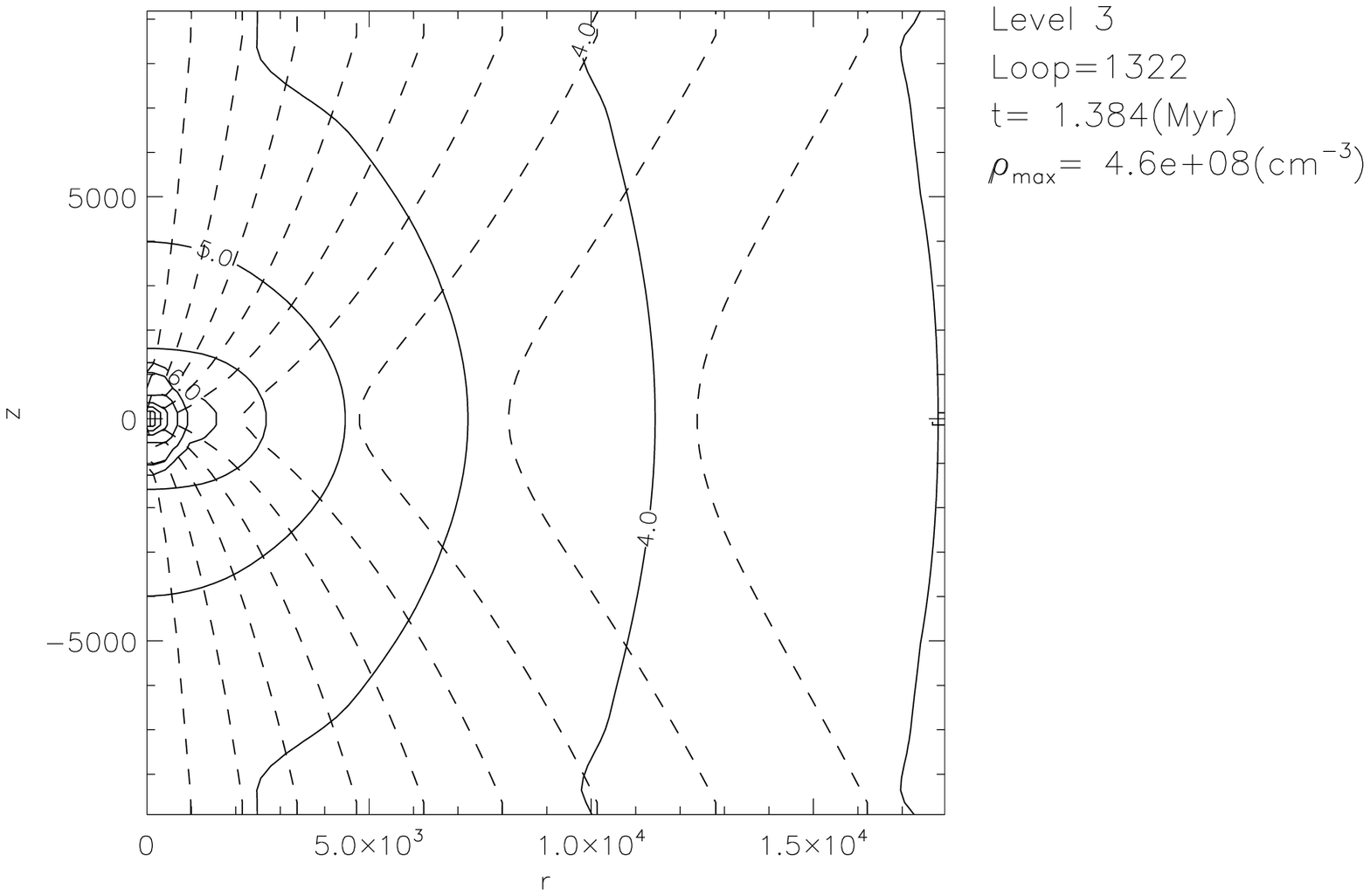}
      \one{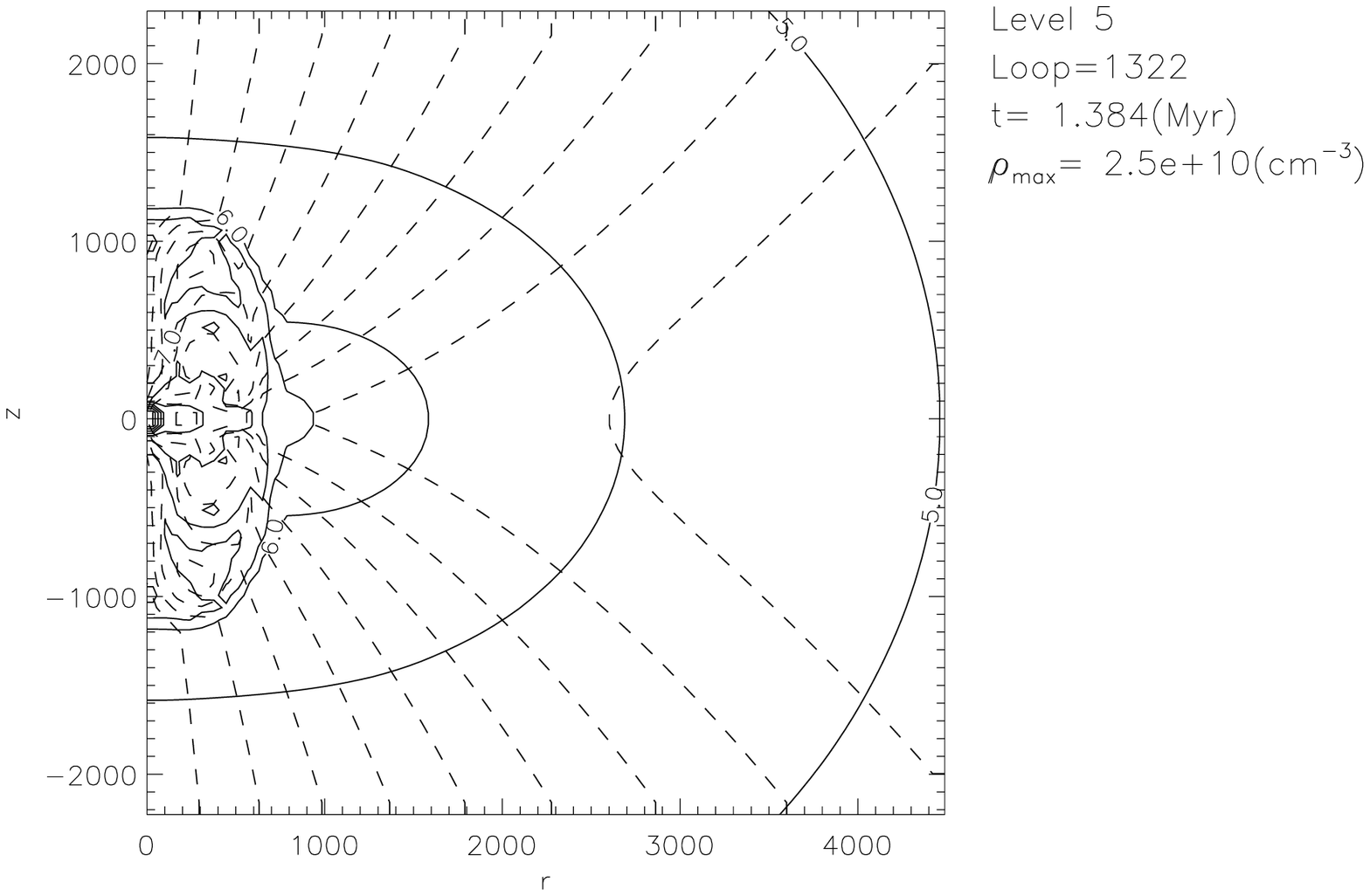}
      \one{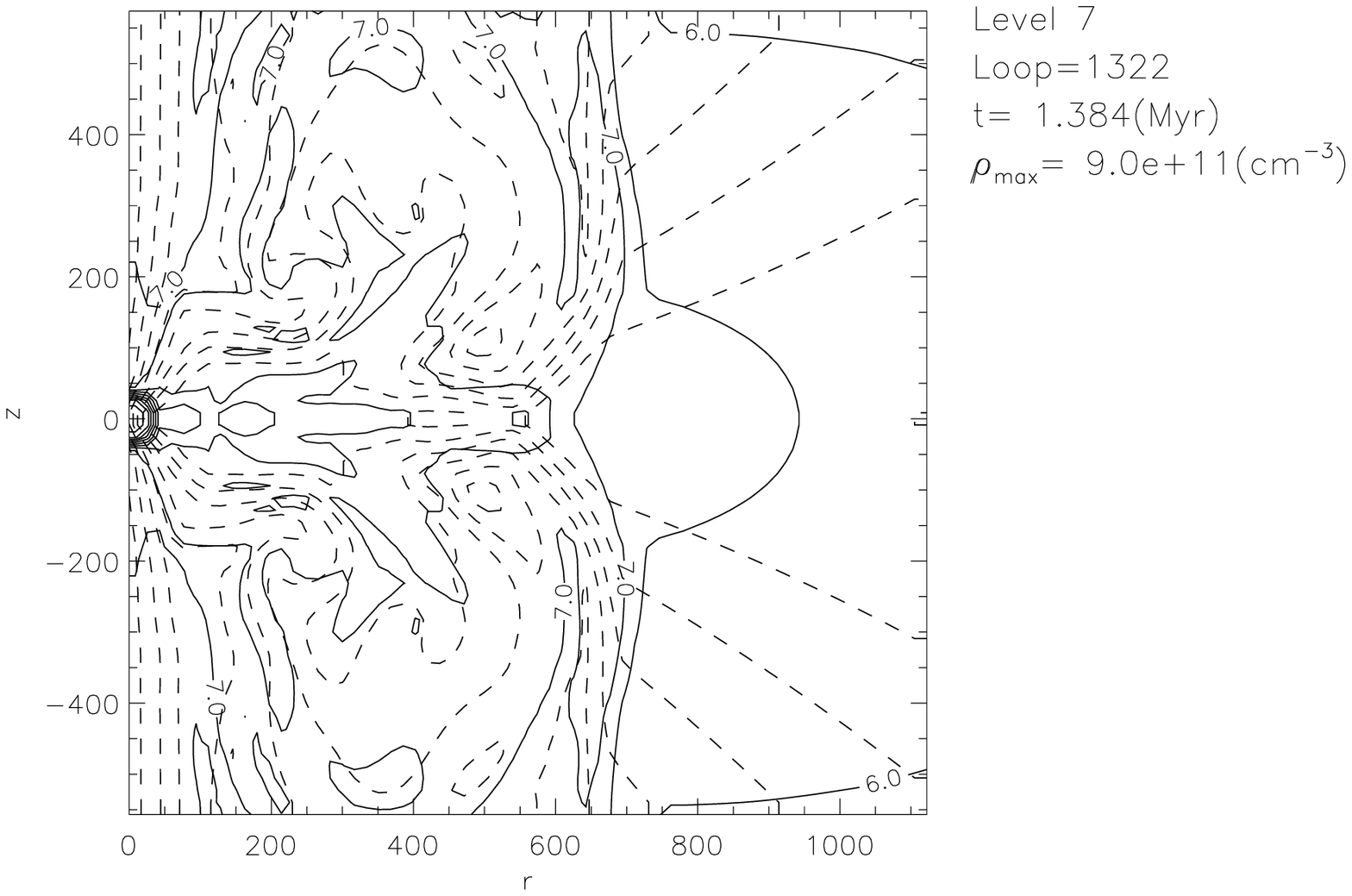}
      \one{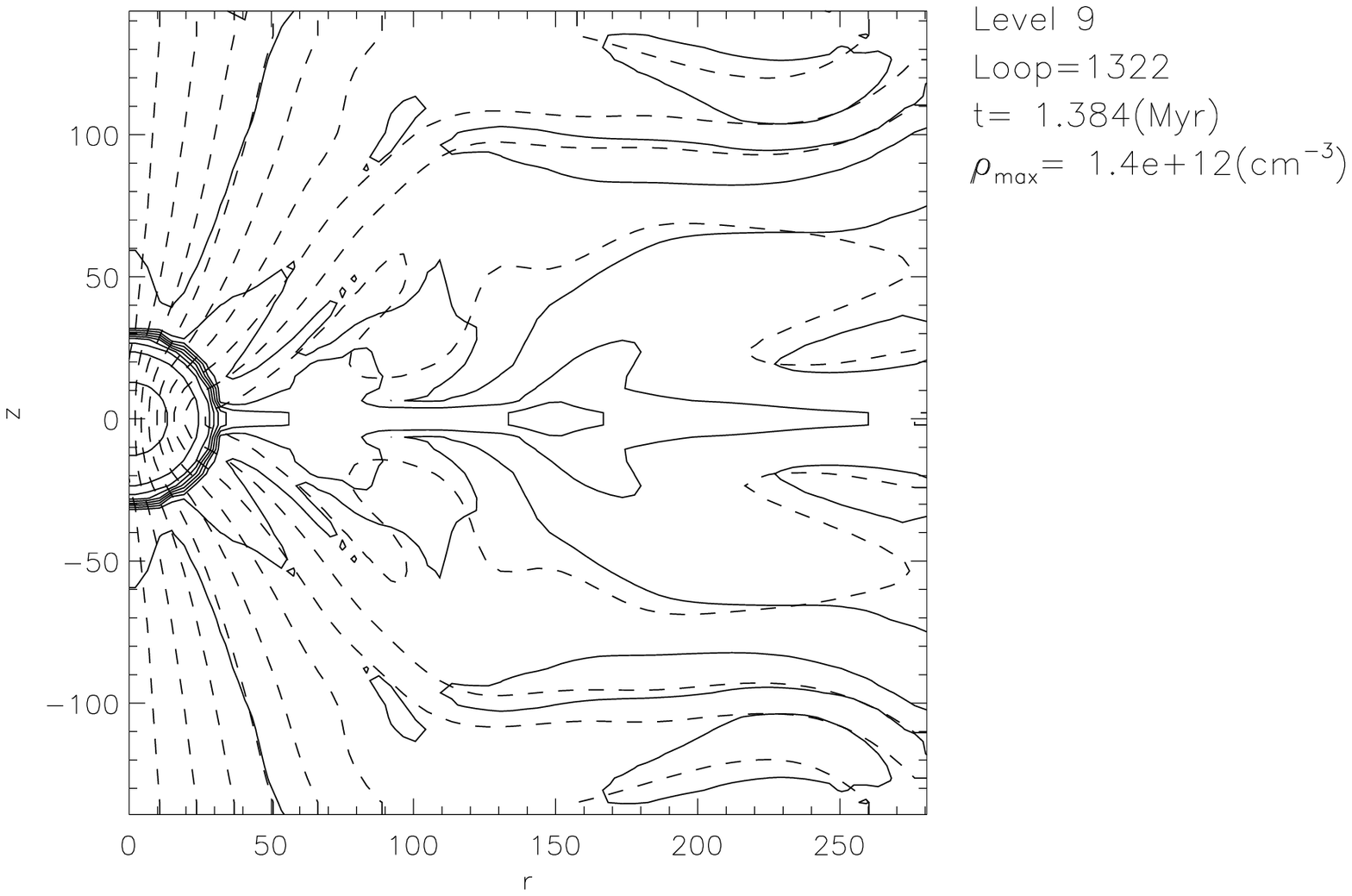}}\\

$\theta=0^\circ$\hspace*{12mm}
      \raisebox{-20mm}{\one{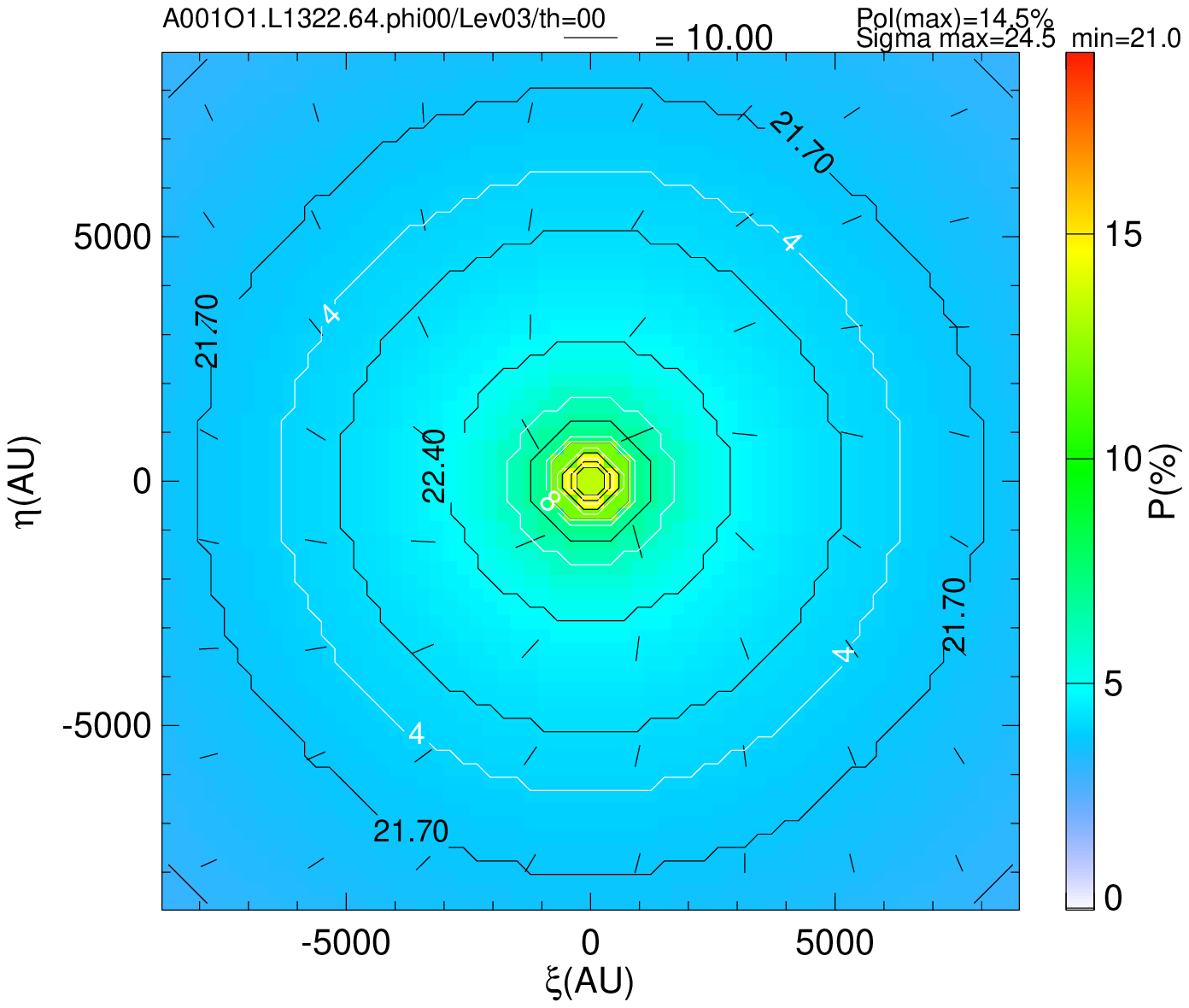}
      \one{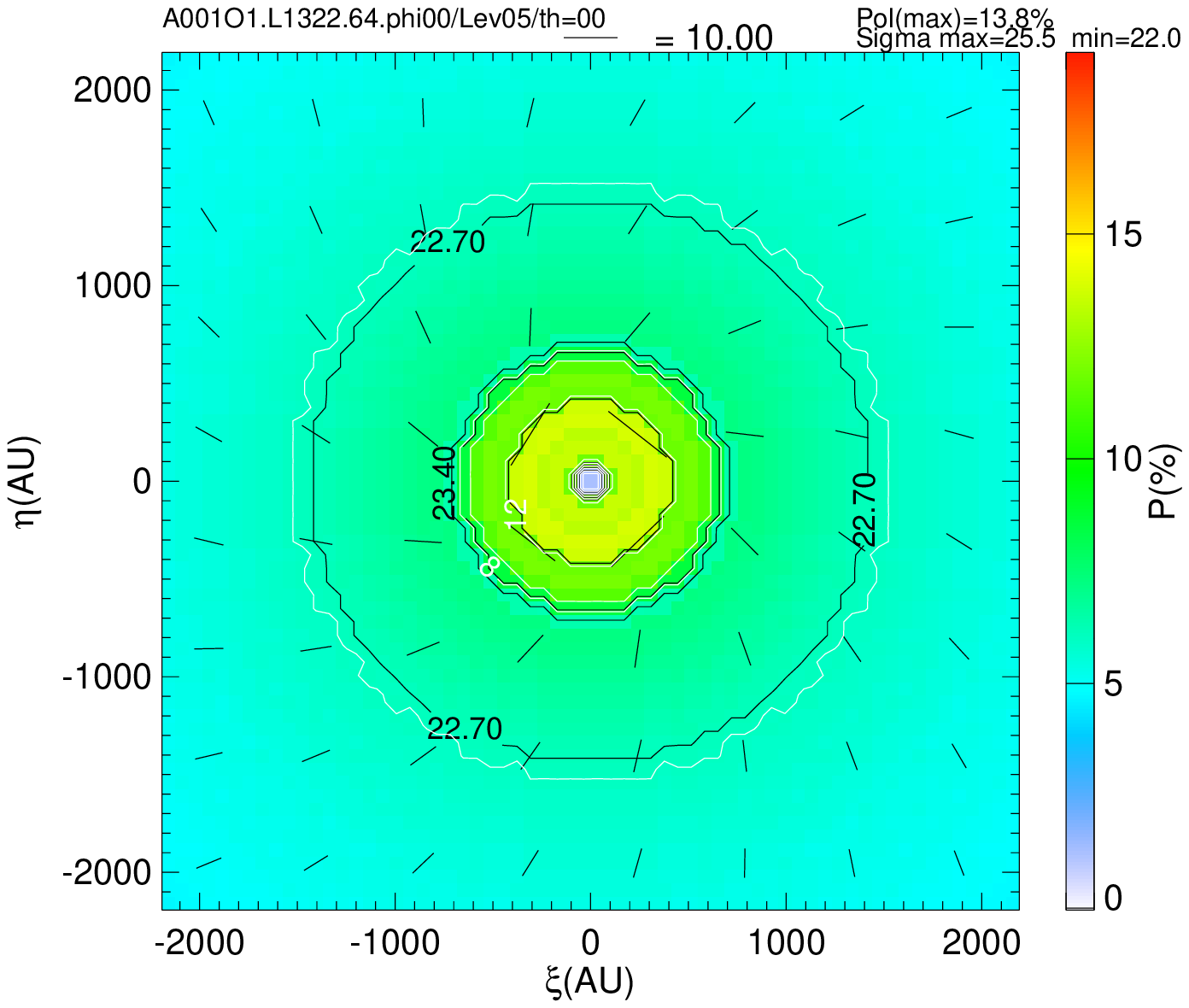}
      \one{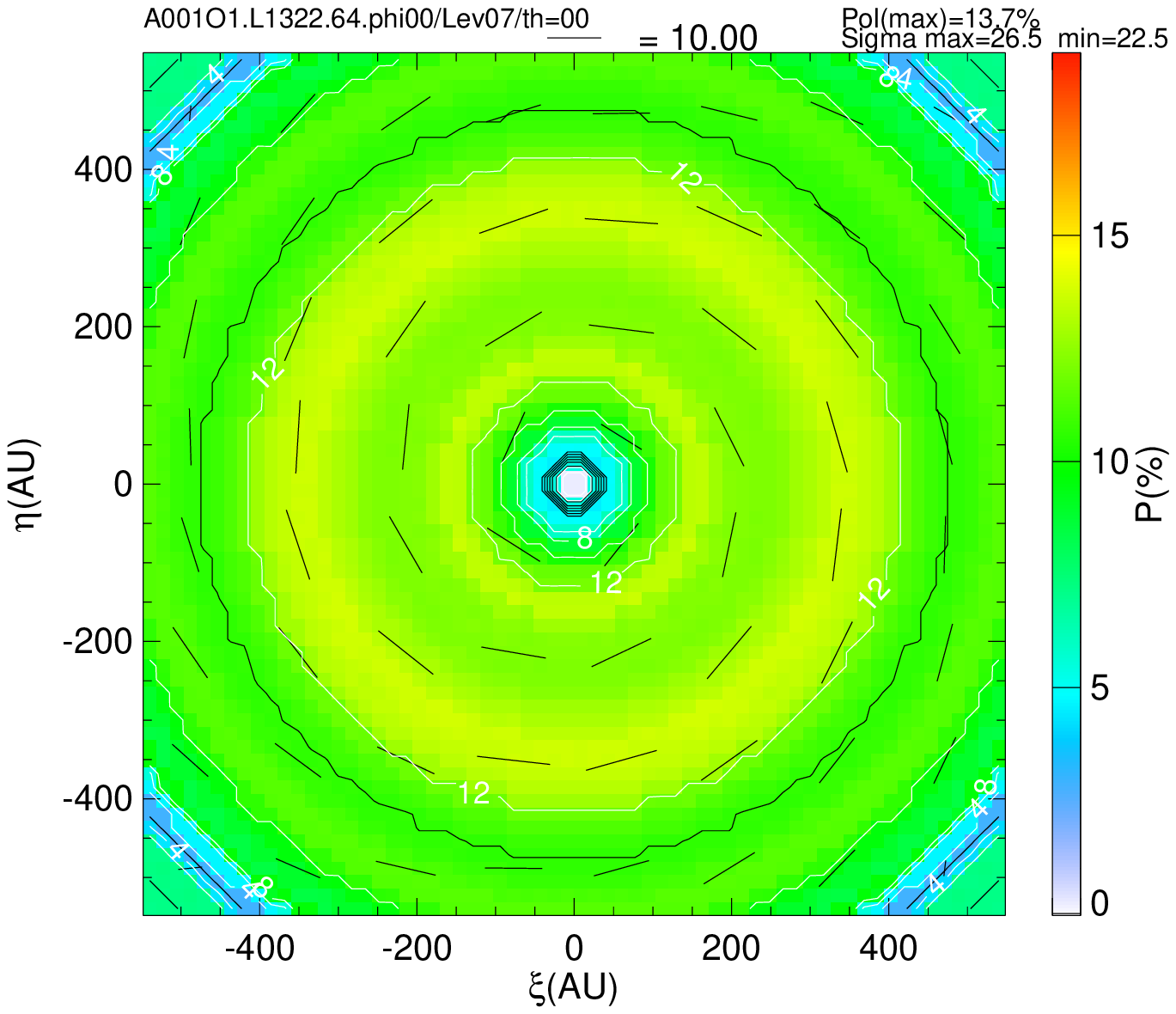}
      \one{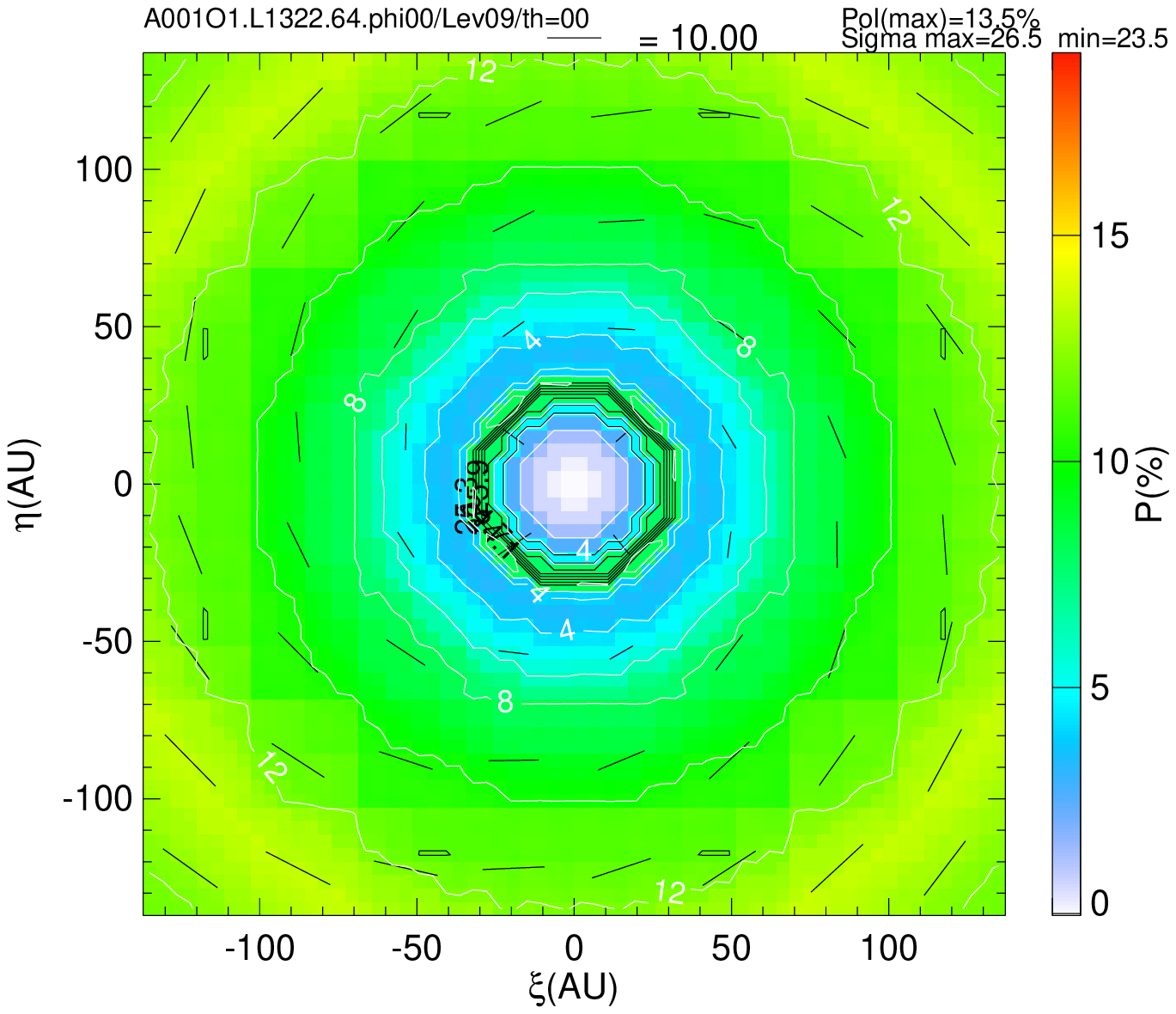}}\\
$\theta=30^\circ$\hspace*{10mm}
      \raisebox{-20mm}{\one{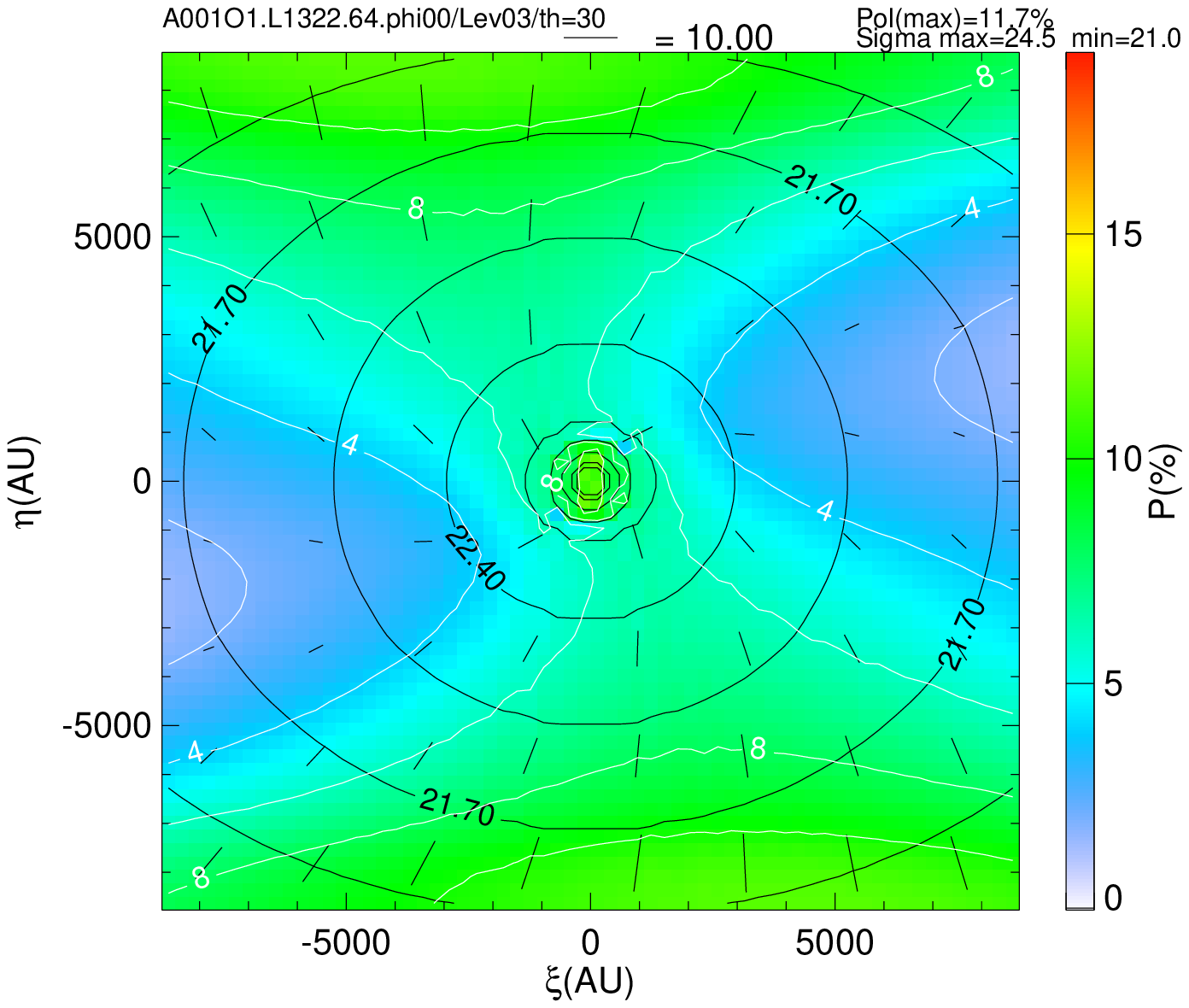}
      \one{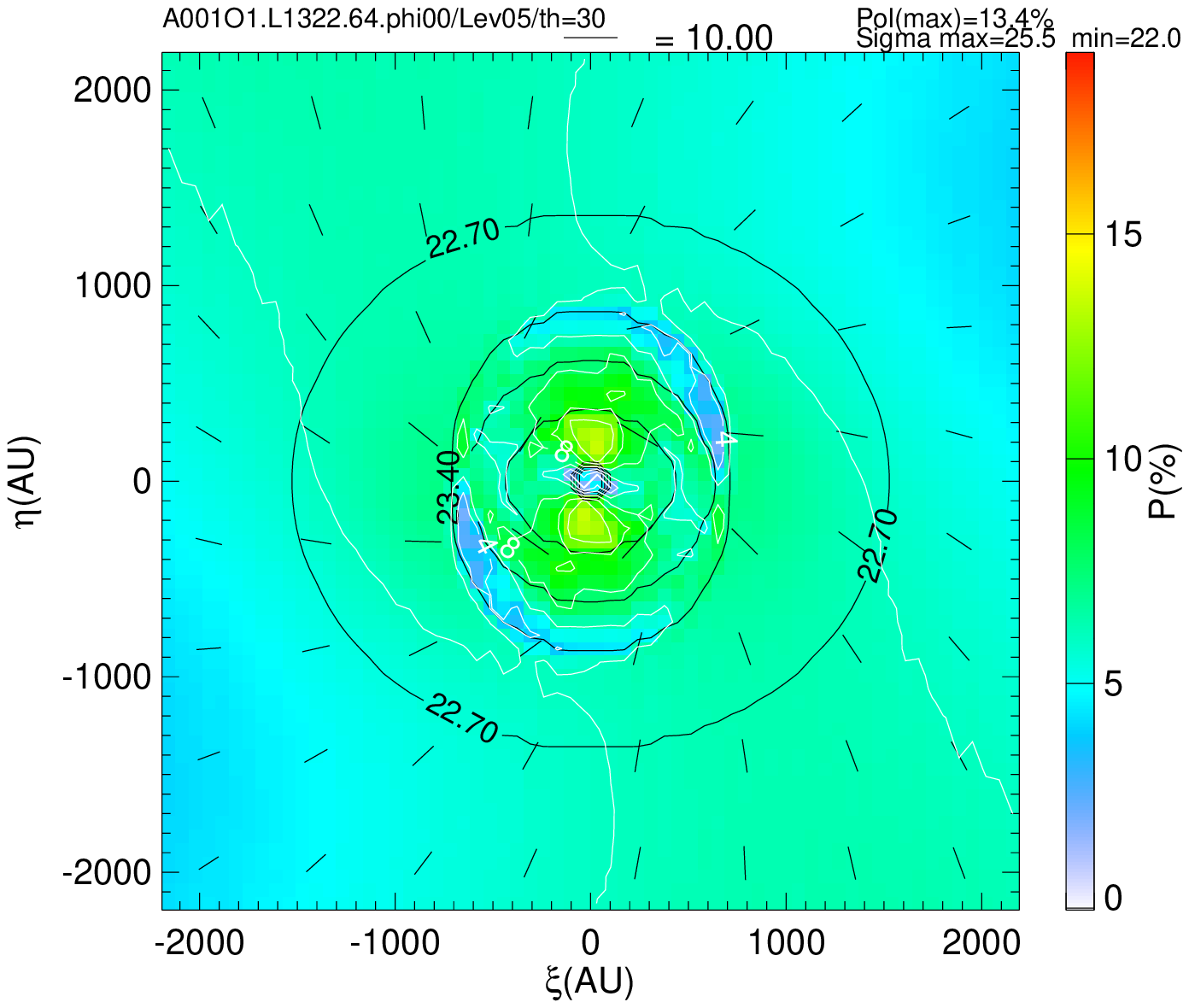}
      \one{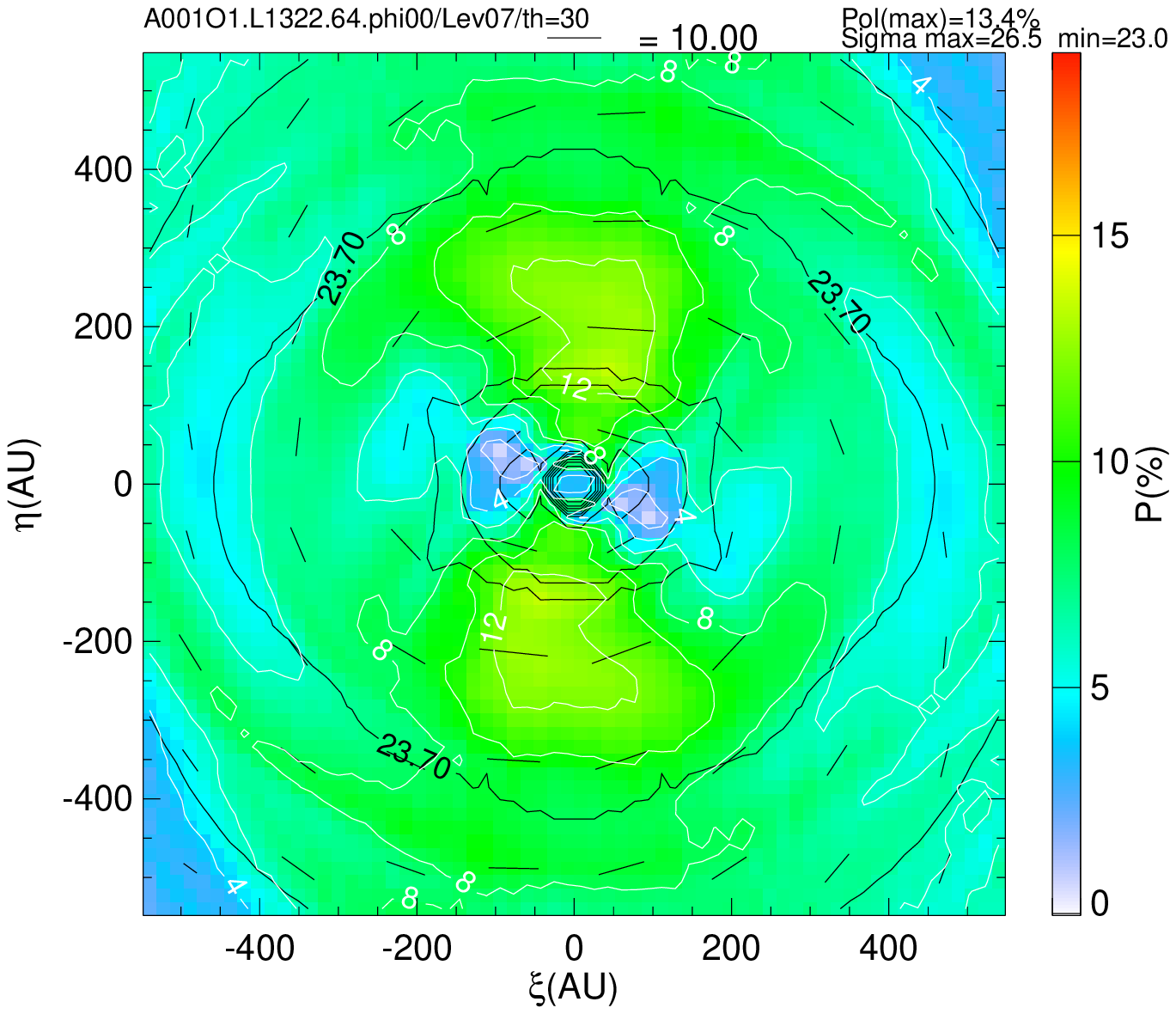}
      \one{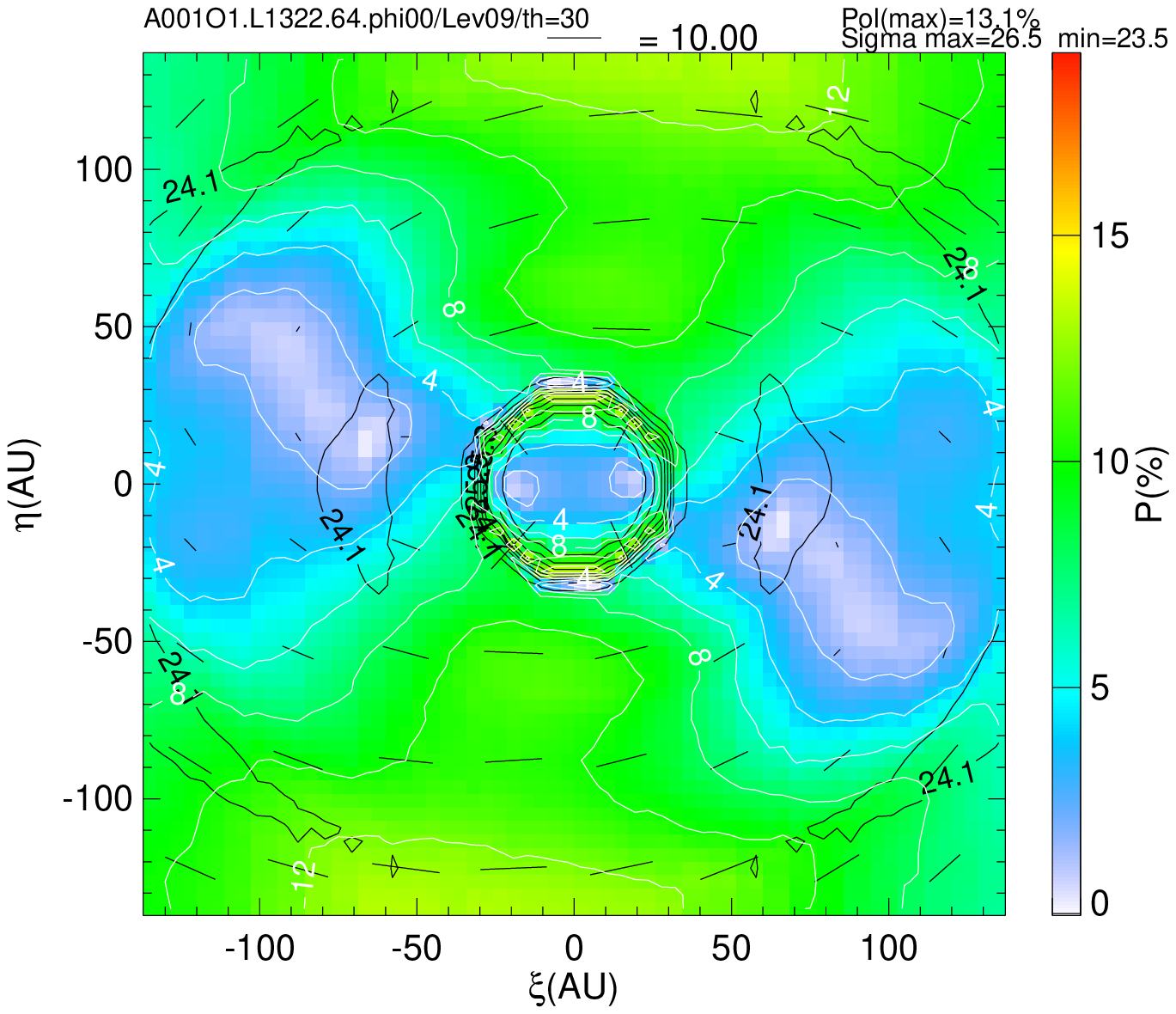}}\\
$\theta=45^\circ$\hspace*{10mm}
      \raisebox{-20mm}{\one{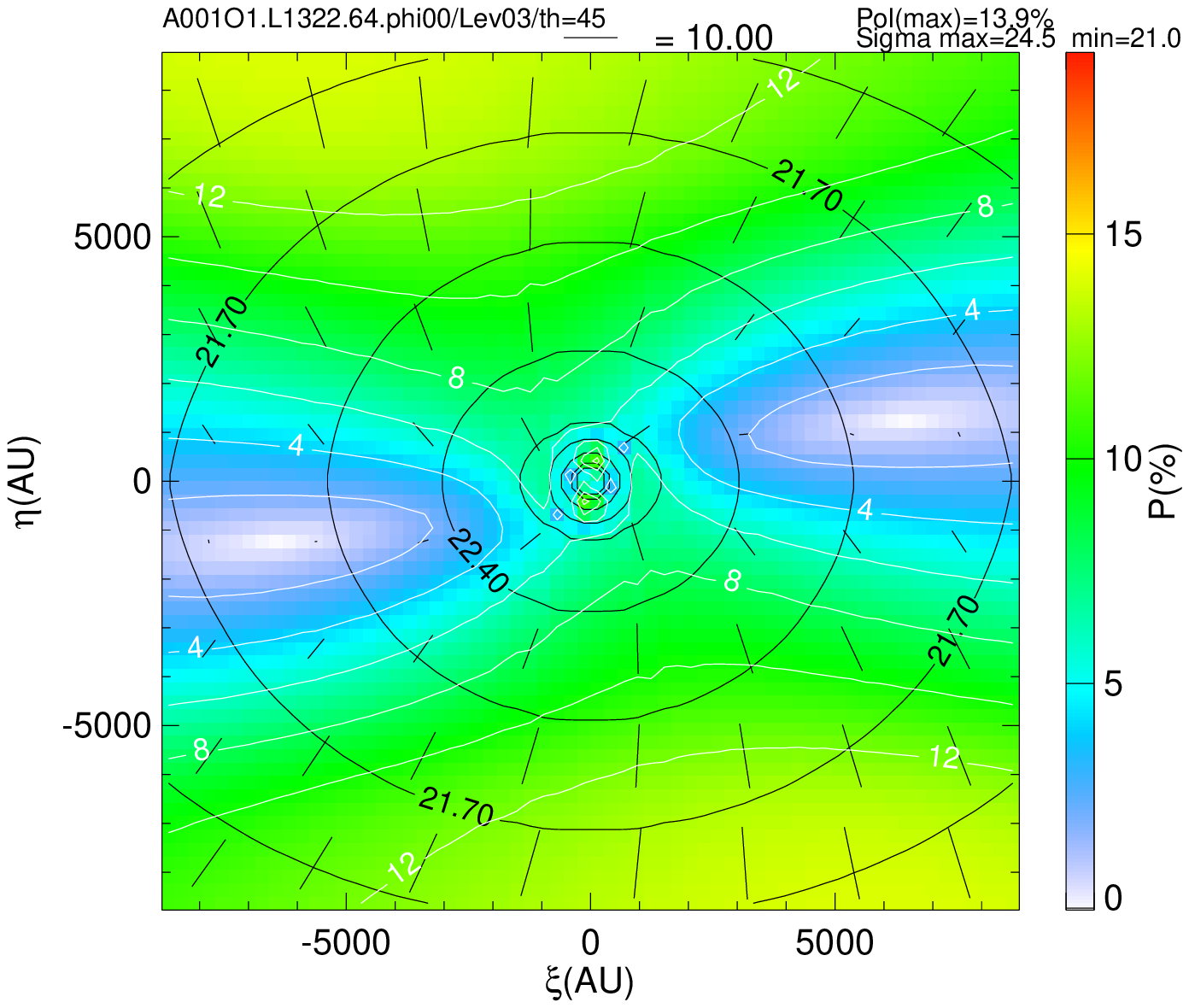}
      \one{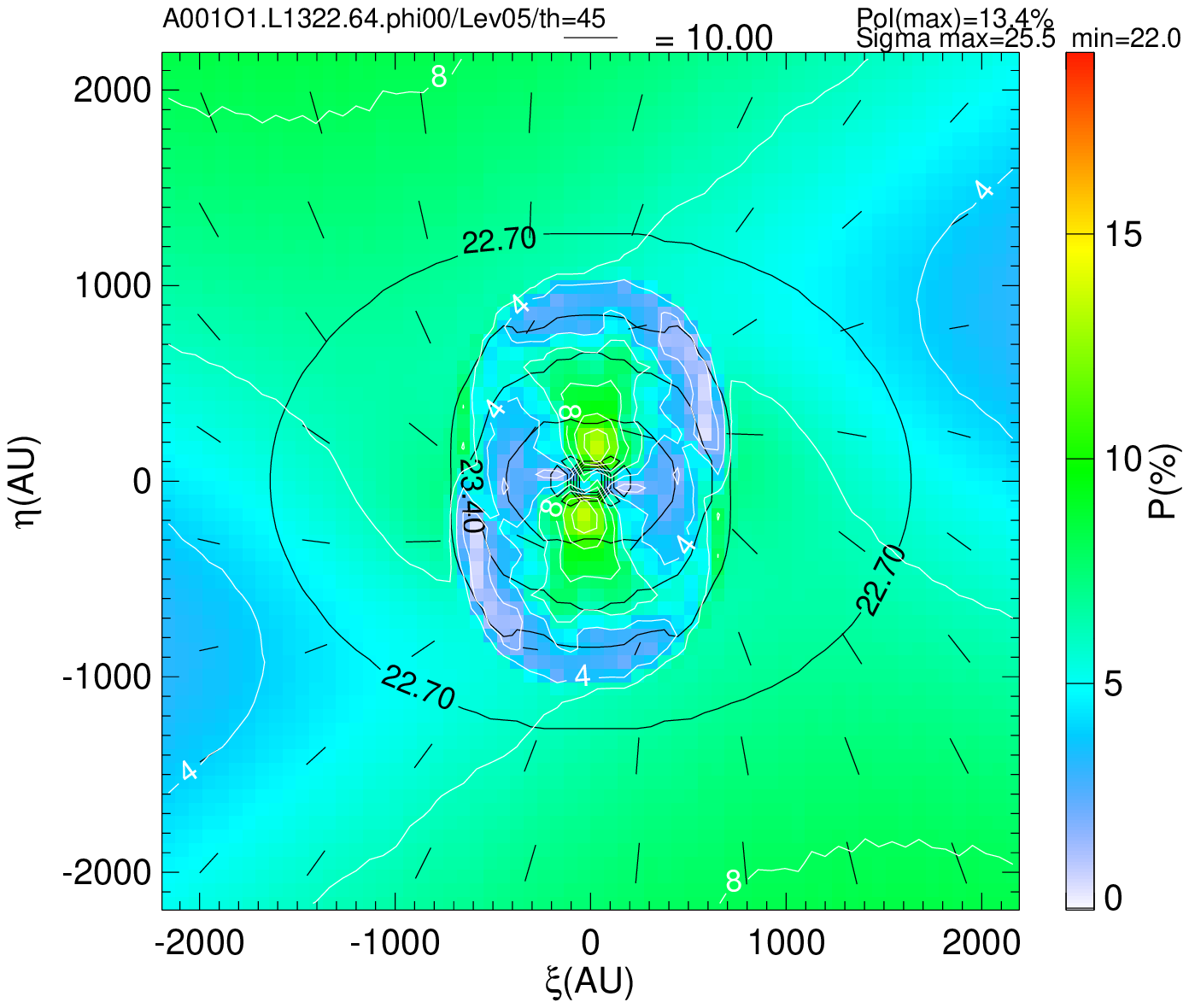}
      \one{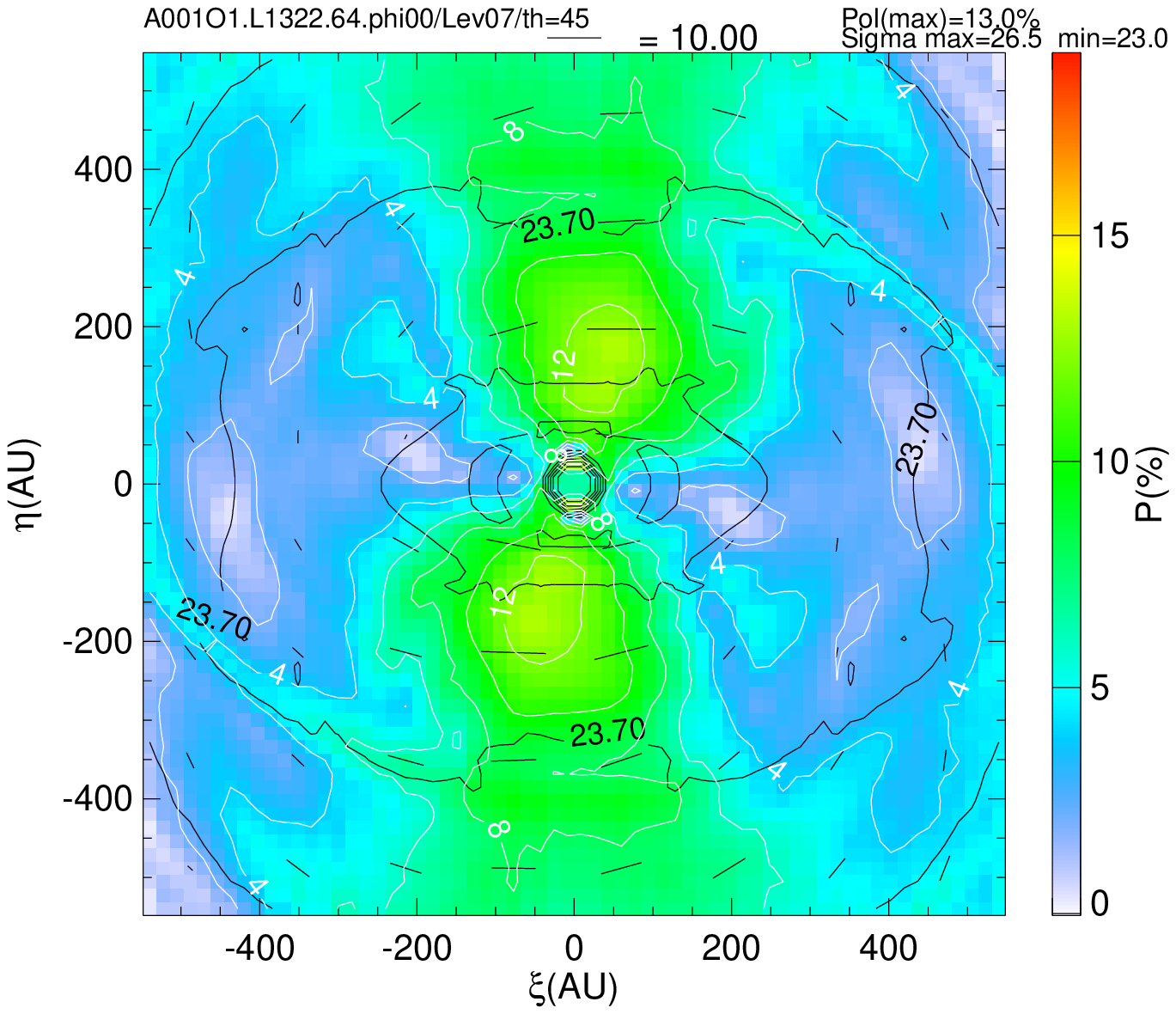}
      \one{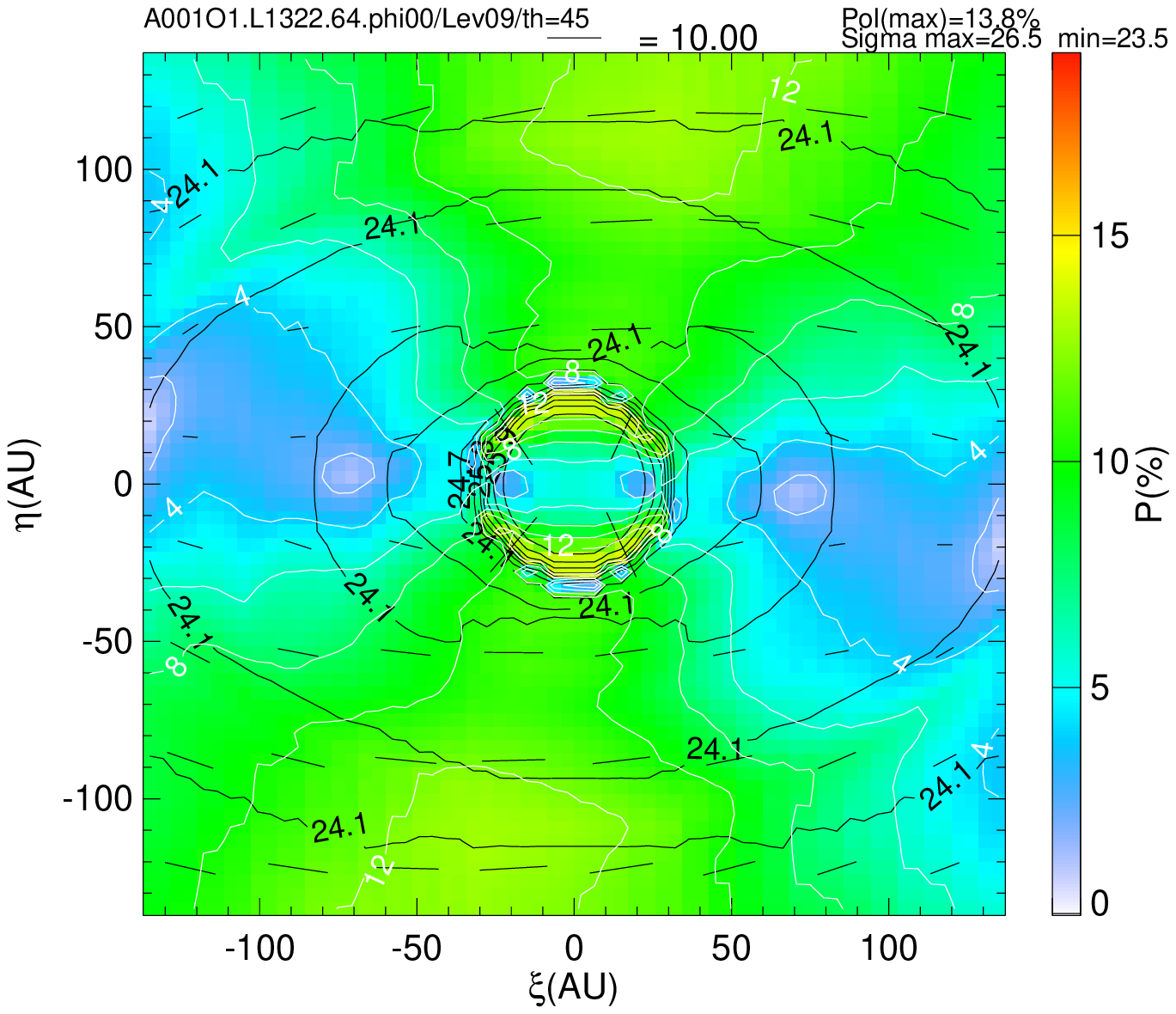}}\\
$\theta=60^\circ$\hspace*{10mm}
      \raisebox{-20mm}{\one{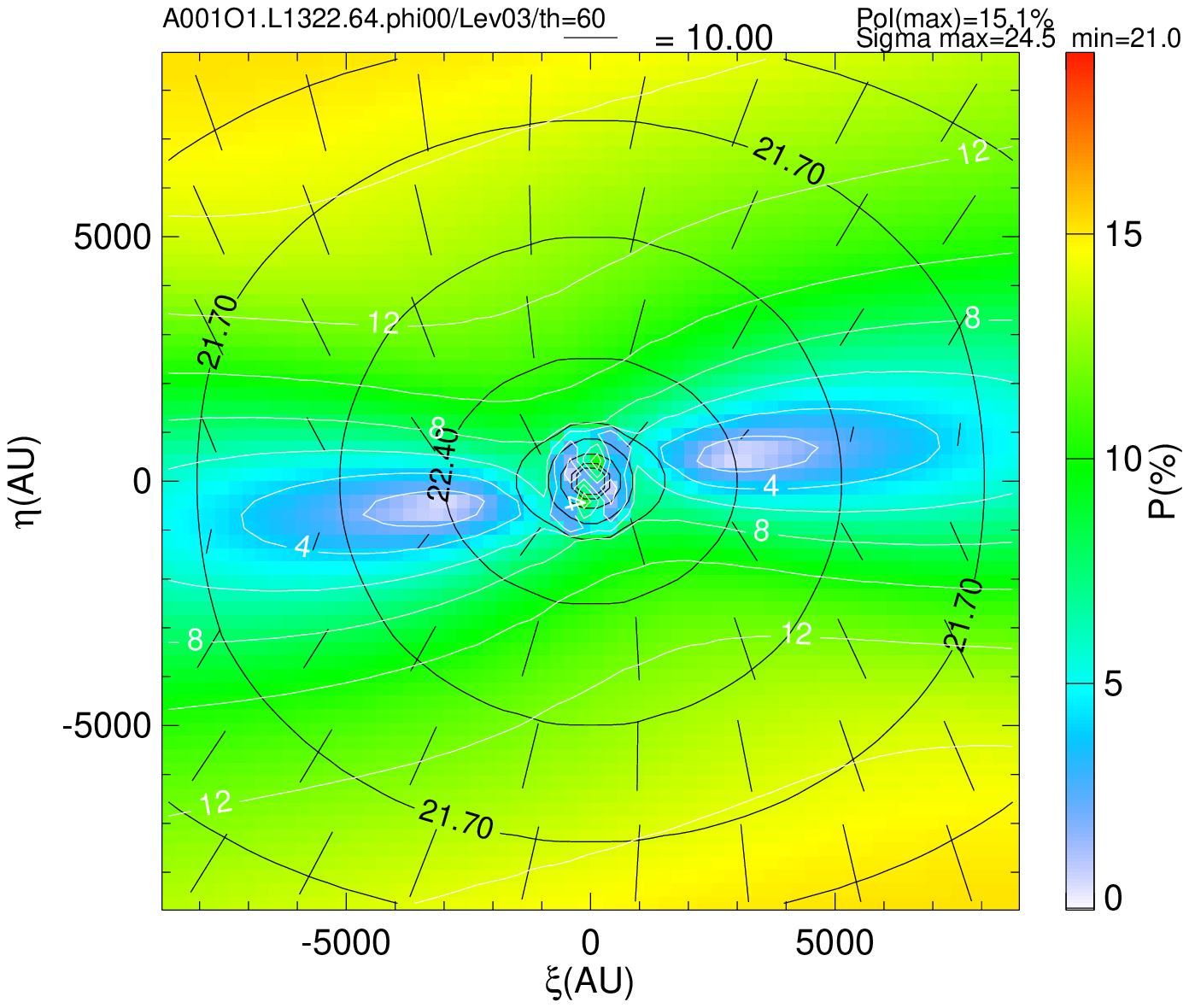}
      \one{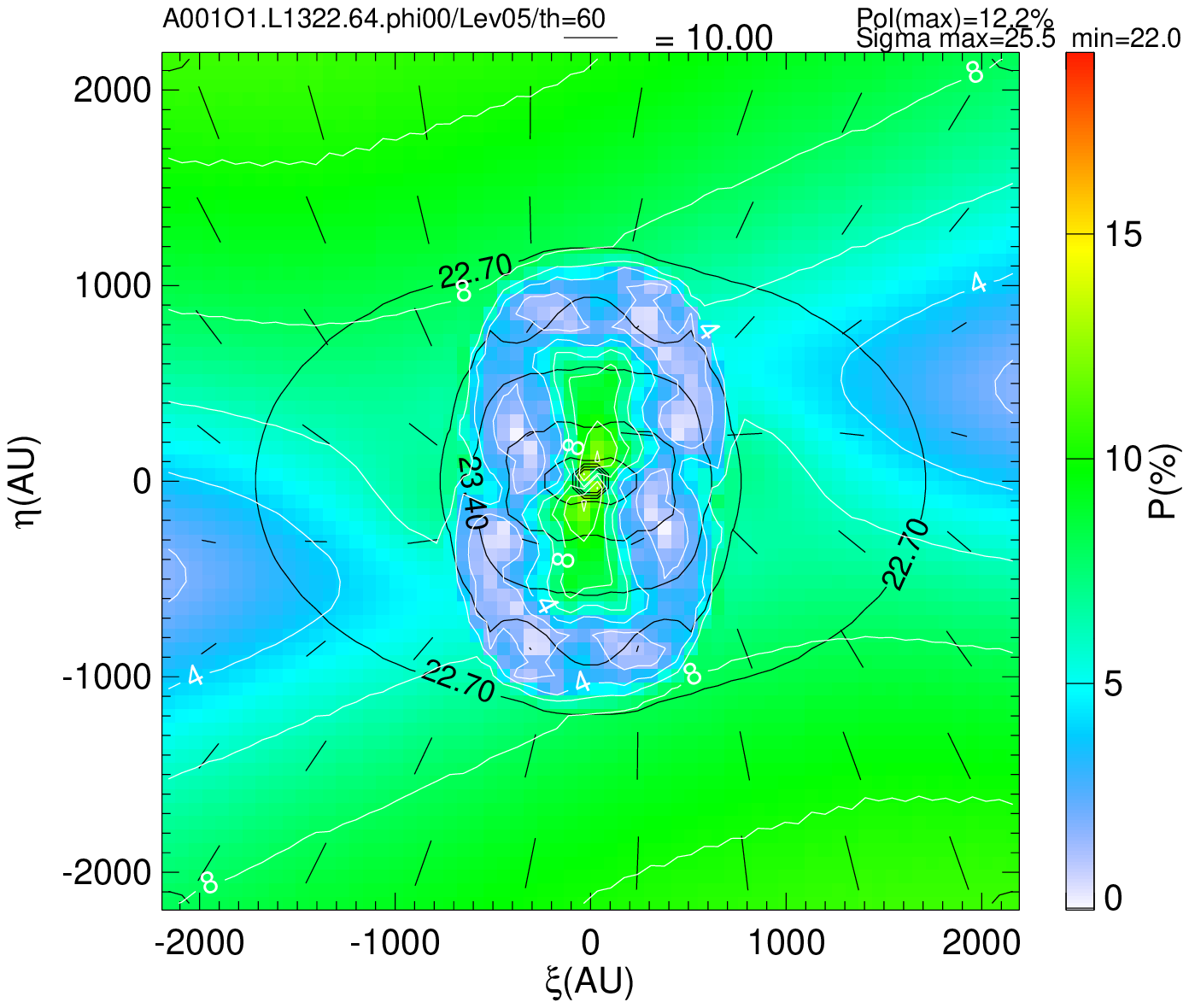}
      \one{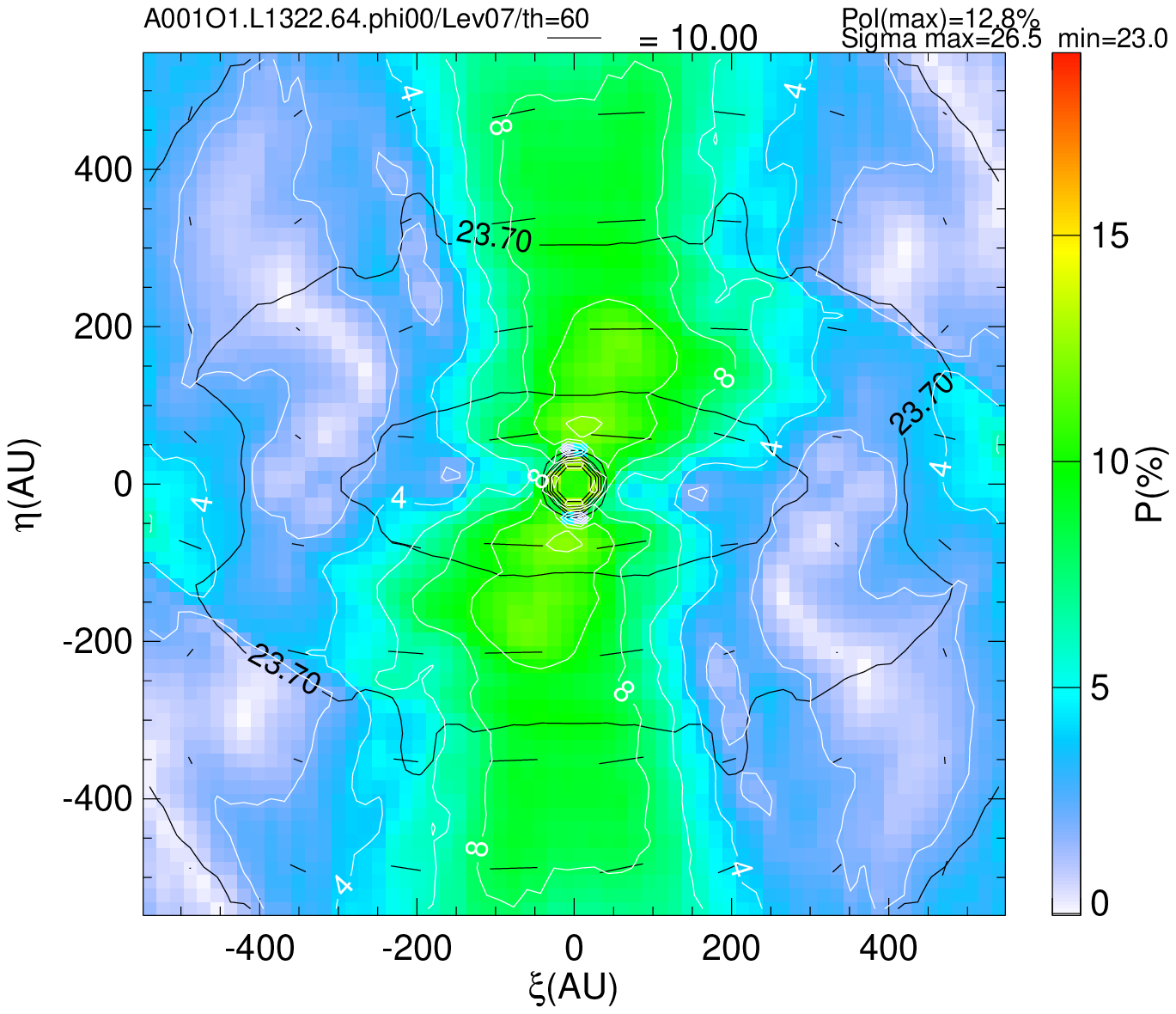}
      \one{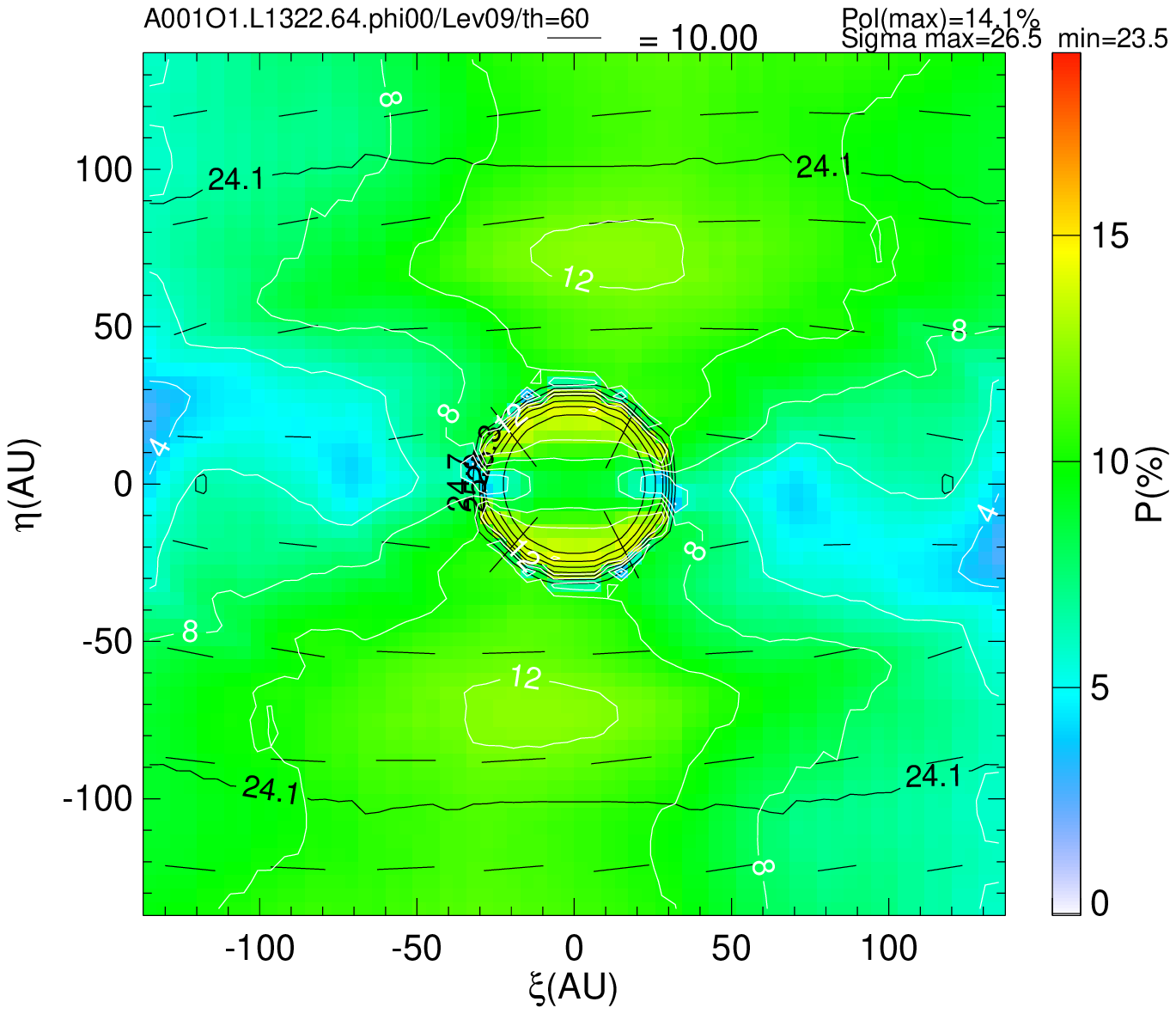}}\\
$\theta=80^\circ$\hspace*{10mm}
      \raisebox{-20mm}{\one{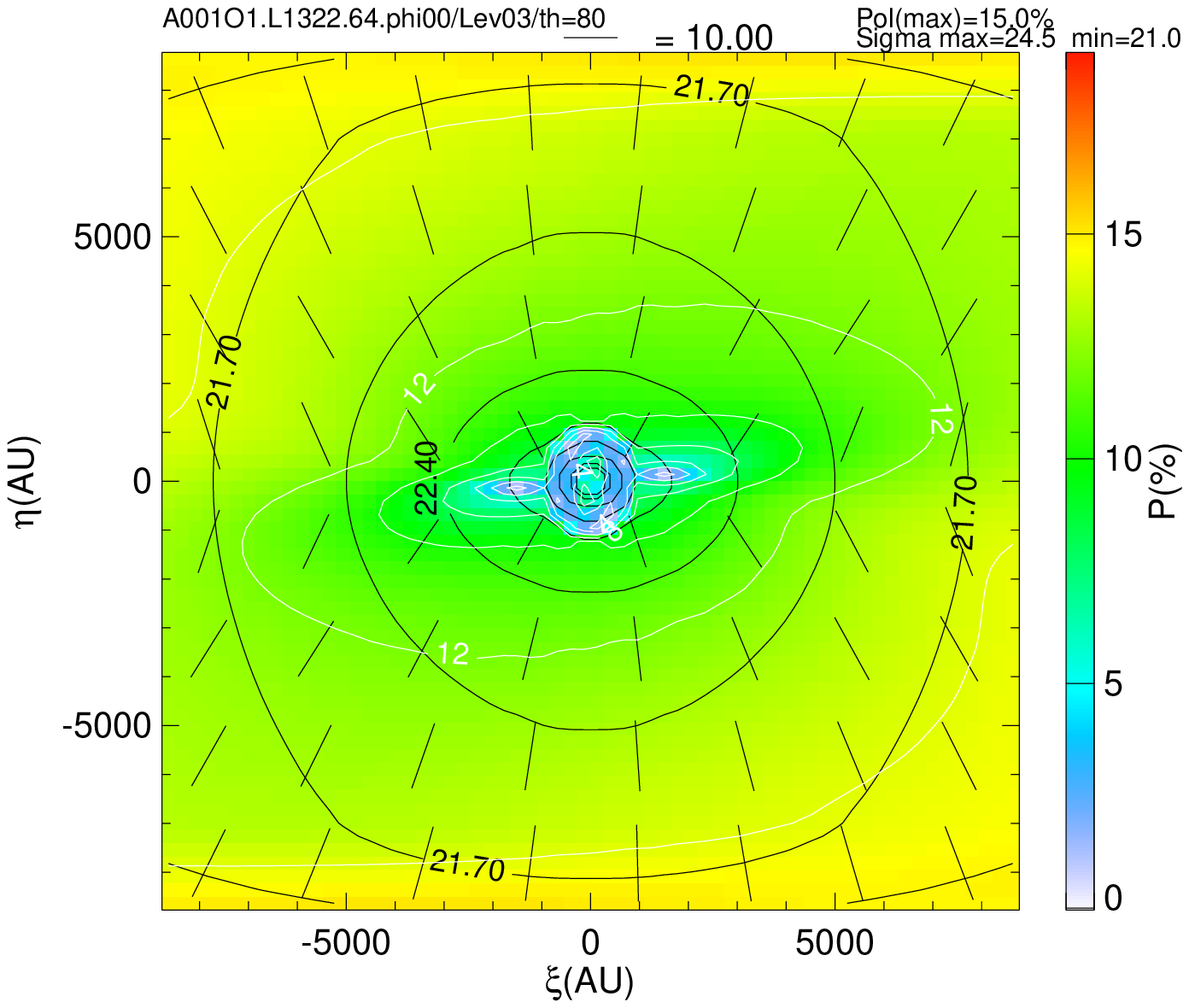}
      \one{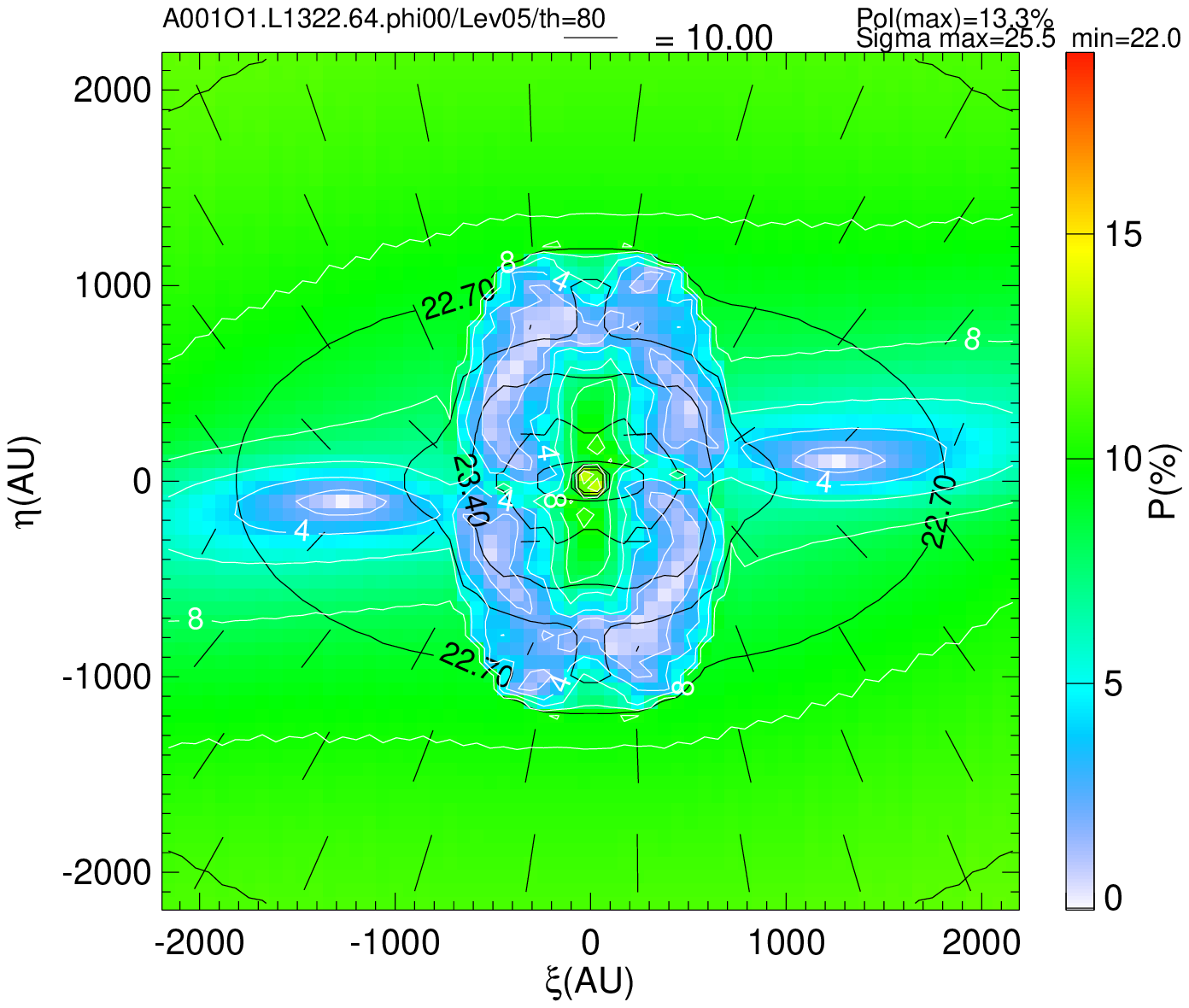}
      \one{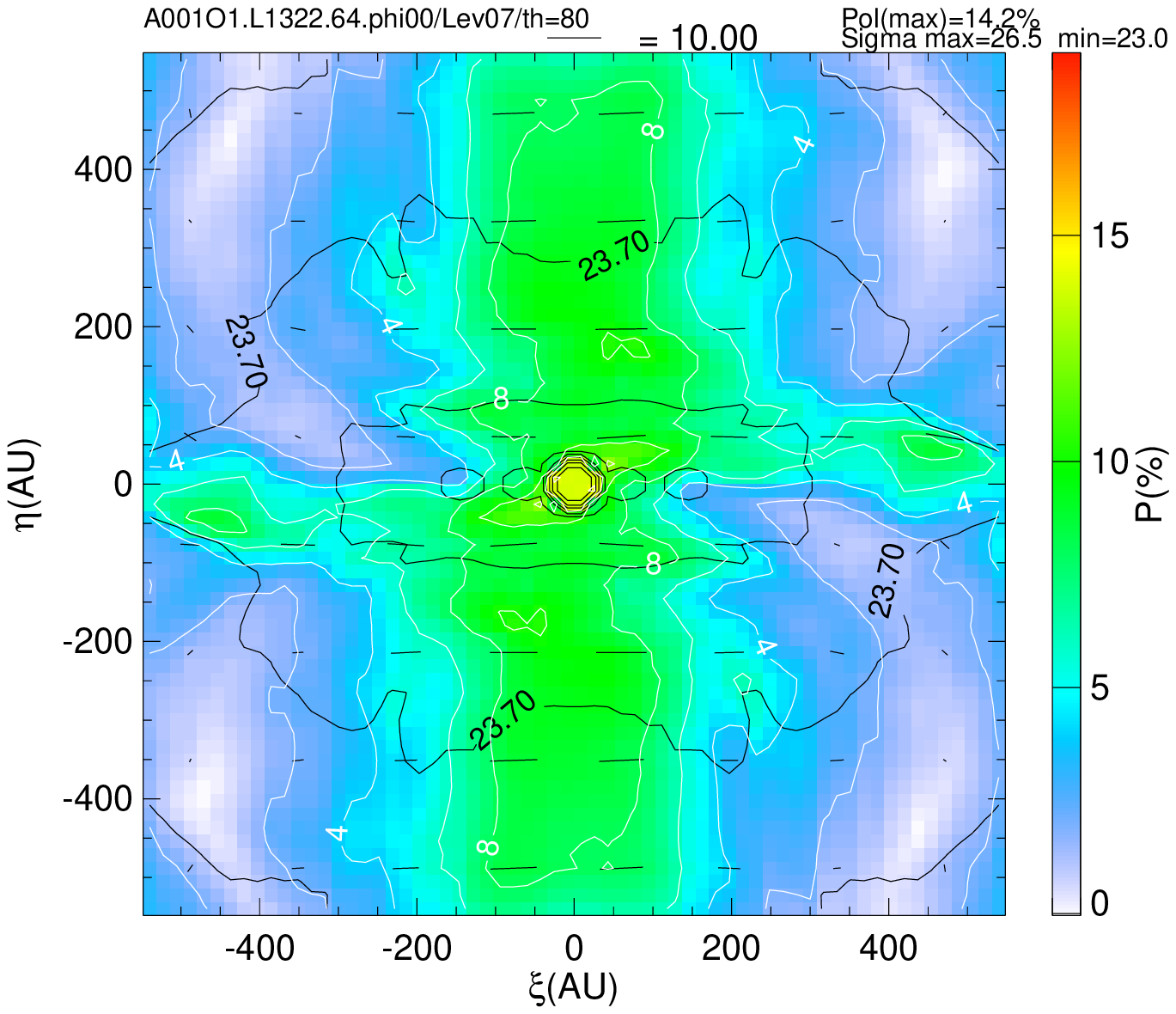}
      \one{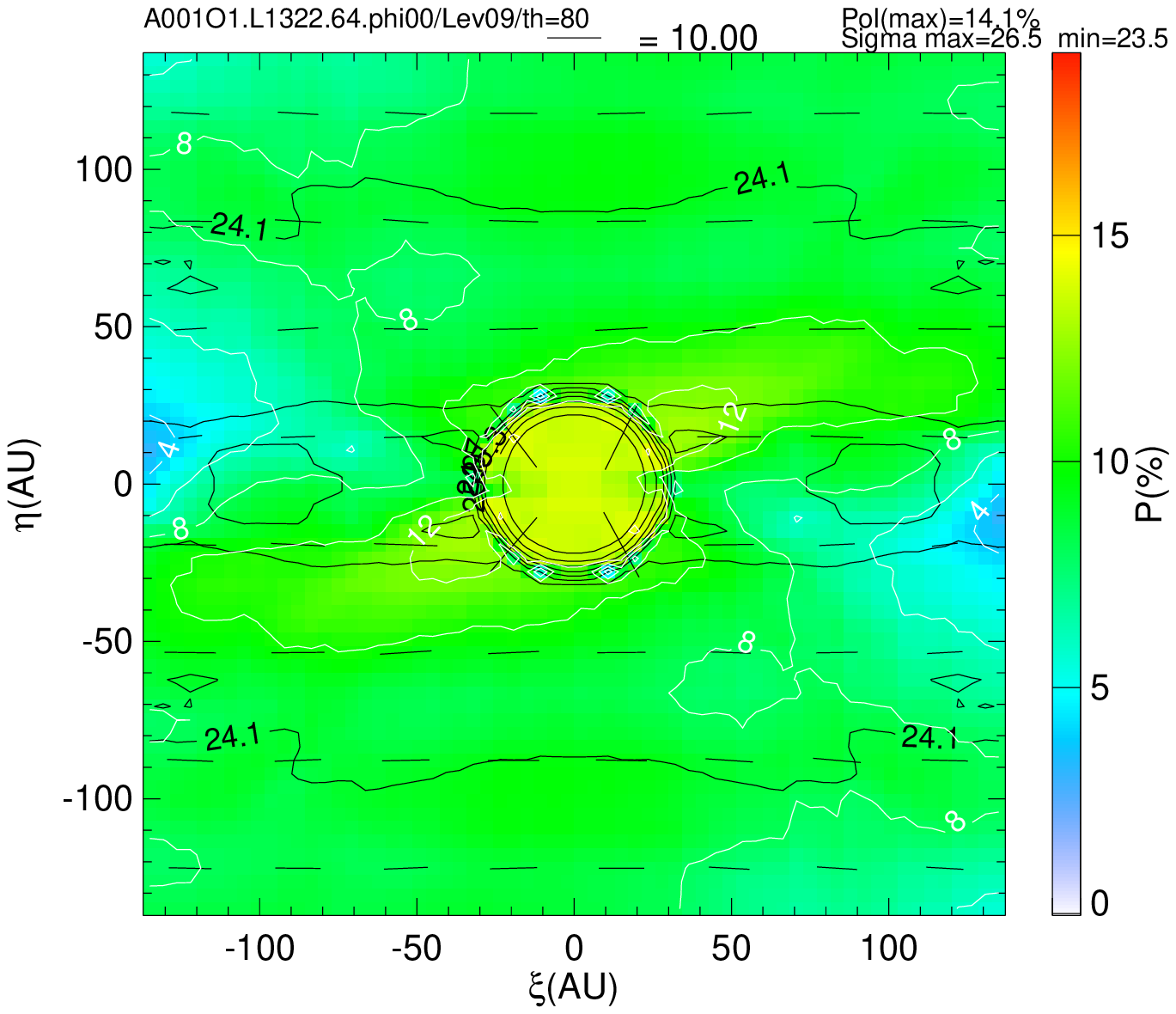}}\\
$\theta=90^\circ$\hspace*{10mm}
      \raisebox{-20mm}{\one{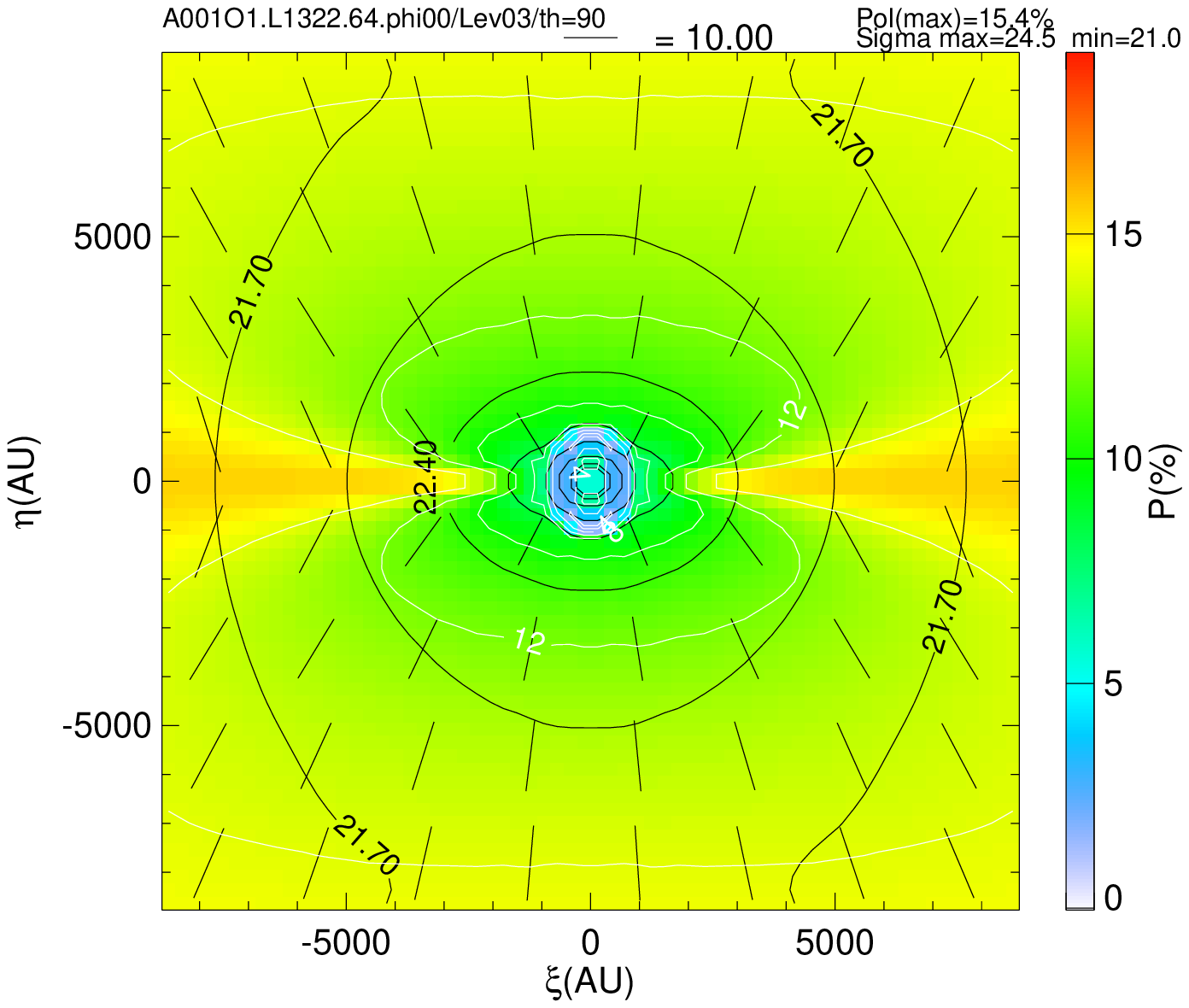}
      \one{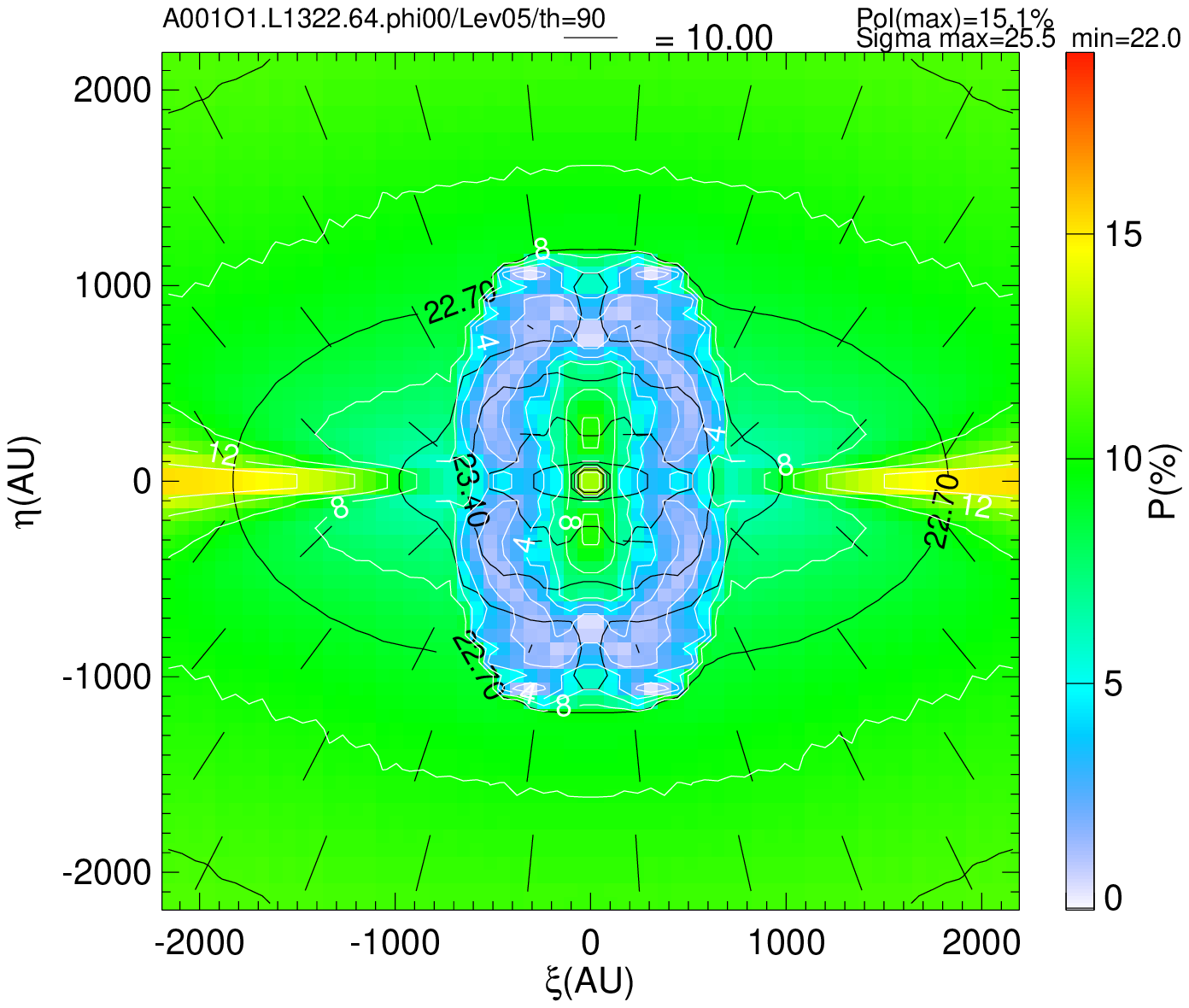}
      \one{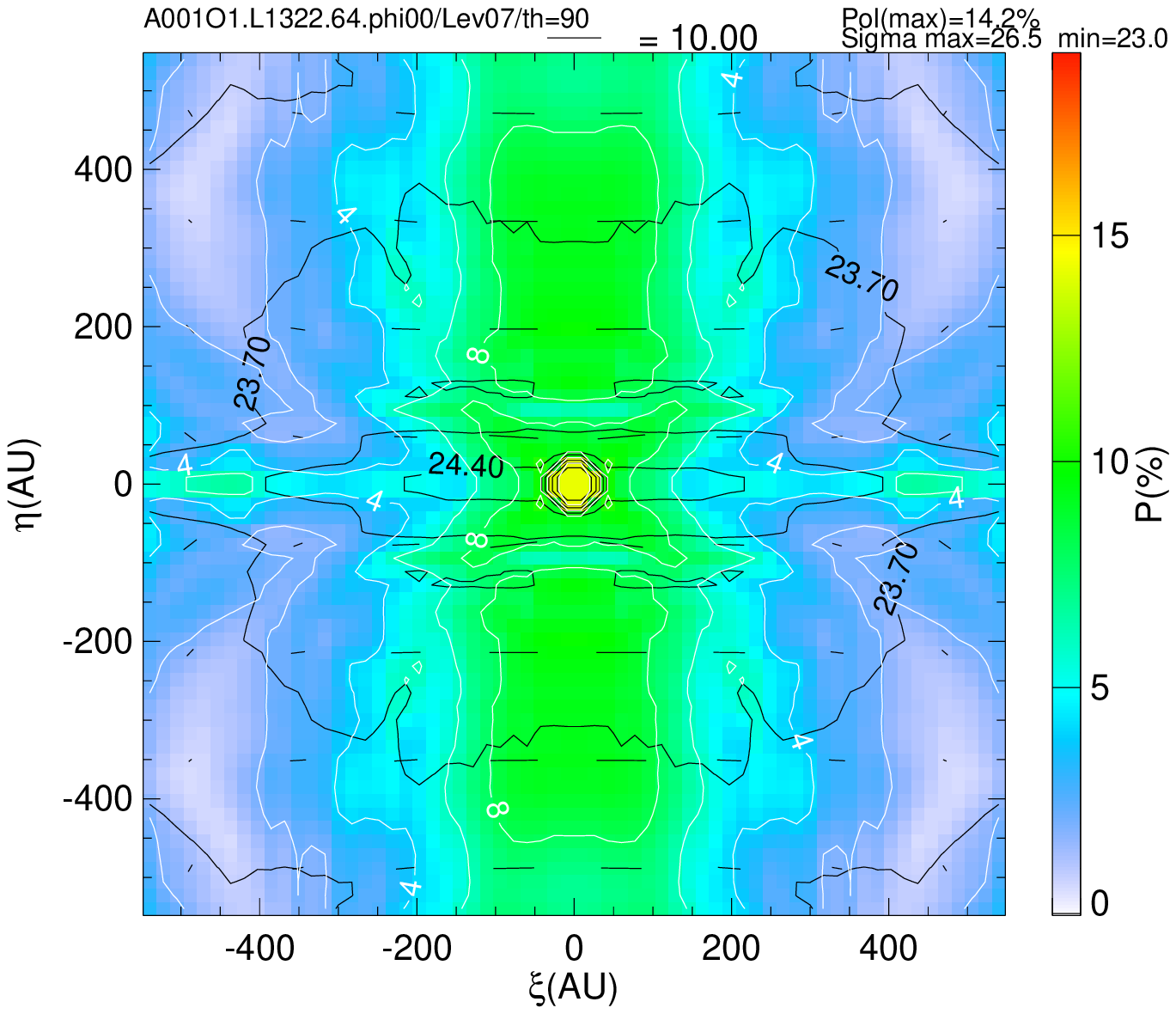}
      \one{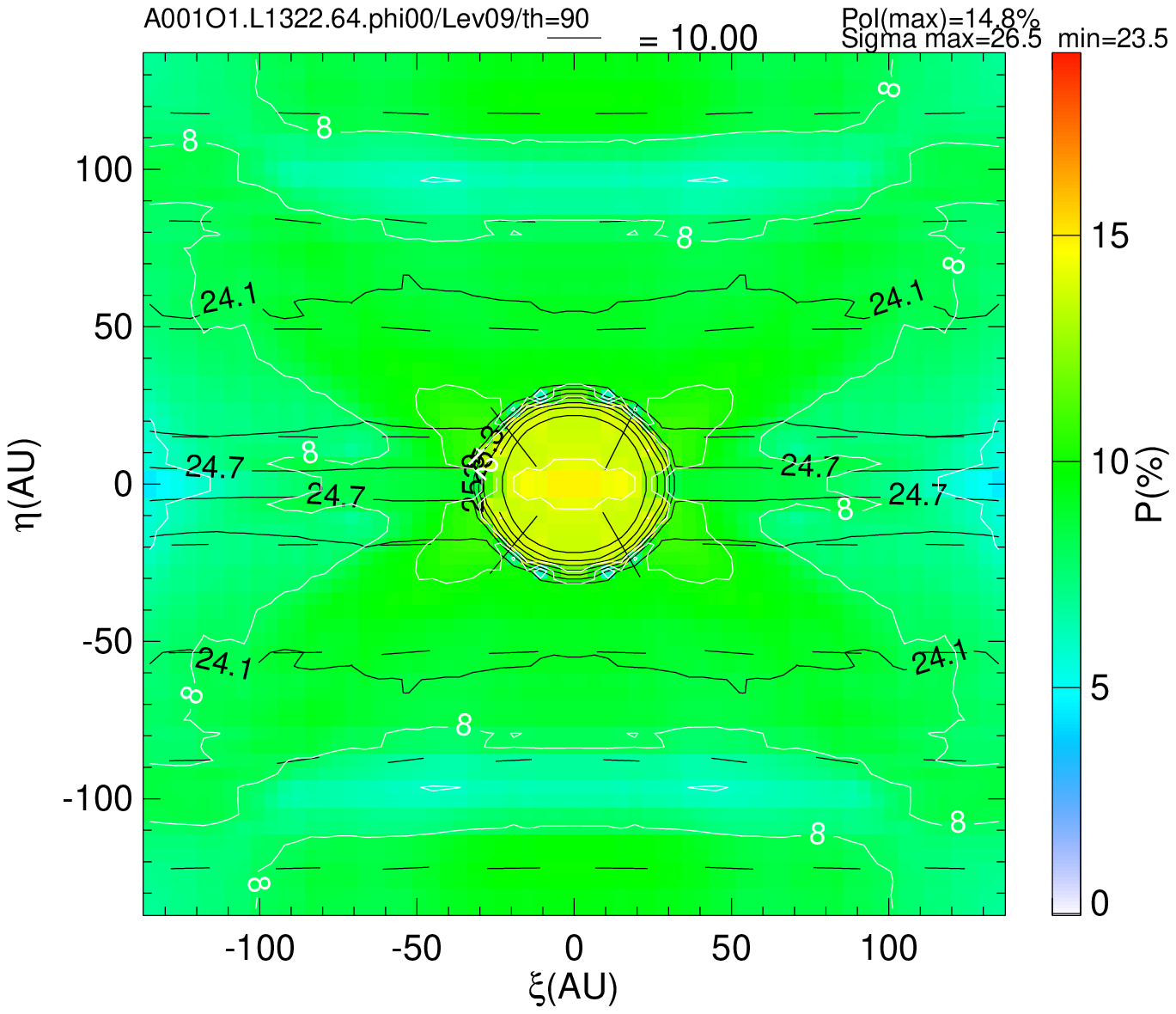}}
   \end{center}
\caption{\label{fig:A001O1L1322}As for Fig.\ref{fig:A1O5L1001}
 but for model EH with $\alpha=0.01$ and $\Omega'=1$.}
\end{figure}

\begin{figure}[h]
   \begin{center}
  \hspace*{45mm}(poloidal)\hspace*{30mm}(toroidal)\\
$\theta=0^\circ$\hspace*{10mm}\raisebox{-20mm}{\one{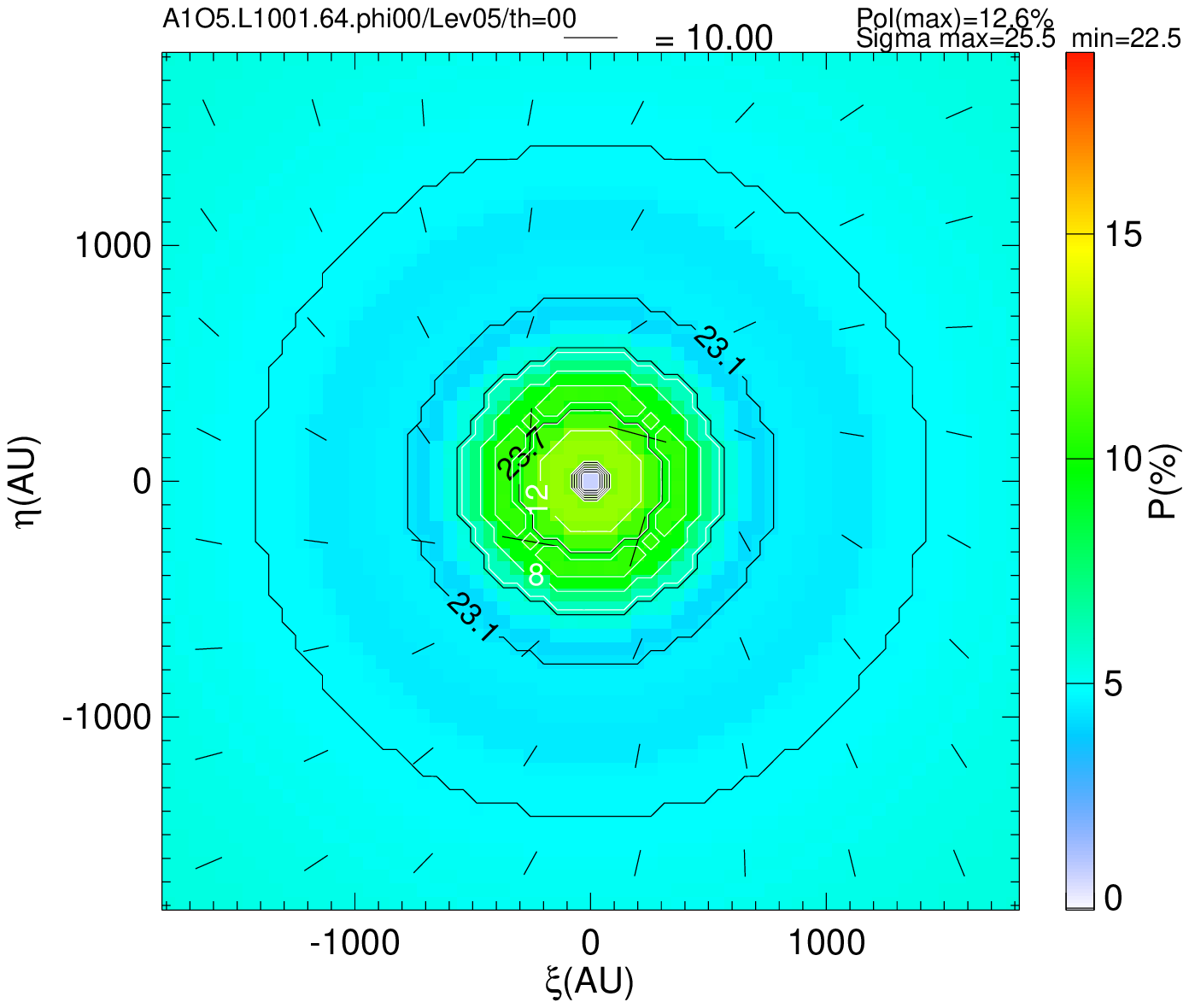}
      \hspace*{5mm}\one{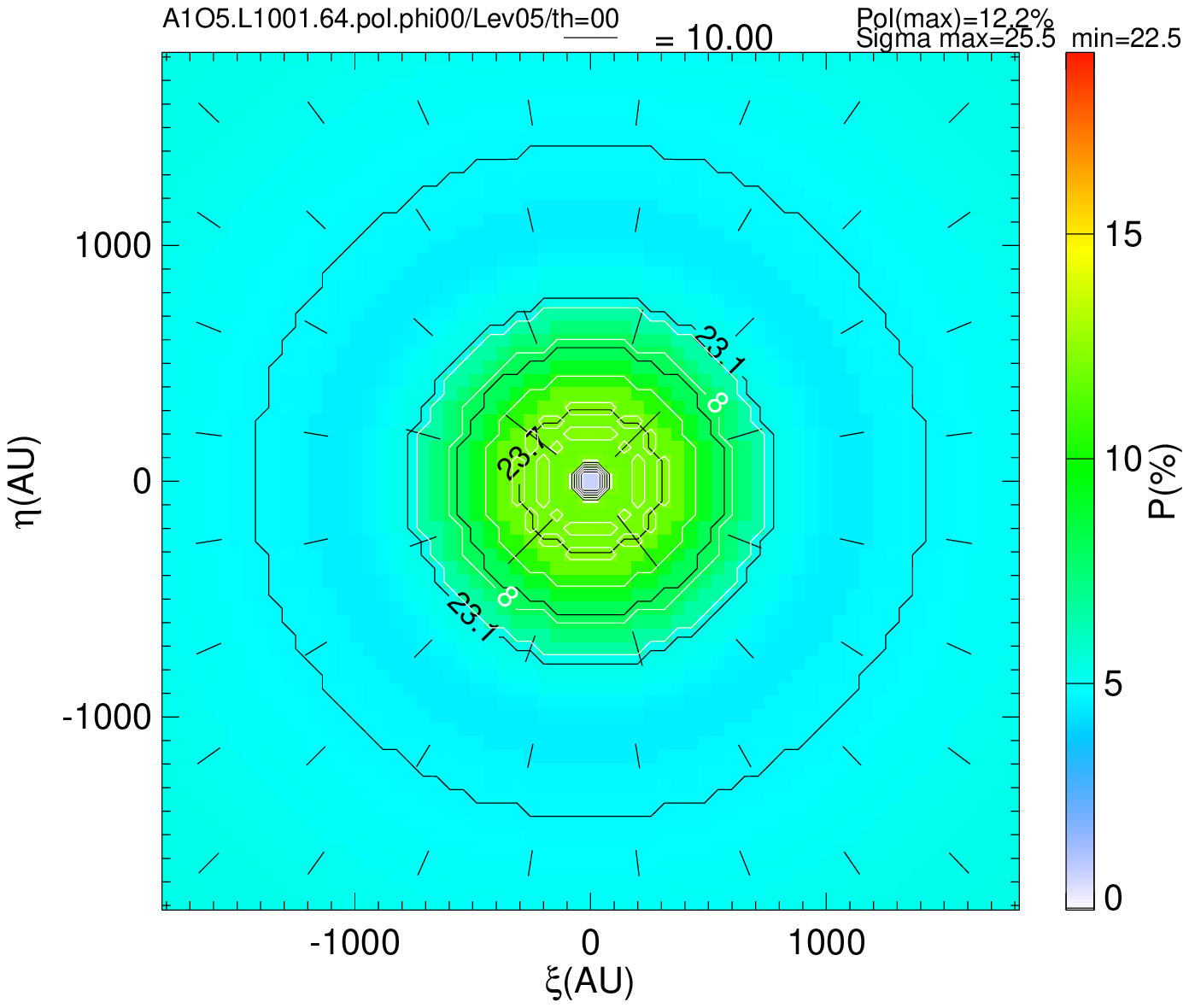} 
      \hspace*{5mm}\one{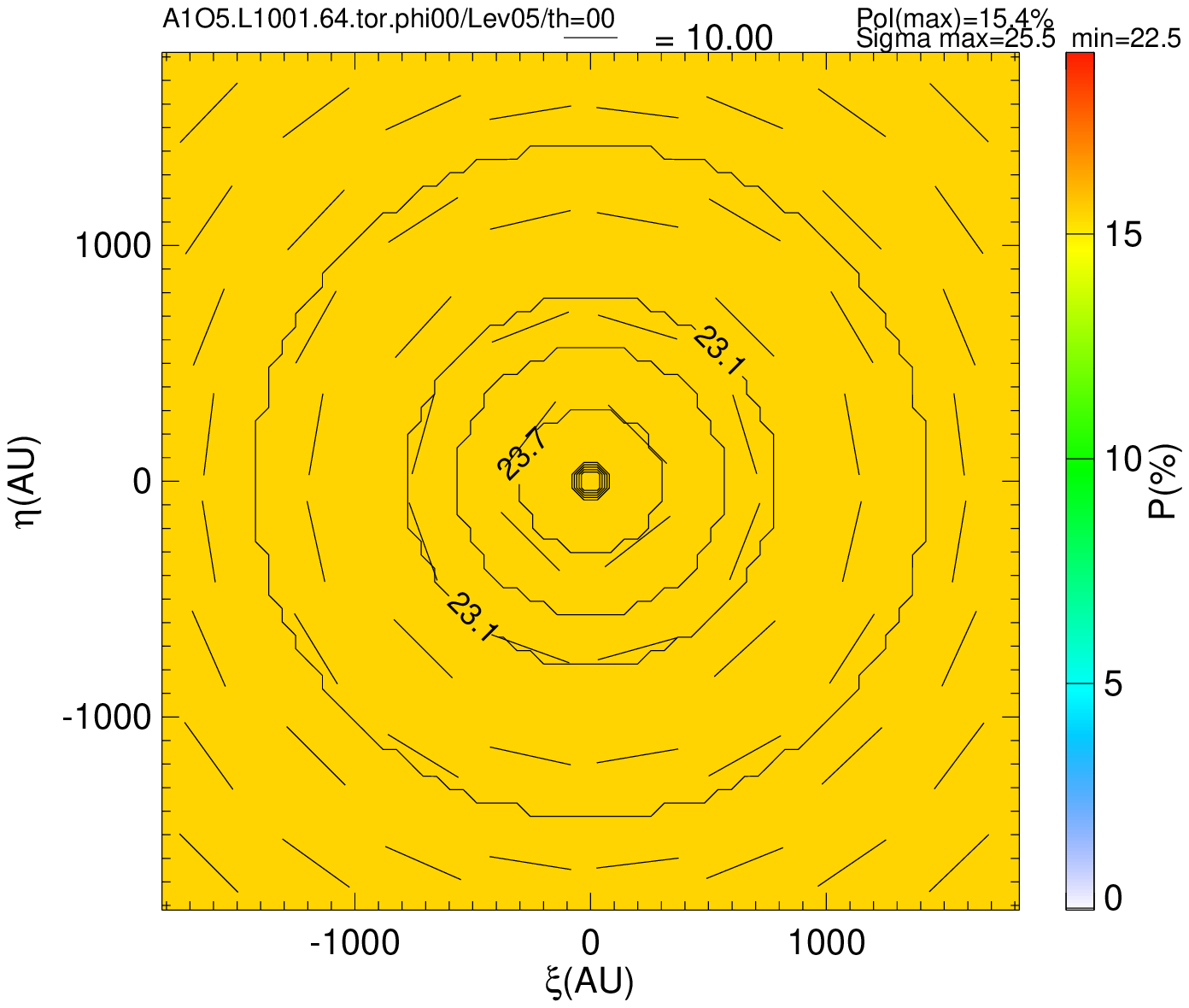}}\\[-0mm]
$\theta=45^\circ$\hspace*{10mm}\raisebox{-20mm}{\one{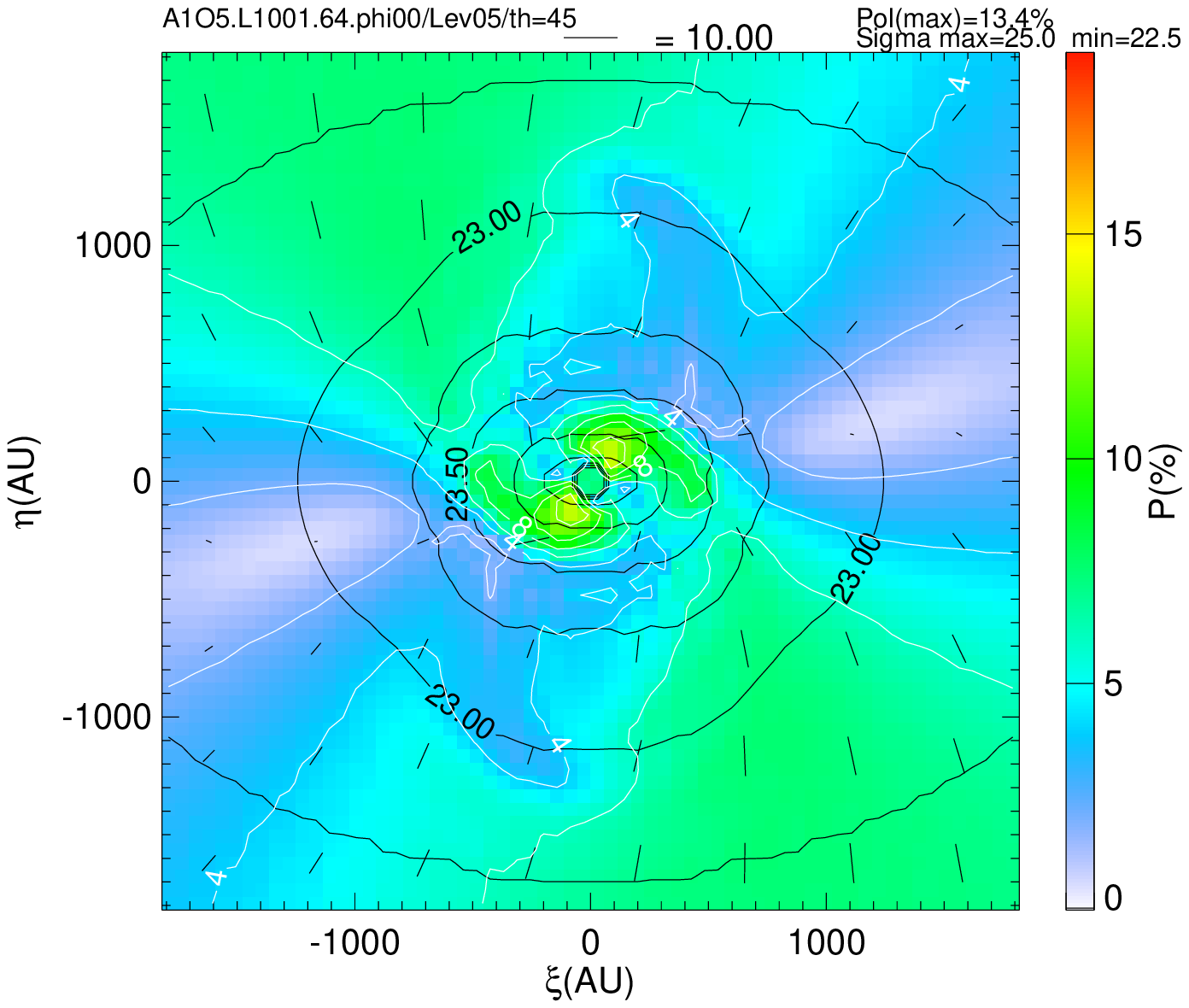}
      \hspace*{5mm}\one{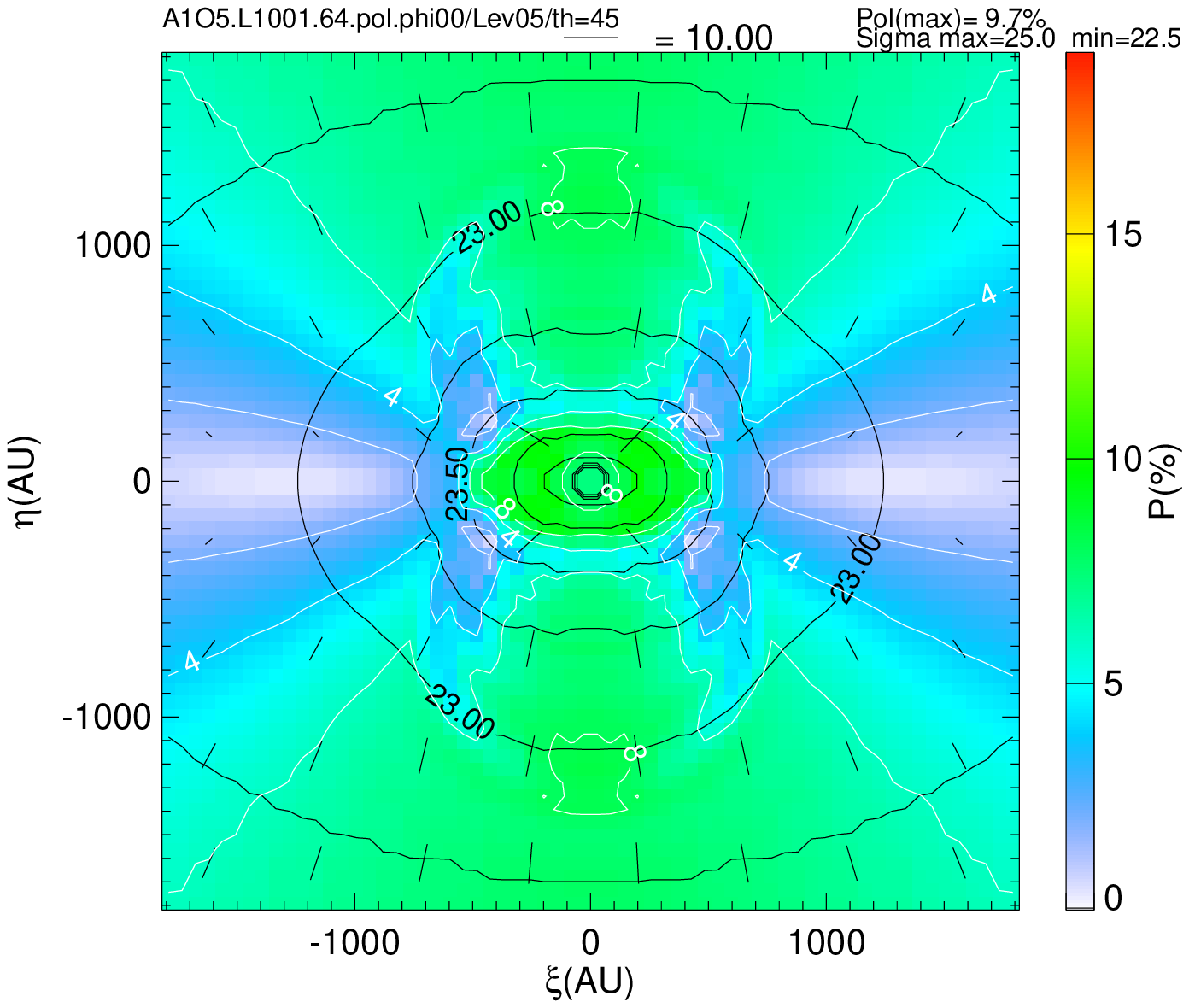}
      \hspace*{5mm}\one{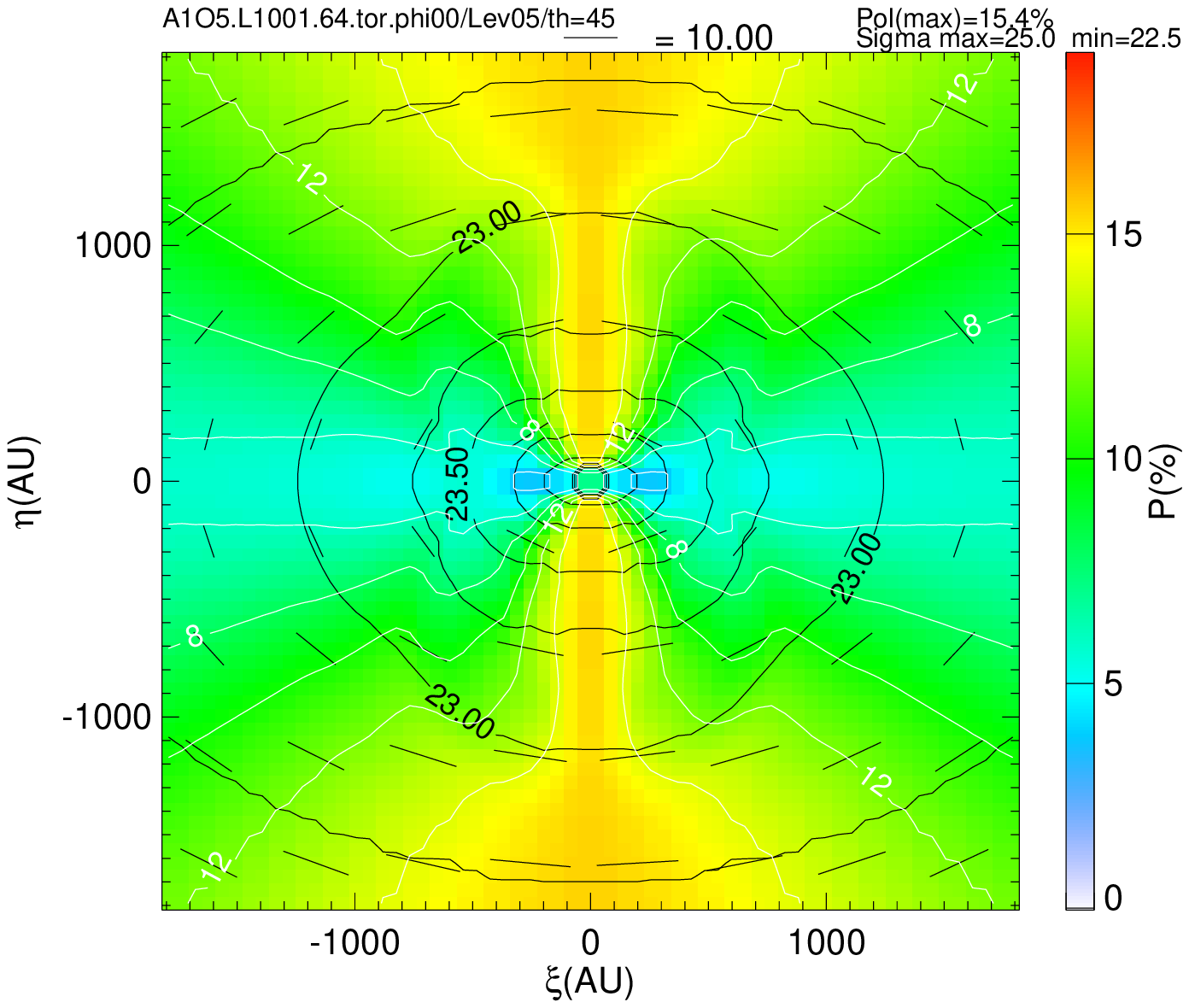}}\\[-0mm]
$\theta=90^\circ$\hspace*{10mm}\raisebox{-20mm}{\one{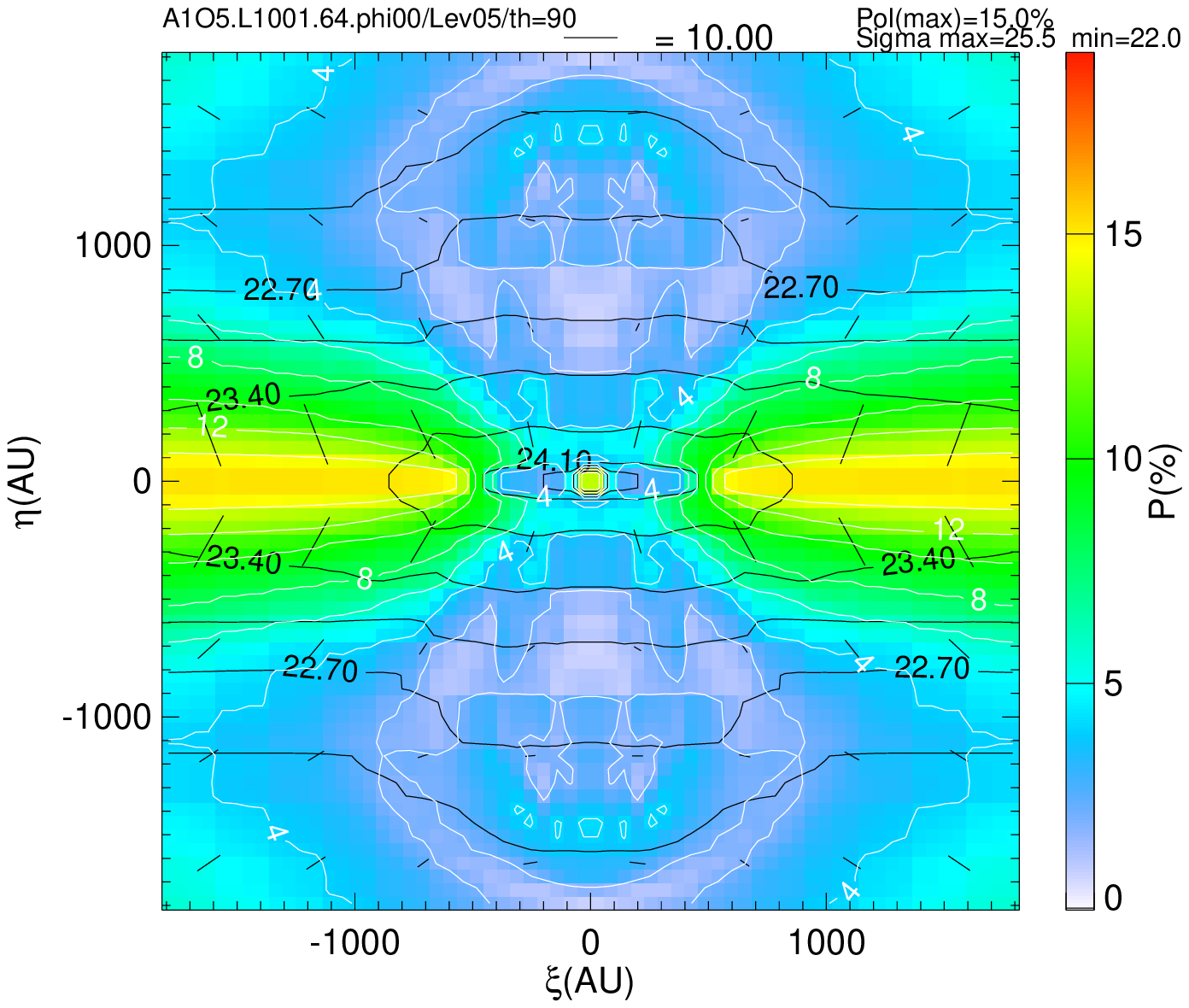}
      \hspace*{5mm}\one{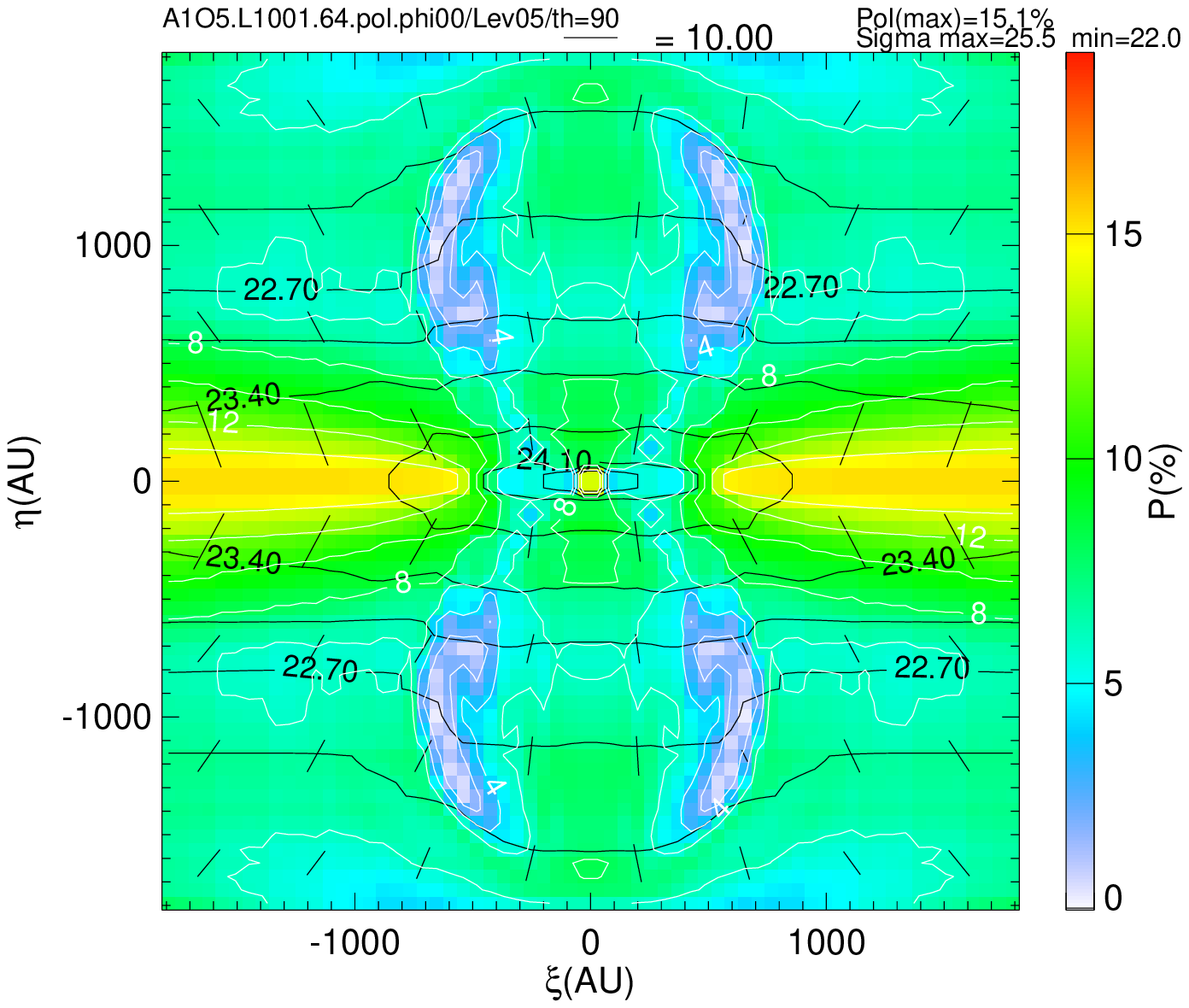}
      \hspace*{5mm}\one{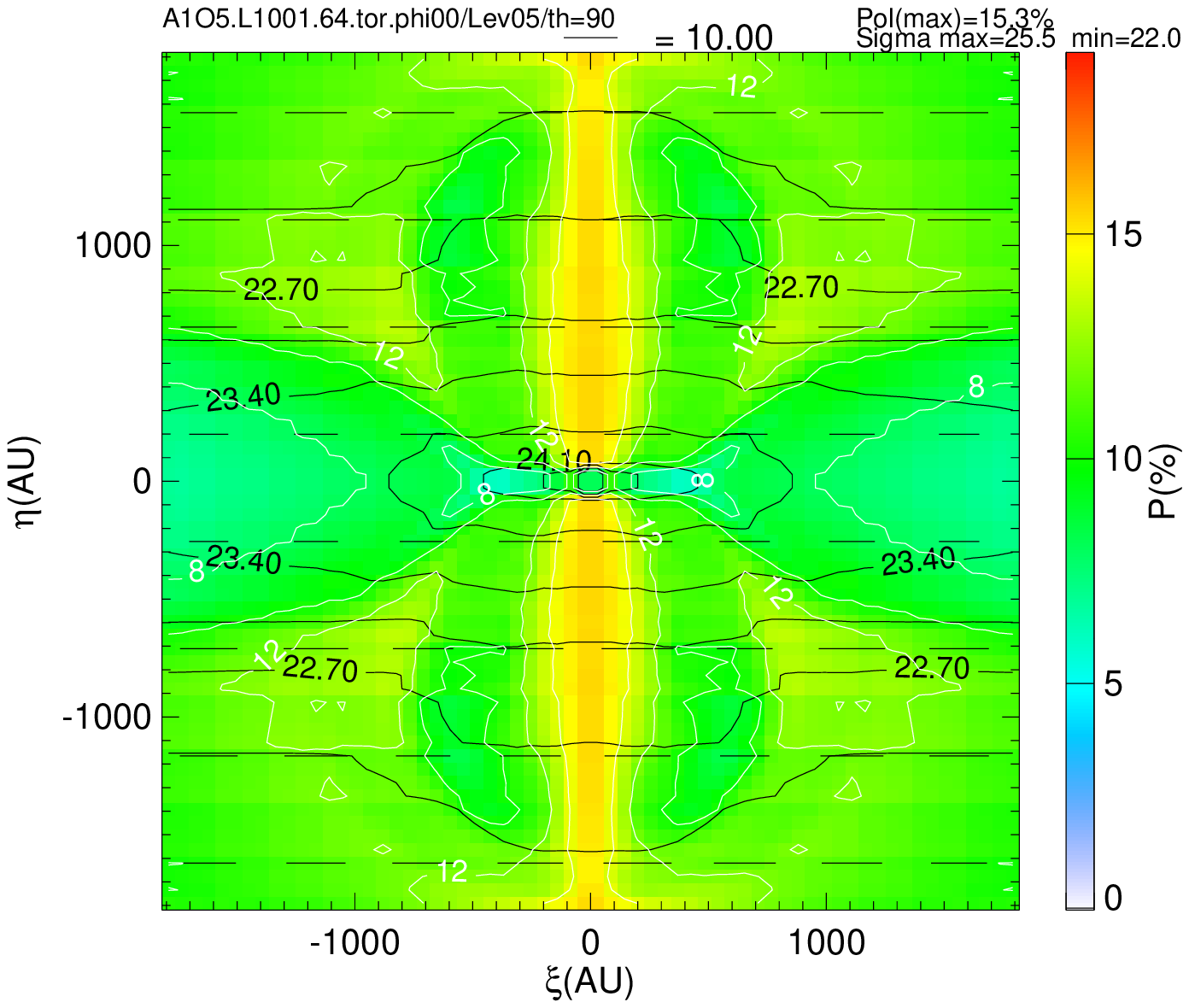}}
   \end{center}
\caption{\label{fig:1001A1O5L5poltor}As for Fig.~\ref{fig:A1O5L1001} 
  (protostellar phases).
 However, the middle and right columns are results
 for artificial data consisting only of poloidal and toroidal magnetic fields,
 respectively.
 Level 5 of model AH1 with $\alpha=1$ and $\Omega'=5$.}
\end{figure}

\begin{figure}[h]
   \begin{center}
\hspace*{45mm}(poloidal)\hspace*{30mm}(toroidal)\\
$\theta=0^\circ$\hspace*{10mm}\raisebox{-20mm}{\one{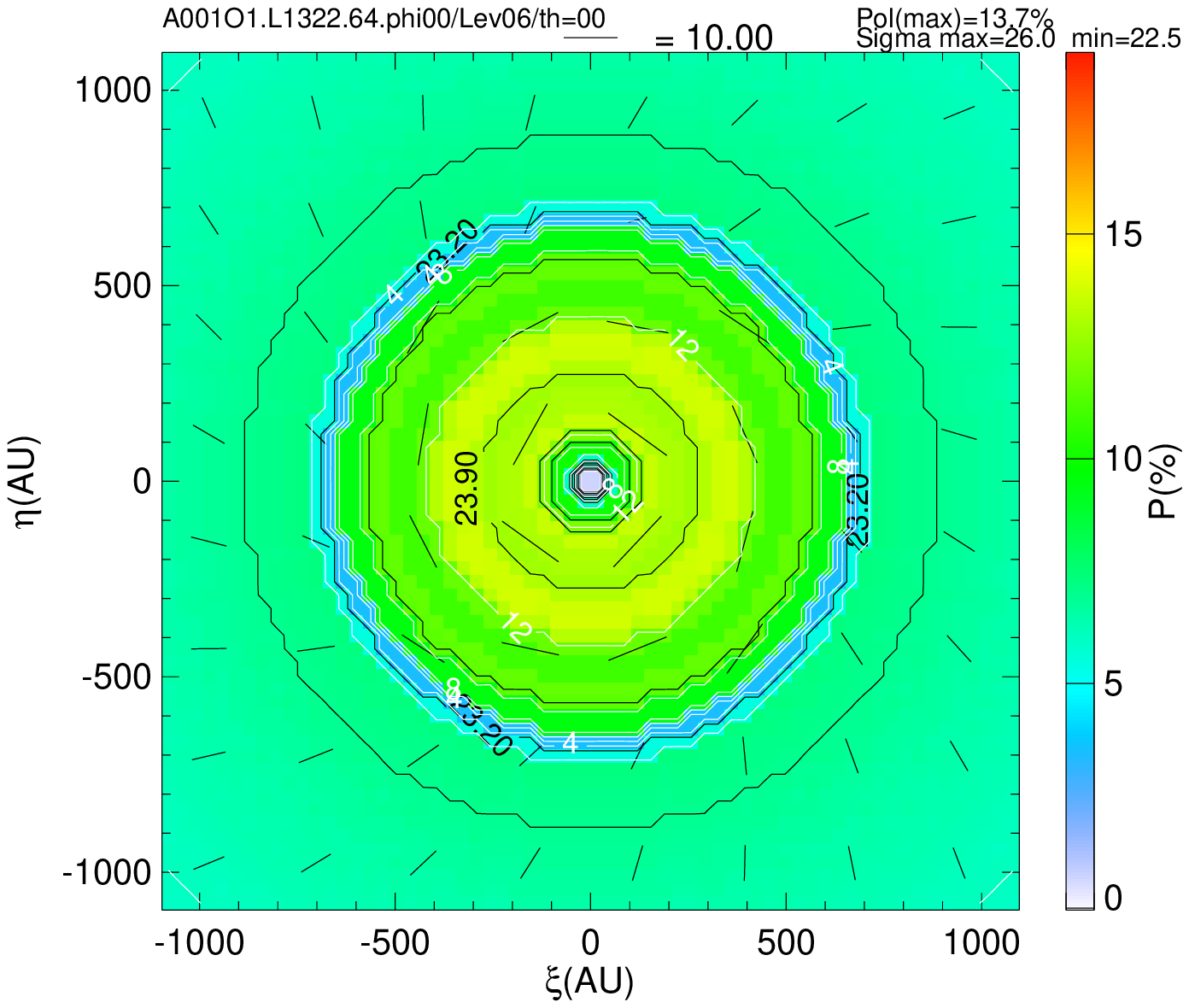}
      \hspace*{5mm}\one{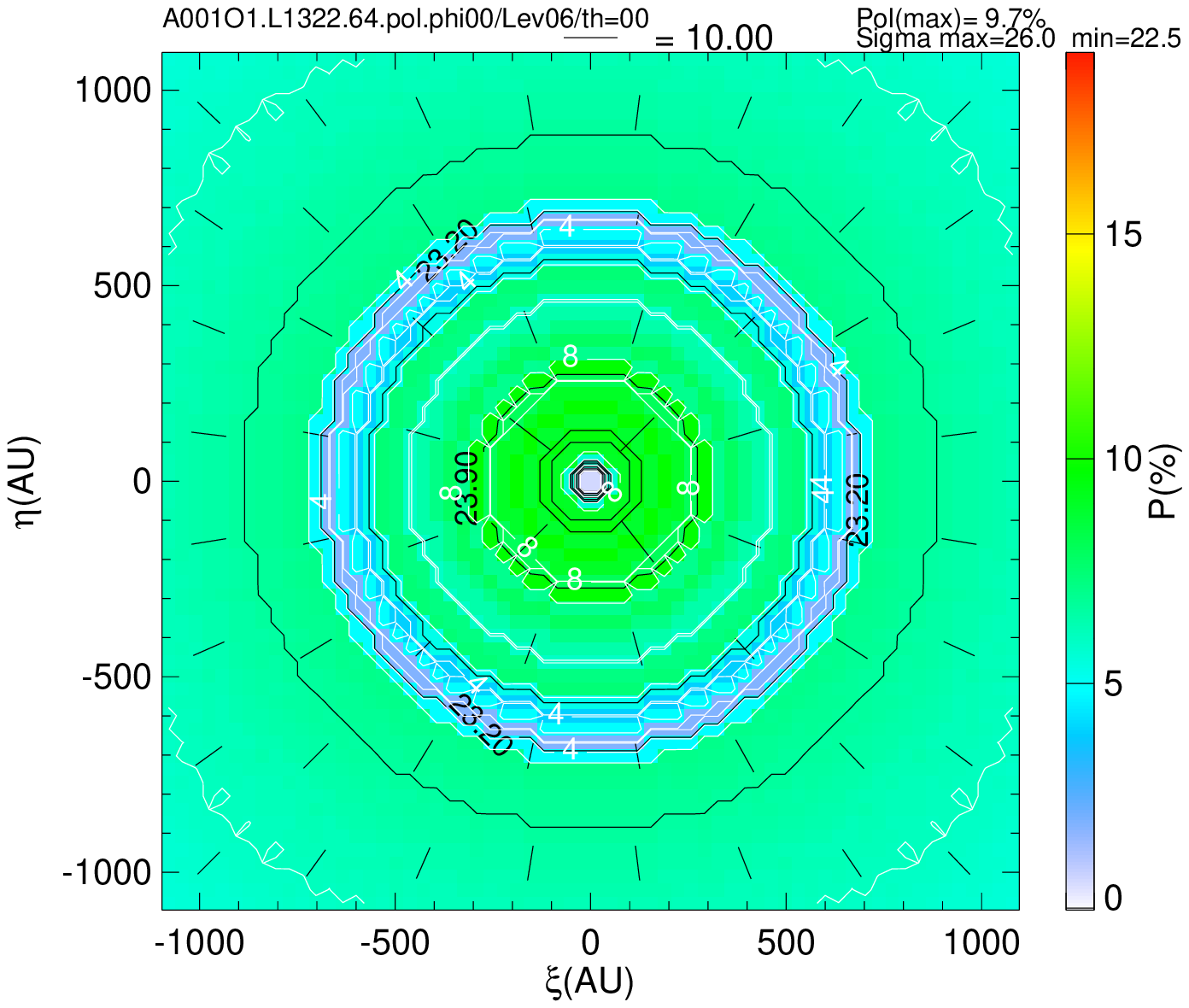} 
      \hspace*{5mm}\one{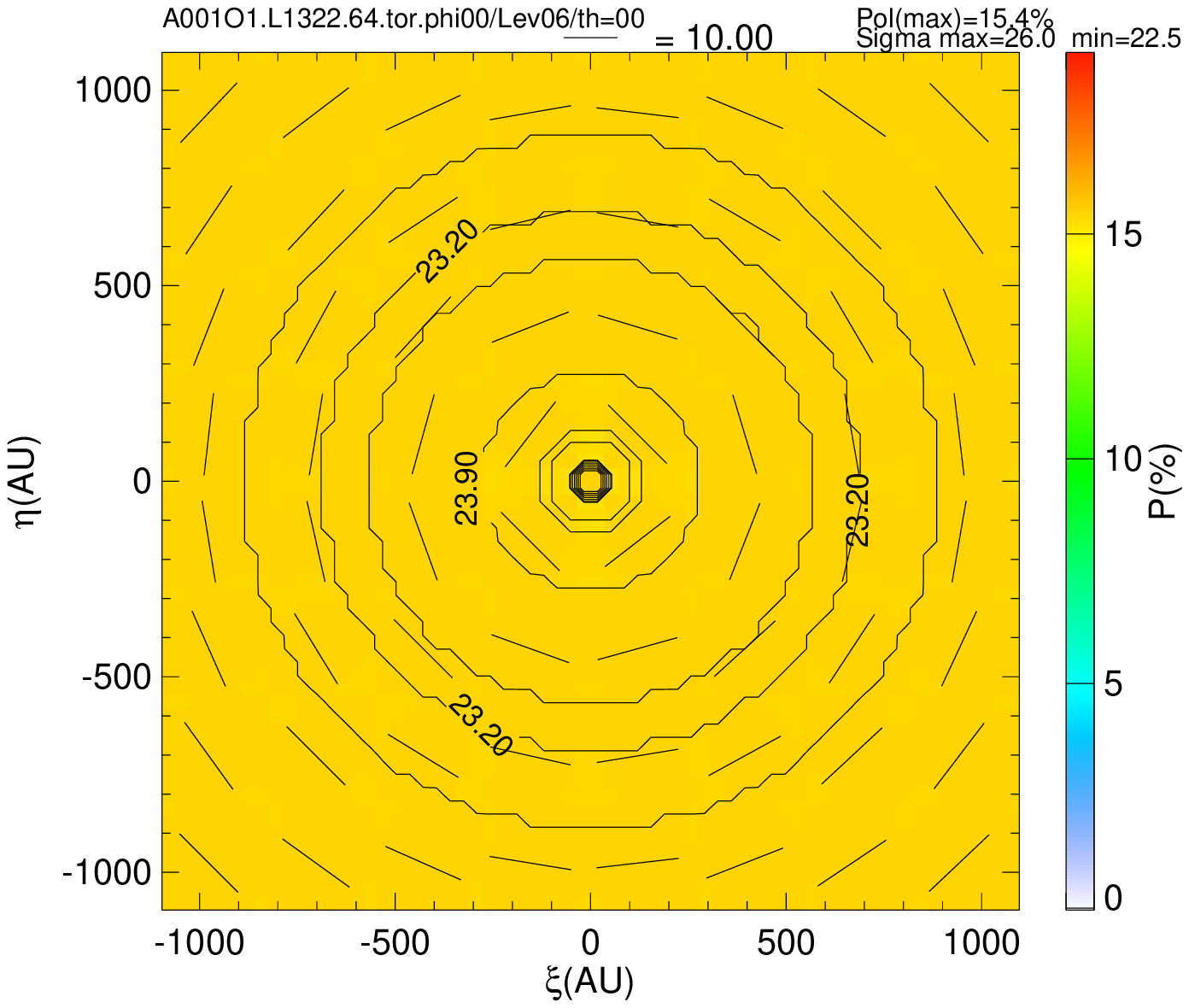}}\\[-0mm]
$\theta=45^\circ$\hspace*{10mm}\raisebox{-20mm}{\one{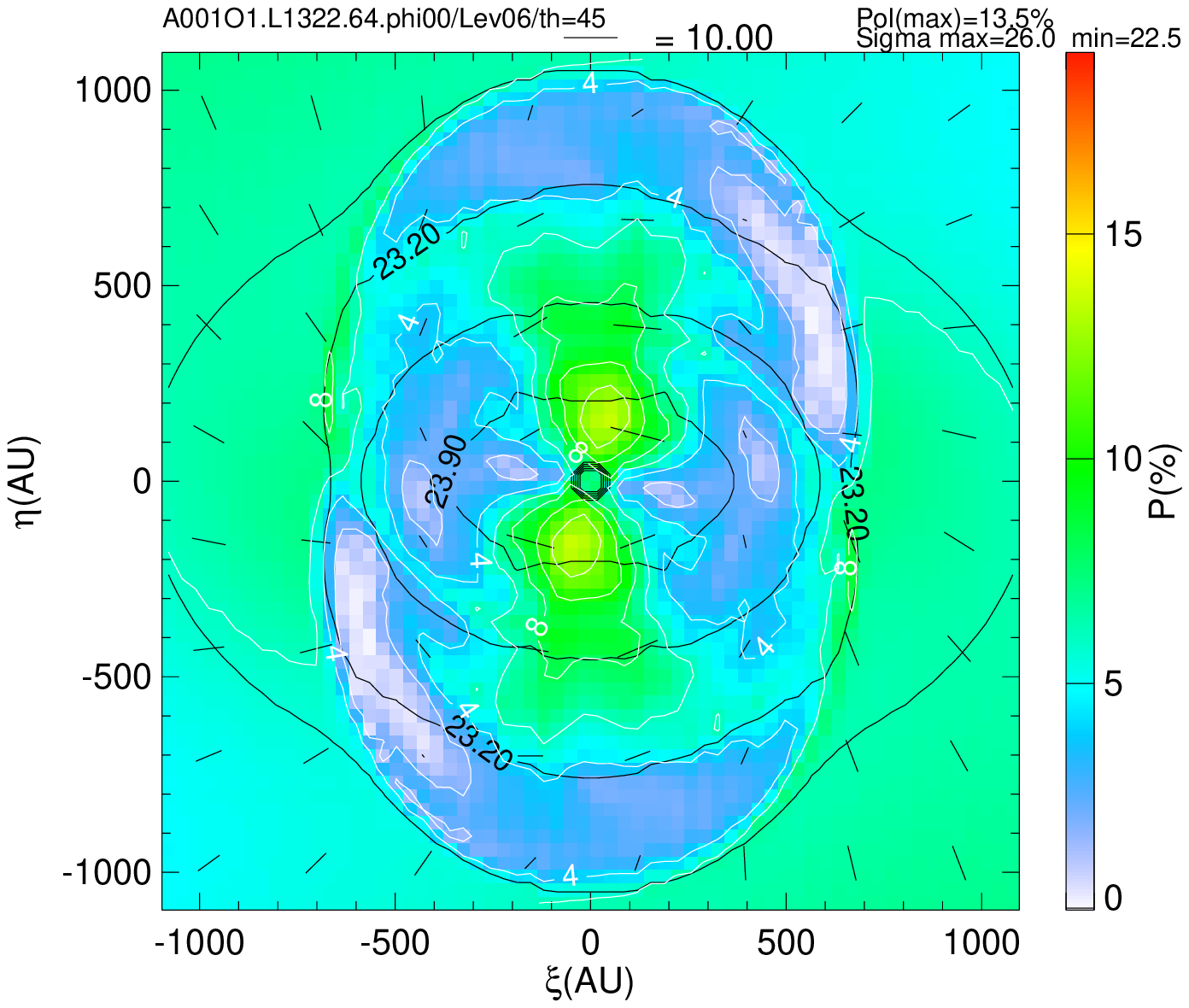}
      \hspace*{5mm}\one{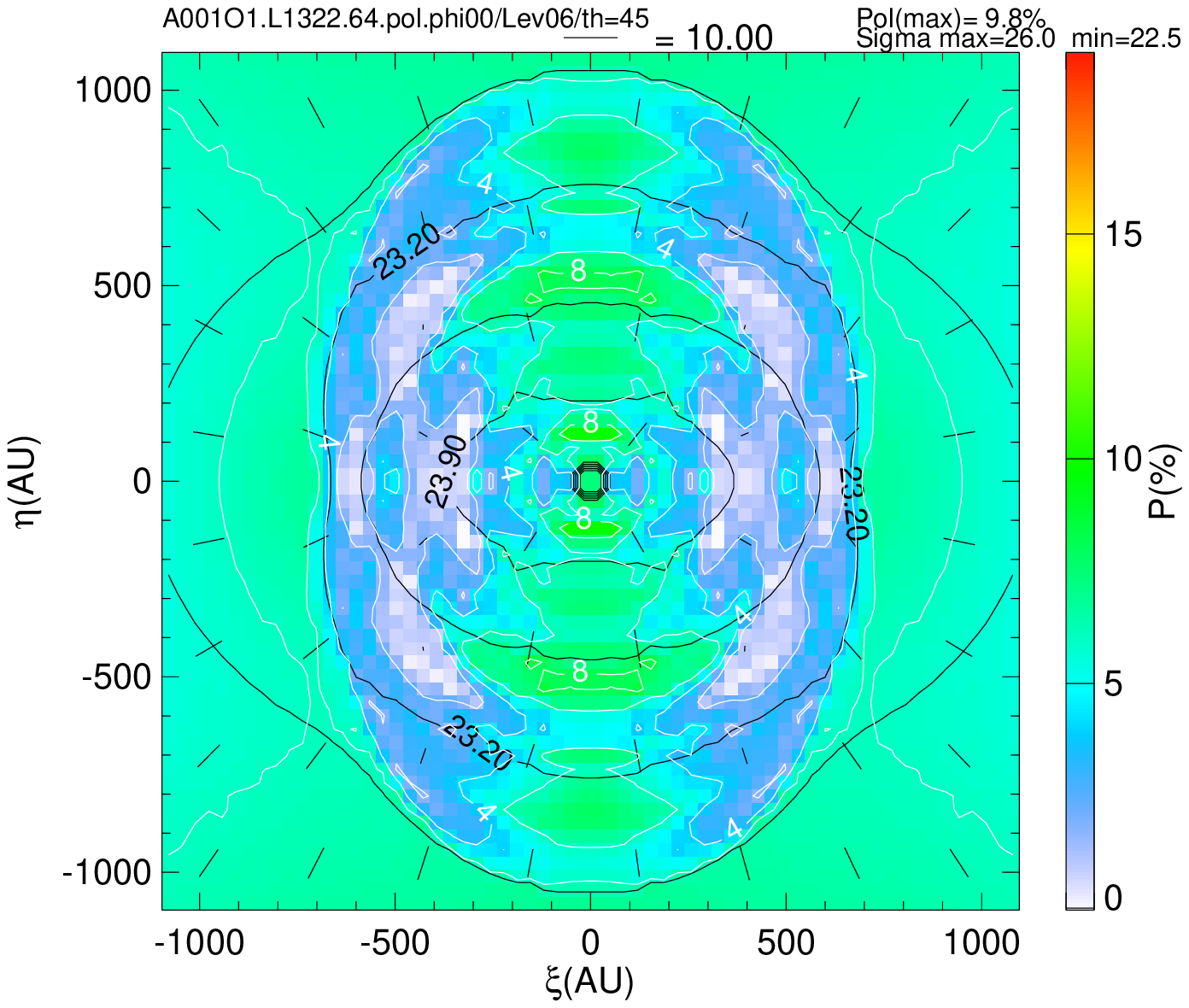}
      \hspace*{5mm}\one{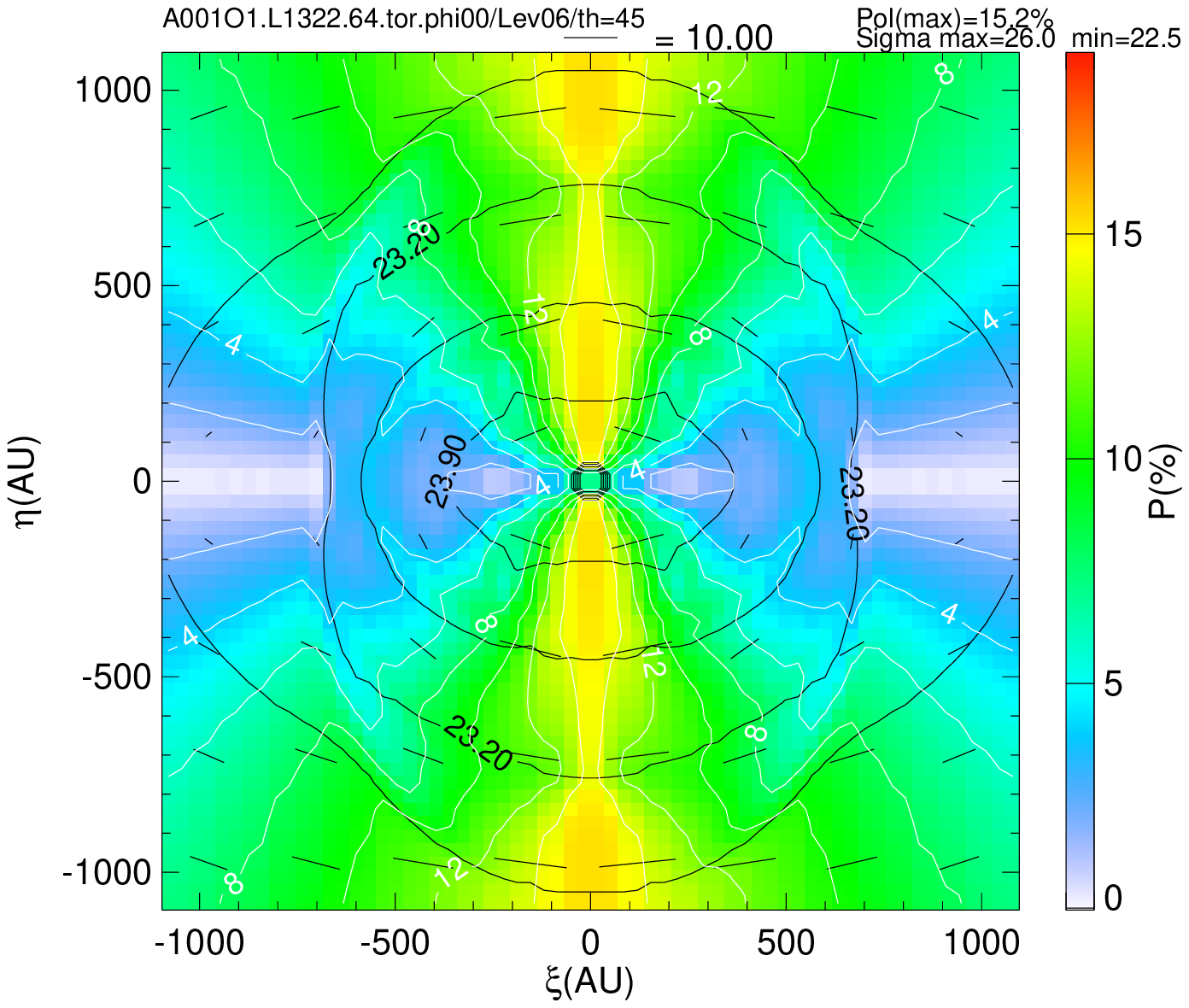}}\\[-0mm]
$\theta=90^\circ$\hspace*{10mm}\raisebox{-20mm}{\one{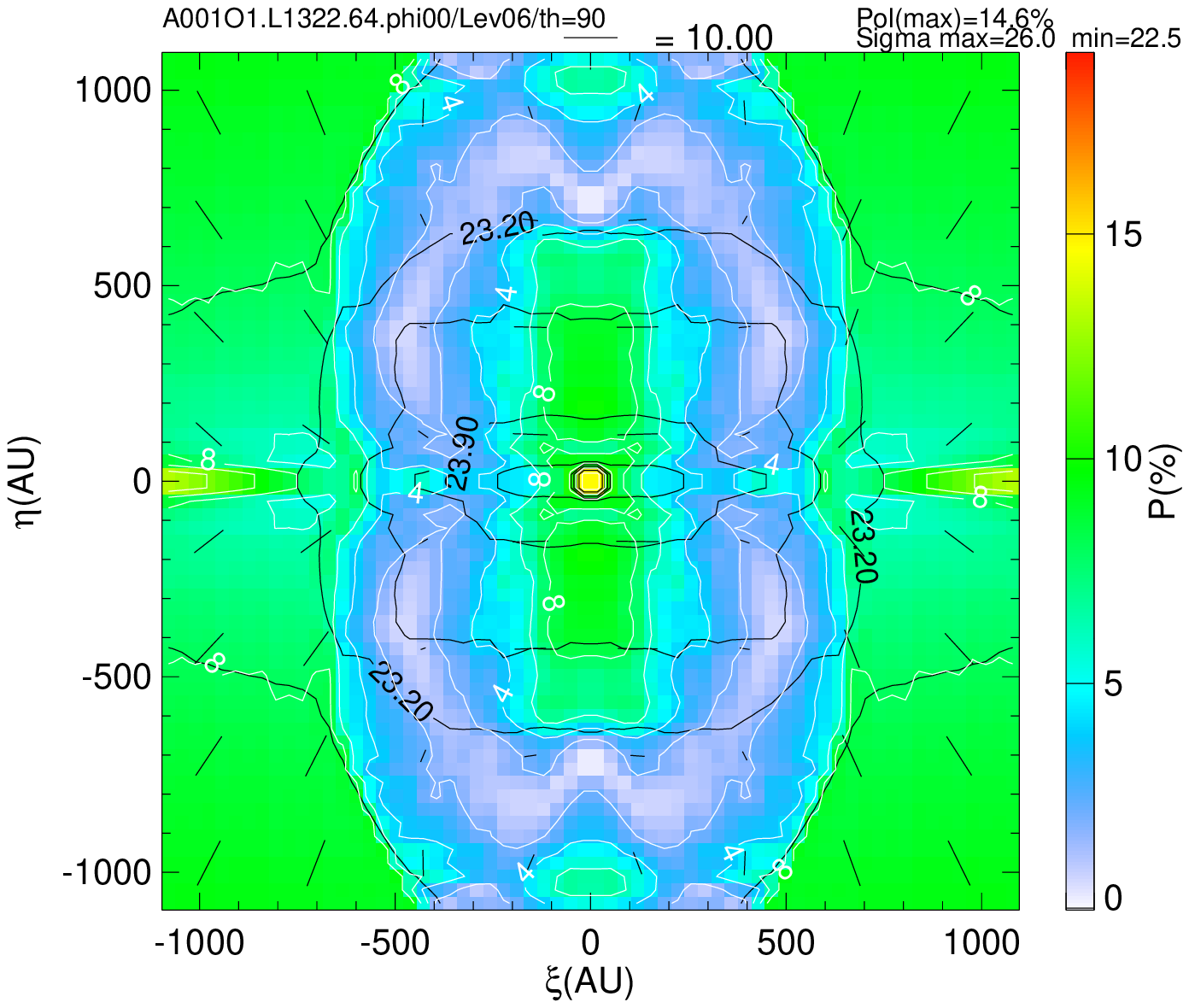}
      \hspace*{5mm}\one{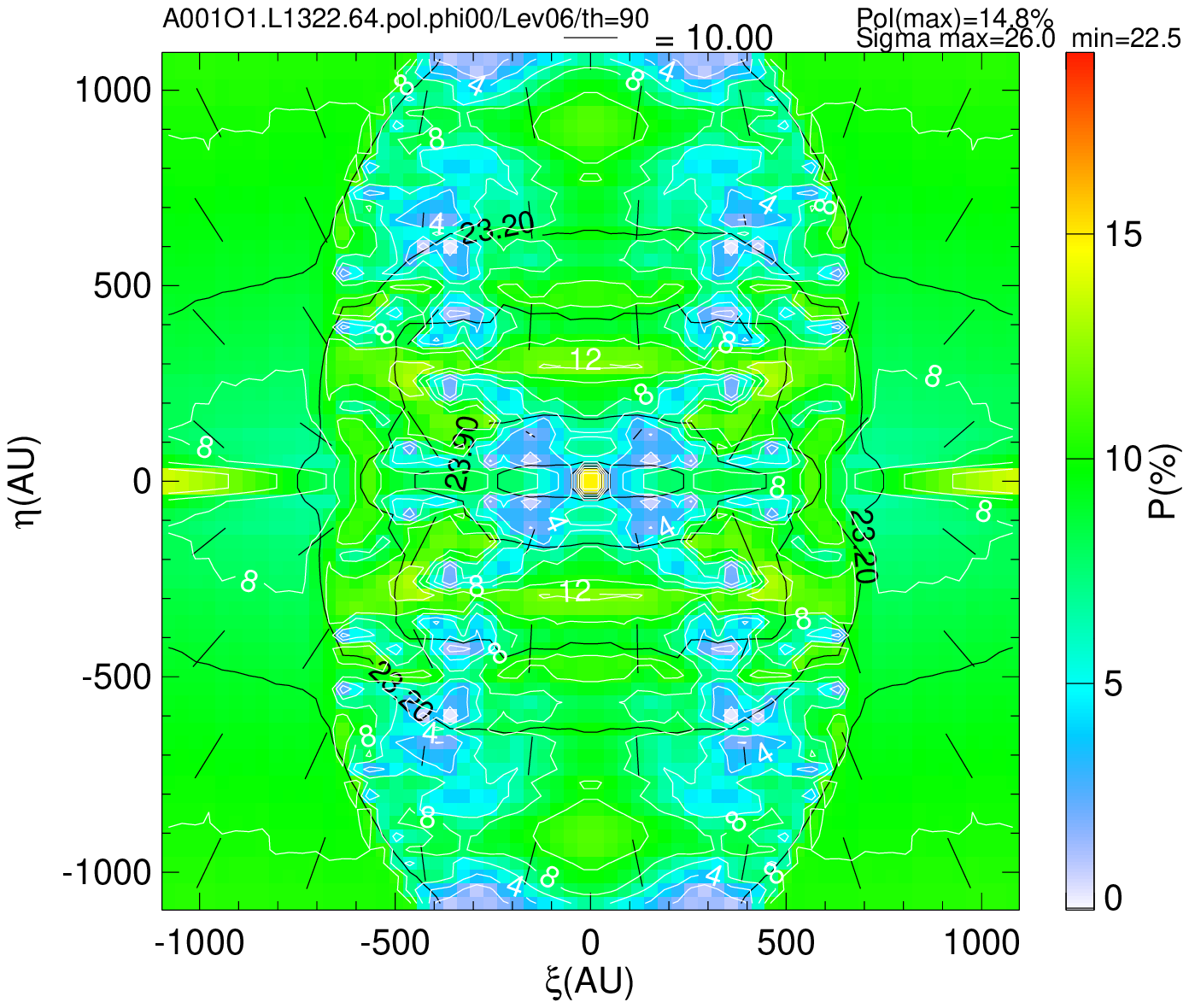}
      \hspace*{5mm}\one{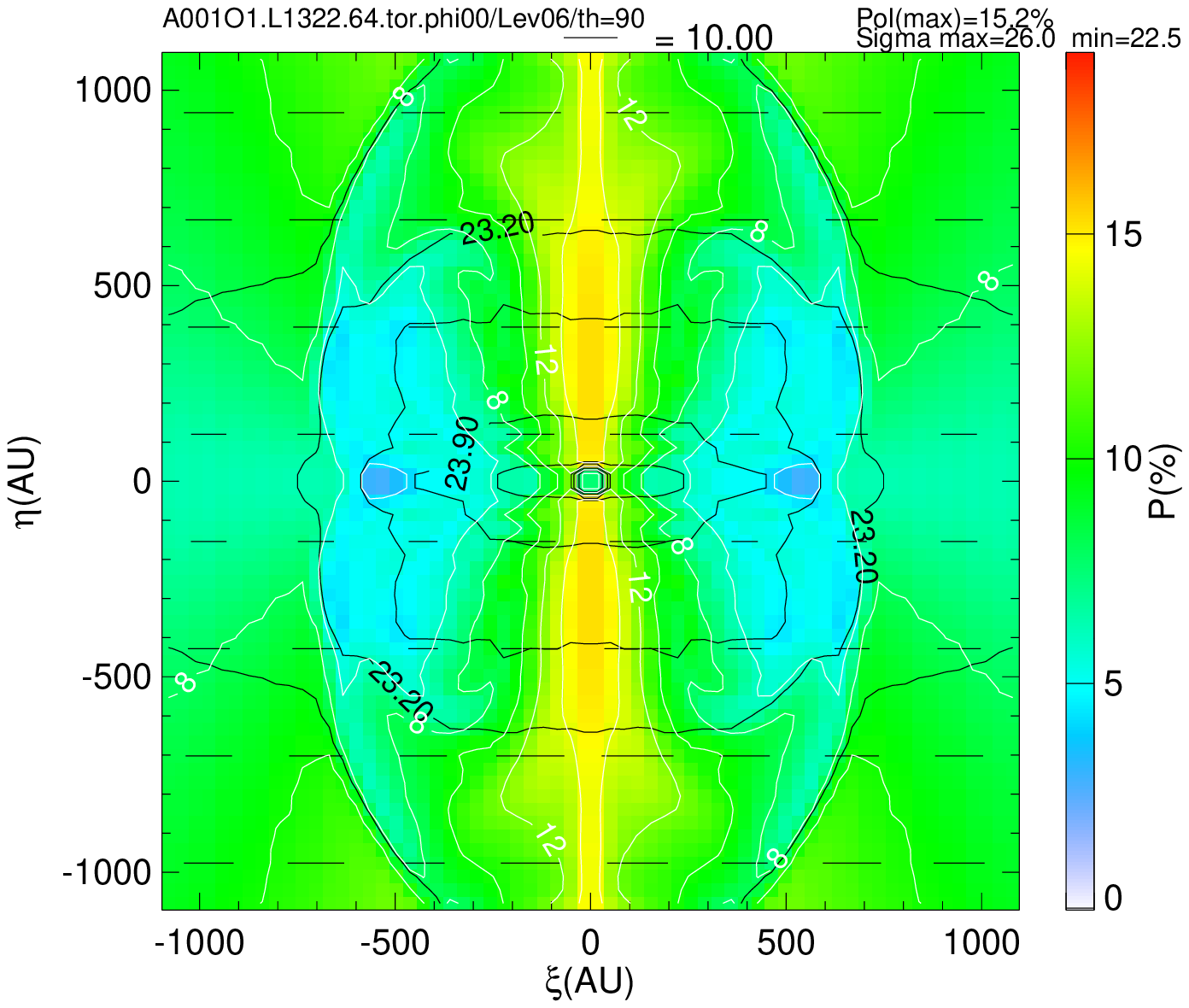}}
   \end{center}
\caption{\label{fig:1322A001O1L6poltor}As for Fig.~\ref{fig:A001O1L1322}.
 However, the middle and right columns are results
 for artificial data consisting only of poloidal and toroidal magnetic fields,
 respectively.
 Level 6 of model EH with $\alpha=0.01$ and $\Omega'=1$.}
\end{figure}

In the previous section we obtained the expected polarization map
 of thermal dust emission toward the prestellar core 
(Fig.~\ref{fig:A1O5L332}, during the runaway collapse) and protostellar phase
(Figs.~\ref{fig:A1O5L1001} and \ref{fig:A001O1L1322}).
To distinguish the origin of the molecular outflow,
 it is essential to observe the toroidal magnetic field.
The magnetically driven molecular outflow model
 induces a toroidal magnetic field with a strength at least comparable
 to the poloidal component,
 at that point and time where the molecular outflow is accelerated.
This is obtained in a cloud core with a relatively strong magnetic field,
  $\alpha \sim 1$ shown in Figure \ref{fig:A1O5L1001}.
The magnetocentrifugal wind acceleration mechanism applies. 
In a molecular core with a weak magnetic field, $\alpha \ll 1$, the outflow
 is accelerated by the magnetic pressure gradient as shown in Figure
 \ref{fig:A001O1L1322}. 
In this case, the strength of the toroidal component dominates
 the poloidal component in the molecular outflow.   
That is, the configuration of the magnetic field 
 $|B_\phi| \gtrsim (B_r^2+B_z^2)^{1/2}$ exhibits a characteristic signature
 of magnetically driven molecular outflow.

A toroidal dominant magnetic field gives a point-symmetric
 polarization distribution rather than a mirror-symmetric one.  
As we have found, a purely poloidal magnetic field recovers a mirror symmetry
 with respect to the $\eta-$axis.
Thus, the predominance of a toroidal magnetic field is imprinted in the disk
 around the protostar.
However, the direction of the major axis of the total intensity
 of the disk and the extension of the region with low polarization degree
 should first be compared.
If the disk has a point symmetric polarization distribution rather than 
 a mirror symmetric one, this indicates that a relatively strong 
 toroidal magnetic field has been generated in the gaseous disk.

Another signature is a low polarization degree in the molecular outflow.
If the coupling between dust and magnetic field is complete,
 this indicates a predominance of a toroidal magnetic field in the molecular
 outflow.
Even if the dust alignment in the contracting envelope is the same as
 in the dark cloud,
 the alignment in the molecular outflow might be incomplete.
Even in this case,
 the molecular outflow has to have a low polarization degree.  

\section{Summary}
Based on two-dimensional axisymmetric MHD simulations,
 we calculated the polarization pattern expected in observations
 of thermal dust emission.
We developed a procedure to calculate Stokes' parameters
 on a nested grid hierarchy.
The distribution of the polarization degree has an apparent signature
 indicating a toroidal component dominated magnetic field
 in the acceleration region of the molecular outflow.
The outflow must have a lower polarization degree
 than the envelope.
Another signature is imprinted on the disk whose rotation amplifies
 the toroidal magnetic field and thus accelerates the gas.
A point-symmetric rather than a mirror-symmetric distribution of 
 the low polarization degree region is another signature of a toroidal dominated
 magnetic field.
If these characteristic features are observed, such a molecular
 outflow has toroidal-dominated magnetic field and is likely to be  
 driven by the magnetic Lorentz force rather than the entrainment mechanism.

\section*{Acknowledgement}
This work was supported in part by JSPS 
 Grant-in-Aid for Scientific Research (A) 21244021 in 2009--2010. 
Numerical computations were in part carried out on Cray XT4 at 
 the Center for Computational Astrophysics, CfCA, of the National Astronomical 
 Observatory of Japan.

\clearpage
\section*{Figure Captions}

\clearpage

\clearpage
\begin{thebibliography}{}
\bibitem[Bacciotti et al.(2002)]{bacciotti02}
 Bacciotti, F., Ray, T. P., Mundt, R., Eisl\"{o}ffel, J., \&  Solf, J.
 2002, \apj, 576, 222
\bibitem[Banerjee \& Pudritz(2006)]{banerjee06}
 Banerjee, R., \& Pudritz, R. E. 2006, \apj, 641, 949
\bibitem[Blandford \& Payne(1982)]{blandford82}
 Blandford, R. D., \& Payne, D. G. 1982, \mnras, 199, 883
\bibitem[Cabrit, Raga, \& Gueth(1997)]{cabrit97}
 Cabrit, S., Raga, A., \& Gueth, F. 1997, in Herbig-Haro Flows and the Birth of Stars,
 ed. B. Reipurth, \& C. Bertout (Kluwer Academic Publishers: Dordrecht)  p.163.
\bibitem[Coffey et al.(2007)]{coffey07}
 Coffey, D., Bacciotti, F., Ray, T. P., Eisl\"{o}ffel, J., \& Woitas, J.
 2007, \apj, 663, 350
\bibitem[Chrysostomou et al.(2000)]{chrysostomou00}
 Chrysostomou, A., Hobson, J., Davis, C. J., Smith, M. D., \& Berndsen, A.
 2000, \mnras, 314, 229
\bibitem[Commer\c con et al.(2010)]{commercon10}
 Commer\c con, B., Hennebelle, P., Audit, E., Chabrier, G., \&
 Teyssier, R. 2010, \aap, 510, L3
\bibitem[Davis et al.(2000)]{davis00} 
 Davis, C. J., Berndsen, A., Smith, M. D., Chrysostomou, A., \& Hobson, J. 
 2000, \mnras, 314, 241
\bibitem[Davis \& Greenstein(1951)]{davis51}
 Davis, L., \& Greenstein, J. L. 1951, \apj, 114, 206
\bibitem[Fiege \& Pudritz(2000)]{fiege00}
 Fiege, J. D., \& Pudritz, R. E.  2000, \apj, 544, 830
\bibitem[Galli \& Shu(1993)]{galli93}
 Galli, D., \& Shu, F. H. 1993, \apj, 417, 220
\bibitem[Kudoh, Matsumoto, \& Shibata(1998)]{kudoh98}
 Kudoh, T., Matsumoto, R., \& Shibata, K. 1998, \apj, 508, 186
\bibitem[Larson(1969)]{larson69}
 Larson, R. B. 1969, \mnras, 145, 271
\bibitem[Launhardt et al.(2009)]{launhardt09}
 Launhardt, R., Pavlyuchenkov, Ya., Gueth, F., Chen, X.,
 Dutrey, A., Guilloteau, S., Henning, Th., Pi\'{e}tu, V., Schreyer, K., \&
 Semenov, D. 2009, \aap, 494, 147
\bibitem[Lazarian(2007)]{lazarian07}
 Lazarian, A. 2007, \jqsrt, 106, 225 
\bibitem[Lazarian, \& Draine(1999)]{lazarian99}
 Lazarian, A., \& Draine, B. T. 1999, \apjl, 516, L37 
\bibitem[Lee \& Draine(1985)]{lee85}	
 Lee, H. M., \& Draine, B. T. 1985, \apj, 290, 211 
\bibitem[Lee et al.(2000)]{lee00}
 Lee, C.-F., Mundy, L. G., Reipurth, B., Ostriker, E. C., \& Stone, J. M.
 2000, \apj, 542, 925 (erratum \apj 549, 1231)
\bibitem[Machida, Inutsuka \& Matsumoto(2007)]{machida07}
 Machida, M. N., Inutsuka, S.-I., \& Matsumoto, T. 2007, \apj, 670, 1198
\bibitem[Masson \& Chernin (1993)]{masson93}
 Masson, C. R., \& Chernin, L. M. 1993, \apj, 414, 230
\bibitem[Matsumoto, Nakazato, \& Tomisaka(2006)]{matsumoto06}
 Matsumoto, T., Nakazato, T., \& Tomisaka, K. 2006, \apjl, 637, L105
\bibitem[Raga \& Cabrit (1993)]{raga93}
 Raga, A., \& Cabrit, S. 1993, \aap, 278, 267
\bibitem[Raga et al.(1993)]{raga+93}
 Raga, A. C., Canto, J., Calvet, N., Rodriguez, L. F., \&
 Torrelles, J. M. 1993, \aap, 276, 539
\bibitem[Shu et al.(1994)]{shu94}Shu, F., Najita, J., Ostriker, E.,
 Wilkin, F., Ruden, S., \& Lizano, S. 1994, \apj, 429, 781
\bibitem[Stahler(1993)]{stahler93}
 Stahler, S. W., 1993 in Astrophysical Jets, ed. D. Burgarella, M. Livio, 
 C. O'Dea (Cambbridge: Cambridge University Press)
\bibitem[Tomida et al.(2010)]{tomida10} Tomida, K., Tomisaka, K.,
 Matsumoto, T., Ohsuga, K., Machida, M. N., \& Saigo, K. 2010,
 \apjl,714, L58
\bibitem[Tomisaka(1998)]{tomisaka98} Tomisaka, K. 1998, \apjl, 502, L163 
\bibitem[Tomisaka(2000)]{tomisaka00} Tomisaka, K. 2000, \apjl, 528, L41
\bibitem[Tomisaka(2002)]{tomisaka02} Tomisaka, K. 2002, \apj, 575, 306
\bibitem[Woitas et al.(2005)]{woitas05}Woitas, J., Bacciotti, F., Ray, T. P., 
 Marconi, A., Coffey, D., \& Eisl\"{o}ffel, J. 2005, \aap, 432, 149 
\end{thebibliography}
\end{document}